%% file: THESIS_MAY09.tex
\input{Definitions.tex}

\documentclass[10pt, a4paper]{mscthesis} 
\usepackage{graphicx}
\usepackage[cmex10]{amsmath}
\usepackage{amssymb}
\usepackage{setspace}
\degree{Doctor of Philosophy}

\author{Jack Robert Raymond}
\title{\input{Title.tex}}
\submitmonth{November} \submityear{2008} 
\begin{document}
\maketitle
\begin{acknowledgements}
\input{Acknowledgements.tex}
\end{acknowledgements}

\begin{abstract}
\input{Abstract.tex}
\end{abstract}
\tableofcontents
\listoffigures

\onehalfspacing
\KEEPNOTE{PAGES IN COLOUR 9,11,12,15, 16,20,21, 22,27,28, 29,32,40, 44,53,55,78,80, 92, 109, 141,148}
\KEEPNOTE{MUST BE/BETTER IN COLOUR 76,77,88,97,98,99, 102,103,104,105 119,120,122,123,124,125,127}
\KEEPNOTE{ESSENTIAL IN COLOUR}
\chapter{Introductory section}
\label{chapter:introduction}
\input{INTRODUCTORYSECTION.tex}

\chapter{UCP analysis of Exact Cover}
\label{chapter:1inkSAT}
\input{HUCPANALYSISOF1INK.tex}
\chapter{Sparse CDMA}
\label{chapter:sparseCDMA}
\input{SPARSECDMA.tex}

\chapter{Composite spin models}
\label{chapter:composite}
\input{COMPOSITEMODEL.tex}
\chapter{Composite CDMA}
\label{chapter:compositeCDMA}
\input{COMPOSITECDMA.tex}

\chapter{Conclusion}
\input{Conclusion.tex}
\UPDATEBIBFILEYES{
\bibliographystyle{unsrt} 
\bibliography{Bibliography}
}
\UPDATEBIBFILENO{

\input{THESIS_MAY09.bbl}
}
\appendix
\input{APPENDICES.tex}

\end{document}

%% file: Definitions.tex
\newcommand\UPDATEBIBFILEYES[1]{}
\newcommand\UPDATEBIBFILENO[1]{#1}

\def\solid{\relbar}

\def\dotted{\cdots}

\newcommand{\cut}[1]{}
\newcommand{\KEEPNOTE}[1]{}

\newcommand{\<}{\left\langle}
\renewcommand{\>}{\right\rangle}

\def\factorial{!}
\def\rmd{{\mathbf d}}
\def\rmD{{\mathbf D}}

\def\rmi{{\mathbf i}}
\def\Reals{{\mathbf R}}
\def\Complexs{{\mathbf C}}

\def\Ham{\mathcal{H}}

\def\quenched{{\cal Q}}
\def\ensemble{{\cal E}}

\def\partitionfunction{Z}
\def\localpartitionfunction{{\cal Z}}
\def\localreplicaprobability{{\cal P}}

\def\freeenergy{{\cal F}}

\def\freeenergydensity{f_\quenched}
\def\safed{f_\ensemble}
\def\saed{e}
\def\sasd{s}
\def\repZ{{\langle \partitionfunction^n\rangle_{\quenched}}}
\def\Gone{{\cal G}_1}
\def\Gtwo{{\cal G}_2}
\def\Gthree{{\cal G}_3}
\def\Gi{{{\cal G}_i}}
\def\gone{g_1}
\def\gtwo{g_2}
\def\gthree{g_3}
\def\gi{g_i}
\def\UCPtransitioner{{\mathbb F}}
\def\mB{\mathbb{B}}
\def\SpinGlassSusceptibility{{\chi_{SG}}}
\def\Susceptibility{{\chi_{Lin}}}
\def\CDMA{CDMA}
\def\FDMA{FDMA}
\def\TDMA{TDMA}
\def\MAP{MAP}
\def\MPM{MPM}

\def\SK{SK}
\def\VB{VB}
\def\UCP{UCP}
\def\HUCP{HUCP}

\def\MAI{MAI}
\def\PSD{PSD}
\def\KL{KL}
\def\KLfunction{\mathrm{KL}}
\def\BP{{BP}}

\def\RSB{{RSB}}
\def\oRSB{{1RSB}}
\def\FRSB{FRSB}
\def\RS{{RS}}
\def\SNR{SNR}

\def\SNRmath{{\mathrm{SNR}_b}}
\def\ASK{ASK}
\def\MUEmath{{\mathrm{mue}}}

\def\SEmath{\mathrm{se}}
\def\BER{{BER}}
\def\BERmath{{\mathrm{BER}}}
\def\EC{EC}
\def\kSAT{kSAT}
\def\SAT{SAT}
\def\UNSAT{UNSAT}

\def\Extr{{\mathrm{Extr}}}

\newcommand{\sign}{{\mathrm{sign}}}
\newcommand{\erfc}{{\mathrm{erfc}}}
\newcommand{\argmax}{{\mathrm{argmax}}}
\newcommand{\argmin}{{\mathrm{argmin}}}
\newcommand{\determinant}[1]{\left| #1 \right|}
\def\atanh{\mathrm{atanh}}
\def\arbitraryfunction{G}
\def\arbitraryfunctiontwo{g}

\def\eokSAT{\epsilon\mbox{-1-in-kSAT}}
\def\okSAT{1\mbox{-in-kSAT}}
\def\otSAT{1\mbox{-in-3SAT}}
\def\eotSAT{\epsilon\mbox{-1-in-3SAT}}
\def\eofSAT{\epsilon\mbox{-1-in-4SAT}}

\def\Jij{J_\ij}
\def\load{\chi}
\def\modulationsymbol{V}
\def\sparsematrixsymbol{A}
\def\couplingsymbol{J}
\def\randomnumber{r}
\def\randomfield{z}

\newcommand{\realvectornotation}[1]{\vec{#1}}
\def\vmodulationsymbol{{\realvectornotation{\modulationsymbol}}}
\def\vy{{\realvectornotation{y}}}
\def\vu{{\realvectornotation{u}}}
\def\vm{{\realvectornotation{m}}}
\def\vp{{\realvectornotation{p}}}

\def\vY{{\realvectornotation{Y}}}
\def\vH{{\realvectornotation{H}}}

\def\vL{{\realvectornotation{L}}}
\def\vC{{\realvectornotation{C}}}

\def\vU{{\realvectornotation{U}}}
\def\vZ{{\realvectornotation{Z}}}
\def\vY{{\realvectornotation{Y}}}
\def\vb{{\realvectornotation{b}}}
\def\vtau{{\realvectornotation{\tau}}}
\def\vsigma{{\realvectornotation{\sigma}}}
\def\vs{{\realvectornotation{s}}}
\def\vomega{{\realvectornotation{\omega}}}

\def\vh{\realvectornotation{h}}

\def\vones{\realvectornotation{1}}
\def\vzeros{\realvectornotation{0}}
\def\vx{\realvectornotation{x}}
\def\vA{\realvectornotation{A}}
\def\vnu{\realvectornotation{\nu}}
\def\vlambda{\realvectornotation{\lambda}}
\def\vzeta{\realvectornotation{\zeta}}
\def\vsigma{\realvectornotation{\sigma}}
\def\vtau{\realvectornotation{\tau}}
\def\vS{\realvectornotation{S}}
\def\vrandomfield{\realvectornotation{z}}

\newcommand{\replicavectornotation}[1]{{\hbox{\boldmath{$#1$}}}}
\def\rvlambda{{\replicavectornotation{\lambda}}}
\def\rvsigma{{\replicavectornotation{\sigma}}}
\def\rvtau{{\replicavectornotation{\tau}}}
\def\rvS{{{\replicavectornotation{S}}}}
\def\rvx{{{\replicavectornotation{x}}}}

\def\mx{{\mathbb X}}

\def\ms{{\mathbb S}}

\def\mJ{{{\mathbb \couplingsymbol}}}
\def\mA{{\mathbb \sparsematrixsymbol}}
\def\mxi{{\mathbb \modulationsymbol}}

\def\mW{{\mathbb W}}
\def\QuadraticForm{{\mathbb T}}
\def\ScalarForm{{T}}
\def\RotationMatrix{{\mathbb R}}

\newcommand{\intZ}[1]{\oint {\rmD_{#1} Z_k}}

\newcommand{\vintlambda}[1]{\left[\int \rmD_{#1} \rvlambda \right]}

\def\GENOP{{\Phi}}
\def\GENOPconj{{{\hat \GENOP}}}
\def\GENOPsp{{\GENOP_b(\rvsigma)}}
\def\GENOPspconj{{\GENOPconj_b(\rvsigma)}}
\def\RSOP{{\pi}}

\def\RSOPconj{{\hat{\pi}}}

\newcommand{\Set}{W}
\def\Histogram{W}

\def\factorG{G(V_v,V_f,E)}

\def\ij{{\langle i j \rangle}}

\newcommand{\orderedtwo}[1]{\<#1_1,#1_2\>}
\newcommand{\orderedthree}[1]{\<#1_1,#1_2,#1_3\>}
\newcommand{\orderedfour}[1]{\<#1_1,#1_2,#1_3,#1_4\>}
\newcommand{\orderedL}[1]{{\langle {#1}_1, \ldots, {#1}_{L} \rangle}}

\def\alal{{\orderedtwo{\alpha}}}
\def\alalal{{\orderedthree{\alpha}}}
\def\alalalal{{\orderedfour{\alpha}}}
\def\qal{{q_\alpha}}
\def\qalal{q_\alal}

\def\hq{{\hat q}}
\def\bq{{\bar q}}
\def\qhal{{\hq_\alpha}}
\def\qbal{{{\bq}_\alpha}}

\def\qhalal{{\hq_\alal}}
\def\qbalal{{\bq_\alal}}

\def\qbalalal{{\bq_\alalal}}
\def\qbalalalal{{\bq_\alalalal}}

\def\True{\mbox{True}}
\def\False{\mbox{False}}

%% file: Title.tex

Typical case behaviour of spin systems in random graph and composite ensembles

%% file: Acknowledgements.tex

This thesis was completed with the help of many people. Most importantly my thesis supervisor David Saad who was available regularly for discussions and guidance on all topics regardless of other responsibilities. The majority of subjects presented in the thesis were explored as suggestions by David as potential sources of interesting phenomena. I was also grateful to be given time exploring some more speculative research directions, in particular consideration of random circuit structures.

Throughout the three years of my PhD it has been of importance to me to work in a research group undertaking a wide variety of high quality research beyond the range of topics I had encountered as a physics undergraduate student. The pattern analysis and neural networks course, undertaken in the first few months as a PhD student, was an eye opener on various applied mathematics topics. During my PhD I was inspired by a number of researchers and topics I encountered as a result of travel to conferences organised by the European statistical physics research community. I found of particular significance in developing my ideas the Les Houches summer school of 2006 on the topic of complex systems.

I also appreciate the two opportunities I was given to undertake research in the year between graduating from Edinburgh University and beginning my PhD. The first was presented to me by Peter Andras, who allowed me visitor status at the computer science department of Newcastle University and provided resources to study a problem on syntactic dependency networks. The second opportunity was presented by Martin Evans, who provided funding so that I could return to Edinburgh to write up my Masters' project on a model of flocking as a research paper. Both opportunities provided a good basis by which to gauge research opportunities and practice.

The financial support of my PhD by an Aston University scholarship, EU Evergrow funding and other sources found by my supervisor were critical in maintaining a focus on research and professional development throughout the last three years. Furthermore I have benefitted greatly from an exemplary administrative system within my department, I thank particularly Vicky Bond.

For the work developed in chapter \ref{chapter:1inkSAT} I am grateful to my collaborators Andrea Sportiello and Lenka Zdeborov\'{a}, who checked many calculations I produced and provided context for the results in pursuing other types of analysis. I should thank my colleagues Mike Vrettas, Eric Casagrande and my supervisor for proof-reading at various stages of thesis preparation. Finally I thank my examiners Ton Coolen and Juan Neirotti for their constructive criticisms that in application have improved on my initial concept, as well as their flexibility in allowing an early examination date enabling me to pursue my post-doctoral research at Hong Kong University of Science and Technology.

%% file: Abstract.tex

This thesis includes analysis of disordered spin ensembles corresponding to Exact Cover, a multi-access channel problem, and composite models combining sparse and dense interactions. The satisfiability problem in Exact Cover is addressed using a statistical analysis of a simple branch and bound algorithm. The algorithm can be formulated in the large system limit as a branching process, for which critical properties can be analysed. Far from the critical point a set of differential equations may be used to model the process, and these are solved by numerical integration and exact bounding methods. The multi-access channel problem is formulated as an equilibrium statistical physics problem for the case of bit transmission on a channel with power control and synchronisation. A sparse code division multiple access method is considered and the optimal detection properties are examined in typical case by use of the replica method, and compared to detection performance achieved by iterative decoding methods. These codes are found to have phenomena closely resembling the well-understood dense codes. The composite model is introduced as an abstraction of canonical sparse and dense disordered spin models. The model includes couplings due to both dense and sparse topologies simultaneously. Through an exact replica analysis at high temperature, and variational approaches at low temperature, several phenomena uncharacteristic of either sparse or dense models are demonstrated. An extension of the composite interaction structure to a code division multiple access method is presented. The new type of codes are shown to outperform sparse and dense codes in some regimes both in optimal performance, and in performance achieved by iterative detection methods in finite systems.

%% file: INTRODUCTORYSECTION.tex

\section{Disordered binary systems}
\label{introduction.spods}
\input{INTRODUCTION/Disordered.tex}
\section{Graphical models}
\label{introduction.graphicalmodels}
\input{INTRODUCTION/Graphical_model.tex}
\section{Algorithmic methods in disordered systems}
\label{introduction.algorithmicmethods}
\subsection{Belief propagation}
\label{introduction.BP}
\input{INTRODUCTION/Belief_propagation.tex}
\subsection{Branch and bound}
\label{introduction.branchandbound}
\input{INTRODUCTION/Branch_and_bound.tex}
\section{Exact cover}
\label{introduction.EC}
\input{INTRODUCTION/Exact_cover.tex}
\section{Multi-user detection models}
\label{introduction.MD}
\input{INTRODUCTION/Multi_user_detection.tex}
\section{The Viana-Bray model}
\label{introduction.VB}
\input{INTRODUCTION/Vb.tex}

%% file: INTRODUCTION/Disordered.tex

Understanding how the macroscopic properties of large assemblies of interacting objects arise from a microscopic description is at the basis of many fields of science. The question arises naturally in physics, where an understanding of elementary particles has become well developed. A concrete understanding of the microscopic systems (atoms/quarks/strings) and their interactions, would seem to be a good basis from which to verify and develop macroscopic theories. Many physical theories, such as thermodynamics, describe the macroscopic dynamics and interactions of large assemblies of particles with great accuracy based on an incomplete description of the microscopic details. Statistical physics connects the macroscopic theories with the microscopic description. Unimportant microscopic degrees of freedom can be marginalised according to some assumed or exact probabilistic description, to give a description of a macroscopic behaviour.

With strong interactions amongst variables sparse and dense graphical models often provide a necessary or insightful simplification of interactions. For point to point interactions each variable is represented by a vertex, and each interaction by an edge. In many classical theories based on simplified structures, such as lattices, the type of order observed at the macroscopic level reflects the microscopic symmetries of interactions. Classical and quantum magnetic spin systems, where each microscopic state take only two values, are a particularly successful application of classical theories based primarily on simplified lattices and fully connected graphical structures~\cite{Pointon:ISP,Stanley:IPT,Yeomans:SM}.

Seminal works, especially in the 1980s, developed the classical theories of statistical mechanics to systems with inhomogeneous interactions. In some cases it was discovered that correlations in the macroscopic order were non-trivial extensions of the
microscopic description. The spin-glass phase of matter became an archetypal case~\cite{Fischer:SG,Mezard:SGT}. Spin glasses are a class of materials in which the microscopic states exhibit both anti-ferromagnetic and ferromagnetic couplings with neighbours. The low temperatures phase for these materials are described by a novel magnetic behaviour, which was initially difficult to formulate within classical exact and mean-field methods. Many realistic models of the microscopic interactions remain unsolved by exact methods~\cite{Edwards:TS}, although simplified models have been developed and solved to correctly describe phenomena consistent with experiment.

In spin glasses the statistical description can be quantified by representing a particular instance of the disordered interactions as a sample from an ensemble. Working in the large system limit self-averaging may be assumed or proved. Self-averaging, a term coined by Lifshitz, is the intuitive phenomena that the macroscopic features of different samples converge as the size of the assembly increases. Therefore, in the large assemblies, the average value of some interesting macroscopic property is statistically identical to the value in any typical sample; the samples breaking this rule being atypical and statistically insignificant.

The separation of microscopic and macroscopic scales is apparent in other fields of science: humans and societies, neurons and brains, bits and codewords. Within the wider scientific community there is an effort to understand the steady state and equilibrium behaviors of complex systems~\cite{Mezard:LH}. Complex systems are characterised by some statistically robust macroscopic features, in spite of strong inhomogeneities in space (and/or time) at the microscopic level. Since the microscopic interactions in these systems are often described by discrete properties of the objects such as left/right, on/off and true/false, the theories of 2-state (spin) physics may often be applied.

\subsection{The Sherrington-Kirkpatrick model}
\label{Introduction.SK}
\begin{figure}[htb]
\centering{
\includegraphics[width=\linewidth]{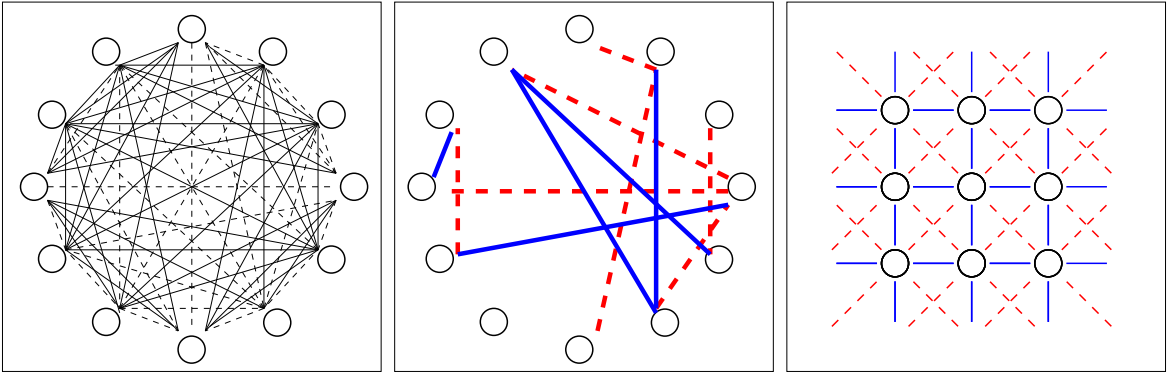}
\caption[Examples of the SK, VB and EA graphical models.]{\label{fig:Introduction.SK} Left figure: The graphical models for the Sherrington-Kirkpatrick model is fully connected, with randomly distributed ferromagnetic (solid) or anti-ferromagnetic (dashed) couplings. Centre figure: The Viana-Bray model has only a small random subset of couplings active per variable. Right figure: The Edward-Anderson model has a regular lattice structure, with nearest neighbour ferromagnetic couplings and next-nearest neighbour anti-ferromagnetic couplings.}}
\end{figure}
The Sherrington-Kirkpatrick (SK) model was developed as a
mean-field model to allow a better understanding of the spin glass phase, and is the model for which the replica method was originally developed~\cite{Sherrington:SMSG}. In the SK model all spin states are coupled through heterogeneous couplings, as shown in figure~\ref{fig:Introduction.SK}.

The model describes a systems of $N$ spins, $\vS$, so that the state space is $\left\lbrace -1,+1\right\rbrace^N$. The interactions for each spin are point to point and described by couplings $\Jij$. A positive value of $\Jij$ will promote alignment of spin $i$ and $j$, whereas a negative value promotes an antiparallel alignment. The equilibrium properties of the system are described by the Hamiltonian
\begin{equation}
 \Ham(\vS) = - \sum_\ij \Jij S_i S_j - \sum_i h_i S_i \;.
 \label{eq:Introduction.SK}
\end{equation}
The formulation is motivated by problems in real spin glasses, and the success of related mean-field models in describing the equilibrium properties of ferromagnets.

\subsection{The replica method}

The replica method is the main tool used in this thesis to
determine equilibrium properties of typical cases described by a statistical ensemble. The SK model can be taken as an example, but the principles of the calculation are quite general. The equilibrium statistics may be determined from a variational form for the free energy, and the typical case behaviour is determined by averaging over the possible samples. Each sample is distinguishable by a set of parameters, called quenched variables in physics, which are either static or slowly evolving (so that equilibrium in some dynamical variables is achieved on a timescale over which the variables can be assumed to be static). Most cases in this thesis involve quenched variables that describe a particular sparse graph, combined with some edge modulation properties, whereas the dynamic variables are bits/spins which adopt particular states subject to this fixed structure. The dynamical variable average, used to calculate properties for the equilibrium configuration, is implicit in the definition of a partition function,
\begin{equation}
Z = \sum_{\vS} \exp\left\lbrace -\beta \Ham(\vtau) \right\rbrace \label{eq:Introduction.Z} \;.
\end{equation}
The quenched disorder average is over the free energy, a generating function for macroscopic statistics,
\begin{equation}
\freeenergy = \frac{-1}{\beta} \log \partitionfunction \;.\label{eq:Introduction.f}
\end{equation}
By averaging over the free energy each quenched sample is given an equal weight in the calculation of every statistic, which is the desired interpretation for typical case analysis.

The free energy average is not directly tractable for general
strongly coupled systems, but the following transformation, the replica identity, can always be applied
\begin{equation}
\<\log \partitionfunction\>_{\quenched}= \lim_{n \rightarrow 0} \frac{\partial}{\partial n}\repZ \;, \label{eq:Introduction.ri}
\end{equation}
and this form allows the average to be taken. An analytic expression in $n$ is required to take the limit, but the problem is normally solved for positive integer $n$, from which an analytic continuation is possible to positive non-integer value. The integer $n$ framework allows an interpretation for the free energy as a calculation of the average partition function for an assembly replicated $n$ times. To each replica is associated a set of dynamical variables $\{\vS^1, \ldots,\vS^\alpha, \ldots, \vS^n\}$, which are conditionally independent given the shared set of quenched variables. The average over quenched variables in the replicated partition function is technically similar to the dynamical variable average except in the $n$ dependence. This allows the quenched average to be taken before the dynamical average in the replicated model.

An exponential form may be derived for the replicated partition function. The relevant terms in the exponent are determined by inter-replica correlations $\sum_i S_i^{\alpha_1} S_i^{\alpha_2}$. The form of correlations can be quite complicated, but take a simplified form in the SK model owing to a central limit theorem in the large system limit, but more general frameworks exist without these feature~\cite{Monasson:OP}. Reasoning on the form of interactions suggests a hierarchy of candidate solutions~\cite{Mezard:SGT}, different levels of Replica Symmetry Breaking (RSB). The simplest non-trivial case is called Replica Symmetric (RS), where all the inter-replica correlations are identical.

With the hypothesis on correlations introduced the average over replicated dynamical variables can be taken, and the dependence on $n$ analytically continued to the real numbers. A variational form for the free energy can be produced from which the appropriate values for the correlations can be inferred by an extremisation procedure.

The RS approximation proves not to be a sufficient description of replica correlations in the SK model. The limiting case in an RSB hierarchy is applicable, and is tractable in the SK model. The results of the equilibrium analysis predict a complicated fragmented phase space, not easily accessible in either real systems or in numerical evaluation of the model, and with dynamical features similar to glasses.

The replica method, combined with the hierarchy of RSB variational solutions, proved to be a major breakthrough in the study of disordered models of much wider importance than in the field of solid state physics~\cite{Nishimori:SP,Hertz:ITNC}. The correctness of the replica solution for the Sherrington-Kirkpatrick model is now accepted after much effort to verify the consistency of all steps~\cite{Talagrand:SG,Parisi:SS}. The limitations of the replica method as applied to other important classes of models, and its relevance for finite dimensional assemblies, including the physical spin glass, and finite size assemblies, remain active areas of research.

%% file: INTRODUCTION/Graphical_model.tex

Graphical models are used to describe dependencies between states in a model, and are valuable in gaining intuition for a problem~\cite{Jensen:ABN}.

The graphical model for SK takes the form of a fully connected graph of binary interactions as shown in figure~\ref{fig:Introduction.SK}. A more general representation of interactions is provided by factor graphs~\cite{Richardson:MCT,Kschischang:FG}. A factor graph is a bipartite graph with dynamical variables associated to circles, and functional dependencies associated to squares. The factor graph $\factorG$ includes a labeled set of variable nodes $V_v$, factor nodes $V_f$, and edges $E$. Microscopic states are associated to the variable nodes, which interact only when connected through some factor node(s). To each factor node is associated some function on the variables, which in this thesis is either a logical constraint or probabilistic relation.

\begin{figure}[htb]
\begin{center}
\includegraphics[width=1.0\linewidth]{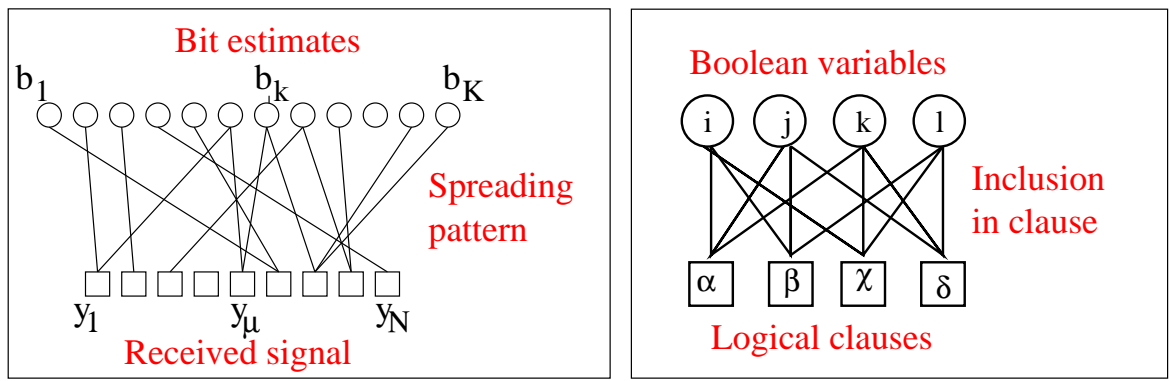}
\caption[Two simple factor graphs.]{\label{fig:Introduction.Simplefactorgraph2} Left figure: A factor graph in the case of source detection is represented. This includes the source bit variables (upper circular nodes) and the evidence, which is the signal spread over some discretised bandwidth. Each factor node (lower square nodes) may label, for example, the received power on some frequency band during a short interval. The source bits are assumed to have a probabilistic relation with the received signal, dependencies are indicated by the links. Properties of the source bits might be inferred by various methods, depending on the structure of the graph. Right figure: A set of logical clauses (squares) on Boolean variables is represented. In each clause exactly one variable is true, where inclusion in a clause is demonstrated by a link. From the factor graph consistent logical assignments may be found.}
\end{center}
\end{figure}

Figure~\ref{fig:Introduction.Simplefactorgraph2} demonstrates two models characteristic of problems studied in this thesis. The left figure describes a source detection problem. A received signal, discretised on some bandwidth, is known to represent a set of source bits, with different sections of in the signal being dependent on different sources. The aim is to estimate the source bits given the evidence (signal) and assumed dependencies, represented in the graphical structure.

In the second graphical model of figure~\ref{fig:Introduction.Simplefactorgraph2} a Constraint Satisfaction Problem (CSP) is represented. A set of logical statements (clauses) on some variables is represented. Each clause is encoded by a factor, and edges imply inclusion of that Boolean variable in the clause. Determining if any assignment to the variables satisfies all clauses simultaneously is the question of interest.

A useful feature of the sparse graphical model is the explicit representation of conditional dependencies of states. It is convenient to define the local quantities $\partial_\mu$ to describe the set of variables on which factor node $\mu$ depends, the set connected through an edge. Similarly $\partial_k$ describes the factors relevant to determination of a particular variable, again associated to through a local edge set. These sets may be much smaller than the full sets of factors or variables, and can be used to identify sub-problems.

%% file: INTRODUCTION/Belief_propagation.tex

\subsubsection{Calculating marginals}
\begin{figure}[htb]
\centering{
\includegraphics[width=\linewidth]{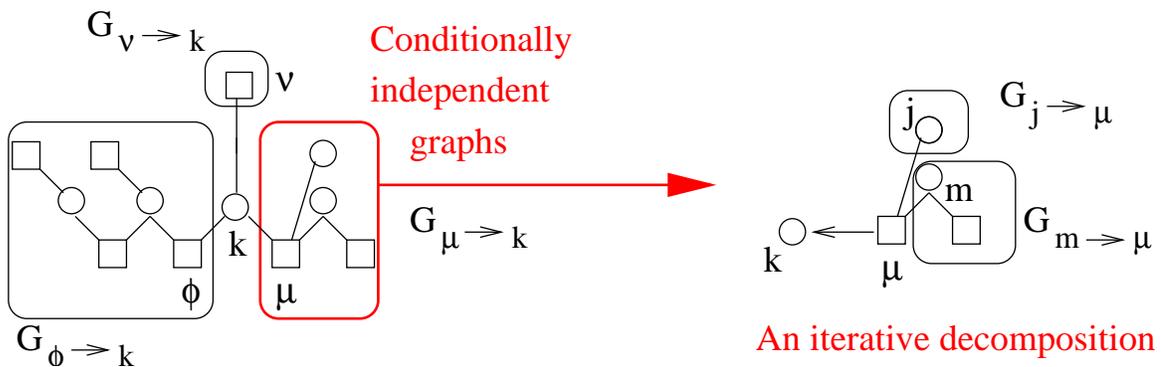}
\caption[Iterative calculation of marginals on a tree.]{\label{fig:Introduction.IterativeDecoding} The source detection problem of figure~\ref{fig:Introduction.Simplefactorgraph2} is a tree on which marginals can be calculated efficiently, providing a basis to estimate source bits. The marginal probability of bit $k$ can be determined by considering sub-problems defined on cavity graphs, $G_{*\rightarrow*}$. The marginals can be calculated on leaves and iterated inwards through probabilistic relations.}}
\end{figure}

This section describes a problem pertinent in Bayesian networks, graphical models in which the factors encode some probabilistic relationship amongst variables; the problem of calculating marginal probability distributions for the states. A general method exists for calculating marginals, this is to marginalise over all states excluding the state of interest
\begin{equation}
P(b_i | G) = \sum_{\vb \setminus b_i} P(\vb |
G)\label{eq:Introduction.exhaustivemarg}\;.
\end{equation}
This process is unfortunately computationally expensive when $N$, the number of states, is large. Belief Propagation (BP), also called the sum-product algorithm~\cite{Kschischang:FG}, provides a more efficient method, which can be applied iteratively and process estimates a distributed manner. The computation complexity is dependent on the cost of evaluating factor node relations, and the number of edges in the model. Unfortunately the uniqueness of any solution produced, or the convergence of messages to any solution, is not guaranteed except in some special cases.

Calculation of marginals is a useful process in the case of the source detection problem (figure~\ref{fig:Introduction.Simplefactorgraph2}), since this provides a basis for determining the most likely value for any source bit. Belief propagation (BP) is a message passing method, whereby messages, associated to directed edges (2 for each undirected edge) in the factor graph, obey some coupled set of equations determined by probabilistic rules. Two types of messages represent solutions to problems defined on subgraphs of the full graphical model ($G$). Evidential messages represent the likelihood of state $i$ on the subgraph $G_{\mu \rightarrow i}$, a sub-graph of $G$ with all dependencies between $i$ and $\partial_i$, except $\mu$, removed. Variable messages represent a posterior distribution of state $i$ on $G_{i \rightarrow \nu}$, a sub-graph of $G$ with all edges attached to variables in $\partial_\mu$, except $i$, removed.

In the case of a tree it is possible to determine the exact value of messages on the sub-problems corresponding to leaves, and these may be iterated inwards to determine marginals at any point in the graph through a combination of the evidential and variable messages, as shown in figure~\ref{fig:Introduction.IterativeDecoding}.

Some graphical models with loops may also be solved exactly and efficiently by BP~\cite{Weiss:CLP}. Small loops can be handled by replacing the non-tree like dependencies implied by a loop with a generalised factor node connecting all the variables in a star like configuration~\cite{Yedidia:GBP}. A tree like structure is formulated at the cost of some potentially more complicated functional relationships. Other cases with only a single loop may be solved, and it is possible in some cases to show convexity relationships, which guarantee convergence in apparently complicated models (e.g.~\cite{Weiss:MELP}).

In this thesis BP is applied to finite loopy graphs, where it is a heuristic rather than exact method~\cite{Weiss:CLP}. Leaves are either absent, or the messages defined on leaves cannot be iterated to determine all messages uniquely. A heuristic guess is used to initialise messages. These initial guesses can be refined by the iteration of the BP relations. In these cases the factor graphs are only one link deep, including the root variable, and messages arriving from other attached nodes.

\subsubsection{Belief propagation algorithm}

The BP algorithm can be defined in a general manner for a variety of problems in statistical physics, where the probability distribution for dynamical variables ($\vb$) is
\begin{equation}
P(\vb) = \frac{1}{Z}\exp\{ - \beta\Ham(\vb)\} \label{eq:Introduction.Pb}\;,
\end{equation}
described by a partition function $Z$, inverse temperature $\beta$ and Hamiltonian $\Ham$. A marginal on the spins, can be conveniently represented as a log ratio, for example
\begin{equation}
H_i = \frac{1}{2 \beta} \sum_{\tau_i} \tau_i \log \left( \sum_{\vtau\setminus \tau_i} \exp\{ - \beta\Ham(\vtau)\}\right)\label{eq:Introduction.Hi}\;,
\end{equation}
describes the marginal for a single spin
\begin{equation}
P(b_i) =\sum_{\vb\setminus b_i}P(\vb) = \frac{\exp(\beta H_i b_i)}{2\cosh(\beta H_i)} \label{eq:Introduction.Pbi}\;.
\end{equation}
These log-likelihoods quantities will be manipulated rather than full probability distributions where possible.

The Hamiltonian can be decomposed as a summation of the energies at factor nodes, where $\partial_\mu$ are the set of variables in a factor
\begin{equation}
\Ham(\vb) = \sum_\mu\Ham_\mu(b_i | i \in \partial_\mu)\;.
\end{equation}
Many relevant graphs include binary couplings so that all possible factors are labeled uniquely by edges $\mu=\ij$, and a ferromagnetic or anti-ferromagnetic interaction may determine the energy $\Ham_\ij(b_i,b_j)= -J_\ij b_i b_j$, as in equation (\ref{eq:Introduction.SK}). In more general scenarios each factor may include many variables, so that $\ij$ does not provide a sufficient labeling of factors, and probabilistic dependencies at a factor may be arbitrary. The representation through a factor graph has the interactions (factor nodes, $\mu$) and dependencies (edges $\mu i$) treated separately.

In BP an estimate for (\ref{eq:Introduction.Pbi}) is achieved by first finding the fixed point for a message passing procedure, each message representing a probability on a subgraph which is initialised (time, $t=0$) through some special insight, or more generally by guesswork. Two types of messages are passed from factors to variables (evidential messages) and from variables to factors (variable messages). The relation can be written as a recursive one, in time, so that each iteration is based on previous estimates and converges, in some cases, on a fixed point.

The variable messages define estimates to posterior probability distributions on a factor graph with factor node $\mu$ removed, which can be encoded by a log-posterior ratios
\begin{equation}
h^{(t)}_{i \rightarrow \mu} = \frac{1}{2 \beta} \sum_{\tau_i} \tau_i \log P^{(t)}(b_i=\tau_i |G \setminus V_{\mu}) = \sum_{\nu \in \partial_i \setminus j} u^{(t-1)}_{\nu \rightarrow i}\label{eq:Introduction.convergedvariable}\;,
\end{equation}
with $\setminus$ used to denote exclusion. The message $h^{(t)}_{i \rightarrow \mu}$ is an estimate to the quantity
\begin{equation}
h_{i \rightarrow \mu}= \frac{1}{2 \beta} \sum_{b_i} b_i \log \left( \sum_{\vb\setminus i} \exp\left\lbrace - \beta \sum_{\nu\setminus \mu}\Ham_\nu(b_i | i \in \partial_\mu) \right\rbrace\right) \;,
\end{equation}
it is the probability on a graph where the dependency $\mu$ is removed from the Hamiltonian (equivalently the node $\mu$ removed from the graph).

Similarly $u_{* \rightarrow *}$ are factor messages, which estimate log-likelihood ratios (\ref{eq:Introduction.convergedfactor}). In the case of a tree removing node $\mu$ creates $|\partial_\mu|$ independent trees, each of which may be described by an independent probability distribution, and this is the reason for the second decomposition in (\ref{eq:Introduction.convergedvariable}), the factor messages are independent and the probabilities factorised, this becomes an approximation in loopy graphs.

The recursion is completed through an update for factor messages in terms of variable messages, and some initial condition. The evidential messages are log-likelihood ratios
\begin{equation}
\begin{array}{lcl}
u^{(t)}_{\mu \rightarrow i} &=& \frac{1}{2 \beta}\sum_{b_i} b_i \log \sum_{\vb \setminus b_i} P^{(t)}(\vb\setminus b_i|b_i, G \setminus V_\mu) \;;\\
&=& \frac{1}{2 \beta} \sum_{b_i} b_i \log \left(\prod_{j\in \partial_\mu \setminus i}\left[\sum_{b_j}\exp \{\beta h_{j \rightarrow \mu} b_j\}\right] \exp \{-\beta\Ham_\mu(b_k|k\in\partial_\mu)\} \right) \label{eq:Introduction.convergedfactor}\;;
\end{array}
\end{equation}
where the product applies only to the term in square brackets. The marginalisation in the calculation is simplified by use of the variable messages (\ref{eq:Introduction.convergedvariable}), which are treated as independent priors so that the marginalisation need be carried out over only one factor node, rather than the entire Hamiltonian. In the case of the simple Hamiltonian with anti-ferromagnetic couplings the marginalisation is straightforward, using $\mu\equiv\ij$
\begin{equation}
u^{(t)}_{\mu \rightarrow i} = \frac{1}{\beta} \atanh\left(\tanh(\beta J_\mu)\tanh(\beta h_{j \rightarrow \mu})\right)\;.
\end{equation}
If $\beta J_\mu$ is large then $u$ is correlated with $h_{j\rightarrow \mu}$, indicating the spins are similarly aligned on the cavity graph (as expected for a ferromagnetic interaction). Weak coupling gives a message which is nearly zero, indicating only a small bias in the variable. The scaling with $\beta$ is chosen so that the messages are always $O(J_\mu)$ when $J_\mu$ is large, which is convenient numerically.

Again the assumption is that the priors for incoming messages are independent, this is trivially true in the $\mu\equiv\ij$ case since there is only one incoming message, but for hyper-edges this is not true except on trees.

From the messages an estimate of the posterior distribution (\ref{eq:Introduction.Pbi}) is given by a product of likelihoods (\ref{eq:Introduction.convergedfactor}) originating in the attached factors. Assuming independence of the factor messages the BP estimate at iteration $t$ is produced
\begin{equation}
H^{(t)}_i = \frac{1}{2\beta} \sum_{\tau_i} \tau_i \log {\hat
P}^{(t)}(b_i \tau_i| G) = \sum_{\mu \in \partial_i} u^{(t)}_{\mu \rightarrow i} \label{eq:Introduction.LLR}\;.
\end{equation}
Other marginal quantities may also be calculated in a simple manner, given a converged set of messages.

\subsubsection{BP and statistical physics}
\begin{figure}[htb]
\begin{center}
\includegraphics[width=0.5\linewidth]{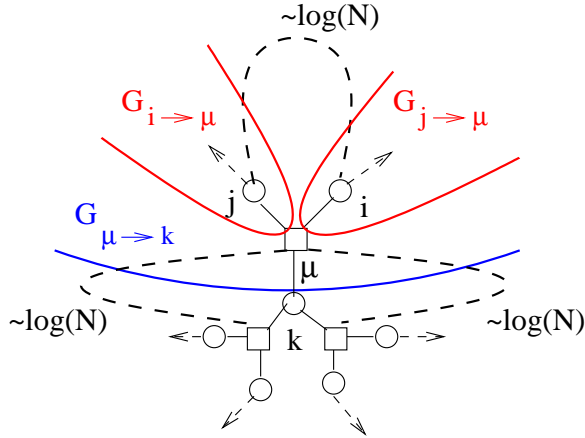}
\caption[Locally tree like structure in random factor graphs, and consequences for belief propagation.]{\label{fig:Introduction.Randomfactorgraph} Sparse graphical models may be characterised by locally tree like structures when the number of variables, $N$, becomes large. Above the percolation threshold two cavity graphs rooted in some variable or factor are not independent, since the priors in the cavity graphs depend on a common set of variables and are connected through (many) loops, each loop containing $O(\log(N))$ links. However, if dependencies are weak then the priors may, at a statistically significant level, depend only on local variables, and these are in the vicinity of the root and not shared by the two cavity graphs. The statement that the posterior probability of variable $i$ is independent of the posterior on $j$ in the absence of factor node $\mu$, may then be correct to leading order in $N$, and the probabilistic recursions implied by BP will be correct.}
\end{center}
\end{figure}

There is a close connection between BP and statistical physics methods. The solution to the extremisation procedure of the replica method, in the RS assumption, produces a set of relations with a structure often equivalent to a special case of BP. Whilst BP represents a dynamical process of messages on a particular graph, the analogous equations in the saddle-point method represent mappings of density in a function. Aspects of dynamics in the former would seem to be unrelated to equilibrium properties of the latter except at fixed points (in the case of convergence), but the similarity of processes are not superficial and conclusions drawn in one framework can be used to form hypotheses on the other.

Sparse random graphs~\cite{Bollobas:RG}, above the percolation threshold, are used throughout the thesis. The topology of interactions in these graphs converge in the large system limit to a locally tree like structure. An example is shown for a regular connectivity graph in figure~\ref{fig:Introduction.Randomfactorgraph}. In this case any two messages are correlated only through (many) long loops. Information may decay exponentially along each of these paths, and if correlations between the paths are weak then the messages arriving at a particular node may be effectively independent. The decay of information has an analogy in physical models as the decay of connected correlation functions, as arises in a pure state~\cite{Mezard:SGT}.

Decay of correlations may be tested within an statistical mechanics framework by posing the problem on a tree and considering properties of the boundary in the large system limit~\cite{Rivoire:GM,Mezard:RT}. This method can provide a proof of the convergence of BP in asymptotic samples, equivalent to the stability of RS solutions at equilibrium. If the convergence is to a unique fixed point, then BP will correctly reconstruct the marginals.

%% file: INTRODUCTION/Branch_and_bound.tex

\begin{figure}[htb]
\centering{
\includegraphics[width=\linewidth]{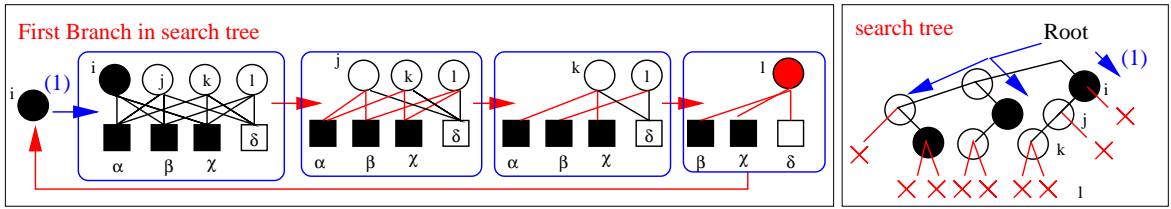}
\caption[Branch and bound on a simple constraint satisfaction problem.]{\label{fig:Introduction.branchandbound} The graphical model introduced in figure~\ref{fig:Introduction.Simplefactorgraph2} can be searched exhaustively for a satisfying solution by branch and bound methods. Left figure: No variable is initially implied, but assuming variable $i$ to be true creates a new branch in which all the other values may be iteratively implied by trivial unit (single-variable) clauses. However, it is found that two unit clauses are in contradiction so this branch is invalid and removed. A new branch is explored choosing the alternative assignment to $S_i$. Right figure: The solution space can be searched efficiently by considering all branches by a combination of implication and guess work, UNSAT is proved efficiently with only two heuristic steps being necessary.}}
\end{figure}

Graphical models for CSPs involve constraints rather than probabilistic relations, and the central question is of satisfiability (SAT), determining if any assignment to variables violates no clauses. In the search for a solution it is convenient to use a branch and bound decimation algorithm that involves a guided search through the solution space. For CSPs it is typical to consider variations of the Davis-Putnam (DP) algorithm~\cite{Davis:MP}. The state space $\{\True,\False\}^N$ can be represented as a regular tree of depth $N$. Each possible assignment to states is represented by a unique path between a leaf and the root, the value of state labeled $i=1 \ldots N$ is determined by the direction of branching (left/right) at level $i$.

This state space is searched from the root, by decimating (assigning) values first according to simple localised constraints, and, in the absence of such constraints, by some heuristic rule. With each assignment the nature of the factor graph is modified, so that some simple non-degenerate clause statements might appear. For example a problem might include a statement on two variables with neither variable being uniquely implied. However, when one variable is decimated the other variable is logically implied. In this way a sequence of heuristic steps might be followed by logical implication steps, with all ambiguity stemming from the heuristic steps.

Either a leaf of the tree is reached from the root by decimating $N$ variables, proving SAT, or else unsatisfiability (UNSAT) is shown on that particular search branch, by a logical contradiction. If a logical contradiction is encountered then the most recent heuristic step is reevaluated to the opposite state and a new branch searched. In the case that both branches evolving from a heuristic step are exhausted it is necessary to consider the next most-recent heuristic step. Each time a contradiction is encountered a new branch is explored. If all branches stemming from heuristic evaluations lead to contradictions then the problem is proved UNSAT.

The process is represented for a small example with Exact Cover clauses in figure~\ref{fig:Introduction.branchandbound}. There are two heuristic steps, the other evaluations being implied by logical constraints, therefore to search the tree only three paths are explored, in a state space of sixteen ($2^4$) possible branches. Every branching leads to a contradiction so there is no SAT solution.

In worst case there are an exponential (in number of variables) number of branches, which must be explored before a solution is found, or the absence of a solution is proved. Branch and bound methods are complete solvers, always terminating with a solution if one exists, but they are not always efficient. Nevertheless, they form the basis for solving hard constraint satisfaction problems.

For problems with a random logic structure much progress has been made in the development of efficient heuristic decoders through the statistical mechanics frameworks~\cite{Mezard:AAS,Montanari:SCS,Measson:MC}. Many of these methods are based on BP, and an abstraction of the CSP to a probabilistic framework, and some outperform the best branch and bound methods for random graphical structures.

The basis of success in these algorithms relates to statistical reasoning, which is important in determining an optimal heuristic rule in branch and bound. To minimise the number of branches searched it is ideal to choose the state maximising the number of SAT solutions in the branched tree. If twice as many assignments contain the decimated variable set to true than false, then this can form the basis for a greedy branching strategy. Statistical arguments may be made concrete in the case of samples from known ensembles.

%% file: INTRODUCTION/Exact_cover.tex

\begin{figure}[htb]
\centering{
\includegraphics[width=\linewidth]{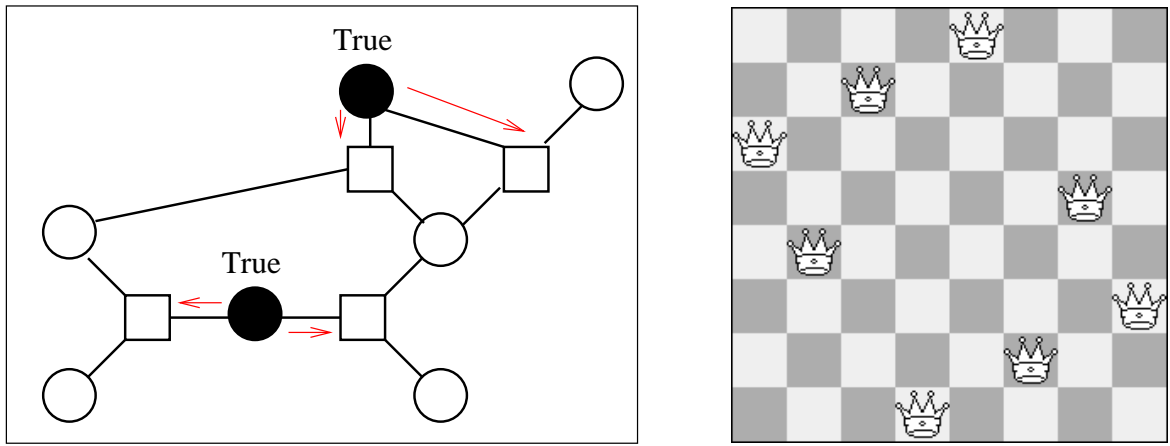}
\caption[A Satisfiability-certificate in a simple Exact Cover
problem.]{\label{fig:Introduction.UCP_SATassignment} Left figure: In a factor graph the interactions amongst a set of Boolean variables (circles) are prescribed by factors (squares). In the ECk decision problem establishing the existence or non-existence of a variable assignment so that each clause (factor) is covered by (connected to) exactly one true variable (circle) is sought. The assignment shown to variables, with black as true, and white as false, indicates one solution amongst several to this problem. The true (black) variables are distributed in such a way that all clauses can be uniquely identified with one variable (the cover). Right figure: The 8 Queens problem is popular realisation of the Exact Cover problem. A solution is shown satisfying the constraints that exactly one queen covers every row and column, and that at most one queen covers each diagonal.}}
\end{figure}

Exact Cover (EC) is a well known CSP encountered in computer science and optimisation~\cite{Rosenthal:ES}. The problem is defined by a set of $N$ Boolean variables and $\alpha N$ logical constraints. Each EC clause represents the statement that one included variable is $\True$, and all others are false.

A familiar example of Exact Cover is the $N$ queens problem in chess, whereby one must choose the positions of 8 queens such that each row and column are covered by exactly one queen, and each diagonal must covered by at most one queen (a slight variation on EC clause). The 64 variables (squares) must be assigned to either true/false depending on whether a queen is present/absent to meet the 46 row, column and diagonal constraints. An Exact Cover solution to this problem is demonstrated in figure~\ref{fig:Introduction.UCP_SATassignment}, alongside the solution to problem where all clauses are in 3 variables.

The standard $k$ variable Exact Cover problem (ECk) is defined by a set of parameters $\{N,M,\alpha\}$. In the typical case formulation of ECk the decision problem may be phrased: Given a set of $N$ Boolean (2-state) variables $S_1,\ldots,S_N$, and a set of $M$ logical clauses, each containing exactly $k$ distinct variables selected at random from the full set, does there exist an assignment to the variables such that exactly one variable is true in every clause. Any assignment of variables that exactly covers the clauses is called a SAT-certificate and is a sufficient proof. The negative version of this decision problem is also interesting: given a sample taken as above, do there exist no satisfying assignments. Again one might have some proof of unsatisfiability, this would be an UNSAT-certificate.

Since the set of candidate SAT-certificates is finite (of size $2^N$) a simple way to find a SAT-certificate is to run through the list of $2^N$ different candidate certificates until a SAT-certificate is found. Suppose, however, that $N$ and $M$ are both large and proportioned so that the state space is much bigger than the solutions space. In this case testing an exponential number of configurations might be required, and the problem is computationally expensive. The question of algorithmic complexity naturally arises, does a fast algorithm, requiring few logical evaluations, exist that can always demonstrate an Exact Cover, if it exists.

Any scalable algorithm must work for arbitrary $N$, and it is usual to classify complexity in terms of the asymptotic (large N) scaling of the algorithm time: the number of elementary logical operations required to find a SAT-certificate. A useful distinction is between fast ($O(N^x)$) and slow ($O(\exp N)$) methods, although distinctions within the fast set such as linear $O(N)$ are also important. To find fast algorithm for the worst imaginable sample from the ECk ensemble is improbable, since ECk (with $M$ polynomial in $N$) is in the class of Non-deterministic Polynomial complete (NP-complete) problems. Completeness is a statement of algorithmic equivalence~\cite{Cook:CTP}, and implies that a fast algorithm for ECk would also be a fast algorithm for a large and important range of combinatorial problems~\cite{Garey:CI}. Unlike Exact Cover, ECk is not in standard lists known by the author, but demonstrating worst case equivalence of ECk to other standard forms such as k-satisfiability is straightforward~\cite{Karp:RACP}.
It is widely assumed, but not proven, that only slow algorithms might work for NP-complete problems on practical computing machines. The NP part implies amongst other things, that if a solution is known it can be validated by a fast algorithm.

In worst case producing an UNSAT certificate might be demonstrated by slow methods only, it is at least as hard as producing a SAT certificate. Other interesting questions within the random ensemble framework include determining existence of a solution with fewer than $E$ constraints violated, determining the number and correlations amongst solutions, or the optimisation problem in which the question asked is 'what is the minimal number of constraints that must be violated in any assignment?'. Pessimistic complexity results also apply to these decision (yes/no) and optimisation questions~\cite{Garey:CI}. However, one reason for recent interest in ECk was apparently excellent performance attained by a quantum adiabatic algorithm~\cite{Farhi:QAEA}, but only for small instances.

Given that no fast algorithm has been shown to exist for worst case, the benchmark by which to judge efficiency of practical algorithms is not obvious. Much work undertaken in studying CSPs before, and since, the interest arose in statistical physics has been in developing algorithms based on refined branch and bound methods (complete solvers that produce results but may work slowly) and incomplete algorithmic methods (solvers that work fast, but may fail to show a result).

Worst case of Exact Cover, even when restricted to clauses with only $k=3$ variables is unsolvable by fast methods, but what of typical samples from the ECk ensemble? Within such an ensemble it may be that there exist hard to solve instances, but these may be unrepresented in sampling a large set.

The interest from the statistical physics community in decision questions for CSPs is a recent phenomena~\cite{Monasson:SMRK,Mertens:CC,Hartmann05}, and is based on the observation that ensemble descriptions provide a benchmark for exploring algorithmic complexity questions. Typical cases are considered to be samples from an ensemble with some concise parameterised description, for example ECk. Statistical physics methods are able to demonstrate detailed parameter ranges for SAT and UNSAT, and the nature of correlations amongst solutions. A second reason for interest from physics is the close relationship between some parity check based channel coding methods and random constraint satisfaction problems~\cite{Richardson:MCT}.

A statistical physics reinterpretation of CSPs is achieved through considering a set of spins (Boolean variables) and interactions (present between variables attached to the same clause). The interactions are defined so that an energetic penalty is paid locally whenever a clause is not exactly covered. The ground state(s) of the system then become SAT certificates, when the ground state energy is zero, or otherwise proved UNSAT. Descending in the energy landscape from some point represents a greedy local optimisation method. More generally insight into the properties of algorithms can be gained by considering the topology of the phase space, and attractors in the energy landscape~\cite{Mezard:AAS,Montanari:SCS}.

The strongest results attained by statistical physics methods are for typical case. The methods have been particularly successful in analysing the properties of CSPs restricted to random graph ensembles with homogeneous clause types. In this thesis algorithmic properties of a set of ensembles closely related to Exact Cover are considered. These ensembles demonstrate an unusual variety of behaviours that are examined and contrasted with equilibrium analysis and insight.

%% file: INTRODUCTION/Multi_user_detection.tex

\subsection{Wireless communication}

\begin{figure}[htb]
\begin{center}
\includegraphics[width=0.5\linewidth]{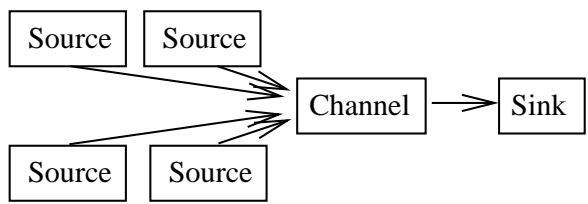}
\caption[A multi-access channel schematic.]{\label{fig:Introduction.manypointtopoint} The multi-access communication channel involves a set of independent sources communicating through a shared noisy channel. The multi-user detection problem involves inference of the sources given the signal received at the sink.}
\end{center}
\end{figure}

Multi-user detection is the problem of detecting source information within a multi-access communication channel~\cite{Verdu:MD}. In a multi-access channel a set of $K$ users (sources) transmit independent information, to a single base station (sink), through a shared noisy channel, as shown schematically in figure~\ref{fig:Introduction.manypointtopoint}. This problem is a natural generalisation of the single user noisy channel, which is the seminal channel coding problem~\cite{Shannon:MTC}. A dual scenario to the multi-user detection problem is that of broadcasting, one to many communication, but the terminology of transmission is used.

The main practical application of multi-user detection is in wireless communication. The bandwidth (frequency $\times$ time) is the medium on which information is transmitted, and may be considered as broken into discrete resolvable blocks (chips), with each chip subject to some environmental noise during a transmission. On this bandwidth each user transmit information to a base station according to some protocol on bandwidth access.

\begin{figure}[htb]
\centering{
\includegraphics[width=1.0\linewidth]{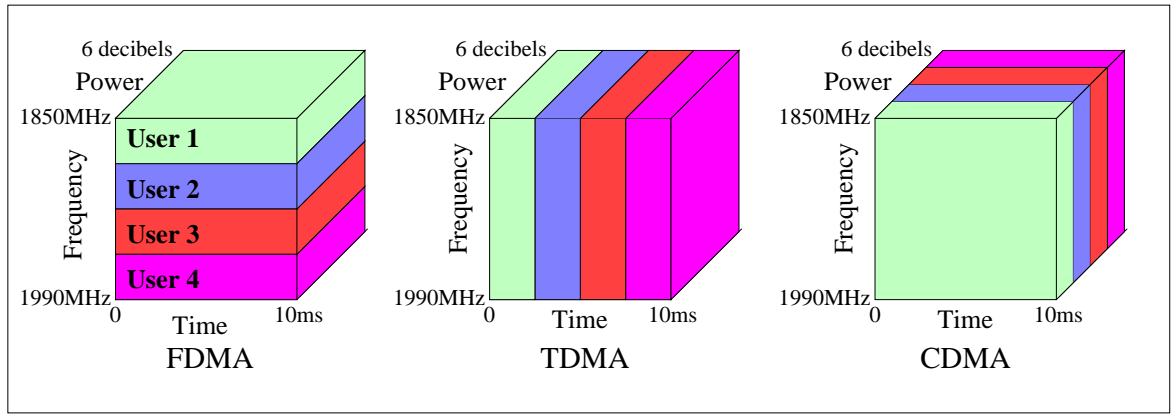}
\caption[An FDMA, TDMA and CDMA schematic for four
users.]{\label{fig:Introduction.FDMATDMACDMA} Each user (source) transmits with some power on the bandwidth, which is described by a time-frequency interval. There are 4 users in the above example distributing power according to some paradigm across the bandwidth. Users can concentrate power on small frequency (FDMA) or time (TDMA) intervals, or else can distribute transmission power across the bandwidth (CDMA). In the final diagram each user transmits with uniform power on all time-frequency blocks, although interference in the channel means that there is not a clear delineation of power sources in the received signal -- which is at the root of the inference problem. The total power is preserved in expectation, but there is signal interference. The labeling in the first figure shows some scales for the components in realistic wireless phone communication, decibels being a measure relative to environmental noise.}}
\end{figure}

There are various ways in which to spread user signals across the bandwidth and achieve successful source detection, a standard method is Code Division Multiple Access (CDMA). CDMA is a method allowing the benefits of wide-band communication to all users simultaneously, as shown in figure~\ref{fig:Introduction.FDMATDMACDMA}, which has a number of attractive theoretical and practical properties over communication on a scalar channel~\cite{Ipatov:SS}. These include the ability to reduce power, increase robustness and resolve scattering effects.

\subsubsection{A realistic model for wireless phone communication}

A range of complicated phenomena are inherent to wireless multi-user communication in realistic environments. Amongst the most important are distance and frequency dependent fading of signals, multi-path effects~\cite{Rappaport:900}, Multi-Access Interference (MAI), Inter-Symbol Interference~\cite{Kavcic:BII} and random environmental noise. As well as this assignment of users between base-stations must be determined through a hand-off process and protocols must exist for a range of different communication scenarios. A separation of these effects is in some cases artificial.

The received signal, in a general case, might include environmental noise along with a superposition of delayed and faded paths from each user. The amplitude at a given chip (frequency/time, $(f,t)$) might be represented, in some cases, by a superposition of discrete paths
\begin{equation}
y(f,t) = \omega(f,t)+\sum_{k=1}^K \sum_{p(k)} F(p) b_k(f(p),t(p)) \label{eq:Introduction.generallvc}\;.
\end{equation}
There are many parameters, the simplest being $\omega(f,t)$, the channel noise, which is local to the receiver. For each user $k$ the set of paths ($p$) along which information arrives must be considered: to each path received on chip ($f,t$) corresponds to a source frequency $f(p)$ and time $t(p)$. There may be paths along a direct line of site preserving the frequency and timing (up to a delay) of the transmitted signal, but there may also be scattered paths. The signal received from each user is dependent on the symbol transmitted by user $k$, which is $b_k(f,t)$ along with some path specific fading $F(p)$.

In the detection problem an estimate for the source bits is desired, under some model approximating the generative process (\ref{eq:Introduction.generallvc}). The transmitted symbols may represent the source information (bits) through a redundant description to allow robust detection even in the presence of noise, or when the detection model is not identical to the generative process.

In practice fading can often be controlled, by appropriate amplitude modulation by the transmitter. Similarly there might be ways to resolve dominant paths either directly from the signal (using for example a Rake correlator~\cite{Verdu:MD}) or from some independent information on the channel. Estimation of detection model parameters, such as the noise variance in the case of Additive White Gaussian Noise (AWGN), may also form part of the inference process~\cite{Neirotti:IMP}, or be determined by independent information. A synchronisation of user transmissions may also be possible, which may be useful in reducing MAI.

MAI is degradation of a user signal caused by overlap between user signals, by contrast with random noise interference. MAI may occur either because the user transmissions are poorly synchronised, or as a result of random processes in the channel. MAI is inherent to multi-user detection, but absent from single user detection problems.

\subsubsection{A model with perfect power control and
synchronisation}
\begin{figure}[htb]
\includegraphics[width=1.0\linewidth]{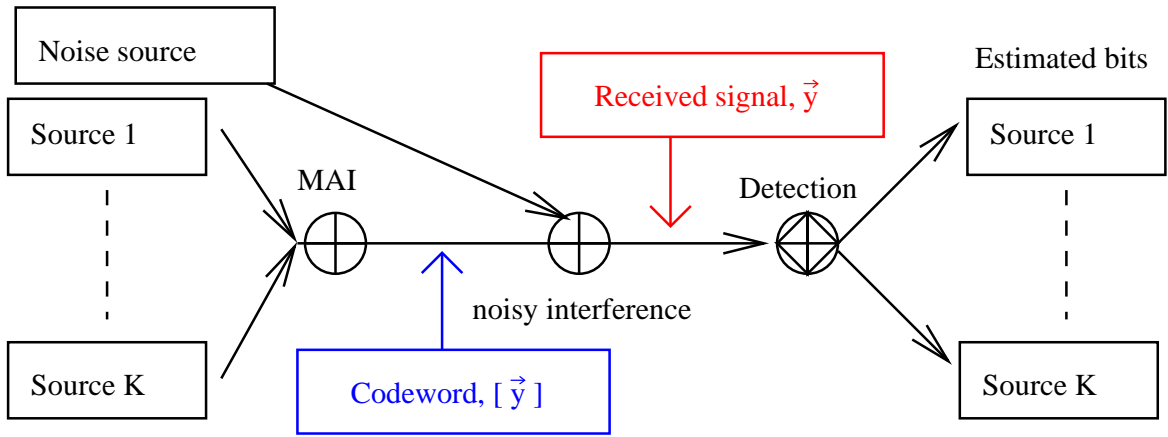}
\caption[The multi-access linear vector channel.]{\label{fig:Introducion.lvc} The multi-access linear vector channel takes as input a set of independent sources ($b_k=\pm 1$). These inputs combine additively to create a codeword and are subject to additive white Gaussian noise within the channel. Detection of bits occurs at the sink.}
\end{figure}

The model analysed is a simpler one than (\ref{eq:Introduction.generallvc}). Detection occurs on some discrete bandwidth of $M$ chips, called a bit interval. In the bit interval each user ($k=1\ldots K$) transmits a single bit ($b_k=\pm 1$), which is modulated according to a real vector spreading pattern ($\vs_k$) on the bandwidth. The modulation can be considered physically as occurring by Binary Phase Shift Keying (BPSK), combined with some amplitude modulation. Two in phase symbols interfere constructively, whereas two out of phase symbols interfere destructively, hence the additive nature of interference. The received signal ($\vy$) is a linear sum of the modulated spreading patterns from every user and random channel noise
\begin{equation}
\vy = \sum_{k=1}^K b_k \vs_k + \vomega \label{eq:lvc}\;.
\end{equation}
A schematic is shown in figure~\ref{fig:Introducion.lvc}. There are no explicit fading, inter-symbol interference or multi-path effects and perfect synchronisation of the users is assumed so that bit intervals are non-overlapping. In the detection problem the powers of different users are controlled by the base station, and it is assumed the receiver has full knowledge of the spreading patterns $\{\vs_k\}$ for all users. The detection problem is complicated by MAI and channel noise.

\subsection{Optimal detection}

\begin{figure}[htb]
 \includegraphics[width=\linewidth]{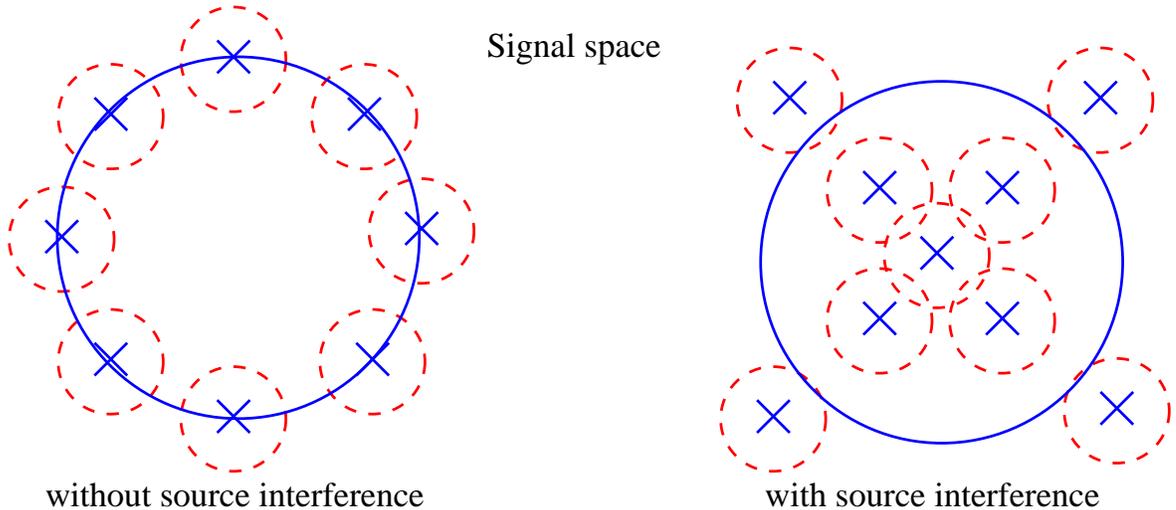}
\caption[Detection of codewords in a vector channel with and
without multi-access interference.]{\label{fig:Introduction.codewords} The detector must establish a hypothesis on source bits based on an $M$ dimensional signal space. The signal in the noiseless channel is detected as a set of at most $2^K$ distinct points (codewords), with noise the signal is determined in the space $\Reals^M$ concentrated at some fixed amplitude (power level). Left figure: With coordination of bit transmission it is possible to separate codewords so that detection is robust again moderate noise levels. Right figure: Without coordination typical codewords are at a smaller distance in signal space, and less robust against noise. The dashed lines represent a distribution on potential received signals, codewords distorted by noise.}
\end{figure}

The detection represents an inference problem in a high dimensional vector space as illustrated in figure~\ref{fig:Introduction.codewords}. A probabilistic detection framework is a principled method of estimating source information. This is achieved through construction of a posterior probability distribution, $P(\vb | \vy)$, many properties of which can be determined by statistical physics and algorithmic methods.

In the large system limit $M,K\rightarrow \infty$ the typical value for performance statistics, describing accurately almost all samples of channel noise and MAI, are the quantities of interest. The spread-spectrum (and many user) limit is a standard benchmark, and statistical mechanics methods are established tools in analysis of such cases~\cite{Guo:MDSP}. Often systems of practical size reflect strongly the properties inferred from the large system result; however, finite size effects may be significant in preventing practical applications.

Normally, a sufficient description of the probability distribution for the purposes of detection is a bit sequence meeting some optimisation criteria. The Marginal Posterior Mode (MPM) detectors~\cite{Tanaka:SMA} are a class of detectors determining bit sequence solutions that maximise the posterior distribution, and hence are optimal in a probabilistic sense. Similarly the Maximum-A-Posteriori (MAP) detector returns a state of the system (bit estimate) consistent with a maximum probability, which may be unique or one of several degenerate states. For the general case of non-zero MAI the optimal detection of source bits is a Non-deterministic Polynomial Hard (NP)~\cite{Verdu:CC}. That is to say there is no algorithm, efficient (polynomial) in running time, guaranteed to determine an optimal bit sequence. As in the previous chapter less pessimistic results can be expected for ensemble descriptions.

\subsubsection{The MPM detector}
The MPM detector returns an individually optimal estimate of bits
\begin{equation}
\tau_k^{(MPM)} = \argmax\left\lbrace \sum_{\vb\setminus b_k} P(\vb | \vy)\right\rbrace \;.
\end{equation}
A common measure of success for this and other detectors is the bit error rate, which is the proportion of errors in the marginal description
\begin{equation}
\BERmath(\vtau) = \frac{1}{2}\left(1 - \frac{1}{K}\sum_{k=1}^K b_k \tau_k\right) \label{eq:Introduction.BER} \;,
\end{equation}
where $\vtau$ is the estimate to the transmitted bits $\vb$.  The BER is minimised in expectation, averaging over $\vb$ consistent with the signal, when the model parameters exactly match the generative process and $\vtau=\vtau^{MPM}$.

The MPM is a special detector in that it is provably optimal amongst all detectors when the generative model and detection model are equivalent, the detection model is said to be at the Nishimori temperature/parameterisation. Many properties of the detection process become simpler in this scenario~\cite{Iba:NL}.

\subsubsection{The MAP detector}
The MAP detector determines a jointly optimal estimate of bits, which is
\begin{equation}
\vtau^{(MAP)} = \argmax \left\lbrace P(\vb | \vy)
\right\rbrace\;,
\end{equation}
where $\argmax$ returns the unique, or one of a degenerate number of bit sequences maximising the posterior. In the case of no prior knowledge on the bit sequence this result is equivalent to maximum likelihood detection. The MPM detector becomes equivalent to the MAP detector in some special models.

\subsection{The case for random codes}

Random spreading patterns/codes offer flexibility in managing bandwidth access by allowing code assignment by independent sampling for each user, and also have robust self-averaging performance for large system sizes. Furthermore the unstructured nature of codes makes them less susceptible to certain attacks and structured noise effects.

Random codes, sampled independently for each user, interfere in the channel. Optimal encoding of sources would involve a correlation of codes so as to minimise MAI. It has been shown that standard dense and sparse spreading patterns can achieve a bit error rate comparable to optimal transmission methods in the AWGN vector channel with only a modest increase in power. Optimal being by comparison with transmission in the absence of MAI, the single user case, with comparable energy per bit transmitted. The small increase in power required to equalise performance is often a tolerable feature of wireless communication.

CDMA methods can be formulated so as to reduce or remove MAI subject to synchronisation and power control of users; for example orthogonal codes ($(\vs^*_k)^T \vs^*_{k'}=\delta_{k,k'}$) can be chosen for sparse and dense systems, whenever the ratio of users to bandwidth $\load=K/M \leq 1$, which achieves the single user channel performance. Codes meeting the Welch Bound Equality minimise the cross-square correlations beyond $\load=1$~\cite{Rupf:OSM}, where some unavoidable MAI is present, for the BPSK case Gold codes achieve minimal MAI~\cite{Gold:OB}. A sparse orthogonal code is achieved by Time or Frequency Division Multiple Access (TDMA/FDMA), whereby each chip is accessed by at most a single user, for $\load>1$ sparse optimal codes may also be formulated.

However, in many cases only limited coordination of codes might be possible, so that MAI is an essential and irremovable feature. The random coding models, with a little elaboration, may also approximated different scenarios other than ones corresponding to deliberately engineered code. Consider for example a TDMA code, which is a sparse orthogonal coding method under good operating conditions, with each transmitted signal uniquely associated to a chip (time slot). In a practical environment the signal may not arrive perfectly but might have a significant power component delayed by random processes, contributing to unintended chips. This may occur in practice by way of multi-path effects. In terms of the optimal detection performance, the properties may then more closely resemble sparse CDMA, rather than an MAI-free TDMA method. Depending on how scattering occurs different random models may be relevant. If the paths are more strongly scattered across a significant fraction of the bandwidth a random dense inference problem is implied. Finally, a scenario with a few strong paths and many weak paths may apply, then the detection problem might involve inference with both sparse and dense spreading considerations.

\subsubsection{Sparse and composite random codes}

The particular focus in this thesis is on CDMA detection problems involving a sparse random component, extending the theory of densely spread codes. Sparse codes might allow more efficient detection methods by connection with sparse inference methods such as BP. There may also be some hardware constraints or adverse channel conditions (such as jamming), which would make a sparse pattern preferable over a uniform power transmission across the bandwidth. Complexity of detection algorithms and power of transmission are key constraints in realistic wireless communication, that could benefit from a sparse formulation. At the same time there are a number of wide-band benefits, which are lost in a sparse description, most importantly the reduced ability to detect scattered signal paths by filtering methods. A more exotic code involving a combination of both sparse and dense processes might preserve some of the practical advantages of the dense codes.

The large $M$ (spread-spectrum) scenario is an efficient multi-user transmission regime~\cite{Ipatov:SS}, and one in which we expect typical case performance of different codes drawn from the sparse or composite ensembles to converge. The properties of dense, sparse and composite random codes are distinguishable and may be calculated from a free energy density.

%% file: INTRODUCTION/Vb.tex

The seminal magnetic spin model is the Ising model, which is a lattice model for a ferromagnet. Lattice models have formed the basis for studying many physical materials, and the first and most realistic graphical models for spin glasses also take this form. The Edwards-Anderson model is a lattice model of spin glasses that captures the spatially dependent combination of
ferromagnetic and anti-ferromagnetic couplings~\cite{Edwards:TS}. A two dimensional model is demonstrated in figure~\ref{fig:Introduction.SK}. Although the EA model contains a number of realistic features of the material it proved not to be easily solved by exact methods.

As a means to understand features of the EA model through exact methods the SK model was proposed. The SK model is a mean field approximation to the EA model, each spin is assumed to interact according to some simple statistics with all neighbours, without spatial considerations. Analysis of the SK model in the large system limit is achieved by the replica method. The relationship between the SK model and EA model is unfortunately less transparent than corresponding mean-field methods in ordered systems. The existence of an upper critical dimension for lattice models above which SK may apply exactly is not known for example.

Between the SK and EA models, in terms of approximation to realistic spin glasses, is the Viana-Bray (VB) model~\cite{Viana:PD}, also shown in figure~\ref{fig:Introduction.SK}. This model includes the dilution effects relevant in the EA model, but there is no finite dimensional topology; couplings are sampled at random according to an ensemble without spatial considerations. In the simplest ensemble the couplings may be represented by an Erd\"{o}s-R\"{e}nyi random graph of mean connectivity $C/N$. There are two important sources of disorder in the model - the graph, and the couplings. The properties of the VB model are dependent on the topology as well as higher order moments of the marginal coupling distribution, by contrast with the SK model.

The VB model can also be analysed exactly by the replica method, although so far there is not a complete description of the spin-glass phase except by perturbative methods near ferromagnetic and paramagnetic phase boundaries, where behaviour is similar to the SK model~\cite{DeDominicis:RSB}, and near the percolation threshold~\cite{Mottishaw:SRF}. The lack of dimensionality is a significant omission from the model, although the VB model has applications in other scenarios where dimensionality may not be an important feature, such as graph partitioning~\cite{Wong:GB,Mezard:MFT}.

The VB and SK models are today viewed as being more useful as prototypes in the development of statistical physics theories for disordered systems, the simple ensembles each allows frameworks suitable for experimental methods. In this thesis a new spin model is studied, a composite model, which contains simultaneously features of both models and is also amenable to an exact analysis.

%% file: HUCPANALYSISOF1INK.tex
\section{Introduction}
\label{1inkSAT.introduction}

\begin{figure}[htb]
\centering{
 \includegraphics[width=0.5\linewidth]{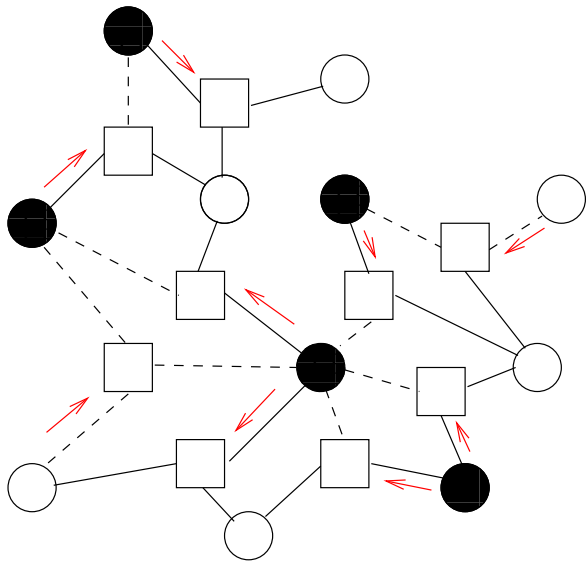}
\caption[A SAT-certificate in a simple $\okSAT$ problem.]{\label{fig:1inkSAT.okSATexample} A small $\okSAT$ problem is represented as a factor graph. Each factor represents an exact cover clause, each circle a variable, and each link the inclusion of a variable in a particular clause as either a positive (solid line) or negative (dashed line) literal. If a variable is set to false, but interacts through a negated literal, then the clause is covered. Arrows demonstrate an {\em exact cover} where black/white circles indicate variables assigned to true/false.} }
\end{figure}

This chapter demonstrates results developed in studying the $\eokSAT$ problem~\cite{Raymond:PD,Maneva:OK}, a generalisation of the k Exact Cover (ECk) Constraint Satisfaction Problem (CSP) outlined in section \ref{introduction.EC}. In ECk a set of Boolean variables interact in a set of ECk clauses. Each ECk clause is a logical constraint on $k$ variables, exactly one variable must be true in any clause. When many clauses exist complicated correlations in the assignments of variables are created. The satisfiability (SAT) question asks if there exists any assignment to variables, which violates no constraints.

In a generalisation, one in k SAT ($\okSAT$), Boolean variables interact indirectly in clauses as either positive or negative literals. A positive literal is identical to the variable, a negative literal takes the opposite logical value to the variable. A $\okSAT$ clause, on a set of Boolean literals, implies that only one of the literals is true. When all literals are positive $\okSAT$ is equivalent to ECk. The $\okSAT$ can be contrasted with the better known 3SAT clause, for which at least one literal must be true. $\okSAT$, like ECk, can be represented by graphical models, an example of a problem, along with an assignment to variables satisfying all clauses (a SAT-certificate), is shown in figure~\ref{fig:1inkSAT.okSATexample}.

The SAT question for logical CSPs is important in a wide range of fields~\cite{Garey:CI} and generating efficient and scalable algorithms to determine SAT is of great importance to computer science research. One standard algorithm employed to determine SAT is the Davis-Putnam-Logemann-Loveland algorithm (DPLL)~\cite{Davis:MP}, which is a complete branch and bound algorithm. DPLL generates certificates (proofs of SAT or UNSAT) by assigning variables in an iterative manner, and backtracking once a particular search pathway is shown not to contain any viable solutions. In the absence of simple logical deductions, variables are fixed by some heuristic rule, and these free steps determine a branching process on the space of feasible configurations. DPLL is complete, always returning a correct answer to the SAT/UNSAT question.

Unit Clause resolution is a simple logical deduction step employed in DPLL that is vital in making the branch and bound algorithm efficient for some CSPs. A partial assignment on variables might imply a necessary assignment to others. A unit clause is a clause in one variable, the constraint being that the variable is either true or false, a structure making explicit deductive reasoning. If unit-clauses are generated in an algorithm they constrain the branching to take a particular direction, which may reduce substantially the search space. Unit Clause Propagation (UCP) is the recursive application of the resolution, it is possible that in resolving some unit clauses others are generated, and this is the propagation effect.

\begin{figure}[htb]
\includegraphics[width=\linewidth]{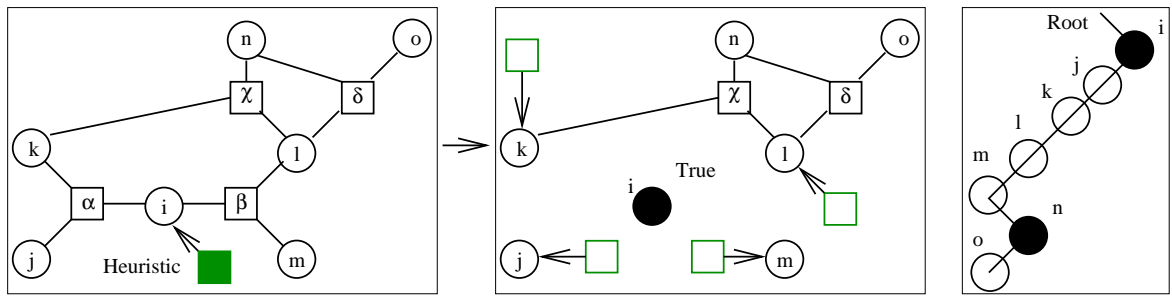}
\caption[SAT-certificate generation by UCP for a small ECk problem.]{\label{fig:1inkSAT.UCPsmallset_joined} Left figure: The graphical model, with factors labeled in Greek and variables labeled in Latin, can be searched by recursively resolving unit clauses combined with an initiating guess. Selecting a variable at random, $i$, and setting this to True [black] covers all attached clauses. Middle: This implies all variables in these clauses are False (white), as indicated by the unit clauses. Resolving these unit clauses then implies the final two variables. Right figure: In order to find a solution it is necessary to search only one branch of the search tree.}
\end{figure}

\begin{figure}[htb]
\includegraphics[width=\linewidth]{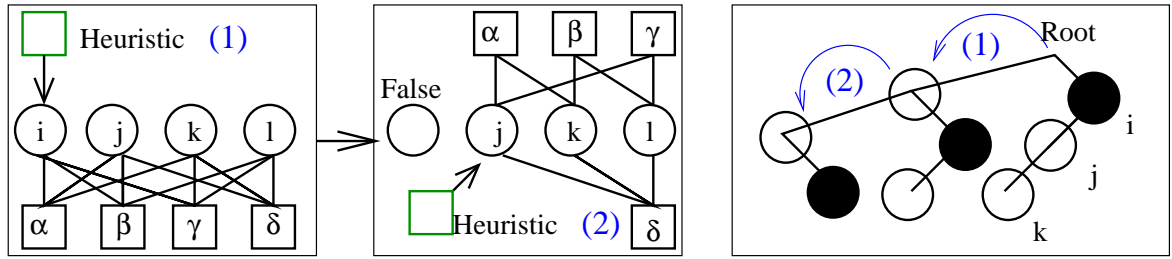}
\caption[UNSAT proof by UCP for a small ECk
problem.]{\label{fig:1inkSAT.UCPsmallset_contr_joined} The graphical model of figure can be searched by decimation. Decimating variable $i$, setting it to false, implies the reduction of 3-clauses to 2-clauses in several cases. A further heuristic step is required to generate the first unit clauses, but resolving the unit clauses leads to a contradiction: two unit clauses, which dictate opposite values to some variable. The first branching is unsuccessful. By backtracking each path in the search tree leads may be proved to lead to a contradiction at some depth demonstrating UNSAT.}
\end{figure}

Figure~\ref{fig:1inkSAT.UCPsmallset_joined} demonstrates some steps in applying branch and bound to a case of ECk. There exist 7 logic variables in the problem, therefore an exhaustive state space search requires evaluating $2^7$ configurations. The space of solutions can be search by first assigning variable $S_i$ to $\True$. This generates 4 unit clauses $\left\lbrace S_j=S_k=S_l=S_m=\False\right\rbrace$ and no contradictions. Resolving these unit clauses generates first a clause $\left\lbrace S_n=\True\right\rbrace$, and finally $\left\lbrace S_o=\False\right\rbrace$ is implied so that a SAT instance is found without testing a large number of assignments, only one branch of the search tree need be considered.

In figure~\ref{fig:1inkSAT.UCPsmallset_contr_joined} there is a different outcome to a branch and bound search. Many assignments are tested, but all searches result in contradictions. All branches are searched to the depth at which a contradiction is apparent, the tree itself constitutes a certification of unsatisfiability.

\subsubsection{Computational complexity}

For small systems DPLL will work fine, as may other methods. However, in larger systems it will, in worst case, require $O(\exp(N))$ evaluations to determine SAT, where $N$ is the number of variables. Typical large samples may not correspond to this worst case performance, and an ensemble description of large instances provides a statistical definition of complexity by which to test algorithm viability~\cite{Mertens:CC,Hartmann05}. Amongst the simplest ensembles includes all structures consistent with a fixed number of clauses ($M$) and variables ($N$). Since the number of variables included in a clause is three $\gamma$, the mean number of clauses per variable, is a convenient intensive parameter to describe the ensemble ($k M = \gamma N$). The structure of interactions in typical samples from this ensemble, with either ECk or $\okSAT$ clauses, has a sparse random graph structure.

It has been observed that many large random ensembles exhibit phase transitions similar to those in thermodynamics. As $\gamma$ increases there is often a transition from a phase in which typical samples have many satisfying variable assignments (SAT phase) to one in which there are no solutions (UNSAT phase). SAT phases almost surely (a.s.) contain a satisfying (SAT) assignment of variables, a.s. implies with probability asymptotically at least $1-O(1/N)$. In the UNSAT phase there is a.s. no SAT assignment. There is a SAT-UNSAT transition which is discontinuous in the probability of SAT.

It is also observed that there are other transitions relating to algorithmic performance. The Easy-SAT phase is a portion of the SAT phase for which an algorithm exists that a.s. finds a
satisfying assignment in polynomial time (quickly). In the
Easy-UNSAT phase the unsatisfiability can also be determined quickly. The Hard-SAT and Hard-UNSAT phases are implied only by negative results, the failure to find some efficient algorithm, although there are various hypotheses on the origins of hardness in random CSPs relating to the structure of the solution space~\cite{Krzakala:GS,Zdeborova:Thesis}.

In this chapter I shall concentrate on the algorithmic analysis which was my contribution to~\cite{Raymond:PD}, and some unpublished work produced in support of this paper. The performance of a simplified DPLL algorithm is analysed with respect to random graph ensembles parameterised by $k$ - the number of variables in each clause of the ensemble, $\gamma$ - the mean connectivity of any variable in the ensemble, and $\epsilon$ - the probability of a literal being in a negated form. The large system limit is studied where the problems of computational complexity are acute and well formulated.

\subsection{Summary of related results}

The possibility to examine typical case properties of large CSPs through DPLL has been considered, and numerical work undertaken for ensembles including ECk varieties. Special cases of DPLL have also been developed recently allowing exact analysis~\cite{Deroulers:CUU}, including UCP and some heuristic features.

A symmetric $\okSAT$ ensemble parameterised by $\gamma$ was examined by Achlioptas et al~\cite{Achlioptas:PT}, and it was shown that the SAT question could be determined at all $\gamma$ by a simple version of DPLL. A SAT/UNSAT transition was demonstrated, without any Hard-SAT/UNSAT phases. This is unusual in the study of typical case Boolean CSPs, usually there exists a range for $\gamma$, close to the transition from SAT to UNSAT, in which all fast local search algorithms fail.

An ECk ensemble has also been investigated by a DPLL method, resulting in a lower bound for Easy-SAT~\cite{Kalapala:PTEC}. Hard-SAT/UNSAT phases exist about the SAT/UNSAT transition for a range of $\gamma$ in this ensemble. Although UCP proves a strong upper bound in the case of $\okSAT$, the method fails in ECk. Approximating the $\okSAT$ clauses by XOR clauses, which have more degrees of freedom but can be exactly analysed, is one alternative constructive proof method.

A rigorous upper bound for SAT may be determined by an annealed approximation~\cite{Knysh:AST}. Non-rigorous exact results for the SAT transition have also been developed through the cavity method~\cite{Raymond:PD}. These results demonstrate the existence of a sharp SAT/UNSAT threshold in agreement with analysis of complete solvers~\cite{Kalapala:PTEC,Knysh:AST}.

A parameter $\epsilon$ may be introduced to interpolate between standard ECk ($\epsilon=0$) and $\okSAT$ ($\epsilon=\frac{1}{2}$) ensembles. The $\eokSAT$ ensemble has been examined and it was demonstrated that for small $\epsilon$ behaviour with variation of $\gamma$ is similar to ECk~\cite{Raymond:PD}. As $\epsilon$ increases the range of $\gamma$ corresponding to Hard SAT/UNSAT behaviour about the transition decreases continuously to zero at a critical parameterisation $\epsilon^*<\frac{1}{2}$, so a range of $\eokSAT$ ensembles also behave similarly to $\okSAT$, without Hard phases. This approach is akin to methods used to understand the emergence of algorithmic hardness in 3SAT using mixtures of clause types~\cite{Monasson:TP,Achlioptas:RR}.

\subsection{Chapter outline and result summary}

Section \ref{1inkSAT.typicalcasealgorithm} defines the $\eokSAT$ ensemble studied and the dynamics of the simplified DPLL algorithm considered, as well as introducing relevant notation. Marginal transition probabilities within the ensemble are determined and a simplified statistical description developed sufficient to determine typical algorithmic properties.

In section \ref{1inkSAT.UB} the upper bound $\gamma_{\rm
UCP}(\epsilon)$ is demonstrated proving an Easy-UNSAT phase for a range of connectivity ($\gamma> \gamma_{UCP}(\epsilon)$) for the $\eokSAT$ ensemble. This is demonstrated by showing super-critical UCP.

In section \ref{1inkSAT.LB} an exact lower bounds $\gamma_{\rm
H}(\epsilon)$ for the connectivity below, which an Easy SAT phase exists is demonstrated. The lower bound is determined by UCP analysed in a subcritical regime combined with several heuristic (H) rules. If $\gamma < \gamma_{\rm SCH}(\epsilon)$, by fixing variables according to a heuristic, short clause (SCH) being the optimal choice amongst those investigated, one can find a solution with finite probability on any run.

In section \ref{1inkSAT.ESU} the upper and lower bounds for $\eotSAT$ are shown to coincide on the interval $\epsilon \in [0.2726, 1/2]$. This fact indicates that there exists a range of $\epsilon$ for which typical instances of the ensemble are Easy for all $\gamma$.

The algorithmic results for $k=3$ are placed in the context of statistical mechanics analysis obtained by the cavity method in section \ref{1inkSAT.PD}. With large $\epsilon$ the position of the boundary is reproduce by Belief Propagation (BP), and the phase space is shown to be a simple one even very close to the transition. However, it is possible to identify a region in which the state space near the transition is described correctly only with Replica Symmetry Breaking (RSB) effects and yet the DPLL continues to work efficiently. The ability of algorithms to work beyond the RSB threshold in typical case has been established by numerical studies of heuristic algorithms, such as Walk-SAT~\cite{Selman:LSS}, but the analytical proof is an exceptional case.

The case of $k=4$ is examined in section \ref{1inkSAT.k4}. Some features are repeated for these ensembles including the tightness of the upper and lower bounds over a wider range of $\epsilon$. It is demonstrated that as $\epsilon$ decreases there are discontinuous transitions in the minimum amount of variables, which must be revealed to solve the problem. This can be understood by considering properties of UCP; it is argued that $k=3$ is the exception.

\section{Typical case analysis}
\label{1inkSAT.typicalcasealgorithm}
\subsection{The $\eokSAT$ ensemble}

The $\eokSAT$ ensemble describes a problem of $N$ variables, each appearing in expectation $\gamma$ times in clauses. Each clause is a function of $k$ literals, literals are negative with probability $\epsilon$ and positive otherwise. The clause is in all cases $\okSAT$, that one literal is true and all others are false. Each literal is determined by a variable sampled uniformly from the set of $N$ Boolean variables.

Defining an $i$-clause as a clause containing $i$ literals let $C_i(X)$ be the number of clauses containing $i$ literals, and $C_{i j}(X)$ be the number of $i$-clauses with $j$ negative literals, after $X$ variable decimations. The $\eokSAT$ ensemble is defined before the decimation process begins as
\begin{equation}
 C_{i j}(0) = \frac{N k}{\gamma} \delta_{i k} \frac{i\factorial}{(i-j)\factorial j \factorial} \epsilon^j (1-\epsilon)^{i-j}\label{eq:1inkSAT.IC}\;.
\end{equation}
With this definition the special cases $\epsilon=\frac{1}{2}$ corresponds to symmetric $\okSAT$, and $\epsilon=0$ corresponding to the ECk (also called positive $\okSAT$).

\subsection{Heuristic driven unit clause propagation dynamics}

\begin{figure}[htb]
\includegraphics[width=\linewidth]{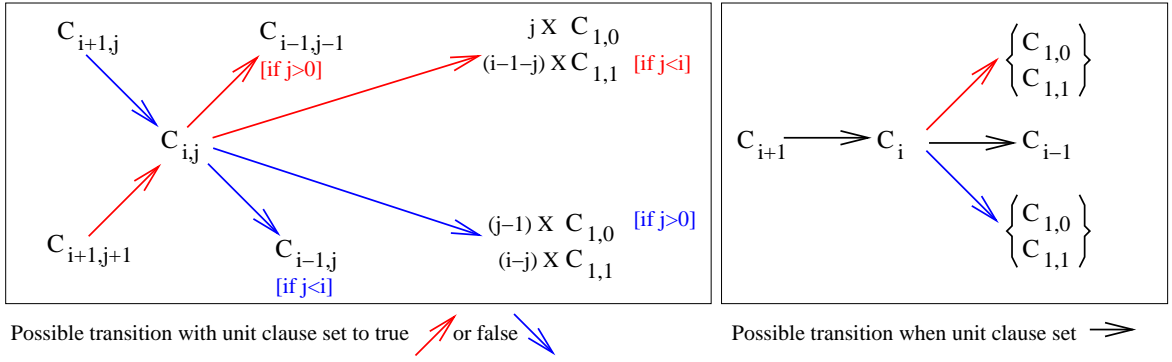}
\caption[Transitions amongst clause populations in response to unit clause resolution.]{\label{fig:1inkSAT.ClauseDynamics} The population of clauses containing i variables and j negations changes as unit clauses are resolved. The transitions are from $i$-clauses to $(i\!-\!1)$-clauses, or else where clauses are covered $i$-clauses can become $(i\!-\!1) \times$ unit clauses. Left figure: Resolving a negative unit clause results in the downward set of transitions amongst clauses, and conversely for resolving a positive unit clause. Right figure: At a statistical level, and for simple heuristic rules on resolving clauses, the distribution of negations within clauses of size 2 and greater depends only on $\epsilon$.}
\end{figure}

All $i$-clauses with $i>1$ allow ambiguity in the value of the literals. Unit clauses are the exception, whether positive $C_{1,1}$, or negative $C_{1,0}$, a unique assignment is implied to the literals, and hence to some variable in the ensemble. In an ensemble containing unit clauses the associated variables can be immediately decimated to leave a reduced problem. In resolving a unit clause variables are decimated from other clauses, and some larger clauses may be reduced to unit clauses, it is possible therefore to have a branching process, this is UCP.

In the presence of only clauses where $i>1$, there is local ambiguity in the value a variable can take. DPLL determines a SAT assignment by {\em branching}, which involves guessing by heuristic the value of a variable, proposing and resolving a unit clause, and if a contradiction does not arise proceeding with the search based on the reduced problem.

The algorithm employed to determine SAT is called Heuristic driven Unit Clause Propagation (HUCP). The dynamics of the back-tracking step, which is a necessary reexamination of heuristic inference when a contradiction is encountered, is not essential in determining results of this chapter. HUCP is a decimation procedures, so that the population of different clause types departs from the initial condition (\ref{eq:1inkSAT.IC}) as the algorithm is run.

It is useful to consider the algorithm as partitioned into rounds that consist of a free step, followed by implied steps (UCP). First variables are assigned by a heuristic, this causes a change to the clause structure as described shortly, and may include creation of unit clauses. Resolution of these unit clauses can then be done recursively. With each resolution of a unit clause further changes occur to the problem structure. A branching process describes UCP, so that decimation of one variable by heuristic rule may result in a substantial UCP cascade. After this round has finished, with no unit clauses remaining, a new round begins with decimation by heuristic.

The dynamics of clause populations in assigning variables by HUCP involves the transfer of mass from larger clauses $C_i$ to smaller clauses, either to $C_{i\!-\!1}$ or to unit clauses $C_1$. The decimation of a variable (as either $\True$ or $\False$) leads to different transitions amongst these populations. The set of transitions is shown in figure~\ref{fig:1inkSAT.ClauseDynamics}.

The following two heuristics are used in this chapter~\cite{Deroulers:CUU}
\begin{itemize}
\item RH$[p]$: A random unassigned variable is selected and assigned to value $\True$ with probability $p$.
   \item SCH: Select at random a literal within a 2-clause, if any exist. Set this to $\True$, and the other literal in the clause to $\False$. If no 2-clause exists apply RH$[p]$.
\end{itemize}
RH$[p]$ makes few assumptions, but the possibility exists to optimise the algorithm with respect to $p$. SCH reduces $C_2$, minimising the number of clauses in the reduced problem, by comparison with the number of variables set.

Variables decimated by RH$[p]$ or SCH, are determined independently of the frequency with which they appear as positive or negative literals, given $\epsilon$. UCP can be applied with a similar independence assumed. Therefore in a round the distribution of literals in all clauses $i>1$ is unchanged. The set of heuristic rules is chosen so that given the initial condition (\ref{eq:1inkSAT.IC}) the identity
\begin{equation}
C_{i j}(X) = C_i(X) \frac{i\factorial}{(i-j) \factorial j\factorial} \epsilon^j (1-\epsilon)^{i-j}
\label{eq:1inkSAT.Cij}\;,
\end{equation}
holds at the level of expected values for the clause populations. The parameter $X$ denotes algorithm time (the number of variables set by HUCP). The dynamics of $C_i$ as variables are decimated determines, in expectation, those for the sub-populations $C_{i j}$. Although it seems likely an optimal heuristic would involve a distinction in $C_{i j}$, these cases are avoided.

Variables are selected for decimation uniformly at random for $RH[p]$ and at random subject to their multiplicity within two clauses in the case of $SCH$. The distribution of variables within clauses are conditionally independent given some shared mean connectivity, and the unit clauses created are therefore uncorrelated with these heuristic rules. The reduced instance is thus assumed to be described by a typical sample from an ensemble characterised by $N-X-1$ variables, the new adjusted set of clause populations $\{ C_{i} \}$ and $\epsilon$. Clause populations and $X$ are sufficient to determine the algorithmic properties of HUCP for the ensemble.

The concentration of clause populations is an important feature assumed~\cite{Deroulers:CUU}. In a sub-critical round the populations of clause types change by small random amounts, an accumulation of these processes is assumed to concentrate on the mean, so that at leading order in $N$ all clauses with at least $2$ variables are described by their mean quantities. It is therefore sufficient to consider mean quantities in determining subcritical branching processes, and subcritical branching processes will be shown as sufficient to determine the mean values. The distribution of negations need not be monitored given (\ref{eq:1inkSAT.Cij}).

\subsection{Sub-critical round dynamics}

In expectation, the number of variables fixed throughout a round of HUCP is described by a transition matrix (${\cal M}(X)$) depending on the clause populations $\left\lbrace C_i(X)\right\rbrace$. The unit clauses generated by heuristic go on to generate other unit clauses and so forth, this can be described by a geometric series in ${\cal M}(X)$. Calling $\boldsymbol{p} = (p_T, p_F)$ the expected numbers of variables fixed to $(\True, \False)$ by heuristics, $\boldsymbol{m} = (m_T, m_F)$ the number of variables set to $(\True, \False)$ in a round, the following relation applies
\begin{equation}
 \vm = \vp + {\cal M}(X)\vp + {\cal M}^2(X)\vp + \cdots = (I-{\cal M}(X))^{-1} \vp \label{geometric_series}\;.
\end{equation}
The matrix inverse description is consistent if the round is subcritical, thus the description is restricted to cases where the modulus of the largest eigenvalue of ${\cal M}$ is smaller than $1$. Since $C_i(X)$ are $O(N)$ these are unchanged during any finite round (to $O(\frac{1}{N})$), so that ${\cal M}(X)$ remains invariant at leading order during a round.

The transition matrix has two components. A first contribution comes from clauses in the population $C_i$, which may be reduced to $i\!-\!1$ unit-clauses if the set variable is present in a clause and the literal is set to $\True$ -- all other literals are then implied to be $\False$. A second contribution comes from 2-clauses (if any) where setting any literal to false implies the other literal is true. These two processes are distinguishable in figure~\ref{fig:1inkSAT.ClauseDynamics}. The probability a unit clause is positive or negative is determined only by $\epsilon$ and $\vp$. Since there are $C_i(x)$ clauses of length $i$, and $N-X$ variables in the reduced problem, these two terms are combined in the expression to give
\begin{equation}
 {\cal M}(X) \!= \!\frac{1}{N-X}\! \left[\!\sum_{i=2}^k i (i-1)C_i(X)
\begin{pmatrix}
\epsilon (1- \epsilon)\! & \epsilon^2\! \\
(1-\epsilon)^2 \! & \epsilon (1-\! \epsilon)\!
\end{pmatrix}
\!+ \!C_2(X)
\begin{pmatrix}
\epsilon (1- \epsilon) \!&\! (1-\epsilon)^2 \\
\epsilon^2 \! & \!\epsilon (1- \epsilon)\!
\end{pmatrix}\right] \label{eq:1inkSAT.transitionmatrix}\;.
\end{equation}
Rounds on a large graph are a simple uncorrelated process, governed by the spectrum of a $2 \times 2$ transition matrix. If all the eigenvalues have $|\lambda_i| < 1$, the process is \emph{subcritical}: the typical size of the rounds is $\sim
1/\min_i (1-|\lambda_i|)$, and their average size concentrates. Conversely, if during the algorithm $1-|\lambda_i|\rightarrow 0$, the percolation threshold for the UCP branching process is reached. The branching process can then only become subcritical once some macroscopic change occurs to the transition matrix (\ref{eq:1inkSAT.transitionmatrix}), or once the loopy structure of the graph emerges to curtail the exponential expansion of the branching process. Both of these processes only occur once the branching process has reached a finite fraction of a graph even in the large $N$ limit.

During a super-critical round the number of unit clauses grows to an extensive value, measurable in $N$. Amongst such a population there is certain to be a contradicting pair, and so the branch searched must be a.s. UNSAT. If all rounds are subcritical then the maximum number of unit clauses present at any time in a round is finite, and the probability of a contradiction is $O(1/N)$. Over the course of an algorithm the probability of creating a unit clauses is a finite fraction, however, in this scenario it can be assumed a random restart will be independent, and so by making only a few random restarts the probability of contradiction occurring is reduced to zero so that the algorithm will a.s. work in linear time.


\section{Unit clause bounds to the SAT/UNSAT transition}
\subsection{SAT upper bound}
\label{1inkSAT.UB}

If, for a variable $i$ selected randomly from the instance at algorithm time zero, both the rounds initiated by setting the literal to $\True$ and $\False$ percolate, there is a finite probability that they result in a certificate of contradiction. Thus the upper bound for the SAT/UNSAT transition comes from the requirement that the rounds are almost entering this regime of criticality. At this point $C_{i<k}=0$ and $C_k= N \gamma/k$ (\ref{eq:1inkSAT.IC}), so that
\begin{equation}
 {\cal M}(0) = 2 \frac{\gamma}{k} \frac{k(k-1)}{2} \begin{pmatrix} \epsilon (1- \epsilon) & \epsilon^2 \\ (1-\epsilon)^2 & \epsilon (1- \epsilon)
\end{pmatrix}\;.
\end{equation}
From this it is clear that a random instance is a.s.~(randomised linear time) provable to be unsatisfiable for $\gamma$ larger than the percolation threshold
\begin{equation}
 \gamma_{UCP}(\epsilon) = \frac{k}{4 \frac{k(k-1)}{2}
\epsilon (1-\epsilon)} \label{eq:1inkSAT.gammaUC}\;.
\end{equation}
Randomised linear/polynomial time complexity is to say that the algorithm runs to completion within linear/polynomial time if a source of random numbers is available. The random numbers are important to guarantee certain assumptions of unbiased selection in the branching step, but in practice bias in standard pseudo random number generators is not crucial.

\subsection{UNSAT lower bound}
\label{1inkSAT.LB}

The differential equations studied here are a generalisation of those found by Kalapala and Moore~\cite{Kalapala:PTEC} for Exact Cover. A heuristic rule for clause or variable selection determines the nature of the free step in our rounds. The two rules examined herein are Random Heuristic (RH$[p]$) and Short Clause Heuristic (SCH).

If at some time $X$, $\boldsymbol{p}$ variables are set to $(\True,\False)$ in expectation, then the expected change in $C_{i}$ may be calculated. If an $(i\!+\!1)$-clause contains the variable just fixed, it is with probabilities $\boldsymbol{\epsilon} = ( \epsilon, 1-\epsilon )$ that the clause is reduced to an $(i\!-\!1)$-clause (rather than to unit clauses). Similarly an $i$-clause is reduced to a set of $(i\!-\!1)$-clause with probabilities $\boldsymbol{1} = (1, 1 )$ (with certainty regardless of the literal type). Still in expectation for cases where $1<i<k$:
\begin{equation}
C_i(X+\boldsymbol{1} \cdot \boldsymbol{p}) \!= \!\left( C_{i}(X) \!-\! \delta_{SCH} \delta_{i,2} (1-\delta_{C_2(X),0})\right) \left( 1 \!- \frac{i (\boldsymbol{1} \cdot \boldsymbol{p})}{N-X} \, \right) \!+\! \frac{i+1}{N-X}(\boldsymbol{\epsilon} \cdot \boldsymbol{p})\, C_{i+1}(X) \label{Cidyn}\;,
\end{equation}
where $\delta_{SCH}=0$ or $1$ respectively in the cases of RH$[p]$ and SCH. The two heuristics are also distinguished in that to initiate the rounds for RH$[p]$, $\boldsymbol{p}_{\textrm{RH}[p]} = (p,1-p)$, while for SCH $\boldsymbol{p}_{\textrm{SCH}} = (1,1)$. Remembering the vector $p$ describes the number of variables set in expectation not the probability distribution, the SCH value can be understood since setting one random literal in a two clause implies setting the other to the opposite value, thus setting variables to either $\pm 1$ is equally likely in expectation and decimation always occurs in pairs.

A round can be described by incorporating the variables set in the forced steps. Suppose that during a subcritical round $\vm$ variables are decimated in expectation (including the free step). To leading order in $N-X$ the variation is
\begin{equation}
C_i(X+\boldsymbol{1} \!\cdot\! \boldsymbol{m}) = \left( C_{i}(X) \!-\! \delta_{SCH} \delta_{i,2} (1-\delta_{C_2(X),0})\right) \left( 1\! -\! \frac{i (\boldsymbol{1} \cdot \boldsymbol{m})}{N-X} \, \right) \! +\! \frac{i+1}{N-X} \UCPtransitioner(X/N,\epsilon) C_{i+1}(X) \label{eq:1inkSAT.Ciround}\;, \end{equation}
with
\begin{equation}
\UCPtransitioner(X/N,\epsilon) = (\boldsymbol{\epsilon} \cdot \boldsymbol{m}) \label{UCPcalF}\;,
\end{equation}
where $\boldsymbol{m}$ is a function of $X/N$.

A final simplification in the clause dynamics is to summarise the behavior by continuous variables $x=X/N$ and $c_i=C_i/N$, which is justified by Wormald's Theorem~\cite{Wormald:DE}. In the hypothesis of sub-criticality, $\boldsymbol{m}/N$ is infinitesimal, and a differential equation description is attained
\begin{equation}
\frac{\rmd}{\rmd x} c_i(x) = - \delta_{SCH} \delta_{i,2} \theta(c_2(x)) \< \frac{1}{\boldsymbol{1} \cdot \boldsymbol{m}} \>+ \frac{1}{1-x} \left(- i c_i(x) + (i+1) \< \frac{\boldsymbol{\epsilon} \cdot \boldsymbol{m}} {\boldsymbol{1} \cdot \boldsymbol{m}} \>
c_{i+1}(x) \right) \label{eq:1inkSAT.ciroundSC}\;,
\end{equation}
where $\theta$ is the step function. The expression corresponds to the SCH, in the case of RH$[p]$ the first term is absent. For both RH$[p]$ and SCH rules, the equation for $c_k(x)$ gives
\begin{equation}
c_k(x) = \frac{\gamma}{k} (1-x)^k \label{eq:1inkSAT.c_k}\;.
\end{equation}
Instead, for $c_{i|i<k}(x)$ the equation is non-linear. In this way the terms $\<m_T\>$ and $\<m_F\>$ are given by the combination of equations (\ref{geometric_series}) (\ref{eq:1inkSAT.transitionmatrix})(\ref{eq:1inkSAT.c_k}), and thus depend on the unknown function $c_i(x)$ (besides, of course, $x$, $\epsilon$ and $\gamma$). Using this expression within (\ref{eq:1inkSAT.ciroundSC}) allows $c_i(x)$ to be determined by numerical integration, and thence $\lambda_{\textrm{max}}(x)$.

Since the aim is to prevent contradictions arising, a greedy choice for the parameter $p$ in RH$[p]$ creating the smallest rounds at a given time seems reasonable. This depends on the largest eigenvalues of ${\cal M}$, which varies between $\boldsymbol{\epsilon}$, when $C_k$ dominates the branching process at early times, and $\frac{1}{2}\vones$ ($\vones$, a vector of ones), when $C_2$ dominates the branching process at later algorithm times. However, minimising the probability of a contradiction locally (in $X$) can cause higher probabilities of contradiction at later times. In particular contradictions will almost surely occur if super-critical branching occurs at later times, and rules should be chosen to mitigate this most important source of contradictions.

If super-criticality occurs at $x=0$ the heuristic rule employed is statistically insignificant, since it only decimates $O(1)$ variables, whereas if the maxima in round size occurs at $x>0$ the heuristic rule plays a role and the best choice is $p=1$ (satisfying as many clauses as possible), which curtails the growth of $C_2$. This reduces the amount of super-critically in ECk and small $\epsilon$ ensembles.

\section{Results for $\eotSAT$}
\subsection{Upper and lower bounds by numerical integration}
\begin{figure}[htb]
\begin{center}
\includegraphics[width=\linewidth]{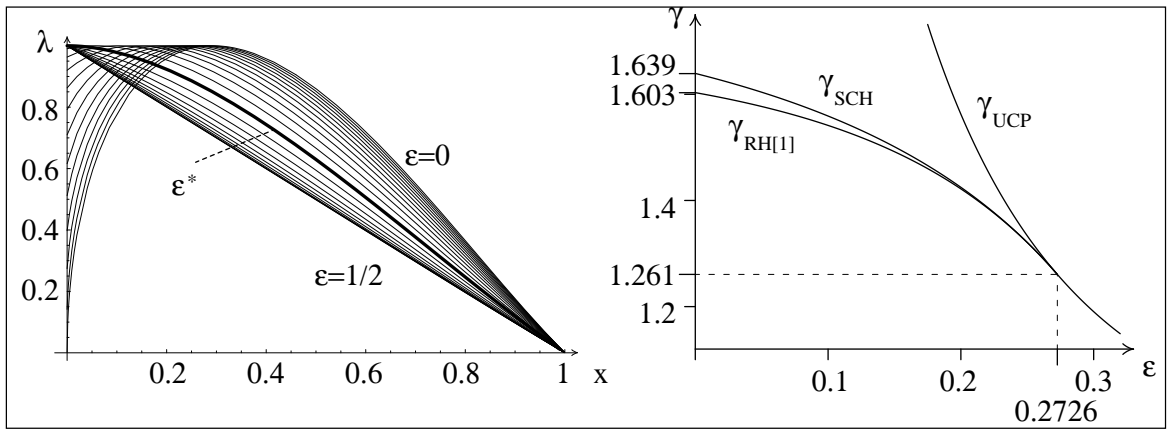}
\caption[Algorithmic criticality in $\eotSAT$.] {\label{fig:1inkSAT.lambdas} Results for $\eotSAT$. Left figure: Profiles of $\lambda(x)$ along decimation time $x$, for RH$[1]$, at various $\epsilon$ and at the corresponding critical value of $\gamma$. In all the cases, the functions $\lambda(x)$ are concave (up to the limit value $\epsilon=1/2$, where $\lambda(x)=1-x$). For $\epsilon$ larger or smaller than the tri-critical value $\epsilon^* = 0.272633$, the maximum of $\lambda(x)$ is achieved respectively at $x=0$ or at $x>0$. Right figure: Critical curves $\gamma_{\rm SCH}(\epsilon)$ and $\gamma_{\textrm{RH$[1]$}}(\epsilon)$, obtained through SCH and RH[1] are shown along with the upper bound $\gamma_{UCP}$.}
\end{center}
\end{figure}

Figure~\ref{fig:1inkSAT.lambdas}~\footnote{Figure taken, with modifications, from a collaborative work~\cite{Raymond:PD}.} shows the results for RH$[1]$ and SCH. The latter is always at least as good as the former, and gives a lower bound of $\gamma_{\rm SCH} = 1.6393$, while RH$[1]$ attains $\gamma_{\rm RH[1]} = 1.6031$, for the case $\epsilon=0$ (EC3). Kalapala and Moore calculated these quantities for EC3, with compatible results for the $k=3$ case (up to perhaps a misprint exchanging RH$[p]$ with RH$[1-p]$).

The numerical integration is a somewhat cumbersome process in spite of the smoothness for the range of parameters and rules chosen. In Appendix~\ref{app:AnalyticBounds} a method of bounding the integration curve, allowing analytical estimations of the maxima, is constructed by which numerical integration can be checked or directed. The analytic bounds are not tight to the result by numerical integration except above a critical value of $\epsilon$, but this result can be used, without requiring a numerical integration, to demonstrate the tri-critical point beyond which UCP cannot solve quickly samples near the SAT/UNSAT transition.

\subsection{Exact SAT/UNSAT thresholds}
\label{1inkSAT.ESU}

This section proves the coincidence of the curves $\gamma_{\rm
SCH}(\epsilon)$ and $\gamma_{UCP}(\epsilon)$ for $\epsilon>0.2726$ when $k=3$. It was shown in the previous section that whenever the rounds remain subcritical Easy-SAT behaviour is realised. The criteria for the rounds to be subcritical at $x=0$ is precisely $\gamma<\gamma_{UCP}(\epsilon)$. It thus suffices to show that the maximum (over the decimation time $x$) of the $\max_i
|\lambda_i(x)|$, is attained for $x=0$. This is indeed what happens in the interval $\epsilon \in [0.2726,1/2]$.

Building on the previous section we will see that, for $\eotSAT$ and our heuristics, $\lambda(x)$ is a concave function. So, the interval on which $\gamma_{UCP}(\epsilon)$ (the upper bound) and $\gamma_{\rm SCH}(\epsilon)$ (the lower bounds) coincide is the one in which
\begin{equation}
 \left. \frac{\rmd \lambda(x; \epsilon, \gamma)} {\rmd x} \right|_{x=0,\gamma=\gamma_{UCP}(\epsilon)} \leq 0 \label{localmaxx}\;,
\end{equation}
the endpoint being determined by the corresponding equality.

It is possible to calculate the characteristic polynomial (and differentiate with respect to $x$). However, the expressions thereby found can only be evaluated exactly at $x=0$. At this value $\{c_i(x)\}$, and their derivatives, are known exactly in terms of the initial conditions and $\boldsymbol{m}$. Restricting attention to the nearly super-critical case (\ref{eq:1inkSAT.gammaUC}), a further simplification is in the eigenvectors of ${\cal M}$, the principal eigenvalue becomes $\boldsymbol{\epsilon}$ and dominates the other process. The criticality of the branching process then becomes independent of $\boldsymbol{p}$, since there must be a component along $\boldsymbol{\epsilon}$, from (\ref{eq:1inkSAT.Ciround})
\begin{equation}
 F(0,\epsilon) = \frac{\boldsymbol{\epsilon} \cdot \boldsymbol{\epsilon}} {\boldsymbol{1} \cdot \boldsymbol{\epsilon}} =1 - 2 \epsilon (1-\epsilon) \label{1inkSAT.F0e}\;.
\end{equation}
Finally, the condition (\ref{localmaxx}) becomes
\begin{equation}
 \left( 1 + \frac{1}{4} \left(\frac{\epsilon}{1-\epsilon} + \frac{1-\epsilon}{\epsilon}\right)^2 \right) (1 - 2 \epsilon (1-\epsilon)) - 2 \leq 0 \label{criteria}\;.
\end{equation}
So that after the change of variable $y=2 \epsilon (1-\epsilon)$, one gets the equation for the endpoint of the interval
\begin{equation}
2 y^3 - 2 y^2 + 3 y - 1 = 0\;,
\end{equation}
whose only real solutions are $\epsilon = 0.272633, 1-0.272633$, the appropriate solution being in the interval $[0,0.5]$.

To show that the properties at $x=0$ are sufficient to determine $\gamma_{\rm SCH}$ it is necessary to show that whenever criteria (\ref{criteria}) is met, and $\lambda(0)<1$, the algorithm is subcritical at all $x$. An analytic proof, not reliant on numerical integration (as in figure~\ref{fig:1inkSAT.lambdas}), is to find a function ${\hat \lambda}(x)$ such that
\begin{equation}
\lambda(x) \leq {\hat \lambda}(x) \leq \lambda(0) \qquad \forall x \label{ubcond}\;,
\end{equation}
establishing the bound. Such an upper bound is also motivated as a variational method in Appendix~\ref{app:AnalyticBounds}.

Since $\lambda(x)$ is a monotonically increasing function of $c_2(x)$ (\ref{eq:1inkSAT.transitionmatrix}), an upper bound ${\hat c}_2(x)>c_2(x)$ implies an upper bound in $\lambda(x)$, which we take to be ${\hat \lambda}(x)$. The bound function ${\hat c}_2$ is defined by replacing the complicated function $\UCPtransitioner(x,\epsilon)$ by the constant value $\UCPtransitioner(0,\epsilon)$ in the expression (\ref{eq:1inkSAT.ciroundSC}), which are then exactly solvable for all $x$ as
\begin{equation}
{\hat c}_2(x) = \gamma x (1-x)^2 \UCPtransitioner(0,\epsilon) = \gamma x (1-x)^2 \big( 1-2 \epsilon (1-\epsilon) \big) \label{eq:1inkSAT.upperbound}\;.
\end{equation}
For RH$[\frac{1}{2}]$ and certain other heuristics this approximation can be shown to produce an upper bound for $c_2(x)$, and yet be exact at $x=0$ in both absolute value and derivative.

This then allows an exact expression for $\frac{\rmd \hat{\lambda}(x)}{\rmd x}$ to be written in terms of $x$. Though the dependency on $x$ remains complicated
\begin{equation}
\frac{\rmd {\hat \lambda}(x)}{\rmd x} < 0 \qquad
\hbox{whenever}\qquad {\hat \lambda}(0) < 1 \;,
\end{equation}
exactly in the same interval of $\epsilon$ in which (\ref{criteria}) holds, by examination of the derivatives of the eigenvalues of the stability matrix. These fact proves that the local analysis at $x=0$ is sufficient for the purpose of identifying the maximum over $x$ of $\lambda(x)$ in this interval.


\subsection{A comparison of algorithmic and statistical physics results}
\label{1inkSAT.PD}
It is possible to place the problem of satisfiability in a statistical physics framework, interpreting results derived for the $\eotSAT$ ensemble~\cite{Raymond:PD}. By representation of the clauses as energetic interactions on spin states $\{\True,\False\}\rightarrow\{-1,1\}$, with satisfied clauses contributing zero energy, and unsatisfied clauses energetically penalised, a standard Hamiltonian can be formulated. Proving SAT is then equivalent to evaluating the ground state energy and determining if this is zero.

\begin{figure}[!htb]
\begin{center}
\setlength{\unitlength}{50pt}
\begin{picture}(5,3.8)
\put(-0.1,-0.03){\includegraphics[scale=1, bb=0 0 270 250]
{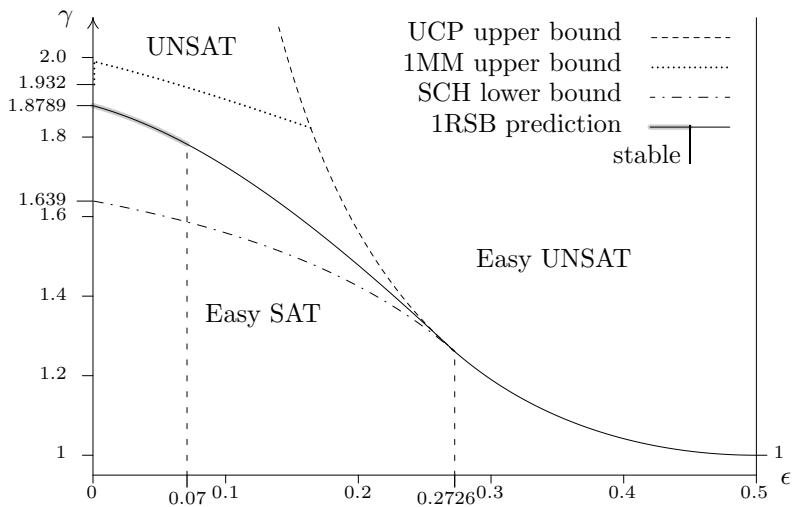}} \put(0.05,0){${\scriptstyle
0}$} \put(1,0){${\scriptstyle 0.1}$} \put(2,0){${\scriptstyle
0.2}$} \put(3,0){${\scriptstyle 0.3}$} \put(4,0){${\scriptstyle
0.4}$} \put(5,0){${\scriptstyle 0.5}$}
\put(2.55,-0.05){${\scriptstyle 0.2726}$}
\put(0.67,-0.05){${\scriptstyle 0.07}$} \put(5.3,0.1){$\epsilon$}
\put(-0.05,3.6){\makebox[0pt][r]{$\gamma$}}
\put(-0.1,0.3){\makebox[0pt][r]{${\scriptstyle 1 }$}}
\put(-0.1,0.9){\makebox[0pt][r]{${\scriptstyle 1.2}$}}
\put(-0.1,1.5){\makebox[0pt][r]{${\scriptstyle 1.4}$}}
\put(-0.1,2.1){\makebox[0pt][r]{${\scriptstyle 1.6}$}}
\put(-0.1,2.7){\makebox[0pt][r]{${\scriptstyle 1.8}$}}
\put(-0.1,3.3){\makebox[0pt][r]{${\scriptstyle 2.0}$}}
\put(5.25,0.3){\makebox[0pt][l]{${\scriptstyle 1 }$}}
\put(-0.1,2.217){\makebox[0pt][r]{${\scriptstyle 1.639}$}}
\put(-0.1,2.925){\makebox[0pt][r]{${\scriptstyle 1.8789}$}}
\put(-0.1,3.105){\makebox[0pt][r]{${\scriptstyle 1.932}$}}
\put(4.1,3.466){\makebox[0pt][r]{UCP~upper bound}}
\put(4.1,3.233){\makebox[0pt][r]
{1MM~upper bound}}
\put(4.1,3.0){\makebox[0pt][r]{SCH~lower bound}}
\put(4.1,2.766){\makebox[0pt][r]{1RSB prediction}}
\put(4.62,2.533){\makebox[0pt][r]{stable\ \rule{0.25pt}{14pt}}}
\put(3.0,1.75){Easy UNSAT} \put(0.95,1.33){Easy SAT}
\put(0.5,3.35){UNSAT}
\end{picture}
\caption[Phase diagram for
$\eotSAT$.]{\label{fig:1inkSAT.fig_PhD_bounds_hum2} The phase diagram of $\eotSAT$ problem is shown. The parameters $\epsilon$ and $\gamma$ describe the probability of negations and the average variable connectivity. For $\epsilon > 0.2726$, the threshold is rigorously $\gamma^*(\epsilon) = 1/\big(4 \epsilon(1-\epsilon) \big)$ (drawn as a solid line), since the UCP upper bound 
and SCH lower bound 
coincide in that region. For $\epsilon < 0.2726$, the dot-dashed, dashed and dotted line denote respectively the SCH lower bound, the UCP upper bound, and an alternative algorithmic bound based on the first moments method (annealed approximation)~\cite{Knysh:AST}, which improves on UCP at small $\epsilon$. The solid line is the one-step replica-symmetry-breaking (1RSB) prediction for the SAT/UNSAT threshold. For $0 \le \epsilon<0.07$ the 1RSB result is stable (gray shading) and so the threshold is likely to be exact. For $0.07<\epsilon<0.2726$ the 1RSB result is unstable, and expected to be an upper bound).}
\end{center}
\end{figure}

A benefit of a statistical physics analysis is that if the ground state is degenerate then the number of such states can be calculated, and also the correlations in the phase space. The consequences of an analysis up to the first level of
replica symmetry breaking is shown in figure
\ref{fig:1inkSAT.fig_PhD_bounds_hum2}~\footnote{Figure taken from a collaborative work~\cite{Raymond:PD}.}.
In a range of the state space $\epsilon=(0.33,0.5]$ the replica symmetric solution is correct everywhere with a transition from a paramagnetic/liquid (Easy SAT) state to an (Easy UNSAT) state. Such a prediction is consistent with the HUCP result. Between $\epsilon\sim(0.2277,0.33)$ BP equations predict a solution coincident with the transition, but the equations are themselves unstable near the transition point, indicating a failure of the RS assumption. This is surprising since HUCP, a local search method, can reach the transition everywhere above $\epsilon=0.2736$, it is strange for these two local search methods not to coincide, or for BP not to be a stronger method. Between $\epsilon=(0.07,0.33)$ a stable description is provided by 1-step RSB solutions of the free energy. This result indicates there is clustering in the state space, an effect that would indicate ergodicity breaking in many search dynamics. The analytical result that UCP is successful in a phase space described by 1RSB is unexpected, but indicates that the dynamical transition in this model does not coincide with the emergence of RSB in the thermodynamic solution.

The UCP upper bound diverges for small $\epsilon$, but alternative constructive and non-constructive upper bounds can be formulated, such as an annealed approximation. The Hard phase about the transition is bounded by HUCP, algorithmic difficulty is also observed in other algorithms for finite systems in this range of parameters.

\section{Algorithmic bounds for $\eofSAT$}
\label{1inkSAT.k4} The proof of the exact bound for the case $k=3$ is indirectly reliant on the concavity of the curves for all $\epsilon$ (figure~\ref{fig:1inkSAT.lambdas}). For $\eokSAT$ with $k>3$ the curves are not convex for any $\epsilon>0$ and the gradient in the principal eigenvector of the transition matrix, at $x=0$ is negative everywhere that $\epsilon>0$. In spite of this criticality in $\lambda$ appears at $x>0$ for sufficiently small $\epsilon$. On first inspection a rigorous bound appears more challenging to obtain in these cases.

For $k=4$ numerical integration is used to solve the lower bound dynamics and is found to coincide with the upper bound on a larger range of $\epsilon$. This is not surprising when one considers that longer clauses imply tighter constraints leading to a greater number of implications near the start of the algorithm. A second observation is that the size of rounds does not decrease monotonically throughout this regime. Instead the curve of $\lambda$ against algorithm time is bimodal at small $\epsilon$, with a maximum at algorithm time $x=0$ and a second maxima elsewhere. As $\epsilon$ decreases the latter maxima grows to dominate the branching process so that in spite of decreasing round sizes in the initial stages of the algorithm UCP can later become supercritical.
\begin{figure}[htb]
 \centering{
\includegraphics[width=0.8\linewidth]{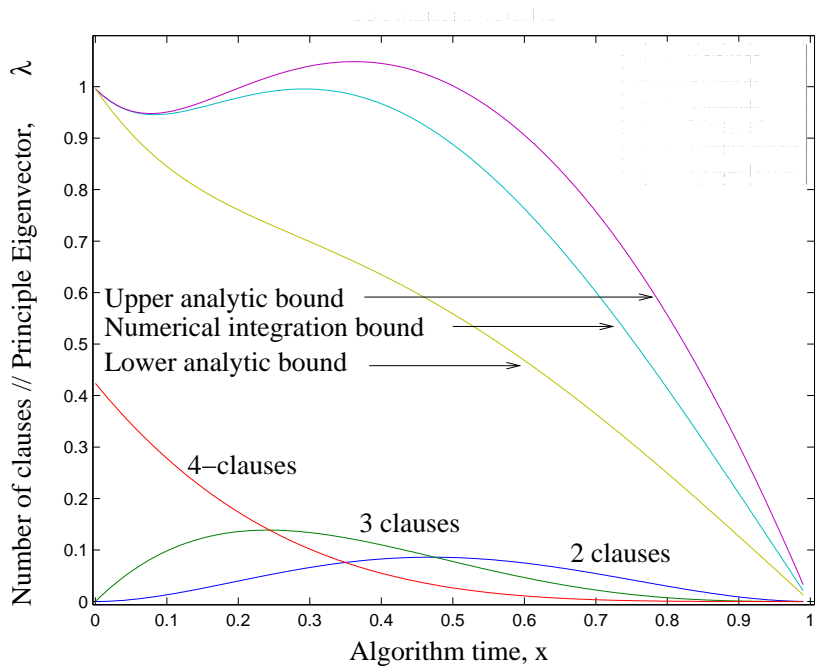}
\caption[Algorithm criticality in $\eofSAT$.]
{\label{fig:1inkSAT.UCP_k4} Using RH[p] the critical point in $\epsilon^*= \argmin \{\gamma_{RH}(\epsilon) = \gamma_{UCP}(\epsilon)\} \lesssim 0.11$, for $k=4$. Numerical integration of the coupled equations shows that for some $k=4$ ensembles there are two algorithm times that represent local maxima in the strength of the branching process. Applying the analytic bounds (\ref{eq:1inkSAT.upperbound}) on the numerical integration process does not predict the critical point accurately, dynamics of RH[$0.5$] are shown in the diagram. Also shown are the clause populations dynamics arising in numerical integration at the critical point. For small $\epsilon$ the 3-clauses have a smaller effect on the size of rounds than do the 4-clauses or 2-clauses. However, 2-clauses are only generated dynamically through 3-clauses leading to the bimodal distribution which is characteristic of all ensembles with $k>3$.}
}
\end{figure}

This effect in $k=4$ and larger clause ensembles can be seen in figure~\ref{fig:1inkSAT.UCP_k4}. The total number of clauses is non-increasing, but in the first small fraction of algorithm time the number of 3-clauses created is proportional the number of 4-clauses decimated $O(x)$, whereas the number of two clauses is proportional to the number of 3-clauses decimated $O(x^2)$. Therefore the number of 2-clauses grows very slowly and is irrelevant to the early dynamics. The initial rounds are largest when decimating $4$-clauses, as these decrease and make way for $3$-clauses the rate of unit clause creation drops. At a later time it is possible that a statistically significant number of 2-clauses is created and begins to dominate the process for some $\epsilon$. There is a gap between these different dominating effects, which becomes wider as $k$ increases but at the same time is restricted to a narrower regime in $\epsilon$ closer to the Exact Cover ensemble.

The upper bound based on taking $\UCPtransitioner$ as a constant remains valid as well as the lower bound, both of, which are derived in Appendix~\ref{app:AnalyticBounds}. The upper bound is exact at $x=0$, but further away is giving a description of $c_2$ too far from the numerical integration result to be useful, as shown in figure~\ref{fig:1inkSAT.UCP_k4} - it is not possible to determine, other than by taking the limit in numerical integration, an exact value for the critical branching point ($\epsilon^*$) when $k>3$.

Unlike $k=3$ the determination of the critical branching point depends on the details of the free step heuristic. The nature of this transition in $k>3$ is fundamentally different to the transition in $k=3$, which is continuous in the order parameter (which may be taken as $x^*$, the algorithm time at which criticality occurs).


%% file: SPARSECDMA.tex

\section{Introduction}
\label{CDMA.Introduction}
\begin{figure}[htb]
\begin{center}
\includegraphics[width=0.6\linewidth]{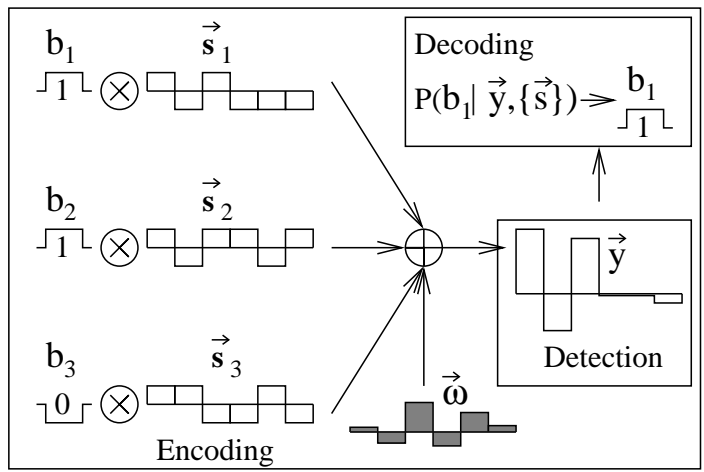}
\caption[CDMA on a linear vector channel, four user example.]
{\label{fig:CDMA.lvc} An instance of communication on the linear vector channel (\ref{eq:CDMA.lvc}) for a single bit interval is displayed. Three users communicate on a bandwidth of five time blocks (spreading factor 5/3). The signal is received and sources approximated through some probabilistic inference.}
\end{center}
\end{figure}

An apparently simple generalisation of the problem of the noisy single user channel is one in which there are several independent, or partly independent, sources communicated to a sink through some shared channel. This is the multi-access noisy channel model~\cite{Verdu:MD}.

A linear vector channel forms a tractable basis for understanding a variety of multi-access channel problems. A linear vector channel is defined as a system in which an input vector of $K$ components, is linearly transformed by an $M$ $\times$ $K$ channel transfer matrix and is additively degraded by noise~\cite{Takeda:SMA}. The channel describes communication on a bit interval, an interval in which each source transmits a single bit as a modulated real valued vector, with $M$ components. The vector signals combine linearly in the channel with some environmental noise, also represented as a real vector, and the detection problem is to identify the most probable values for the transmitted bits from this superposition of signals. An example process is shown in figure~\ref{fig:CDMA.lvc}.

The source information is represented by a set of $K$ bits
($\vb \in \left\lbrace -1,1\right\rbrace^K$), the spreading patterns for users by a code (in matrix form, $\ms=\left\lbrace \vs_{k} \right\rbrace_{k=1\ldots K}$) and the channel noise is $\vomega$; vector notation denotes the spreading on the vector channel. If the transmitted signals are synchronised 
then the received signal on a bit interval is
\begin{equation}
\vy = \sum_{k=1}^K \vs_k b_k + \vomega\label{eq:CDMA.lvc}\;.
\end{equation}
The detection problem is then to infer the source bits from the received signal, based on exact knowledge of the spreading patterns, a noise model and prior assumptions on the source bits.

The linear vector channel is appropriate in the multi user detection problem of wireless communications~\cite{Verdu:MD}, in which a set of users communicate to a single base station over some discretised bandwidth. Each component in the vector can be considered as a chip -- an independent section of the bandwidth such as a time-frequency block. For inference purposes in this thesis a chips are synonymous with a factors (in a factor graph).

Utilising the vector structure of the bandwidth offers a number of practical and theoretical advantages in terms of detection and robustness~\cite{Ipatov:SS}, over communication at equivalent power on a bandwidth without a vector structure. Code Division Multiple Access (CDMA) provides a method of dividing the bandwidth between users so as to achieve a low Bit Error Rate (BER) in communication and maintain some advantageous features of spread spectrum transmission. This is by contrast with Time or Frequency Division Multiple Access (TDMA/FDMA) models, which effectively reduce the transmission/detection problem to a bank of orthogonal scalar channels. In TDMA/FDMA no two users have overlapping spreading patterns, each user transmitting on a separate chip. In CDMA there is overlap between user signals, many chips are accessed by every user, but with lower transmission power on each chip.

Spreading codes/patterns described by some randomised structure have recently become a cornerstone of multi-access channel research. The random element in the construction is particularly attractive in that it provides robustness and flexibility in application, whilst not making significant sacrifices in terms of transmission power efficiency. The extension of standard dense spreading codes to sparse codes can be motivated by the success of sparse ensembles and iterative decoding methods in related coding problems, such as low density parity check codes ~\cite{Vicente:LDPC,Richardson:MCT}. Understanding the sparse CDMA problem also provides a basis for understanding sparse scattering processes arising from more general channel phenomena, such as multi-path scattering and signal fading.

With communication over the channel subject to perfect control over timing, scattering, and power, the possibility exists to develop structured codes that will outperform random codes. However, the random models offer some robustness and flexibility in application, and the difference in performance may be mitigated by a small increase in power. The random code paradigm also offers insight into more general sources of Multi-Access Interference (MAI).

\subsection{Summary of related results}
\label{CDMA.summaryrelatedresults}
This study follows several papers, in applying a typical case analysis based on the replica method to randomly spread CDMA with discrete inputs~\cite{Guo:MDSP}. The paper by Tanaka established many of the properties of random densely-spread CDMA~\cite{Tanaka:SMA}, with respect to several different detection methods including Marginal Posterior Mode detectors, maximising some measures of probability. Sparsely-spread CDMA differs from the conventional CDMA, based on dense spreading sequences, in that any user only transmits on a small number of chips (by comparison to transmission by all users on all chips in the case of dense CDMA). The sparse nature of this model facilitates the use of methods from statistical physics of dilute disordered systems for studying the properties of typical case transmission~\cite{Mezard:SGT,Nishimori:SP}.

The study of dense random codes is a well developed field, some relevant work includes improved iterative methods for detection based on message passing~\cite{Neirotti:IMP,Kabashima:SMA,Bickson:GBP}. Combining sparse encoding (LDPC) methods with CDMA is one way to improve detection properties beyond a single bit interval~\cite{Tanaka:SMALDPC}.

The feasibility of data transmission by sparse random CDMA, at a comparable rate to dense models, was first considered for the case of real (Gaussian distributed) input symbols~\cite{Yoshida:ASS}, the equilibrium problem was solved by a variational approach. A number of results were reported including near equivalence of the dense and sparse codes even where the number of accessed chips as a fraction of the bandwidth goes to zero (in the wide-band limit). In a separate recent study, based on the Belief Propagation (BP) inference algorithm and a binary input prior distribution, sparse CDMA has also been considered as a route to rigorously proving results in the densely spread CDMA~\cite{Montanari:ABP}, some sparse models achieve the dense performance.

There have also been many studies concerning the effectiveness of BP as an optimal detection method~\cite{Montanari:BPB,Guo:MDSS}. However, many of these papers consider the {\it extreme dilution} regime -- in which the number of chip contributions is large but not $O(M)$. In these models the information carried by the channel is identical to a dense random CDMA model.

The theoretical work regarding sparsely spread CDMA remained lacking in certain respects when this thesis began. As pointed out in~\cite{Yoshida:ASS}, spreading codes with Poisson distributed number of non-zero elements, per chip and across users, are systematically failing in that each user has some probability of not contributing to any chips (transmitting no information). This problem was address in ``user regular'' codes~\cite{Montanari:ABP} (where each user transmits on the same number of chips), but an understanding of how inhomogeneous bandwidth usage effects transmission remained poorly understood. Furthermore, the statistical physics analysis of codes with fixed finite connectivity has been solved only by approximation for the special class of code ensembles~\cite{Yoshida:ASS}.

Other theoretical problems are under study with comparable structures to the linear vector channel with MAI. Inter-symbol interference channel models, where a signal from a single user is self-interfering is closely related to MAI~\cite{Kavcic:BII}. A generalisation of the linear vector channel is many input many output (MIMO) channels which have also become an important area of research within statistical physics~\cite{Takeda:SMA}.

\subsection{Chapter outline and results summary}
\label{CDMA.MainContribution}

Section~\ref{CDMA.probability} describes the probabilistic framework, which is used to analyse the CDMA multi-user detection problem, a suitable Hamiltonian is thereby defined. The sparse code ensembles and channel model are presented.

Some special cases and exact results are identified for sparse codes in section~\ref{CDMA.exact}. This includes the identification of the Nishimori temperature, and development BP as an exact method on trees, and Unit Clause Propagation as an exact method in noiseless channels for some loopy ensembles.

Section~\ref{CDMA.marginals} presents a marginal description of the Hamiltonian allowing an understanding of MAI at a microscopic level, and contrasting sparse and dense ensembles~\cite{Raymond:RM}. Adopting a single chip detection model allows an upper bound on information transmission to be identified~\cite{Raymond:SS}.

An analysis of the equilibrium behaviour is constructed for the general case in section~\ref{CDMA.replica} by the replica method~\cite{Raymond:SS}. The Replica Symmetric (RS) saddle-point equations and free energy are constructed, the limitations of RS are explored and a stability analysis of the saddle-point equations is constructed.

Section~\ref{CDMA.results} demonstrates solutions of the saddle-point equations including thermodynamic and metastable cases. Both are shown to be locally stable at the Nishimori temperature. Freezing of the metastable solutions is identified. The dynamical importance of the metastable solution is demonstrated in finite systems, with some moderately sized examples examined using BP, the max-product and multi-stage detection algorithms. The performance of decoders in finite systems matches the results predicted by the equilibrium analysis.

Further discussion of results exists alongside the analysis of composite CDMA in chapter~\ref{chapter:compositeCDMA}.

\section{Probabilistic framework and code ensemble description}
\label{CDMA.probability}

\subsection{Probabilistic framework}
\begin{figure}[htb]
\includegraphics[width=\linewidth]{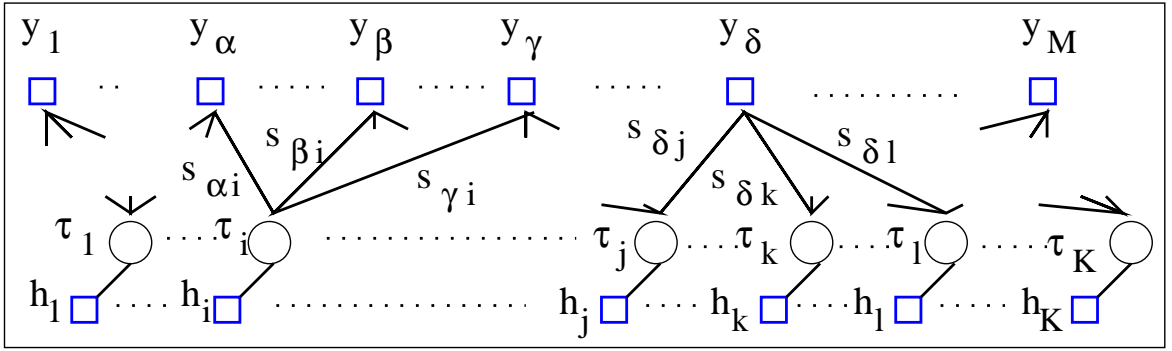}
\caption[Factor graph for a CDMA multi-user detection problem.]
{\label{fig:CDMA.graphicalmodel} A factor graph $\factorG$ for the CDMA detection problem consists of: a set of variable/user vertices $V_v$, which label the dynamical variables $\vtau$; factor vertices $V_f$, labeling the evidence ($\vy$); and edges $E$, encoding the probabilistic dependencies ($\ms$). A user node $i$ is known to have transmitted on three chips ($\partial_i=\{\alpha,\beta,\gamma\}$). The factor nodes are determined through a similar neighborhood ($\partial_\delta=\left\lbrace j,k,l\right\rbrace$). The interaction at each factor ($\mu$) is conditioned on neighbouring gain factors $s_{\mu k}$, and evidence $y_\mu$. The prior on bits, external fields, are represented by the lower set of factor nodes, but these are taken to be zero or infinitesimal in analysis.}
\end{figure}

A probabilistic framework forms the basis for a principled detection methods, and analysis of theoretical channel limits. This may be encoded in a graphical model as shown in figure~\ref{fig:CDMA.graphicalmodel}, which is based on received signal $\vy$, a modulated set of access patterns for each user $\vs_k$ and some prior on the source bits. The channel includes interference between users (MAI), and interference from an independent noise source. The detection model is based on an (assumed) random generative framework for the source signals and channel noise
\begin{equation}
{\hat P}(\vy | \vb, \ms) = \int \rmd\vnu {\hat P}(\vomega) \prod_{\mu=1}^M \left[\delta\left(y_\mu - \sum_{k=1}^K s_{\mu k} b_k + \omega_\mu\right)\right] \label{eq:CDMA.Pvyvbms}\;,
\end{equation}
where ${\hat P}$ is used to distinguish model probability
distributions from the true (generative) ones.

By working with a white noise model, assuming no correlations between the noise on each chip, a factorised form for ${\hat P}(\vomega)$ is taken, and hence (\ref{eq:CDMA.Pvyvbms}) is totally factorised with respect to $\mu$, the chip index. Supposing the power spectrum of the noise to be parameterised by $\beta$ (the inverse temperature), then the Gaussian noise model of variance $\beta^{-1}$ might be appropriate
\begin{equation}
{\hat P}(\vomega)= \sqrt{\beta/(2\pi)}^M \prod_{\mu=1}^M \exp\left\lbrace -\frac{\beta}{2} \omega_\mu^2\right\rbrace \label{eq:CDMA.Phatomega}\;.
\end{equation}
If the true noise is weakly correlated between chips, and does not have bursty (large variance) behaviour, then detection based on this AWGN model may still be useful as a variational estimate.

The quantity from which a principled inference can be drawn is the posterior ${\hat P}(\vb | \vy, \ms)$. From this, the model estimates to marginal probabilities for the bits can be constructed, and most probable bit sequence inferred. The probability can be rewritten using Bayes theorem in terms of our model likelihood and prior
\begin{equation}
{\hat P}(\vb | \vy, \ms) \propto {\hat P}(\vy | \vb, \ms) {\hat P}(\vb) \label{Pvbgivenvyms}\;.
\end{equation}
The code is assumed to be known by the detector. The probability distribution over $\vb$ encodes prior belief on the source bits, independent of the received signal. Under the assumption that the prior is conditionally independent for each user, then the distribution is
\begin{equation}
{\hat P}(\vb) = \prod_{k=1}^K \frac{\exp \{- \beta \randomfield_k b_k \}}{2 \cosh (\beta \randomfield_k)}\;.
\end{equation}
If $\randomfield_k = 0$ then no bias is assumed in the source bit towards either $1$ or $-1$. If $\randomfield_k = z b_k$ the detector has some knowledge of the source bit, and if $\randomfield_k>0$ then there is an assumed bias in the source bit towards $1$. The analysis it is convenient to consider that $\randomfield_k$ might be uniform, or could take some discrete set of values. However, the only case evaluated in detail corresponds to cases where $\randomfield_k\rightarrow 0$ ultimately, and this should be assumed in all expressions. However, in the calculation of the free energy (only) it is useful to make explicit an external field, since the derivative in the limit $\randomfield_k\rightarrow 0$ has an important physical interpretation, as explored in Appendix \ref{app:ConjugateFields}.

A natural quantity to consider in terms of the viability of the channel and detection model is the spectral efficiency, which is defined as the mutual information between the signal and source bits, rescaled to the bandwidth
\begin{equation}
\SEmath = \frac{1}{M} \sum_{\vb}\int \rmd\vy' P(\vy,\vb|\ms) \log {\hat P}(\vy|\vb,\ms) - \frac{1}{M}\int \rmd\vy P(\vy|\ms)\log{\hat P}(\vy|\ms) \label{eq:CDMA.se}\;.
\end{equation}
The log term measures model specific surprise at the samples of the signal (and bits in the first term), given the model used. These samples are marginalised over according to the true distribution of bits and signals, rather than the model estimates, hence the combination of two probabilities. If the detection and generative model are identical the conventional mutual information is recovered.

The first term appears superficially to be the more complicated, but is normally the simpler. This part is not relevant in the detection problem for a static model description since it measures model specific surprise at the signal given that the source bits are revealed. The second part, by contrast, measures surprise at the signal, without revealing the source bits. Minimisation of the second term, for a fixed model, by determination of ${\hat P}(\vb|\ms,\vy)$ within the model framework is the objective. When searching the space of models, to fit the data, both parts are relevant.

In the detection problem the code is a random object so that the spectral efficiency is a random variable, but when the number of users is sufficiently large, self-averaging of samples is expected for the sparse ensembles. The average over the instances of the codes allows construction of the non-random self-averaged spectral efficiency.

\subsection{Statistical mechanics framework}

A Hamiltonian that describes the same joint probability distribution of signal and source bits (\ref{eq:CDMA.Pvyvbms}), and allows a determination of many information theoretic properties, is
\begin{equation}
\Ham(\vtau) = - \sum_k \randomfield_k \tau_k + \frac{1}{2}\sum_\mu \left( y_\mu - \sum_k s_{\mu k} \tau_k\right)^2 \label{eq:CDMA.Hamiltonian}\;,
\end{equation}
where $\quenched$ is an abbreviation for the quenched variables ($\ms$,$\vy$,$\vrandomfield$), which are sampled from an ensemble $\ensemble$, and $\vtau$ are the dynamical variables. This defines the estimated posterior distribution
\begin{equation}
P(\vtau) = \frac{1}{Z} \exp\left(-\beta\Ham(\vtau) \right) \label{eq:CDMA.Ptau}\;,
\end{equation}
where $Z$ is the partition function. It is useful to decompose the signal and code according to
\begin{equation}
y_\mu = \omega_\mu + \frac{1}{\sqrt{C}}\sum_{k=1}^K A_{\mu k} \modulationsymbol_{\mu k} b_k \label{eq:CDMA.yAxi}\;,
\end{equation}
in some analysis, $\vomega$ is the source noise and $\mA$ is a sparse spreading matrix
\begin{equation}
A_{\mu k} = \left\lbrace \begin{array}{l l}
1 &\hbox{If user $k$ transmits on chip $\mu$}\;; \\
0 &\hbox{otherwise}\;;
\end{array}\right.
\end{equation}
$\mxi$ is a dense modulation pattern, and $\vb$ is the source bit sequence.

An alternative Hamiltonian relevant to some sections is obtained from (\ref{eq:CDMA.Hamiltonian}) by expansion of the square, up to constant terms
\begin{equation}
\Ham(\vtau)= \sum_{\ij} J_{\ij} \tau_{i} \tau_{j} - \sum_k h_k \tau_k \;,\label{eq:CDMA.HamiltonianJh}
\end{equation}
where the binary couplings $J_\ij$ and fields are given by:
\begin{equation}
 J_\ij = - \sum_\mu s_{\mu i} s_{\mu j}\;;\qquad
 h_k = \randomfield_k + \sum_\mu s_{\mu k} y_\mu \label{eq:CDMA.Jh}\;;
\end{equation}
with $\ij$ indicating the ordered pair (each edge is labeled uniquely with $i<j$). The Gaussian noise model implies a special Hamiltonian case in which a quadratic form is possible. For general marginal noise models a polynomial of degree ${l_e}$ is required to describe a chip of connectivity ${l_e}$.

The partition function for any model Hamiltonian is
\begin{equation}
\partitionfunction(\quenched) = \sum_{\vtau} \exp \left\lbrace -\beta\Ham(\vtau)\right\rbrace\;,
\end{equation}
and the self-averaging free energy density is given by
\begin{equation}
\beta \safed = \lim_{K\rightarrow \infty}\< - \frac{1}{ K} \log \partitionfunction(\quenched))\>_{\quenched} \label{eq:CDMA.safed}\;,
\end{equation}
which is affine to the spectral efficiency (\ref{eq:CDMA.se}), when averaged over codes. The free energy density is dependent on a particular sample/instance of the codes and channel noise, quenched variables ($\quenched$), whereas the self-averaged quantity is dependent only on the ensemble parameterisation ($\ensemble$). The relation to spectral efficiency is given by
\begin{equation}
\SEmath = -S_\beta(\beta_0) + \left(\load\log(2) + \frac{1}{2}\log(2\pi/\beta) + \load \beta f \right) \;,
\end{equation}
where $S_\beta$ is the signal entropy assuming variance $\beta$ in the detection model
\begin{equation}
S_\beta(\beta_0) = \frac{\beta}{2\beta_0} + \frac{1}{2}\log\left(2 \pi/\beta\right) \label{eq:CDMA.noiseentropy} \;,
\end{equation}
where $(\beta_0)^{-1}$ is the true variance of the noise, defined shortly (\ref{eq:CDMA.SNR}). The notation $\load=K/M$ is the spreading factor, instead of $\beta$ from the information theory literature, and $\alpha$ in some of my published papers.

\subsection{Bit sequence ensemble}

The source bits are assumed to be independently generated. The bit transmitted by any user is then controlled by a probability distribution parameterised by $\randomfield_0$
\begin{equation}
 P(\vb)= \prod_k \frac{\exp\left\lbrace \randomfield_0 b_k\right\rbrace}{2 \cosh \randomfield_0}
 \label{eq:CDMA.Pb}\;.
\end{equation}
The rate of transmission is the entropy of the probability distribution (\ref{eq:CDMA.Pb}), which is an upper bound on the amount of information that might be extracted from the channel. The rate is maximum when $\randomfield_0=0$, which is considered without exception in this thesis. A reduced rate involves a bias in the users transmissions towards $\pm 1$, or correlations amongst user transitions.

\subsection{Noise ensemble}
The model used to explore CDMA is an AWGN model. This is a reasonable model for realistic wireless communication, and is also easy to work with analytically. The instance of quenched noise is drawn independently for each chip according to a distribution parameterised by variance $\beta_0^{-1}$
\begin{equation}
P(\vomega) = \prod_\mu \frac{1}{\sqrt{2 \pi/\beta_0}} \exp
\left\lbrace-\frac{\beta_0}{2}(\omega_\mu^2)\right\rbrace\;.
\end{equation}
This is the same form as assumed in the detection model (\ref{eq:CDMA.Phatomega}), the discrepancy between the model is quantified by $\beta/\beta_0$. The Signal to Noise Ratio (SNR) per bit is defined as
\begin{equation}
\SNRmath = \beta_0 \left(\frac{1}{K}\sum_k \sum_\mu (s_{\mu k})^2 \right)/2 = \beta_0/2 \label{eq:CDMA.SNR}\;,
\end{equation}
the user codes are normalised to one either exactly or in expectation. The entropy of the additive white noise is given by $S_{\beta_0}(\beta_0)$ (\ref{eq:CDMA.noiseentropy}) hence a logarithmic scale is appropriate to describe variability (decibels, dB are used).

\subsection{Spreading pattern ensembles}
\label{code_ensembles}

Sparse codes share the common feature that if the connectivity of user $k$ is $C_k$, then $C_k/M\rightarrow 0$ in the wide-band (large $M$) limit. The particular case considered in this thesis has $C_k$ finite, as opposed to other studies where $C_k$ might scale with $M$, e.g. $C_k\sim M^\delta$~\cite{Guo:MDSS}.

The spreading pattern ensembles defines $\ms$, the set of codes, through a distribution on the connectivity matrix $\mA$ and modulation pattern $\mxi$ (\ref{eq:CDMA.yAxi}). The distribution on matrices $\mA$ is parameterised by a marginal chip connectivity profile of mean value $L$, and a marginal user connectivity distribution of mean value $C$, constrained by the spreading factor
\begin{equation}
\load = \frac{L}{C}=\frac{K}{M}\;.
\end{equation}
The modulation pattern $\mxi$, has components which are independent and identically distributed (i.i.d). The modulation pattern distribution is constrained to be of mean square value one, non-zero, and of finite higher order moments. In the limit $C\rightarrow \infty$ all ensembles described in this way converge to a standard dense code ensemble~\cite{Tanaka:SMA}.

\subsubsection{Sparse connectivity ensemble}
\label{CDMA.connectivityensembles}

Sparseness implies the probability that a user $k$ makes a transmission on some chip is small, implying a prior distribution
\begin{equation}
P(A_{\mu k})= \left(1-\frac{C}{M}\right)\delta(A_{\mu k}) + \frac{C}{M}\delta(A_{\mu k}-1)\label{eq:CDMA.PsMarg}\;.
\end{equation}
The simplest ensemble is the irregular ensemble defined shortly, based only on this constraint.

A generalised ensemble is usefully described by a pair of marginal distributions $\left\lbrace P_C(C_k),P_L(L_\mu)\right\rbrace$, where $C$ and $L$ are the mean connectivity for the user and chip connectivities. As argued in Appendix~\ref{app.Sparsematrix} the probability distribution can then be used in a form given by
\begin{equation}
P(\mA| P_C,P_L) \propto \prod_{\mu} \<\frac{{l_e}\factorial}{L^{l_e}}\delta\left(\sum_k A_{\mu k} - {l_e} \right)\>_{l_e} \prod_{k} \<\frac{c_f\factorial}{C^{c_f}} \delta\left(\sum_\mu A_{\mu k} - c_f \right)\>_{c_f} \prod_{\mu,k} P(A_{\mu k}) \label{eq:CDMA.GENSPARSEENSEMBLE}\;,
\end{equation}
where $c_f$ and ${l_e}$ are sampled from the $P_C$ and $P_L$, and the prior is sparse.

Four types of sparse ensemble are defined as special cases. The ensembles irregular and chip regular ensembles are unconstrained in user connectivity. Some fraction of users, $\exp\left\lbrace -C \right\rbrace$ fail to communicate at all which places a strict limit on the recoverable information; but power is distributed more uniformly on the bandwidth and the mean excess connectivity is reduced, which can be shown to reduce MAI. The excess degree distributions for the chip and user are defined as conditional probabilities:
$E({l_e})=P(L_\mu={l_e}-1|L_\mu>0)$, $E(c_e)=P(C_k = c_e-1|C_k>0)$ respectively.

The irregular and user regular ensembles have a fraction, $\exp\left\lbrace -L\right\rbrace$, of the bandwidth unused. It seems likely that a better use of channel resources would be to have a uniform distribution of power in expectation. Chip regular ensembles use the bandwidth more uniformly, but this spreading can only be realised with a coordinated sampling of codes for different users.

\subsubsection{The irregular ensemble}

In the irregular ensemble the joint code connectivity distribution is a product of the marginal distributions. The ensemble is described by
\begin{equation}
P(\mA) = \prod_k \prod_\mu P(A_{\mu k}) \label{eq:CDMA.irregular}\;,
\end{equation}
and represents a good null model for sparse effects, it was first considered in~\cite{Yoshida:ASS}. The marginal chip connectivity distribution is described by a Poissonian distribution $P(L_\mu={l_e}) = P_L({l_e})$, as is the marginal variable connectivity distribution $P_C$. Where
\begin{equation}
P_x(z) = \frac{\exp\left\lbrace -x\right\rbrace x^z}{z\factorial}\label{eq:Poisson}\;.
\end{equation}

\subsubsection{User regular ensemble}
\label{CDMA.userregular}

A special case of a spreading pattern without a disconnected component is found by constraining all users to transmit on exactly $C$ chips, which has been frequently studied (e.g.~\cite{Montanari:BPB}). The probability distribution is
\begin{equation}
P(\mA)=\prod_k P(\vA_k | C_k = C)\;;\qquad P(\vA_k|C_k=C) = \left(\frac{M\factorial}{(M-C)\factorial C\factorial}\right)^{-1} \delta\left(\sum_\mu A_{\mu k} - C \right) \label{eq:CDMA.userregular}\;.
\end{equation}
The chip connectivity distribution is described by $P_L$ (\ref{eq:Poisson}). The encoding method used by each user is independent given C, the generation of codes may be undertaken independently for each user. Furthermore, with a uniform modulation pattern distribution (\ref{eq:CDMA.BPSK}), the user signal powers are equalised.

\subsubsection{Chip regular ensemble}
\label{CDMA.chipregular}

The number of users is constrained to be $L$ for all chips in this model
\begin{equation}
P(\mA |L_\mu=L)=\prod_\mu P(\vA_\mu|L_\mu=L) \;;\qquad P(\vA_\mu | L_\mu=L)=\left(\frac{K\factorial}{(K-L)\factorial L\factorial}\right)^{-1}\delta\left(\sum_k A_{\mu k} - L\right)\label{eq:CDMA.chipregular}\;.
\end{equation}
This ensemble implies a homogeneous power spectral density across all chips in expectation. However, the model allows a consideration of sparse processes with a homogeneous power spectrum, and it is also an ensemble for which the study of the noiseless channel is simplified. With only the chip regular constraint applied the user connectivity is described by a distribution $P_C$.

\subsubsection{The regular ensemble}
\label{CDMA.regular}

Amongst choices for the marginal chip and user connectivity distributions it would seem a model, which is doubly regular might be most efficient~\cite{Raymond:RM,Raymond:SS}
\begin{equation}
P(\mA |L_k=L,C_k=C) \propto \prod_\mu \delta\left(\sum_k A_{\mu k} - L\right)\prod_k \delta\left(\sum_\mu A_{\mu k} - C \right) \label{eq:CDMA.regular}\;.
\end{equation}
This description implies a homogenous power spectral density on a microscopic scale with respect to the users and bandwidth.

\subsubsection{Modulation patterns}
\label{CDMA.modulationpatterns}

The sparse access patterns determines the existence of links between different users and nodes. A non-zero modulation strength may be assigned to each non-zero user chip pair independently through a distribution
\begin{equation}
P(\mxi)=\prod_\mu \prod_k P(\modulationsymbol_{\mu k})\;;\qquad P(\modulationsymbol_{\mu k}=z)=\phi(z) \label{eq:CDMA.mxi}\;.
\end{equation}
The distribution $\phi$ has mean square value $1$ and finite higher order moments, and no measure on $0$ (zero quantities are encoded through $\mA$). The physical interpretation on a wireless channel is as Binary Phase Shift Keying (BPSK) and/or Amplitude Shift Keying (ASK).

The standard implementation, BPSK, involves no amplitude modulation and a phase shift of $\pm 1$ with equal probability. In the sparse ensemble it is also possible to transmit information without any modulation shift keying at all, the disorder implicit in the structure of the problem is sufficient to extract information. For the sake of generality a simple Gaussian ASK distribution can also be considered. These methods are described by
\begin{equation}
\phi(z)=\left\lbrace
\begin{array}{ccc}
\frac{1}{2}\left(\delta(z-1) + \delta(z+1)\right) \qquad
 &\hbox{(symmetric)} &\hbox{BPSK} \;; \\
\delta(z-1) \qquad & \hbox{(asymmetric)} &\hbox{unmodulated}\;; \\
\frac{1}{\sqrt{2\pi}}\exp\left\lbrace -z^2/2\right\rbrace \qquad &
\hbox{(symmetric)}&\hbox{Gaussian ASK}\label{eq:CDMA.BPSK}\;.
\end{array}\right.
\end{equation}
In examining the generic properties all three methods are useful and span a range of behaviour in the sparse code. The BPSK code is used primarily, but the unmodulated code tests the effect of symmetries and highlights some subtleties in the methods~\cite{Raymond:RM}.

The ASK case is worthy of considering for two reasons. Firstly, it breaks a codeword degeneracy problem in the noiseless channel case, therefore it may be a better choice for high SNR. Secondly, theoretical model of a linear channel with a single characteristic power scale would seem to identify the Gaussian model as a null model. If a random process is responsible for generating the sparse spreading pattern, rather than deliberate coding, then Gaussian amplitudes may be representative.

\section{Exactly solvable sparse ensembles}
\label{CDMA.exact}

For some special ensembles it is possible to calculate exactly
many quantities through a statistical mechanics treatment.
Furthermore, the constructive multi-user detection problem, of
finding an optimal decoding, might be solved exactly in typical case in spite of unfavourable worst case algorithmic properties of multi-user detection with MAI~\cite{Verdu:CC}. An analogy will be made with the 1 in 3 SAT problem in the noiseless channel, extending results of chapter~\ref{chapter:1inkSAT}. Many special cases on sparse graphs are solvable and some are outlined in the following subsections.

\subsection{The Nishimori temperature}
\label{CDMA.Nishimori}

The Nishimori temperature in the CDMA problem describes the parameterisation of the detection model that correctly describes the generative model~\cite{Nishimori:ER,Nishimori:CO}. The proofs derived in this framework are a generalisation of the gauge theory for spin glass systems. The role of temperature is taken by the detection model parameters (the noise variance and priors $\beta,\vrandomfield$)~\cite{Iba:NL}. The Nishimori temperature in the proposed model is the parameterisation of the detection probabilities given by $\beta=\beta_0$, and in the case of a uniform prior $\{\randomfield_k=\randomfield_0\}$ (\ref{eq:CDMA.Pb}).

At the Nishimori temperature it is possible to exactly calculate many thermodynamic properties of ensemble including the energy density, which is $\saed=1/(2\load)$. More importantly it is possible to show that the phase space takes a simple connected form. This observation indicates that some simpler types of mean-field approximation and algorithms, such as RS and BP, may be successful in describing the detection problem. These properties are derived in~\ref{app:Nishimori}, and the significance will be considered in the context of an equilibrium analysis in later sections.

\subsection{Trees}
\label{CDMA.Trees}

Some standard bandwidth sharing models, F/TDMA, correspond to trivial one variable trees in the factor graph representation. The graphical model for sparse codes (\ref{eq:CDMA.GENSPARSEENSEMBLE}) below the percolation threshold also corresponds to a forest, with many trees of size at most $O(\log M)$ in the large system limit.

On a tree, the problem of calculating marginal distributions becomes exact by message passing methods. Furthermore a single pure state is guaranteed to exist and so the self-averaging result can be studied by an RS approximation at all temperatures, simplifying analysis.

\subsubsection{Sparse CDMA BP equations}

BP can realise the Marginal Posterior Mode (MPM) and Maximum A Posteriori (MAP) detectors on trees, once the messages have converged. The messages from and to prior factor nodes are trivial, the equations for $z=0$ are presented.

The BP equations, given the Hamiltonian form (\ref{eq:CDMA.Hamiltonian}), are derived as demonstrated in section~\ref{introduction.BP} and include a weighted marginalisation step, determining log-likelihood ratios
\begin{equation}
u^{(t)}_{\mu \rightarrow k} = \frac{1}{2 \beta} \sum_{b} b \log \left(P^{(t)}(y_\mu | b_k=b,\vy\setminus y_\mu )\right) = \frac{1}{2 \beta} \sum_{\tau_k} \tau_k \log \partitionfunction^{(t)}_{\mu \rightarrow k} \label{eq:CDMA.mutok}\;,
\end{equation}
with
\begin{equation}
\partitionfunction^{(t)}_{\mu \rightarrow k} = \prod_{l \in \partial_\mu \setminus k}\left[\sum_{\tau_l} \exp \left\lbrace \beta h^{(t)}_{l \rightarrow \mu} \tau_l\right\rbrace \right] \exp \left\lbrace -\frac{\beta}{2}\left(y_\mu - \sum_{l \in \partial_\mu} s_{\mu l} \tau_l\right)^2\right\rbrace \label{eq:CDMA.Zmuk}\;.
\end{equation}
Combined with a step determining log-posterior ratios
\begin{equation}
h^{(t+1)}_{k \rightarrow \mu} =\frac{1}{2 \beta} \sum_{b=\pm 1} b\log \left(P^{(t)}(b_k=b |\vy\setminus y_\mu )\right) = \sum_{\nu \in \partial_k \setminus \mu} u^{(t)}_{\nu \rightarrow k} \label{eq:CDMA.ktomu}\;.
\end{equation}
From the messages a determination of log-posterior ratios for the source bits is possible
\begin{equation}
H^{(t+1)}_k = \frac{1}{2 \beta} \sum_{b} b \log \left( P^{(t)}(b_k=b |\vy) \right) = \sum_{\mu \in \partial k} u^{(t)}_{\mu \rightarrow k} \label{eq:CDMA.H}\;.
\end{equation}

\subsubsection{The individually and jointly optimal detectors}

Different values of $\beta$, for fixed SNR, define a class of detectors. With $\beta=\beta_0$, the Nishimori temperature, the correct marginal probability distributions are described and exact marginals can be constructed through iteration of BP equations. An individually optimal estimation of bits is equivalent to
\begin{equation}
\tau^{(t)}_k = \sign\left(\sum_{\mu \in \partial_k} u^{(t)}_{\mu \rightarrow k} \right)\label{eq:CDMA.IO}\;,
\end{equation}
after sufficiently many updates (large t). On a tree relatively few updates are required, and optimal ordering is possible, so that MPM estimation is possible in $O(M)$ updates.

Assuming the ground state of the Hamiltonian is unique then the MAP detector result can also be achieved in linear time by the BP in the limit $\beta \rightarrow \infty$. In this limit the algorithm is well defined and called the max-product algorithm; the weighted marginalisation step (\ref{eq:CDMA.Zmuk}) is replaced by a maximisation step. The variable messages are simplified to
\begin{equation}
u^{(t),MP}_{\mu \rightarrow i} = \frac{1}{2} \sum_i \tau_i \max_{\left\lbrace \tau_k\right\rbrace | k \in \partial_\mu \setminus i} \left\lbrace \frac{1}{2}\left(y_\mu - \sum_{k\in \partial_\mu} s_{\mu k} \tau_k\right)^2 + \sum_{j \in
\partial_\mu \setminus i} h^{(t)}_{j \rightarrow \mu} \tau_j \right\rbrace \label{eq:CDMA.maxproduct}\;.
\end{equation}
The jointly optimal bit sequence is determined from converged messages as
\begin{equation}
\tau_k^{MAP} = \sign\left(\sum_{\mu \in \partial_k} u^{(t),MP}_{\mu \rightarrow k}\right)\label{eq:CDMA.MAP}\;.
\end{equation}
The uniqueness of ground states on the tree will not be met in general; for example, in the absence of ASK (\ref{eq:CDMA.BPSK}). In non-unique cases the max-product algorithm will identify a superposition of solutions, and one solution may be picked out by introducing some symmetry breaking.

\subsection{Graphs with many loops}
\subsubsection{Ferromagnetic systems}

If codes are anti-correlated or orthogonal $\vs_k . \vs_{l}\leq 0$, then an efficient MAP detector is realisable for any sparse graph. In these cases the Hamiltonian (\ref{eq:CDMA.HamiltonianJh}) has exclusively ferromagnetic interactions, and so the problem is equivalent to a random field Ising model (RFIM). The MAP detector is realisable by a polynomial time algorithm in these cases by analogy with max-flow algorithms~\cite{Chertkov:EB}.

A special case of the above system is one where all user codes are orthogonal, $\vs_k . \vs_l = 0$. These codes can be optimally decoded by a matched filter
\begin{equation}
\tau^{MF} = \sign\left(\sum_\mu y_\mu s_{\mu k}\right) \label{eq:CDMA.MF}\;,
\end{equation}
since the probability distribution for any two bits are
independent given the signal, MAI is zero. Orthogonal codes can be constructed with power control and synchronisation whenever $K\leq M$.

In overloaded regimes no orthogonal or ferromagnetic codes exist, codes meeting the Welch-Bound Equality are known to maximise capacity~\cite{Rupf:OSM}, but achieving an interference free performance ceases to be possible. Optimisation of codes is an important issue in this regime, and a compromise is often required between optimality and practicality, due to the computational cost of optimisation and inflexibility of optimal code sets. A popular set of codes reducing MAI are Gold codes, which are applicable to BPSK systems~\cite{Gold:OB}.

\subsubsection{Detectors in the noiseless limit, $\SNRmath \rightarrow 0$}

In this scenario each chip is a constraint, which must be met exactly, although the probabilistic framework may be a useful abstraction. The clause structure for sparse ensembles above the percolation threshold implies no simple solution in general. If a unique bit sequence is implied by every chip then the detection of all source bits connected to a factor becomes practical in the sparse case, by a detection on a chip by chip basis. In the absence of amplitude modulation (\ref{eq:CDMA.BPSK}) this will not be the case and additional correlations between chips must be used to infer the source bits.

Attention is restricted to the case without amplitude modulation, and without prior knowledge of the source bits. In this case the signal, on a specific chip of connectivity $L_\mu$ can take values $\sqrt{C} y_\mu = \left\lbrace -L_\mu+2i | i=0\ldots L_\mu\right\rbrace$ when BPSK is employed. The distribution of values is Binomial
\begin{equation}
P(y_\mu=\sqrt{C}^{-1} i) = \sum_{\vb} P(y_\mu=i |\vb) P(\vb) = \frac{1}{2^L_\mu} \frac{L_\mu\factorial}{(L_\mu-i)\factorial i\factorial} \label{eq:CDMA.rescaledy}\;.
\end{equation}
Only the constraints in which $y_\mu=\pm L_\mu$ imply unique values for connected source bits.

Indirect inference of variables might be made by a branch and bound method as explored in chapter~\ref{chapter:1inkSAT}. The chip regular ensemble with $L=3$ (\ref{eq:CDMA.chipregular}) is a special loopy case; The set of constraints implied by $y_\mu$ can be converted into a set of $\otSAT$ statements, for which the methods of chapter~\ref{chapter:1inkSAT} can be applied to produce a MAP bit estimate efficiently for any load. This result is developed in Appendix~\ref{app:CDMAto1ink}, and a consideration of a broader range of ensembles is possible~\cite{Raymond:OD}. More generally it appears the noiseless sparse case is MAP decodable when load $\load$ is sufficiently small, but becomes discontinuously inefficient to decode by decimation above some threshold in $\load$ for a variety of ensembles.

\section{Marginal descriptions}
\label{CDMA.marginals}
The main analysis of codes is through the replica method; however, it is frequently valuable to examine properties at a marginal level. These analysis provide bounds on the equilibrium properties and may allow insight into dynamics and inspiration based on comparable models. .

\subsection{Marginal field and binary coupling description}
\begin{figure}[htb]
\begin{center}
\includegraphics[width=\linewidth]{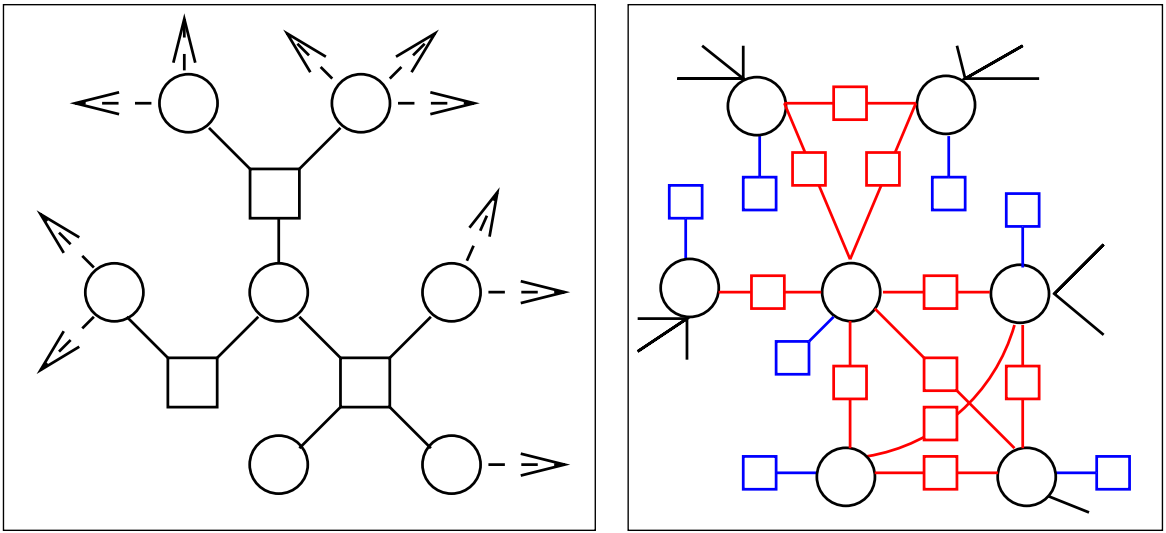}
\caption[Graphical model of CDMA with fields and couplings.]
{\label{fig:CDMA.cliques} Left figure: The truncated locally tree like structure of a CDMA inference problem is shown, with each factor representing the evidence $y_\mu$ (without prior factors, $\randomfield_k=0$). Right figure: The Hamiltonian description in terms of couplings and fields implies a different graph structure, with each chip implying a clique of $L_\mu$ coupled variables, and each variable subject to an external field.}
\end{center}
\end{figure}
One can gain insight into the origins of complexity in detection on the noisy channel by examining the interaction structure, making analogies between this model and the Sherrington-Kirkpatrick (SK)~\cite{Sherrington:SMSG}, Viana-Bray (VB)~\cite{Viana:PD} and other canonical models for disordered systems~\cite{Mezard:SGT}.

Each multi-variable coupling in the Hamiltonian, one for each chip, may be written as a set of binary couplings and fields (\ref{eq:CDMA.HamiltonianJh}). The field referred to in this subsection differs from the external field $\randomfield_k$. This is a standard formulation in physics, where the set of couplings $J_{\ij}$ and fields $h_k$ describe the problem. The coupling term is given by (\ref{eq:CDMA.Jh}), whereas the field term may be expanded in several components according to (\ref{eq:CDMA.lvc})
\begin{equation}
h_k = \randomfield_k \tau_k + \left[\sum_\mu s_{\mu k}^2\right] b_{k} + \left\lbrace \sum_\mu s_{\mu k} \sum_{l\setminus k} s_{\mu l} b_{l}\right\rbrace + \left\lbrace \sum_\mu \omega_\mu s_{\mu k}\right\rbrace \label{eq:CDMA.marginals}\;.
\end{equation}
Since the coupling term has no dependence on the source bits $\vb$, the states induced by the couplings alone must be uncorrelated with the source bits. By contrast, the field term includes a prior term, a bias towards the source bits, and an MAI plus noise term.

The marginal distributions can be evaluated for the symmetric modulation ensembles (\ref{eq:CDMA.BPSK}) to provide insight on the structure. The couplings and fields are strongly correlated through the code. In the case of a dense random code ensemble $C\!\rightarrow\! M$ marginal distributions over couplings and fields are both described by Gaussian random variables according to the central limit theorem. Marginalising over the un-factorised quenched variables gives distributions
\begin{equation}
P(J_{\ij}) = {\cal N}\left(0,\frac{1}{M}\right)\;; \qquad
P(h_k) = {\cal N}\left(b_k + \randomfield_k, \load + \frac{1}{\beta_0}\right)\label{couplings}\;;
\end{equation}
in the case of dense codes, ${\cal N}(a,b)$ indicates the Gaussian distribution of mean $a$ and variance $b$.

For the sparse code the binary couplings occur in cliques of size $\left\lbrace L_\mu\right\rbrace$ as shown in figure~\ref{fig:CDMA.cliques} couplings are $\pm 1/C$ with equal probability in the marginalised case. The field term contains a similar set of terms to the dense case. The MAI term has a non-Gaussian structure, but ignoring for convenience higher order moments allows a Gaussian description
\begin{equation}
P(h_k) = {\cal N}\left( b_k + \randomfield_k, \frac{\<|nn|\>}{C^2} + \frac{1}{\beta_0}\right) \label{Phkfields}\;,
\end{equation}
where $\<|nn|\>$ is the expected number of nearest neighbours (sources sharing a chip with source k), which is dependent on the mean excess chip connectivity ($P(L_\mu-1 |L_\mu>0)$). Chip regular ensembles (\ref{eq:CDMA.chipregular})(\ref{eq:CDMA.regular}) minimise this source of interference, $\<|nn|\>/C^2=\load \frac{L-1}{L}$. In the Poissonian chip connectivity cases the term has an identical value to the dense case, $\load$. This MAI term is the only difference between the dense and sparse terms in the marginal field distribution.

In the dense (sparse) model the marginalised description is consistent with a random field SK (VB) model, as discussed in chapter~\ref{chapter:introduction}. Another analogy may be made with the Hopfield model when considering in detail the form of the couplings (\ref{eq:CDMA.Jh})~\cite{Amit:SG,Kechriotis:HNN}, except that the couplings are reversed (anti-Hebbian). These models are famous for their complicated phase spaces caused by frustration in the couplings.

An intuitive feature of the marginal description is a competition between a mean dominated field promoting source bit reconstruction, and a variance dominated field preventing this. In channels with low SNR the variance dominates and there is only a weak net alignment with $\vb$. With increased SNR the noise term in the field variance becomes negligible, so that MAI is responsible for field misalignment with the source bits. With small load $\load$ the MAI term is reduced and the state will be orderly in the field part. Since the MAI is smallest in chip regular codes, these may demonstrate an improved performance.

When considering the topology of interactions the analogy of dense ensembles with the SK model seems reasonable given that the topology is fully connected and the marginal field and coupling distributions are Gaussian. A comparison of the sparse ensemble to the VB model seems less reasonable given that interactions are correlated within small cliques. In the VB model frustration arises through long loops, but in the sparse CDMA models frustration is implicit to each clique of size at least $3$. The frustration within cliques cannot be gauged from the problem, even within an isolated clique. This frustration within cliques is most explicit in the unmodulated (\ref{eq:CDMA.BPSK}) sparse ensemble, in this case all links are exclusively anti-ferromagnetic $J_{k l} = \frac{1}{\sqrt{C}}$ if variables are not gauged.

\subsection{Information extracted from an ensemble of scalar channels}
\label{CDMA.ScalarChannel}
\begin{figure}[htb]
\begin{center}
\includegraphics[width=\linewidth]{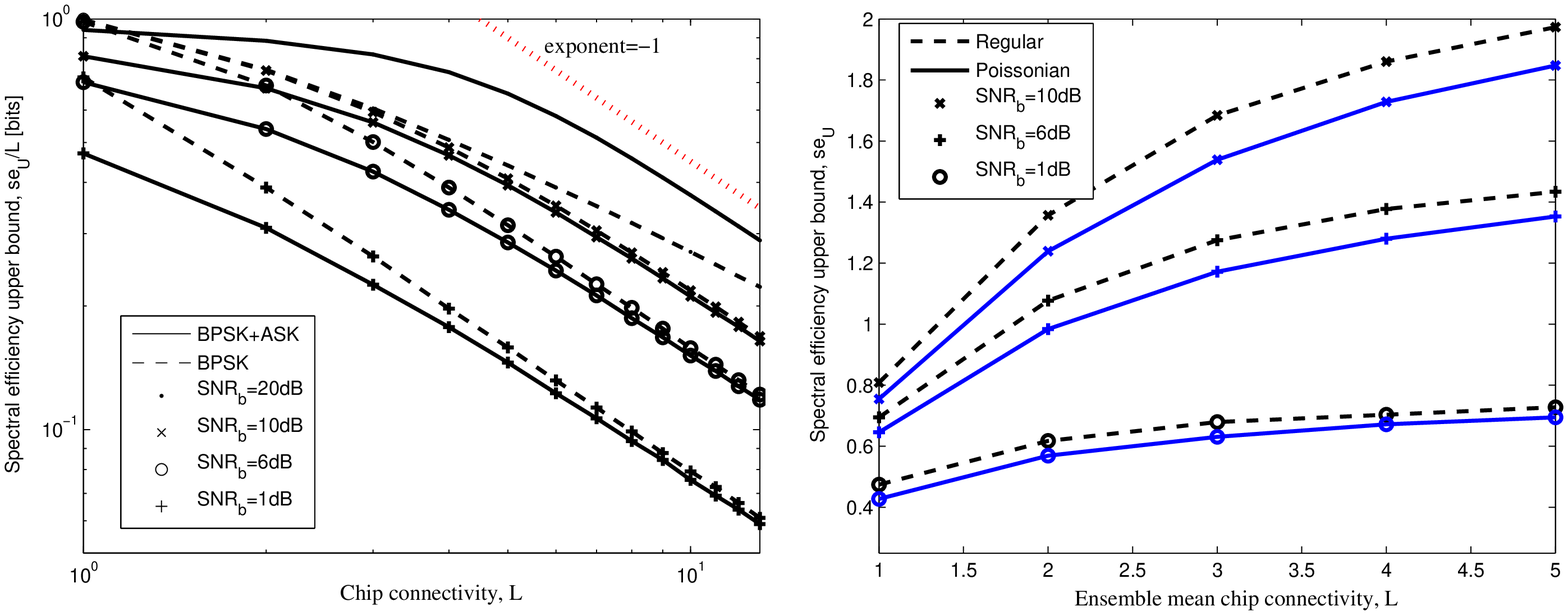}
\caption[Spectral efficiency upper bounds $\SEmath_U$, for sparse ensembles.]
{\label{fig:CDMA.entropybounds} Left figure: The figure demonstrates spectral efficiency for regular and Poissonian chip connectivity ensembles on the Gaussian scalar channel, for various modulation patterns (two line types) and SNRs (mixed symbols). The spectral efficiency per user tends asymptotically to a power law of exponent $-1$ (upper line shows the power law for comparison). Except at very high SNR, BPSK modulation outperforms Gaussian ASK in the noisy channel. The capacity of the channel saturates as $L$ is increases at fixed SNR. Right figure: The ensemble of chips described by Poissonian chip connectivity are compared to those with regular chip connectivity with BPSK of fixed amplitude per transmitted bit. Both converge to the same asymptotic value, but the chip regular ensemble conveys more information for small $L$.}
\end{center}
\end{figure}

The difficult part in calculation of the spectral efficiency for a given model is determining the entropy of the signal $\vy$ (\ref{eq:CDMA.se}), this may be approximated by assuming a factorised dependence on $\vy$, ${\hat P}(\vy)=\prod_\mu {\hat P}(y_\mu)$, and thereby the spectral efficiency may be written
\begin{equation}
\<\frac{1}{M}\log {\hat P}(\vy)\>_\vy = \<\log {\hat P}(y_\mu)\>_{y_\mu} - \left( \frac{1}{M}\KLfunction[P(\vy),{\hat P}(\vy)] - \frac{1}{M} \KLfunction[P(\vy),\prod_\mu{\hat P}(y_\mu)]\right) \;.
\end{equation}
where the Kullback-Leibler (KL) divergence
\begin{equation}
 \KLfunction[P,{\hat P}] = \<\log \frac{P(X)}{{\hat P}(X)}\>_X\label{eq:CDMA.KL}\;,
\end{equation}
is always positive provided the outer average is with respect to $P(X)$.

If the full model is accurate ${\hat P}(\vy)=P(\vy)$ then the first KL term is zero and an upper bound is proved. Otherwise it may be assumed that the unfactorised model is a better estimate, so that the difference between the two KL divergences (\ref{eq:CDMA.KL}) is positive. With the AWGN detection model the upper bound to spectral efficiency ($\SEmath_U$) can be written
\begin{equation}
\SEmath_U = - S_{\beta_0}(\beta) - \frac{1}{M}\sum_\mu \<\log \left[{\hat P}(y_\mu) \right]\>_{y_\mu} \label{eq:CDMA.seU}\;,
\end{equation}
with $S_{\beta_0}(\beta)$ being the channel noise entropy
(\ref{eq:CDMA.noiseentropy}) and the outer average is with respect to the generative distribution, but for a single chip only.

This effectively reduces the problem to a scalar channel, one for each chip. To determine ensemble properties an average over the distribution of scalar channels is needed, parameterised according to chip dependent terms in the ensemble.

The information, which can be extracted from a single chip for various $L_\mu$ determines the difference in spectral efficiency between ensembles for the simplified model. The model is one in which each factor node has an independent set of dynamical variables.

The bound on the self-averaging spectral efficiency is determined from the free energy calculation for a single chip, the spectral efficiency upper bound can be written, averaging over codes
\begin{equation}
\begin{array}{lcl}
\SEmath_U(\beta) &=& L\log(2) - \frac{\beta}{2
\beta_0} + \frac{1}{M}\sum_{\mu=1}^M \<\<-\log \left(
\prod_{l=1}^{{l_e}}\left[\sum_{\tau_l}\right] \right.\right. \right. \\
&\times&\left.\left.\left. \exp \left\lbrace - \frac{\beta}{2}\left(\sum_{k=1}^{l_e} \frac{1}{\sqrt{C}}
\modulationsymbol_{k} (\tau_k - b_k) - \omega
\right)^2\right\rbrace \right) \>_{\left\lbrace
b_i,\modulationsymbol_i\right\rbrace} \>_{{l_e}}
\label{eq:CDMA.generalseU}\;.
\end{array}
\end{equation}
The spectral efficiency is factorised in several parts and $y_\mu$ has been expressed as a combination of source bits, Gaussian channel noise, and a modulation pattern. For fixed modulation patterns (\ref{eq:CDMA.BPSK}), not varying with $L_\mu$, an ensemble of scalar channels handling bit vectors of length $L_\mu$, each with identical SNR.

A special case of the above expression is the noiseless channel with $\beta\rightarrow \infty$, allowing a simplified expression in the cases without amplitude shift keying (\ref{eq:CDMA.BPSK})
\begin{equation}
\lim_{\beta = \beta_0 \rightarrow \infty} \SEmath_U(\beta)
= L\log(2) - \left\langle \sum_{p=0}^{{l_e}} \frac{1}{2^{l_e}} {{l_e} \choose p} \log \left(\sum_i^{\min(p,{l_e}-p)} {{l_e} - p \choose i} {p \choose i} \right)\right\rangle_{{l_e}}
\label{eq:CDMA.noiselessseU}\;.
\end{equation}
This capacity grows asymptotically as $\log(L)$, it is the entropy of a Binomial random variable, which is the channel alphabet. At finite $C$ there is a finite upper bound in $L$ (equivalently $\load$) above which only a fraction of information may be conveyed, even in the noiseless channel. By contrast an asymptotic scaling of $L\log(2)$ is obtained with Gaussian amplitude modulation, indicating non-uniform modulation patterns may provide an improvement at high SNR.

In the limit that $L \rightarrow \infty$ (with $\load=L/C$) the sum over random variables can be reexpressed through the central limit theorem as a Gaussian integral. A calculation for finite $\beta$ and SNR gives a constant value,
\begin{equation}
\lim_{L\rightarrow \infty} \SEmath_U(\beta) = - \frac{\beta}{2 \beta_0} + \frac{1}{2}\log(1 + \beta \load) +\frac{\beta(1+ \beta_0 \load)}{2\beta_0(1+\beta\load)}
\label{eq:CDMA.denseseU}\;.
\end{equation}
In the case of $\beta \rightarrow \infty$ the capacity grows logarithmically, a trend found in a numerical evaluation of (\ref{eq:CDMA.generalseU}), except at small $L$.

Figure~\ref{fig:CDMA.entropybounds} gives a numerical evaluation of expression (\ref{eq:CDMA.generalseU}) for the different sparse ensembles at $\load=1$. Adding more bits to the channel allows more information to be conveyed, but $\SEmath_U$ scales asymptotically as $L^{1-\delta}$, with $\delta$ approaching $1$ so that the total information is bounded at finite SNR. The bound in capacity given by the dense model (\ref{eq:CDMA.denseseU}) is rapidly approached, but nowhere exceeded in the sparse models. With decreasing noise the BPSK curve approaches the curve for the noiseless case even at relatively large $L$. The linear trend for noiseless Gaussian ASK is only seen when the ratio $\SNRmath/L^2$ (the distance between codewords) is large.

The factorised detection model involves determination of marginals on trees, which include only one factor each. The exact marginals calculated on the variables are equivalent to those constructed in the first iteration of BP (\ref{eq:CDMA.ktomu}).

\section{The replica method}
\label{CDMA.replica} The replica method is a mean-field method that determines typical case properties of samples from an ensemble. It can be used to calculate the ensemble average of the free energy, and through the analysis of conjugate variables many macroscopic properties can be determined. Many methods exist for calculating properties on sparse factor graphs~\cite{Mezard:BLSG,Monasson:OP,Wong:GB,Vicente:LDPC}, and standard procedures are employed. The method involves a number of standard analytic continuations and transformations, which are outlined in Appendix~\ref{app:Identities}, these are Cauchy's integral formula, the Fourier transform, the Hubbard-Stratonovich transform, and the saddle-point method.

The replica method employs the following identity with respect to the logarithm of the partition function
\begin{equation}
\<\log \partitionfunction\>_{\quenched}= \lim_{n\rightarrow 0} \frac{\partial}{\partial n} \repZ \label{eq:CDMA.logofZtoZn} \;,
\end{equation}
to solve the self-averaged free energy (\ref{eq:CDMA.safed}) in the limit of large $M$ (wide-band). The values $\quenched$ are the quenched variables and represent samples from the ensemble, $\ensemble$, of signals, codes and bits. The model assumptions are captured by the inverse temperature $\beta$ and detection priors $\vrandomfield$, but $\vrandomfield$ is left from the expressions for brevity.

The problem for general $n$ is solved through an auxiliary formulation where $n$ is integer valued. This allows a decomposition of $\partitionfunction^n$ as a discrete set of spin assemblies, conditionally independent given $\quenched$
\begin{equation}
\repZ = \prod_{\alpha=1}^n \left[\sum_{\vtau^{\alpha}}\right] \<\exp \left\lbrace - \beta \sum_{\alpha=1}^n\Ham(\vtau^\alpha) \right\rbrace \>_{\quenched} \label{eq:CDMA.repZ1}\;.
\end{equation}

\subsubsection{A site factorised form}
Taking the averages requires, in the general ensemble case, a
factorisation of the site dependencies in $A_{\mu k}$ and other quenched variables in the partition functions. The Hamiltonian is already factorised in terms of $\mu$, factorisation with respect to $k$ is achieved by transforming the square for each chip replica pair by the Hubbard-Stratonovich transform.
\begin{equation}
\begin{array}{l}
\exp \left\lbrace - \frac{\beta}{2}\left(\omega_\mu -
\frac{1}{\sqrt{C}}\sum_k A_{\mu k} \modulationsymbol_{\mu
k}(b_k - \tau^\alpha_k)\right)^2 \right\rbrace = \\
\int \rmD_{1}\lambda^\alpha \exp \left\lbrace \sqrt{- \beta} \sum_\alpha \lambda^\alpha\left( \omega_\mu + \frac{1}{\sqrt{C}} \sum_k A_{\mu k} \modulationsymbol_{\mu k} (b_k - \tau^\alpha_k)\right) \right\rbrace \label{eq:CDMA.HS}\;.
\end{array}
\end{equation}
Introducing the notation $D_x$ to mean a Gaussian weighted integral of covariance $x$
\begin{equation}
\int \rmD_x \lambda = \int_\Reals \rmd\lambda \frac{1}{\sqrt{2 \pi x}} \exp\left\lbrace - \frac{1}{2} x \lambda^2\right\rbrace\;.
\end{equation}
In this factorised form it is straightforward to exchange the order of the set of replicated integrals in $\lambda$ with the quenched averages. This leads to a factorised form with respect to $k$ and $\mu$, and all averages except $\mA$ and $\vb$ may be taken directly
\begin{equation}
\begin{array}{l c l}
\repZ &=& \sum_{\vb} P(\vb)\prod_{\alpha=1}^n
\left[\sum_{\vtau^\alpha}\right] \int \prod_\alpha \left[\rmD \lambda_{1}^\alpha\right] \prod_\mu \<\exp \left\lbrace \sqrt{-\beta} \omega_\mu \sum_\alpha \lambda_\alpha \right\rbrace \>_{\omega_\mu} \\
&\times& \<\prod_\mu \prod_k \left[\<\exp \left\lbrace \sqrt{-\beta/C} \modulationsymbol_{\mu k} \sum_\alpha \lambda_\alpha \left(b_k -\tau^\alpha_k \right) \right\rbrace \>_{\modulationsymbol_{\mu k}}\right]^{A_{\mu k}} \>_{\mA} \label{eq:CDMA.simpleaverages}\;. \end{array}
\end{equation}
The averages with respect to $\modulationsymbol_{\mu k}$ and $\omega_\mu$ are left unevaluated for generality, though the site dependent quantities can be replaced by unlabeled integration variables.

In the case of a sparse constrained connectivity matrix, $\mA$, the average does not take a straightforward form, but a series of steps outlined in Appendices~\ref{app.GAmu}-\ref{app.GAk} allows this part of the average to be taken. The calculation for the irregular ensemble (\ref{eq:CDMA.irregular}) is presented in this section, which is finally written in a form inclusive of the general case. The final line of (\ref{eq:CDMA.simpleaverages}) can be written in the form up to corrections of order $\frac{1}{K}$,
\begin{equation}
\exp\left\lbrace - M \right\rbrace \prod_\mu \exp \left\lbrace \frac{1}{K}\sum_{k=1}^K\<\exp \left\lbrace \sqrt{-\beta/C} \modulationsymbol \sum_\alpha \lambda_\alpha \left(b_k -\tau^\alpha_k \right) \right\rbrace \>_{\modulationsymbol}\right\rbrace\;.
\end{equation}
All the $\mu$ dependence is factorised subject to the integral over $\vlambda$. It is possible to exchange the order of marginalisation, first averaging over quenched parameters with chip dependence (noise and modulation patterns), before evaluating in a closed form the Gaussian integral (\ref{eq:CDMA.HS}). A marginalisation over chip connectivity forms part of the quenched averages in a general ensemble (Appendix~\ref{app.GAmu}).

The factorisation of $k$ dependence is achieved by introducing an identity function into the exponent
\begin{equation}
1= \sum_b \delta_{b,b_k} \prod_\alpha \left[\sum_{\sigma^\alpha} \delta_{\tau_k^\alpha,\sigma^\alpha} \right] \;.
\end{equation}
The dynamical variable dependence and quenched dependence on $\vb$ is captured by
\begin{equation}
1 = \int \rmd\GENOPsp \delta\left(\GENOPsp - \frac{1}{K}\sum_{k=1}^K \delta_{b_k,b} \delta_{\rvsigma,\rvtau_k}\right) \label{eq:CDMA.GENOP}\;,
\end{equation}
introduced for all $b$ and $\rvsigma$. This defines the order parameter for the sparse irregular ensemble, ensembles with constraints on variable connectivity require a small modification as demonstrated in Appendix~\ref{app:SparseMatrixAnalysis}. Overhead arrow notation used to indicate a vector with replica indices rather than site indices, and the Kronecker delta function generalised to indicate vector equivalence. The analytic continuation of the $\GENOPsp$ to the real interval $[0,1]$ implied by the integration is self-consistent with the eventual continuation of $n$ from an integer back to a real number. This identity is $\vtau_k$ and $b_k$ dependent and is introduced for all $k$ by inclusion of a trace over all states of $\GENOP$, line 1 of (\ref{eq:CDMA.simpleaverages}), up to the product over $\mu$ is then written
\begin{equation}
\int \prod_\alpha\left[\sum_{\vtau^\alpha}\right] \< \prod_{\rvsigma,b}\left[ \rmd \GENOPsp\delta\left(\GENOPsp - \frac{1}{K}\sum_{k=1}^K \delta_{b_k,b} \prod_\alpha \delta_{\sigma^\alpha,\tau^\alpha_k}\right)\right]\>_\vb \label{eq:CDMA.EntropicPart}\;.
\end{equation}

Finally the Fourier transform of the delta functions (\ref{eq:CDMA.EntropicPart}) is taken so that a form factorised both in terms of variables ($k$) and chips ($\mu$) is achieved, with the introduction of conjugate reciprocal space parameters $\GENOPspconj$,
\begin{equation}
\prod_{b,\rvsigma}\delta\left(\GENOPsp - \frac{1}{K}\sum_{k=1}^K \delta_{b_k,b} \delta_{\rvsigma,\rvtau_k}\right) \propto \int \prod_{b,\rvsigma}\left[ \rmd \GENOPspconj \exp \left\lbrace - \GENOPspconj \GENOPsp \right\rbrace \right] \prod_k \exp \left\lbrace \GENOP(b_k,\rvsigma_k)\right\rbrace\;.
\end{equation}
The quenched bit sequence dependence, and dependence on replicated dynamical variables, is now factorised in the final term allowing the marginalisation of both, and removing the site dependence from the expression. For the more general ensembles a marginalisation over variable connectivity is also required (Appendix~\ref{app.GAk}).

The free energy and its constituent parts are written in such a way as to be inclusive of all the connectivity ensembles. The difference between the ensembles are encapsulated in difference in the averages on connectivity distributions, in the Poissonian case the average can be replaced by exponential function, but not in the general case. Results henceforth are inclusive of all ensembles unless stated otherwise.

The replicated partition function can then be decomposed as an
integral over three factorised terms
\begin{equation}
\begin{array}{lcl}
 \repZ &=& {\cal N}\int \prod_{b,\vsigma} \left[\rmd \GENOPsp \rmd \GENOPspconj\right] \exp \bigg\lbrace - K \Gone(\GENOP,\beta,P(L_\mu),P(\modulationsymbol_{\mu k}))\\
&-&\! K \Gtwo(\GENOPconj,P(C_k),P(b_k))\! - K \Gthree(\GENOP,\GENOPconj)\!\bigg\rbrace \label{eq:CDMA.repZstandard}\;.
\end{array}
\end{equation}
where $\GENOPconj$ are a set of conjugate order parameters introduced in taking the Fourier transform of the identity function. For a given value of the generalised order parameter the term $\Gone$ is dependent on all parameters describing inter-variable factor node properties, and can also be called the energetic part of the free energy. It can be written
\begin{equation}
 \load \Gone = - \log \<\prod_{l=1}^{l_e} \left[\sum_{b_l,\rvsigma_l} \GENOP_{b_l}(\rvsigma_l) \right] \localreplicaprobability_{{l_e}} \>_{{l_e}} \label{CDMA.f_e}\;,
\end{equation}
with
\begin{equation}
\localreplicaprobability_{l_e} = \<\exp\left\lbrace -\frac{\beta}{2} \sum_\alpha \left(\omega + \frac{1}{\sqrt{C}} \sum_{l=1}^{l_e} \modulationsymbol_l (b_l -\sigma_l^\alpha) \right)^2\right\rbrace \>_{\omega,\left\lbrace \modulationsymbol_l \right\rbrace} \label{eq:CDMA.Zchip}\;,
\end{equation}
where the averages are with respect to the true marginal chip noise distribution, the modulation patterns on ${l_e}$ user chip pairs, and the marginal chip connectivity distribution.

The term $\Gtwo$ describes properties of the ensemble attached to sites and takes a form given by
\begin{equation}
\Gtwo = - \log \sum_{\rvsigma}\<\<\left[\GENOPspconj\right]^{c_f} \>_{{c_f}} \>_{b}\label{eq:CDMA.G_2}\;.
\end{equation}
Where $\GENOPconj$ is chosen to be an extensive measure, scaling linearly with $K$. This can be determined retrospectively from the saddle-point equations. The final term $\Gthree$ generates a coupling of these effects in the mean field model. It is determined as
\begin{equation}
\Gthree = C \sum_{b,\rvsigma}\GENOPspconj\GENOPsp \label{eq:CDMA.G_3}\;.
\end{equation}
The normalisation constant, ${\cal N}$ has a contribution from the normalisation of quenched averages, and several constant terms dropped for convenience in the calculation, these do not effect thermodynamic behaviour. The non-trivial part arising from an average over a generic connectivity distribution is calculated in Appendix~\ref{app.normalisation}.

The method has replaced the many spin state problem with quenched couplings by a site factorised form in a complicated state $\GENOPsp$. The generalised order parameter, $\GENOPsp$, describes a lattice gas problem~\cite{Monasson:OP}, determining the occupation density in $[0,1]$ for each point on the lattice defined by $P(b,\vsigma)$ is now the challenge. Since there is no topology each site is effectively unlabeled so only the distribution of occupation densities is meaningful, the distribution of densities is required to be invariant with respect to labeling of replica. A further feature of the partition sum can be utilised to find the correct density distribution dominating thermodynamics, which is the exponential dependence on $M$. Introducing the large $M$ limit for the purpose of calculating the maxima, and assuming $n$ to be finite, the expression will be determined by one, or many, global maxima, which can be evaluated through the saddle-point method.

The lattice gas problem is analysed with $n$ assumed to take an arbitrary integer value. For the Poissonian ensemble the dimension of the space of order parameters ($\GENOPsp,\GENOPspconj$) is $2^{n+1}$, for which an analytic continuation is assumed. The order parameter and its conjugate for general ensembles (\ref{app.GAk}) is defined on the complex plane. In each case the self averaged free energy can only be evaluated by approximation at the saddle-point, which becomes a correct description in the limit of large $K$ (Appendix \ref{app:saddle-pointmethod}).

\subsection{Saddle-point equations}
The saddle-point method determines the replicated partition sum in terms of only one or several extremal values of the integral, corresponding to real-valued saddle-points. The order parameters (integration variables) describing the relevant extrema are assumed not to be on the boundaries of the integration range. Furthermore the search is restricted to real valued integration variables, a justification of this is provided in Appendix~\ref{app:ConjugateFields}. Denoting the exponent (\ref{eq:CDMA.repZstandard}) as ${\tilde f}$, the approximation made is
\begin{equation}
\lim_{K\rightarrow \infty} \frac{1}{K} \log \int \prod_{b,\rvsigma} \left[\rmd\GENOPspconj \rmd\GENOPsp\right] \exp \left\lbrace - K {\tilde f}(\GENOP,\GENOPconj) \right\rbrace = {\tilde f}(\GENOPconj^*,\GENOP^*)\;,
\end{equation}
where $\{\GENOPconj^*,\GENOP^*\}$ are the order parameter values that extremise the exponent. At this point the first order functional derivatives with respect to the order parameters vanish (assuming a maxima exists away from the boundary), the breadth of the maxima (second derivatives) do not contribute at leading order in $K$.

Determination of the saddle-point is achieved by finding the fixed point of the functional derivatives of the exponent. Taking the partial functional derivative with respect to $\GENOPconj$
\begin{equation}
\GENOP^*_{b}(\rvsigma) \propto
\<[\GENOPconj^*_{b}(\rvsigma)]^{c_e} \>_{b,c_e} \label{eq:CDMA.saddle1}\;,
\end{equation}
where the average in $c_e$ is with respect to the marginal excess user connectivity distribution. The derivative with respect to $\GENOP$ gives
\begin{equation}
\GENOPconj^*_{b}(\rvsigma) \propto \<\prod_{l=1}^{{l_e}}\left[\sum_{b_l,\rvsigma_l} \GENOP^*_{b_l}(\rvsigma_l)\right] \delta_{b,b_{L'+1}}\delta_{\rvsigma,\rvsigma_{L'+1}} \localreplicaprobability_{{l_e}+1} \>_{{l_e}}\label{eq:CDMA.saddle2}\;,
\end{equation}
where the average in ${l_e}$ is with respect to the excess chip degree distribution of the ensemble.

In addition, the normalisations for the order parameters must be determined. One constraint on the normalisation is provided by the pair of saddle-point equations, which can be applied to the case that the order parameters are simply constants. The second criteria is either directly from the definition (\ref{eq:CDMA.GENOP}) or by a definition of the free energy density in the limit $\beta\rightarrow 0$, which is $\log(2)$, the entropy with no constraints.

A solution to the saddle-point equations, defines an extrema or point of inflexion in the parameter space. The correct extremum can be labeled $\GENOP^*$ and $\GENOPconj^*$ (if unique). Instability of the fixed point can be tested by examining the Hessian, which is found from the second order functional derivatives. The fixed points might be determined as locally stable by considering the linear stability of the saddle-point equations. To sufficiently test local stabilities it is necessary to consider breaking of symmetries within a particular model for replica correlations, and also latitudinal stability -- the possibility of instability towards a more inclusive model of replica correlations.

Stable solutions are presumed to exist under some parameterisation of $\GENOP,\GENOPconj$, and a solution within the subspace invariant under replica relabeling is developed, the replica symmetric solution. The validity of the solution is considered retrospectively in section~\ref{CDMA.RS}, and in the context of a local stability analysis section~\ref{CDMA.stability}.

Evaluating the identity (\ref{eq:CDMA.logofZtoZn}) at the saddle-point the self averaged free energy density up to constant terms may be written
(\ref{eq:CDMA.safed})
\begin{equation}
\beta \safed = \lim_{n \rightarrow 0} \frac{\partial}{\partial n} \Extr_{\GENOP,\GENOPconj}\left\lbrace - K \left( \Gone(n,\GENOPconj,\GENOP) + \Gtwo(n,\GENOPconj) + \Gthree(n,\GENOP) \right) \right\rbrace\label{eq:CDMA.freeenegyreplica}\;,
\end{equation}
where the extrema requires a solution of the saddle-point equations. By taking instead a limit $n\rightarrow 1$ in (\ref{eq:CDMA.logofZtoZn}) it is possible to generate an annealed approximation. In the case of CDMA the annealed approximation is at almost all interesting parameterisations inaccurate, lower bounding the free energy.

\subsection{Replica symmetric solution}
\label{CDMA.RS}

A tractable form for the saddle-point equations is attained using the RS assumption. The invariance of the order parameters under relabeling implies order parameters dependent only on the sum of replicas $\sum_\alpha \sigma^\alpha$. The order parameters are then simplified as functions of $\RSOP_b(h)$, the dependency on be can take several forms. It is convenient to consider a symmetric and antisymmetric parts with respect to $b$, which provides a general form given $b$
\begin{equation}
\GENOP_{b}(\rvsigma) = \frac{1}{2} \int \rmd h
\pi(h)\frac{\exp\left\lbrace b h \sum_\alpha
\sigma^\alpha\right\rbrace}{2^n\cosh^n \left\lbrace h
\right\rbrace} + b \frac{1}{2}\int \rmd h_A
\pi_A(h_A)\frac{\exp\left\lbrace b h_A \sum_\alpha
\sigma^\alpha\right\rbrace}{2^n\cosh^n \left\lbrace
h_A\right\rbrace}\label{eq:CDMA.RSOP}\;,
\end{equation}
with similarly structured definitions for the conjugate parameters, characterised by distributions $\RSOPconj,\RSOPconj_A$. Odd moments are expected to align macroscopically with the source bit sequence due to the asymmetry of the marginal fields (\ref{eq:CDMA.marginals}), but no other directions are preferred with respect to the Hamiltonian.

For symmetric modulation ensembles (\ref{eq:CDMA.BPSK}) the dependence on the bit sequences can be gauged from the order parameter so that $\pi_A(h_A)=0$. The gauging of the distribution on modulation patterns $\modulationsymbol_{\mu k}$ to bit $b_k$ removes all $\vb$ dependence in the couplings and hence the simplified order parameter description is sufficient, even where the true and assumed priors on source bits are not uniform.

Asymmetric modulation distributions do not allow the same gauging of bit sequences in the free energy. However, a variational approach to the problem might assume $\pi_A(h_A)=0$, call this the 'bit symmetric assumption' in the case of asymmetric modulation patterns, unlike symmetric patterns where it is exact. If this is assumed then the dynamics of the saddle-point equations become identical to a symmetric code with an equivalent amplitude distribution; the free energy determined for the unmodulated code becomes identical to the BPSK code.

The free energy, evaluated according to (\ref{eq:CDMA.freeenegyreplica}) with the RS order parameters becomes, taking only the relevant coefficients in $n$ the chip-centric term in the free energy is
\begin{equation}
\left. \frac{\partial}{\partial n} \right|_{n=0} \Gone^{RS}(n) \!=\! - \frac{1}{\load}\<\int \!\prod_{l=1}^{l_e} \left[\rmd h_l \RSOP(h_l)\right] \<\log(\partitionfunction^{RS}) \>_{\omega,\left\lbrace \modulationsymbol_l\right\rbrace}\>_{l_e} + C \int \rmd h \RSOP(h) \log (2\cosh h) \label{eq:CDMA.G1}\;,
\end{equation}
defining a local cavity-type partition sum as
\begin{equation}
\partitionfunction^{RS} = \prod_{l=1}^{l_e} \sum_{\tau_l} \exp(h_l \tau_l)\exp\left\lbrace -\frac{\beta}{2} \sum_\alpha \left(\omega + \sum_{l=1}^{l_e} \frac{1}{\sqrt{C}} \modulationsymbol_l (1-\sigma_l) \right)^2\right\rbrace \label{eq:CDMA.singlenodeinter}\;,
\end{equation}
where $b_l$ can be taken gauged to the modulation pattern distribution. The user-centric term in the free energy is
\begin{equation}
\left. \frac{\partial}{\partial n}\right|_{n=0} \Gtwo^{RS}(n) = \int \prod_{c=1}^C \rmd u_c \RSOPconj(u_c) \left\langle \log\left(2 \cosh\left(\sum_{c=1}^{c_f} u_c \right)\right) \right\rangle_{{c_f},\modulationsymbol}\!
- \!C \int \rmd u \RSOPconj(u) \log(2\cosh u) \;, \label{eq:CDMA.G2}
\end{equation}
and finally the coupling term is
\begin{equation}
\left. \frac{\partial}{\partial n}\right|_{n=0} \Gthree^{RS}(n)\!=\! -C\int \!\rmd h \RSOP(h)\! \rmd u \RSOPconj(u) \log \left(1 + \tanh(u)\tanh(h)\right) \label{eq:CDMA.G3}\;.
\end{equation}

The functions $\left\lbrace \RSOP,\RSOPconj\right\rbrace$ are chosen so as to extremise the free energy, the RS definition significantly restricts the search space making the problem tractable. The saddle-point equation (\ref{eq:CDMA.saddle2}) becomes
\begin{equation}
\RSOPconj(u) = \<\int \prod_{l=1}^{{l_e}} \left[\rmd h_l \RSOP(h_l) \right]\<\delta\left(u - \frac{1}{2}\sum_\tau \tau \log(\partitionfunction^{RS}_{0}(\tau)) \right)\>_{\quenched_{{l_e}}} \>_{{l_e}} \;, \label{eq:CDMA.cavitybias}
\end{equation}
letting $\quenched_{{l_e}}$ denote the integration variables $\omega, \left\lbrace \modulationsymbol_l \right\rbrace$ relevant in evaluating the Hamiltonian for a chip attached to an edge of excess connectivity ${l_e}$. The local partition sum is defined
\begin{equation}
\partitionfunction^{RS}_0( \tau_0) = \prod_{i=1}^{{l_e}}\left[\sum_{\tau_l} \exp \{h_l \tau_l \} \right]\exp\left\lbrace -\frac{\beta}{2} \left(\omega + \sum_{l=0}^{{l_e}}\frac{1}{\sqrt{C}} \modulationsymbol_l (1- \tau_l) \right)^2 \right\rbrace \label{eq:CDMA.saddlepartitionsum}\;,
\end{equation}
and the simpler saddle-point equation (\ref{eq:CDMA.saddle1}) becomes
\begin{equation}
\RSOP(h) = \<\int \prod_{c=1}^{c_e}\left[\rmd u_c \RSOPconj(u_c) \right] \delta\left(h - \sum_{c=1}^{c_e} u_c \right)\>_{c_e} \label{eq:CDMA.cavityfield}\;.
\end{equation}
The average is with respect to the excess variable degree distribution $c_e$.

A close relation is apparent between the form of updates (\ref{eq:CDMA.cavitybias}),(\ref{eq:CDMA.cavityfield}) and BP equations (\ref{eq:CDMA.mutok})(\ref{eq:CDMA.ktomu}). The order parameters describing the RS saddle-point also describe the distribution of log-likelihood ratios for a fixed point, or steady state (if non-convergent), of iterated BP equations in a sparse random graph once the limit $M\rightarrow\infty$ is taken. Furthermore a quantity describing the distribution of log-posterior ratios, equivalent quantities to (\ref{eq:CDMA.H}), is apparent
\begin{equation}
P(H) = \lim_{K \rightarrow \infty} \frac{1}{K}\<\sum_{k=1}^K \delta(H-\atanh\<b_k \tau_k\>)\>_{\quenched}= \int \<\prod_{c=1}^{{c_f}}\left[\rmd u_c \RSOPconj(u_c)\right] \delta\left( H - \sum_{c=1}^{c_f} u_c\right)\>_{c_f} \label{eq:CDMA.PH}\;.
\end{equation}
Quantities relying on the overlap such as the BER may be calculated from this, this distribution is a collection of real moments which can be established by conjugate fields, as in Appendix \ref{app:ConjugateFields}.

\subsection{Numerical evaluation}
Population dynamics~\cite{Mezard:BLSG} is used to solve the RS saddle-point equations. A pair of order parameter histograms, containing $N$ points, are used to represent the functions $\RSOP$, $\RSOPconj$ (\ref{eq:CDMA.RSOP}),
\begin{equation}
\RSOP\rightarrow
\Histogram = \left\lbrace x_1,x_2,\ldots x_N \right\rbrace\;; \qquad \RSOPconj\rightarrow {\hat \Histogram} =\left\lbrace x_1,x_2,\ldots x_N\right\rbrace \label{eq:CDMA.Histogram}\;.
\end{equation}
The histogram $\Histogram$ is initialised in a random state and the saddle-point equations (\ref{eq:CDMA.cavityfield}) (\ref{eq:CDMA.cavitybias}) are iterated according to samples from the histograms rather than integrals over the distributions. In each iteration the other integration and summation parameters are sampled from the corresponding marginal distributions, alongside fields from the order parameter distribution.

It is useful to constrain fluctuations in the numerical evaluation, by sampling according to micro-canonical distributions; fluctuations in the mean connectivity and noise variance are then order $\frac{1}{N}$, rather than $\frac{1}{\sqrt{N}}$ for independent sampling. Fluctuations can also be reduced by sampling uniformly from $\Histogram$ and ${\hat W}$ in the updates. Sampling in most case was chosen not to preserve perfect symmetries or create artificial structure in the sample points, fluctuations are introduced or numerically unavoidable, and linearly unstable distributions are not found. In some cases exact scalable methods which do preserve symmetries, such as Gaussian quadrature were advantageous but used with caution~\cite{Press:NRC}.

The update scheme employed in solving the saddle-point equations involved parallel updates of all variables in each of the histograms $\{\RSOP,\RSOPconj\}$. Parallel updates create an artificial oscillation in models with anti-ferromagnetic couplings, but this is not a significant effect in the case of sparse CDMA. Histograms of $10000$ points under these conditions provided robust resolution of the fixed point distributions.

Convergence of population dynamics was assumed when two co-evolving histograms, initialised in antipodal states, converged to a unique solution. These two histograms are known to converge towards the unique solution, where one exists, from opposite directions in state space, and their convergence may be used as a halting criteria for the recursions, as well as to test for multiple stable solutions. In the case that they converge to different solutions the solution converged to from the ferromagnetic initial condition (FIC) is termed a {\it good} solution - in the sense that it is of low bit error rate, and that arrived at from random/paramagnetic initial state (PIC) is termed a {\it bad} solution. In the detection problem one cannot in general start with prior knowledge of the state -- knowing the exact solution would of course makes the decoding redundant, although limited prior knowledge ($\randomfield_k \neq 0$) could be an interesting case. It is reasonable to expect that dynamical features observed for PIC may be more characteristic of practical detection methods such as BP, which must start from an unbiased situation.

\subsection{Stability analysis}
\label{CDMA.stability}

\subsubsection{Stability against symmetry breaking in $\vb$}
\label{CDMA.stabilityb}
The stability of the bit symmetric assumption for asymmetric modulation patterns can be tested by considering an arbitrary perturbation in $\GENOPsp$ away from a symmetric description. In evolving the perturbations through (\ref{eq:CDMA.saddle1}), the perturbations in the order parameters are determined by a simple sum of the perturbation in the $\GENOP$. The second recursion from (\ref{eq:CDMA.saddle2}) is more involved giving, after gauging of the summation variables to $b_k$, the equation
\begin{equation}
\sum_b b \delta\GENOPconj_{b}(\rvsigma) = \sum_b b \< \sum_{k=1}^{{l_e}} \sum_{b_k,\rvsigma_k} \delta\GENOPconj_{b_k}(\rvsigma_k) \prod_{l=1\setminus k}^{{l_e}}\left[\sum_{b_l,\rvsigma_l} \GENOP^*_{b_l}(\rvsigma_l)\right] \delta_{b,b_{L'+1}}\delta_{\rvsigma,\rvsigma_{L'+1}} \localreplicaprobability_{{l_e}+1} \>_{{l_e}}\label{eq:CDMA.stabb}\;,
\end{equation}
describing the fluctuation in the antisymmetric part $\pi_A$ when ${l_e}>0$, the case ${l_e}=0$ gives a contribution of zero in the average. The value of this term depends on whether the function $\< \>$ is an even function of $b$, and if it is odd, whether the largest eigenvalue exceeds one.

In the case of a symmetric codes the modulation patterns $\modulationsymbol_l$ (\ref{eq:CDMA.Zchip}) may be gauged to the bits, so that the final term $[\cdots]$ is an even function of $b$, and the perturbations are zero. This applies also to non-linear terms in the expansion and the bit symmetric solution is correct. In the unmodulated code the term is also even with respect to $b$, and hence the fluctuations again go uniformly to zero and the bit symmetric solution is again locally stable. The 'bit symmetric' solution is therefore locally a valid solution. However, if one considers non-linear perturbations then there are couplings between the fluctuations which are $b$ dependent in the order parameter. This may indicates a source of difference between the susceptibility of the modulated codes, and unmodulated codes, with unmodulated codes having susceptibility properties dependent on $RSOP_A$.

\subsubsection{The spin glass susceptibility}

A necessary criteria for the validity of the replica symmetric assumption for all ensembles is that the spin glass susceptibility is finite, as discussed in Appendix~\ref{app:physicsfromconjfields}, which implies a single pure state description~\cite{Mezard:SGT,Fischer:SG}. In the case of instability towards replica symmetry breaking, strong correlations are manifested as microscopic instabilities in the RS saddle-point equation indicating a failure of the pure state criteria.

A convenient way to test this criteria is through the cavity method framework~\cite{Rivoire:GM}. This formulation transforms the direct evaluation of the spin glass susceptibility into a stability test on the form of the order parameter in the cavity (saddle-point) equations.

An equivalent test of stability can be developed by considering a re-weighted connected correlation function. A random external field is applied which is proportional to the coupling strengths of each variable. This description increases the weight from highly connected variables by a constant factor, but does not exclude any non-zero contributions. The external field is defined
\begin{equation}
\sqrt{z}\sum_k \zeta_k \left(\sum_\mu s_{\mu k} \right) \tau_k \label{eq:CDMA.conjfield}\;,
\end{equation}
where $\zeta_k=\pm 1$ is used as a (quenched) random modulation of the external field. Taking $z$ as the positive external field term then some physical quantities may be calculated, as demonstrated in Appendix~\ref{app:ConjugateFields}.

In calculation of the free energy the average over $A_{\mu k}$ must now include the term (\ref{eq:CDMA.conjfield}), after taking the sparse connectivity average the $k$ dependence must be extracted (\ref{eq:CDMA.GENOP}), but this now includes a dependence on $\zeta_k$. This dependence takes the same form as the dependence on $b_k$, with the definition of a pair of order parameters, the analogue to (\ref{eq:CDMA.GENOP}) is
\begin{equation}
1 = \int \rmd\GENOP_{b,\zeta'}(\rvsigma) \delta\left(\GENOP_{b,\zeta'}(\rvsigma) - \frac{1}{K}\sum_{k=1}^K \delta_{b_k,b} \delta_{\zeta_k,\zeta'} \delta_{\rvsigma,\rvtau_k}\right) \label{eq:CDMA.GENOPzeta}\;.
\end{equation}
For simplicity the case of a symmetric code is considered. With this choice the $\vb$ dependence can be gauged from the Hamiltonian. Only the symmetric order parameter with respect to $b$ is required to describe equilibrium properties. However, it remains necessary to define an order parameter with $\vzeta$ dependence, the dependence being of an equivalent form to that on $\vb$ (\ref{eq:CDMA.GENOP}).

This dependence is processed through to the free energy and saddle-point equations in the same way as the dependence on $b$ in the original derivation, with a quenched average over $\vzeta$ in $\Gtwo$ (\ref{eq:CDMA.G2}) in place of the average over $b$, and replacement of the summation variable $b$ by a summation variable $\zeta'$ in (\ref{eq:CDMA.G2}) (\ref{eq:CDMA.G1}). In (\ref{eq:CDMA.G1}) there is an additional energetic term, which has the form $ \exp \sqrt{z} \sum \zeta'_l \tau_l$, but this can be finally taken to be $1$ in the small external field limit. The term $\Gone$ is modified to
\begin{equation}
\load \Gone = - \log \<\prod_{l=1}^{l_e} \left[\sum_{b_l,\zeta'_l,\rvsigma_l} \GENOP_{b_l,\zeta'_l}(\rvsigma_l) \exp \left\lbrace -\beta \sqrt{z} \zeta'_l \sum_\alpha \sigma^\alpha_l\right\rbrace \right] \localreplicaprobability_{{l_e}} \>_{{l_e}} \label{eq:CDMA.f_ezeta}\;,
\end{equation}
which is where the $z$ dependence is preserved.

A symmetric and antisymmetric decomposition of the order parameter is possible,
\begin{equation}
\GENOP_{b,\zeta'}(\rvsigma) = \frac{1}{2}\int \rmd h \RSOP(h)\frac{\exp\left\lbrace b h \sum_\alpha \sigma^\alpha\right\rbrace}{2^n\cosh^n \left\lbrace h\right\rbrace} + \frac{1}{2} \zeta' \int \rmd h_A \RSOP_A(h_A)\frac{\exp\left\lbrace b h_A \sum_\alpha \sigma^\alpha\right\rbrace}{2^n\cosh^n \left\lbrace h_A \right\rbrace} \label{eq:CDMA.RSOPstab}\;,
\end{equation}
assuming a symmetric dependence in $b$, and a similar decomposition is possible in the conjugate order parameter. In the case that $z=0$ there can be no dependence on $\zeta$ and hence $\pi_A=0$ is the correct solution (assuming the RS assumption to be otherwise correct). If the solution is stable it is necessary that $\pi_A$ converges to zero in the small external field limit, which can be tested by a linear stability analysis.

%
%
A symmetric distribution for the quenched parameter $\vzeta=\{\vb,-\vb\}$ is a convenient choice, and allows a test of second order instabilities associated with the susceptibility. Different types of susceptibility are developed in Appendix~\ref{app:ConjugateFields}. The derivative of the free energy (\ref{eq:CDMA.f_ezeta}) with respect to the external field variance $z$, evaluated in the limit $z \rightarrow 0$ can not be well defined unless the anti-symmetric part $\RSOP_A$ tends to zero, which is the known solution for $z=0$. The stability of the description towards $\RSOP_A\neq 0$ is tested by considering the stability of $\RSOP$, which in the context of population dynamics corresponds to linear stability of the elements of the Histogram under mapping.

\subsubsection{Stability equations}

The perturbation on the order parameter is assumed to take a restricted form with each point in the distribution $\RSOP$ subject to a deviation described by some variance, the mean perturbation is zero by symmetry of the problem when the quenched average is taken with respect to $\zeta$. These can be defined for each point in the histograms $\chi^2_h$, $\chi^2_u$ (\ref{eq:CDMA.Histogram}). If some moments of the perturbed distribution are unstable this is observed in an instability of the mean value of the variance
\begin{equation}
\< \chi^2_h \> = \int \rmd h  W(h) \chi^2_h \;.
\end{equation}

The recursion of stability measures in the RS equations can be determined by expanding (\ref{eq:CDMA.cavityfield}) to linear order about the fixed point, for a particular sample of the excess connectivity $c_e$, and a corresponding number of points from $\RSOPconj$
\begin{equation}
{\chi^2_h}^{(t+1)} = \< \prod_{c=1}^{c_e} \int \rmd u_c \RSOPconj(u_c) \delta(h- \sum_{k} u_k)\sum_{c=1}^{c_e} {\chi^2_{u_c}}^{(t)} \>_{c_e} \label{eq:CDMA.stability1}\;,
\end{equation}
and in the linear expansion of equation (\ref{eq:CDMA.cavitybias}), the rule depends on the excess connectivity sample $c_e$, a corresponding set of samples from $\RSOP$, and other quantities analogous to quenched disorder
\begin{equation}
{\chi^2_u}^{(t)} = \left\lbrace \begin{array}{l r} 0 & \hbox{if ${l_e}=0$} \;;\\
\prod_{k=1}^{{l_e}} \int \rmd h_{k} \RSOP(h_k) \delta\left(u - u^{SP} \right) \sum_{l=1}^{{l_e}} {\chi^2_{u_l}}^{(t)} \left( \left. \frac{\partial}{\partial h_l}\right|_{h_l=0} u^{SP}\right)^2 & \hbox{if ${l_e}>0$} \;;\label{eq:CDMA.stability2}
\end{array} \right.
\end{equation}
where the derivative is calculated from
\begin{equation}
u^{SP} = \frac{1}{2}\sum_\tau \tau \log \left(\partitionfunction^{RS}_{0}(\tau)\right)\;,
\end{equation}
using (\ref{eq:CDMA.saddlepartitionsum}).

It is possible to take $\vzeta$ as a constant vector in which case the nature of the instability tested is a linear one comparable to that examined earlier in this section. Although the linear stability of the bit symmetric RS description is correct with respect to a linear perturbation, it may be that some simple non-linear instability may be relevant. There are two potential types of second order instability, one is with respect to a simple symmetry breaking, implying $\pi_A \neq 0$ for the unmodulated code, and one which describes a more complicated Replica Symmetry Breaking (RSB) instability. Only the latter type of instability is relevant for symmetric codes.

\subsubsection{Stability population dynamics}

The variances are expected either to decay exponentially or to grow exponentially with iterations at the fixed point, a description of the stability is determined through the decay exponent. A determination of the exponent can be achieved by an iteration of the pair of equations in parallel with the RS equations. Numerically, this can be achieved by considering histograms of squared linear fluctuations
\begin{equation}
W^S=\left\lbrace \chi^2_1, \ldots, \chi^2_N\right\rbrace\;;\qquad {\hat W}^S=\left\lbrace {\hat \chi}^2_1, \ldots, {\hat \chi^2}_N \right\rbrace \label{eq:CDMA.stabilityHistogram}\;;
\end{equation}
associated (by label) to each of the order parameter histogram points (\ref{eq:CDMA.Histogram}). The discretisation of the histograms causes a replacement of the integrals by sums over samples in (\ref{eq:CDMA.stability1})-(\ref{eq:CDMA.stability2}), for self-consistency the same set of quenched variables applies to equations (\ref{eq:CDMA.cavitybias}) and (\ref{eq:CDMA.cavityfield}), the order parameter and fluctuation histograms are updated concurrently.

Each of the square linear fluctuations is initialised independently as the square of a value drawn from a Normal distribution -- but results were not sensitive to reasonable initial conditions. The iteration provides an estimate to the stability of the pure state description even where the numerical resolution of the saddle-point, or convergence to this point, is not complete in the order parameter histogram.

\subsection{Replica symmetry and the phase space}

At the Nishimori temperature the RS solution is guaranteed to describe correctly the thermodynamically dominant states, the solution is a connected one as indicated by the analysis of Appendix~\ref{app:Nishimori}. Across a range of parameters solutions of two types are found: one corresponding to a locally stable bad solution (bad decoding performance, $\BERmath\gtrsim 10^{-2}$) and one to a locally stable good solution (good decoding performance).

In many regimes the saddle-point equations produce unique solutions. The discussion of good and bad solutions is primarily in the context of meta-stability, but the distinction is also useful in the case of unique solutions. A bad solution has a behaviour characteristic of a liquid/paramagnetic phase in many ways, whereas a good solution is characteristic of an ordered phase, with increasing SNR the characteristics of the good solution is realised through either a continuous or discontinuous transition. In the continuous case many properties, such as the correlation length scales determined by the local stability analysis, undergo changes in behaviour at $\SNRmath\sim 6 $dB, and the more rigid configuration is realised at high SNR. In the discontinuous case (at intermediate SNR and high MAI), both a good and bad solution exist, but one dominates thermodynamics. The locally stable but subdominant (metastable) solution is irrelevant to equilibrium properties. However, in terms of dynamics or local sampling the bad solution can be dominant even as a metastable solution due to ergodicity breaking.

At and above the Nishimori temperature the good or bad thermodynamic solution is guaranteed to have simple phase space structures described by RS, this conclusion and some thermodynamic properties can be calculated without the replica trick or cavity method~\cite{Nishimori:CO}, but the results do not extend to any metastable solutions.

The RS solution obtained at equilibrium appears sufficient to describe the good equilibrium and metastable solutions. These are the local solutions to the free energy that correspond to states clustered about the encoded bit sequence $\vb$. In this case the phase is connected in state space, and so we expect the dynamics of the system to be relatively simple, so that the phase space can be explored by local sampling methods such monte-carlo. BP will be locally stable in the vicinity of this solution, in the absence of competing local minima convergence towards this solution may be expected in typical samples. This solution to the free energy exists when SNR is sufficiently large.

By contrast we expect there to also be a bad equilibrium solution when SNR is small. The marginal field term in the Hamiltonian means the overlap with sent bits is never zero, but we expect there to be a suboptimal ferromagnetic solution which is also connected in state space, and that has similar properties in terms of BP and sampling.

Finally at high MAI and intermediate noise there may be a bad metastable solution. The bad metastable solution emerges continuously from the bad equilibrium solution with increasing SNR and so will be characterised by a connected phase space for some parameters. However, as the noise decreases we might expect this solution to become fragmented and the RS metastable solution to become unstable. An indication of the failure of RS is the negative entropy in some metastable solutions, which is not viable. The problem of negative entropy is resolved by restricting the analysis to a connected phase space, but one in which the entropy remains $0$ (frozen). It is not uncommon for systems with simple connected phase spaces to exhibit negative entropy when the RS ansatz is applied under the assumption of extensive entropy~\cite{Gross:SS}, as is employed in the calculation. A result without negative entropy can be formulated by a minor variation on the RS approach called frozen RSB, which effectively re-scales the temperature. However, it is not certain that this solution will be correct without a local stability analysis towards other forms of RSB, in many other systems negative entropy is one indicator of a failure of the connected phase space assumption~\cite{Mezard:SGT}.

In the bad metastable state, and also in the bad equilibrium phase away from the Nishimori temperature ($\beta>1$), the connected description is possibly incorrect even where the entropy is positive; an RSB formalism may be applicable. The good solution is likely to be well described by RS at all temperatures, since it is an intuitive state clustered around the encoded bit sequence. For $\beta>1$ the RS approximation produces a variational approximation to the thermodynamic behavior, which must be tested against RSB. The RS approximation may also describe exactly the metastable states in some regimes, but this is not the case in general.

The hypothesis of a connected state described by the RS treatment has consequences for dynamics, as do the various hypotheses on the nature of RSB, should it occur either in a search for the ground state~\cite{Krzakala:GS}, or at some intermediate temperature. On typical samples, BP may be expected to converge for parameterisations described by RS in the large system limit. However, in small samples finite size effects may dominant behaviour, so that in the absence of a scaling analysis conclusions cannot be drawn directly from BP simulation results. In cases where BP is unstable, due to RSB or finite size effects, BP may still reach a steady state of the dynamics that is strongly correlated with an optimal solutions, and so remains useful in estimation when combined with a suitable heuristics.

\section{Results for specific ensembles}
\label{CDMA.results}

\subsection{Equilibrium behaviour}

Results are presented here only for the canonical case of BPSK at the Nishimori temperature. This guarantees that the RS solution is thermodynamically dominant, the energy takes a constant value and hence the entropy is affine to the free energy. A comparative lower bound is plotted for BER in some figures, the single user Gaussian channel (SUG), and the results alongside for the equivalently loaded densely spread
ensemble~\cite{Kabashima:SMA}.

Computer resources restrict the cases studied in detail to SNR below about 10dB, and small $L$. In particular, at high SNR a majority of the histogram is concentrated at magnetisations\footnote{magnetisations rather than log ratios are used in the results presented from~\cite{Raymond:SS}, both methods suffer from similar finite size effects, although it is slightly easier to approach the zero temperature, or high SNR limit in the latter case} near one, where finite precision problems are encountered. Systems with large but finite $L$ are known, in any case, to converge quickly to the limiting $L\rightarrow \infty$ result.

\subsubsection{Performance measures}

Several different measures are calculated from the converged histograms $\left\lbrace W,{\hat W}\right\rbrace$, indicating the performance of sparsely-spread CDMA. Sampling from the converged histograms a representative sample of log-posterior ratios (\ref{eq:CDMA.PH}) is found, from which BER is calculated
\begin{equation}
 \BERmath = \int \rmd H P(H) \sign(H)\;.
\end{equation}
Spectral efficiency is calculated along with an ad-hoc measure of the strength of correlations to complement the stability measure $\SEmath_L$, which is a lower bound to the true spectral efficiency,
\begin{equation}
\SEmath_L = \load \left(\log(2) - \int \rmd H P(H) \left[\log(2\cosh(H)) - H \tanh(H) \right] \right) \;.
\end{equation}
The lower bound is constructed by testing the entropy in a model of spins conditionally independent given $P(H)$ the distribution of global magnetisations.

Finally multi-user efficiency is shown in figure \ref{fig:CDMA.normalregimes}. The SNR required to achieve a given BER, is compared to the SNR required to achieve the same rate in the absence of MAI
\begin{equation}
 \MUEmath = \frac{1}{\load \SNRmath}\left[ \erfc^{-1}(\BERmath) \right]^2\;.
\end{equation}
where $\erfc$ is the complementary error function for a Gaussian of variance $1$, and BER is the ensemble thermodynamic result for some SNR. Multi-user efficiency is a measure of power efficiency on the interval $[0,1]$.

The validity of the RS assumption is determined by a stability exponent, determining if the perturbations grow or decay in successive (parallel) updates
\begin{equation}
\lambda^{(t)} = \log \frac{\sum_{i=1}^N {\chi_i^2}^{(t)}}{\sum_{i=1}^N {\chi_i^2}^{(t-1)}}\label{eq:CDMA.lambdaRS}\;.
\end{equation}
It is convenient to renormalise the perturbations at each time step to reduce finite size effects.

\subsection{Single solution regimes}

\begin{figure}[htb]
\includegraphics[width=\linewidth]{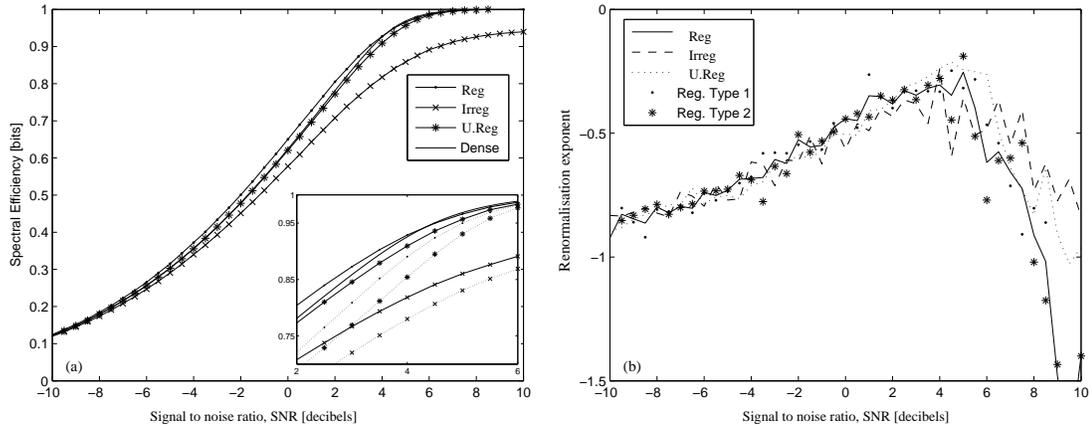}
\caption[Performance of sparse CDMA, single solution regime.]{\label{fig:CDMA.normalregimes} Results for three different connectivity ensembles of section~\ref{CDMA.connectivityensembles} with $C\!\!:\!\!L\!=\!3\!\!:\!\!3$ are shown of Reg(ular), U(ser).Reg(ular) and Irreg(ular) connectivity. All data presented on the basis of 100 runs, error bars are omitted, these are negligible by comparison with symbol size in figure (a), and characterised by the smoothness of the curves in figure (b). (a) The spectral efficiency [$\solid$] indicates a smooth trend, approaching an upper bound of $1$ bit at large SNR in all cases except the irregular code which is limited by a fraction of disconnected users. The gap between $\SEmath$ [$\solid$] and the lower bound $\SEmath_L$ [$\dotted$] is shown in the inset and is everywhere small, indicating weak correlations between variables. (b) The three lines indicating the different ensembles are everywhere noisy, but indicate a comparable trend with a negative stability exponent. All ensembles show a cusp in a range of SNR, but with all solutions being local stable. The two marker types $[*,\cdot]$ are measures of single update variability for the regular code~\cite{Raymond:SS}, the solid line by contrast is an average over $20$ sequential estimations of (\ref{eq:CDMA.lambdaRS}) in the converged state.}
\end{figure}

Figure~\ref{fig:CDMA.normalregimes} demonstrates some general properties of the ensembles parameterised by $C\!\!:\!\!L\!=\!3\!\!:\!\!3$. Equations~(\ref{eq:CDMA.cavitybias}-\ref{eq:CDMA.cavityfield}) were iterated using population dynamics and the relevant properties were calculated from the converged order parameters; the data presented is averaged over 100 runs.

Figure~\ref{fig:CDMA.normalregimes}(a) shows the spectral efficiency and its lower bound, and the trend is a smooth monotonic increase in transmitted information as SNR increases. The effect of the disconnected (user) component is clear in the fact that the irregular code fails to approach capacity at high SNR. At low SNR the reduced MAI in the regular code means this ensemble outperforms the dense ensemble. In all other regimes the ordering of performance is dense, regular, user regular and irregular. In general it appears the chip connectivity distribution is not critical in changing the high SNR trends. It was found in these cases (and all cases with unique fixed points of the saddle-point equations), that the algorithm converged to non-negative entropy ($\SEmath<\load$). The smallness of the gap $\SEmath-\SEmath_L$ is an indication of weak correlations.

The known result that the solution must be RS is verified in the stability exponent, fluctuating about a value less than $0$, as shown in figure~\ref{fig:CDMA.normalregimes}(b). The characteristic cusp near the most correlated point, corresponds also to a gap maximising $\SEmath-\SEmath_L$. The gap in the stability exponent to the neutral stability point ($\lambda^{(t)}=0$) indicates it might be possible to work with the RS assumption at a range of temperatures below the Nishimori temperature, since the stability exponent is expected to vary slowly with $\beta$. The range of SNR at which the RS assumption is likely to break down first is indicated by the cusp.

\begin{figure}[htb]
\includegraphics[width=\linewidth]{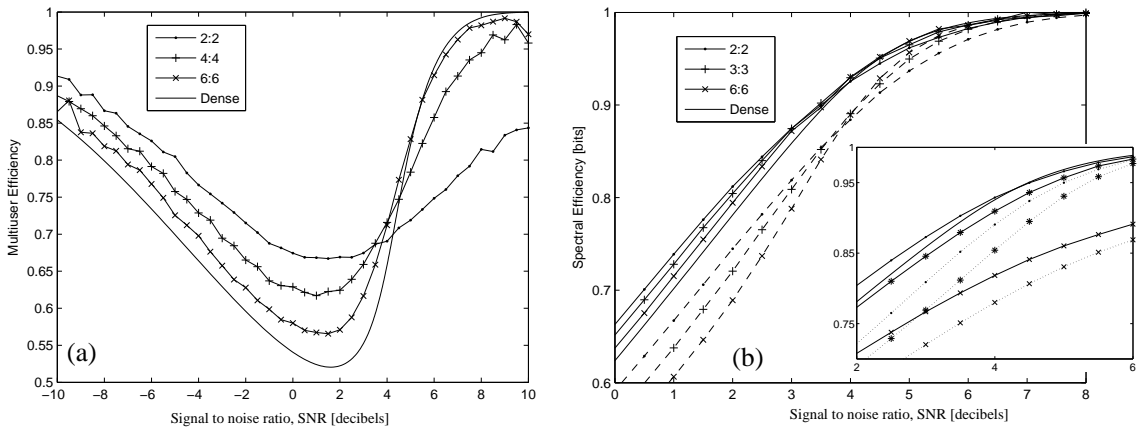}
\caption[Performance of sparse CDMA with connectivity
variation.]{\label{fig:CDMA.density_prop} The effect of increasing density for the regular ensemble is shown, parameterised by $L\!:\!C$. (a) $\MUEmath$ is presented with small error bars are omitted. Below about 4dB the regular ensemble is more efficient than the dense ensemble. As connectivity increases the dense ensemble result is rapidly approached everywhere. The efficiency is worst for all codes at intermediated SNR. As connectivity increases the range of SNR for which the regular code is superior increases slowly. (b) $\SEmath$ [$-$] and $\SEmath_L$ [$--$] demonstrate similar trends to $\MUEmath$.}
\end{figure}

Figure~\ref{fig:CDMA.density_prop} indicates the effect of
increasing density at fixed $\load$ in the case of the regular
code. As density is increased the statistics of the sparse codes approach the dense code in all ensembles tested. For the irregular ensemble performance increases monotonically with density at all SNR. The rapid convergence to the dense case performance was elsewhere observed for partly regular ensembles, and ensembles based on a Gaussian prior input~\cite{Yoshida:ASS,Montanari:ABP}. At all densities for which unique solutions were found violations of the RS assumption were not indicated in the stability exponent or entropy.

Figure~\ref{fig:CDMA.alpha_prop} indicates the effect of channel load $\load$ on performance. Results for codes in which only a single solution was found (no solution coexistence) are first considered. For small values of the load a monotonic increase in BER, and spectral efficiency are observed as $\load$ is increased with $C$ constant, as shown in figures~\ref{fig:CDMA.alpha_prop}(a) and~\ref{fig:CDMA.alpha_prop}(b), respectively. This matches the trend in the dense case, the dense code becoming superior in performance to the sparse codes as SNR increases.

For all sparse ensembles it seems there exist regimes with $\load>1.49$ for which only a single stable solution existed in spite of coexistence of two stable solutions in some range of SNR for dense ensembles~\cite{Tanaka:SMA}, the $L\!\!:\!\!C\!=\!5\!\!:\!\!3$ regular code for example exhibits no metastability. In all single valued regimes positive entropy, and a negative stability exponent were found. However, in cases of large $\load$ many features become more pronounced close to the dense case solution coexistence regime: notably the cusp in the stability exponent, size of $\SEmath-\SEmath_L$ (indicating longer range correlations), and the derivative of BER with respect to SNR.

\subsection{Solution coexistence regimes}
\begin{figure}[htb]
\includegraphics[width=\linewidth]{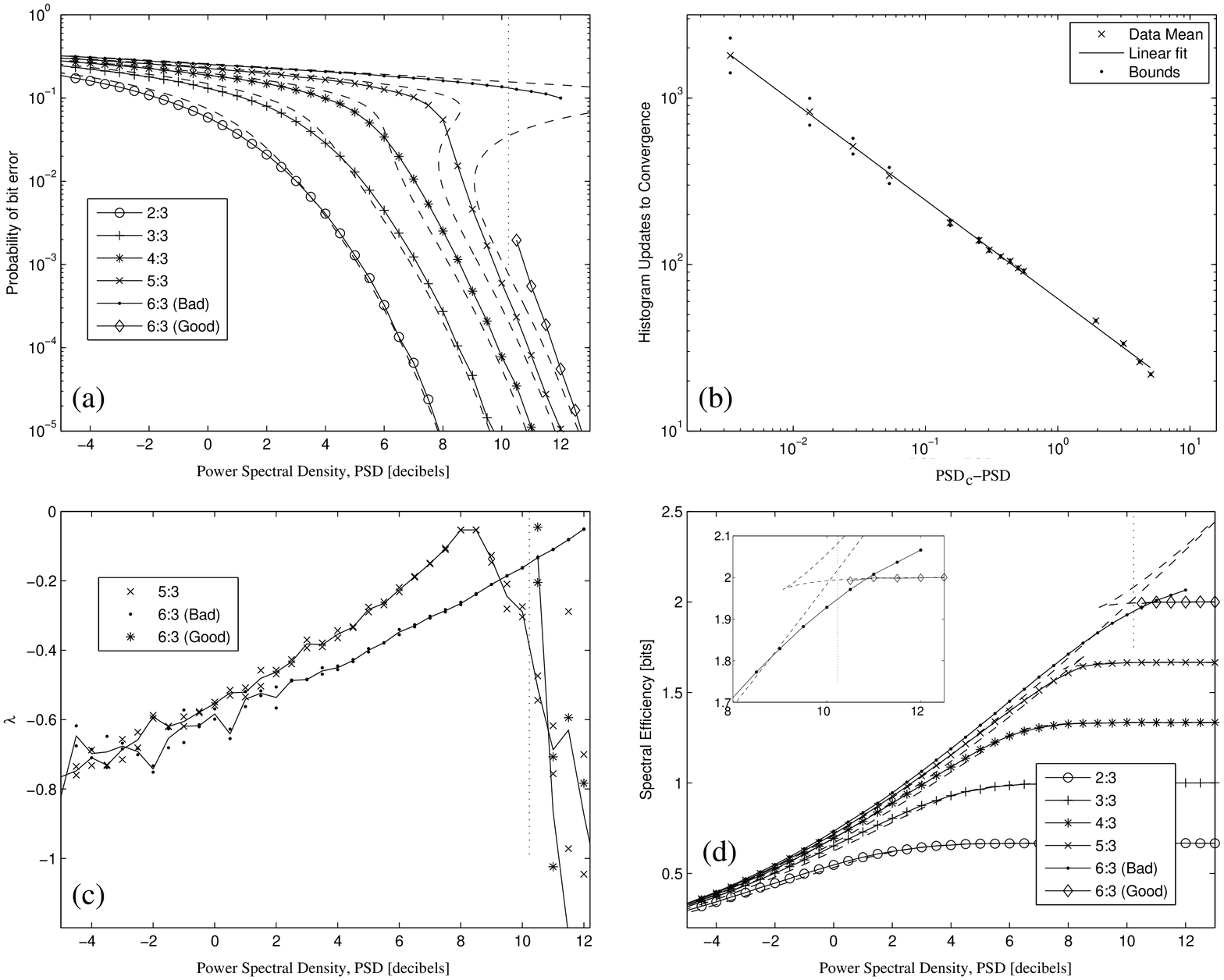}
\caption[Performance of sparse CDMA, multiple solutions regime.]{\label{fig:CDMA.alpha_prop} The effect of channel load $\load$ on performance for the regular ensemble. Data presented are an average of $10$ independent extremisations of the saddle-point equations, error bars are omitted but characterised by the smoothness of curves. Dashed lines indicate the dense code analogues. The vertical dotted line indicates the dynamical critical point beyond which random and ferromagnetic initial conditions failed to converge to the same solution for the $L\!\!:\!\!C\!=\!6\!\!:\!\!3$ ensemble, results for both dynamically stable solutions are shown beyond this point. Power spectral density is plotted, which is the power per chip rather than user PSD ($=\load\SNRmath$) (a) There is a monotonic increase in BER with the increasing load, this is also true at fixed SNR. (b) Investigation of the $6:3$ code ($\load=2$) indicates a divergence in convergence time as PSD$\rightarrow 10.23$dB with exponent $0.59$ based on a simple linear regression of 15 points (each point is the mean of 10 independent runs). Beyond this point different initial conditions give rise to one of two solutions. (c) The stability exponent was found to be negative for all solutions on average (solid lines), indicating the suitability of RS. The stability measures in the case of the good solution are too noisy to provide a firm answer, due in part to under sampling, but also finite precision problems. (d) As load $\load$ is increased there is a monotonic increase in capacity, although the information transmitted per user is a monotonically decreasing function. The spectral efficiency for the 'bad' solution exceeds 2 in a small interval (equivalent to negative entropy), similar to the behaviour reported in the dense ensemble.}
\end{figure}

As in dense CDMA~\cite{Tanaka:SMA}, also here, a regime of parameters were found for which two solutions, of quite different performance, coexist. In order to investigate the coexistence regime the states arrived at from random and ferromagnetic initial conditions (giving bad and good solutions respectively) were examined. Separate heuristic convergence criteria were found for the histograms, and these seemed to work well for the good solution. For the bad solution results are presented based on a conservative runtime of ($500$) histogram updates to ensure counter intuitive features such as negative entropy are correctly captured near the critical points.

Figure~\ref{fig:CDMA.alpha_prop}(a) shows the dependence of the bit error rate on the load, which is also equivalent to $L/C$. There is a monotonic increase in bit error rate with the load and the emergence and coexistence of two separate solutions for a range of Power Spectral Density (PSD$=\load\SNRmath$); in the $L\!\!:\!\!C\!=\!6\!\!:\!\!3$ code the point above which the two solutions coexist is PSD$=10.23$dB as indicated by the vertical dotted line.

The regular code $L\!\!:\!\!C\!=\!6\!\!:\!\!3$ is used to demonstrate the solution coexistence found for a range of SNR in various ensembles. The onset of the bimodal distribution can be identified through the divergence in the convergence time in the single solution regime (the time for the ferromagnetic and random initial condition histograms to converge to a common distribution). The number of updates required for the bit error rate to converge to an identical value is plotted in figure~\ref{fig:CDMA.alpha_prop}(b) as the bimodal regime is approach. By a naive linear regression across 3 decades a power law exponent of $0.59$ and a transition point of PSD$= 10.23$dB (SNR$\approx$ PSD$-3$dB) can be demonstrated, the error implicit in such a fitting is not examined. The evidence indicates the existence of a point at which at least two stable solutions co-exist.

Beyond PSD$\approx12$dB only one stable solution is found from both random and ferromagnetic initial conditions, corresponding statistically to a continuation of the good solution. Thus a second dynamical transition in the region of PSD$=12$dB is found, as might be guessed by comparison with the dense case and observation of the trend in the stability exponent (see figure~\ref{fig:CDMA.alpha_prop}(c)).

The stability results are presented in figure~\ref{fig:CDMA.alpha_prop}(c). Only two stable solutions were found in the region beyond this critical point and up to $12$dB, which are locally stable RS solutions. The bad solution up to 12dB is well resolved. The good solution has a negative value in its mean, but with large error bars, due to insufficient histogram resolution and other numerical issues.

Spectral efficiency monotonically increase with the load as shown in figure~\ref{fig:CDMA.alpha_prop}(d). For the $6:3$ code the dynamical transition point at PSD$= 10.23$dB is indicated by a vertical dotted line and the dashed lines demonstrate behaviour in dense ensembles. The range in which thermodynamic transitions occur is magnified in the inset. A cross over in the entropy of the two distinct solutions, near PSD$\approx 11 $dB, is indicative of a second order phase transition. As in the dense case, only the solution of smallest spectral efficiency is thermodynamically relevant at a given PSD, although the other is likely to be important in decoding dynamics. The trends in the sparse case follow the dense case qualitatively, with the good solution having performance only slightly worse than the corresponding solution in the dense case (and vice versa for the bad solution).

The entropy of the bad solution becomes negative in a small interval (spectral efficiency exceeds $2$ bits), although no local instability is observed. The static and dynamic properties of the histograms appear to be well resolved in this region, but the negative entropy indicates a failure of some assumption in the RS framework as earlier discussed.

The trends in the sparse ensembles match those in the dense ensembles within the coexistence region and $RS$ is locally stable for each of the solutions. The coexistence region is smaller for the sparse codes than the corresponding dense ensembles. In the user regular codes investigated the bad solution of the sparse ensemble outperforms the bad solution of the dense ensemble, and vice-versa for the good solution. Thus regardless of whether sparse decoding performance is good or bad, the dynamical transition point for the dense ensemble would corresponds to an SNR beyond which dense CDMA outperforms sparse CDMA at a particular load. Since our histogram updates mirror the properties of BP on a random graph it is suspected that the bad solution may have implications for the performance of BP decoding in the coexistence region, and that convergence problems will appear near this region.

\subsection{Algorithmic performance in finite systems}
\label{UCICDMATCSP}

\begin{figure}[htb]
 \includegraphics[width=1\linewidth]{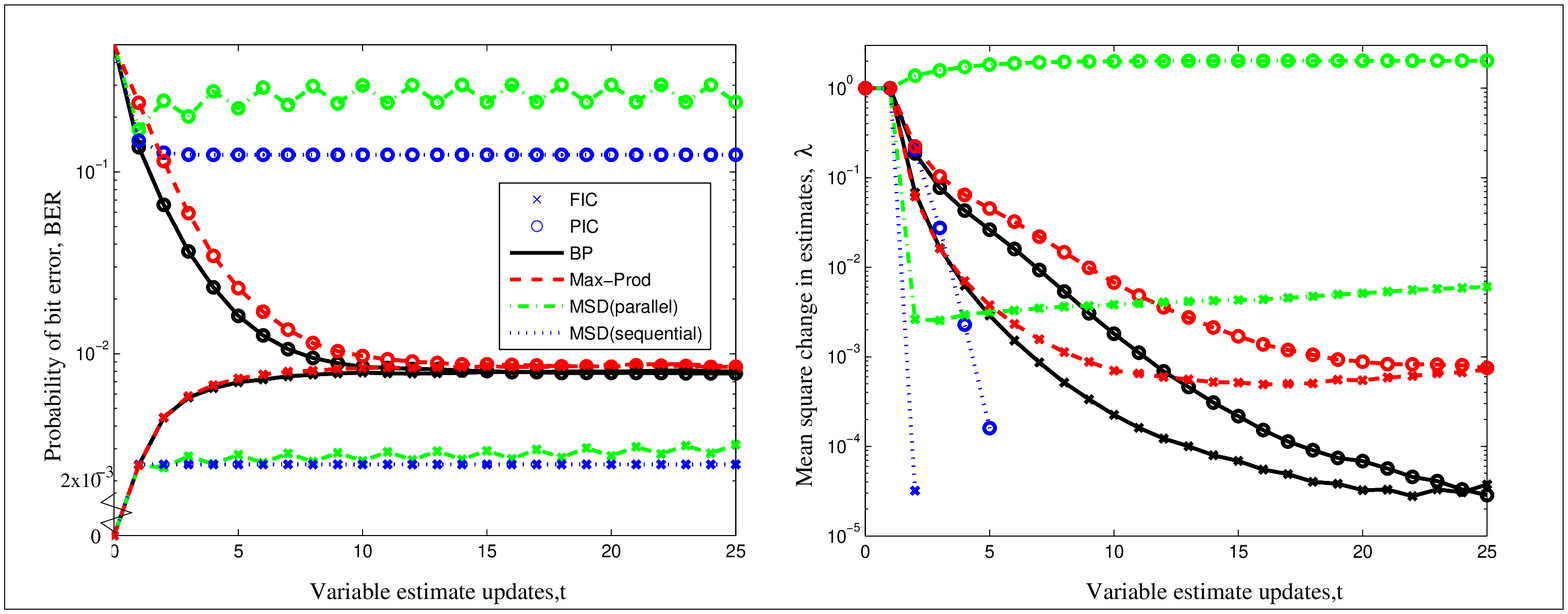}
\caption[Performance of BP and MSD detection for sparse CDMA,
single solution regime.]
{\label{fig:CDMA.regBP} Mean values of $500$ runs, with negligible error bars, are shown for estimate BER and stability for several detectors. The samples are from the regular ensemble ($L\!:\!C\!=\!3\!:\!3$ with $K=500$) at $\SNRmath=6$dB, with BPSK. Decoders are initialised in either a state aligned (FIC) with the source bits, or in unbiased initial condition (PIC). From uninformed initial conditions BP estimates converge exponentially to the lowest BER solution amongst the detectors, this is a unique solution also converged to from FIC in many samples, although convergence is not perfect in some samples, as indicated in the right figure (a curtailing of the exponential decay). The max-product algorithm performs comparably. Sequential MSD is unstable in many samples, from PIC an improved result on matched filter (estimate at time $t=1$) is not typically achieved. Non-sequential MSD is trapped in two attractors, and produces an estimate from PIC often worse than BP.}
\end{figure}
\begin{figure}[htb]
 \includegraphics[width=1\linewidth]{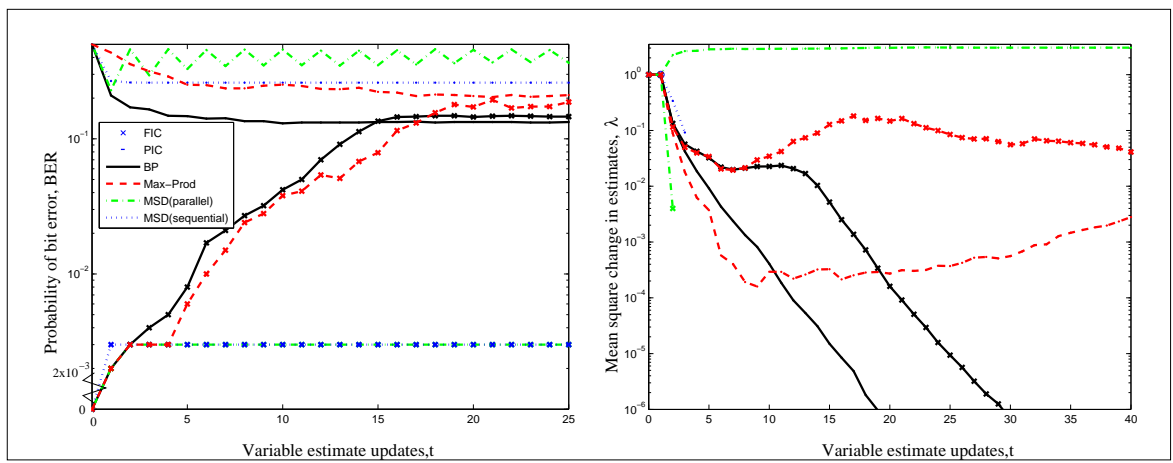}
\caption[Solution properties for BP and MSD, sparse CDMA metastable regime.]
{\label{fig:CDMA.metaBP1} Conditions and symbols as in figure~\ref{fig:CDMA.regBP}, but with more users ($\!L\!:\!C\!=\!6\!:\!3$ with $K=1000$) and only a single typical sample presented, at $\SNRmath=6$dB. Development of a single sample from bad (PIC, no symbol) and good (FIC,$\times$) initial estimates. BP and max-Product achieve BER near the RS thermodynamic prediction from both initial conditions. MSD is trapped in two attractors differing significantly in BER.}
\end{figure}
\begin{figure}[htb]
 \includegraphics[width=1\linewidth]{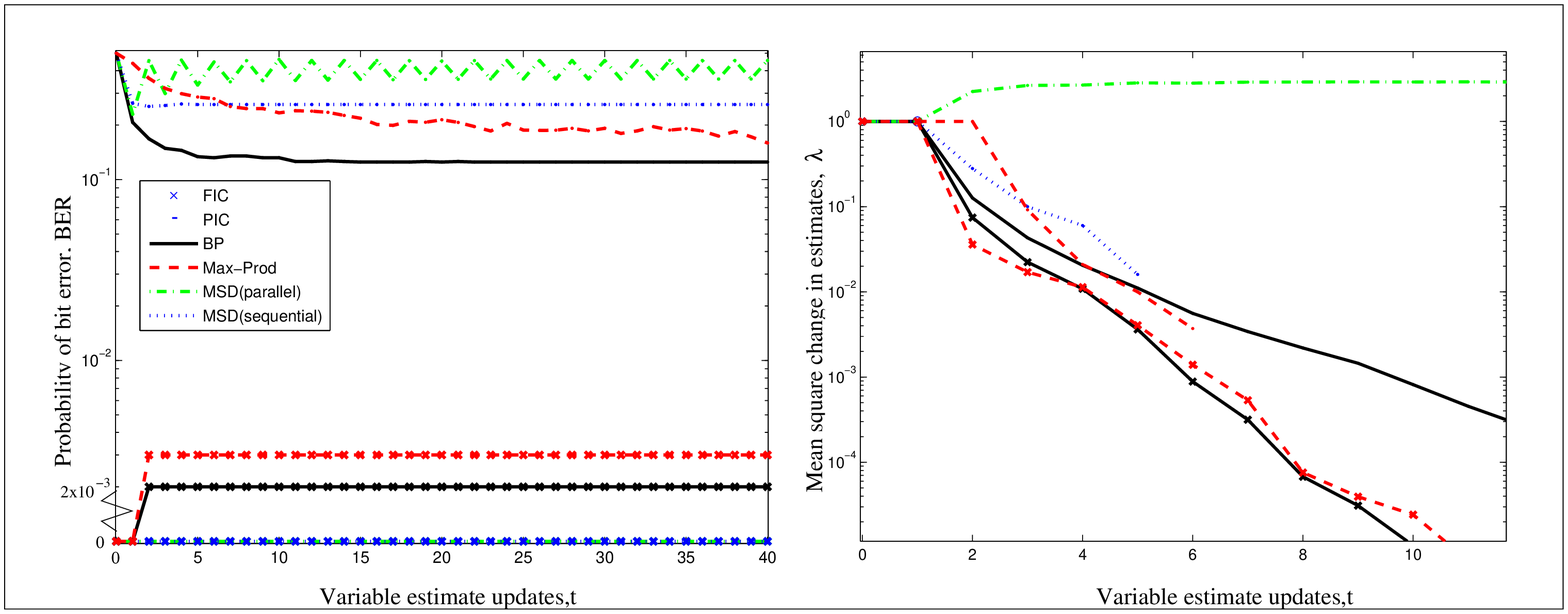}
\caption[Solution properties for BP and MSD, sparse CDMA metastable regime.]
{\label{fig:CDMA.metaBP2} Properties as in figure~\ref{fig:CDMA.metaBP2}, but with decreased noise variance ($\SNRmath=7$dB) on the sample. Local minima emerge close to the FIC which trap the dynamics of all algorithms, in qualitative agreement with the good and bad metastable scenario predicted by the RS thermodynamic solution. BP, MP and non-sequential MSD converge rapidly to one of the two solution types.}
\end{figure}
Results are here demonstrated for a small subset of regular ensembles, some additional examples for the user regular ensemble are demonstrated in chapter~\ref{chapter:compositeCDMA}, along with an elaboration of some of the arguments on the relationship between metastability and decoding. Statistics are presented based on random code and noise samples, the process of code generation for the doubly regular model is outlined in Appendix \ref{app:RandomGraphSamples}, and although some approximations are involved the graphs generated are not expected to be significantly biased. The limited set of algorithms presented here does not represent the great progress that has been made in theoretically guided heuristic decoding methods~\cite{Kechriotis:HNN,Bickson:GBP,Neirotti:IR}. However, the difficulty in detection at high MAI, and ease of detection at $\load<1$ is a feature commonly reported in studies of MAI on linear vector channels.

BP (\ref{eq:CDMA.mutok})-(\ref{eq:CDMA.ktomu}) is applied to small samples, with codes from the regular ensemble at intermediate noise levels. These graphs are loopy and hence BP is not guaranteed to converge. If BP converges to a unique fixed point on a given graph sample then it is guaranteed to be the MPM solution, but this scenario is difficult to prove for particular samples~\cite{Weiss:CLP}. If BP does not converge, or converges to an incorrect local minima, only a suboptimal detection is possible. BP is observed to converge in most parameterisations where a unique RS thermodynamic solution is predicted. In loopy CDMA graphs BP is initiated with the edges set to uninformative values $h_{i\rightarrow\mu}^{(0)}=0$, and then applying parallel updates (which exhibit improved BER over sequential updates in most cases). The results for bit error rate are based on evaluations of (\ref{eq:CDMA.H}) at each time step. Also given is a measure of convergence, the mean square change in estimates, for the BP equations:
\begin{equation}
\lambda^{(t)}=\frac{\sum_k (H^{(t)}_k - H^{(t-1)}_k)^2}{\sum_k \left[(H^{(t)}_i)^2 + (H^{(t-1)}_i)^2\right]} \label{eq:CDMA.stability} \;,
\end{equation}
which is distinguished from (\ref{eq:CDMA.lambdaRS}) by context. $H_k^{(t)}$ is the marginal log-posterior ratio for variable $k$ (\ref{eq:CDMA.H}). If BP converges then this change will decrease exponentially at large time.

Similarly the max-product algorithm may be applied as a heuristic method (\ref{eq:CDMA.maxproduct}), initialising all messages to $h_{i\rightarrow\mu}^{(0)}=0$, and taking intermediate evaluations, with the same stability exponent (\ref{eq:CDMA.stability}). The max-product algorithm is the $\beta\rightarrow 0$ limit of the belief propagation equations.

These results are to be compared to multistage detection (MSD), which is a standard heuristic algorithm~\cite{Tanaka:SMCDMA} that works well in dense codes at low loads. MSD messages are defined as
\begin{equation}
H^{(t)}_k = \sign \left[\vs_k\cdot\vy + \vY_k\cdot\vH^{(t)}\right] \;; \qquad Y_{k l}=(1-\delta_{k,l})\vs_k\cdot\vs_{l} \;.
\end{equation}
Messages are initialised as $H^{0}_k=0$, the first step of multistage detection produces the result of a matched filter detector, subsequent steps refine the estimate based on an ad-hoc calculation of the MAI. This algorithm is very sensitive to the update scheme used. Again, a measure of stability is given by (\ref{eq:CDMA.stability}). Due to the discrete nature of the messages the change in estimates (\ref{eq:CDMA.stability}) does not converge smoothly to zero, the denominator can become exactly zero truncating $\lambda$ curves (figures~\ref{fig:CDMA.regBP}-\ref{fig:CDMA.metaBP2}).

Two timing implementation schemes are considered for MSD. In parallel update schemes all messages $\vH^{(t+1)}$ are simultaneously updated according to $\vH^{(t)}$. The second scheme is random sequential update, which requires a separate timing scheme for updates. The ordering of message updates is randomised and instead of using only messages in generation $t$ to determine generation $t+1$, the most recent updated is always used, so that messages in $\vH^{(t+1)}$ are no longer conditionally independent given $\vH^{(t)}$. This scheme can also be implemented in BP, but results were found to be less striking.

The paramagnetic initial conditions (PIC) defined as ($\{H_i^{0}=h_{*\rightarrow*}^{0}=0\}$) for all message passing methods outlined so far are in no way biased towards the source bit sequence, which is a realistic scenario for detectors. However, it is useful to also consider ferromagnetic initial conditions (FIC) $\vH \propto \vb$, which demonstrates the emergence, or absence, of metastable solutions at low BER.

Figure~\ref{fig:CDMA.regBP} demonstrates some time series for decoding of a $L\!\!:\!\!C\!=\!3\!\!:\!\!3$ regular code at $\SNRmath=6$dB averaged over 100 runs for a system of size $K\!=\!M\!=\!500$. When used at the Nishimori temperature BP is best throughout the algorithm time and converges very quickly. Note that after only one update BP correctly infers $80\%$ of bit values. Max-product algorithm results are similar to BP and is also convergent in all samples through a large number of iterations. In the first step matched filter detection does well, but when using parallel updating the MSD algorithm is clearly unstable and oscillating with period $2$. The MSD algorithm with random sequential updates converges very rapidly, but to a relatively poor estimate. The MSD algorithms with FIC are also trapped in locally stable minima near the encoded solution. For the $3\!\!:\!\!3$ regular codes at low load (and hence low MAI) a variety of simple algorithms perform very well. The BER achieved by BP is comparable to the thermodynamic result (compare to figure~\ref{fig:CDMA.normalregimes}), which is the optimal prediction. Similar near optimal estimates are attainable for the other ensembles at small load.

Figures~\ref{fig:CDMA.metaBP1}-\ref{fig:CDMA.metaBP2} demonstrate regimes of larger load where mean and variance based statistics are not so helpful due to a multi-modal distribution of data, instead decoding of a single typical sample is demonstrated from two different initial conditions. In figure~\ref{fig:CDMA.metaBP1} $\SNRmath=6$dB, estimates from both initial conditions converge towards estimates of comparable BER in most cases. Sequential MSD is the exception which is trapped in a low BER solution from FIC, and oscillates from PIC. The two estimates found by BP are not identical, but represent local minima, and convergence from FIC is slow. The median BER found by BP at large time, over many samples, is close to the RS thermodynamic prediction (figure~\ref{fig:CDMA.normalregimes}). The Max-Product algorithm estimates are unstable in this system, whereas non-sequential MSD converges to a relatively poor estimate by comparison with BP.

In figure~\ref{fig:CDMA.metaBP2} SNR is increased by only $1$dB, and all other aspects of the quenched disorder are identical to figure~\ref{fig:CDMA.metaBP1}. Excluding sequential MSD, convergence occurs for all algorithms, but convergence is towards solutions with a bimodal distribution in BER. An estimate characteristic of a good thermodynamic solution is typically found by algorithms evolving from FIC. However, it is properties of the bad thermodynamic solution which best characterise algorithms evolving from PIC, which is the result achievable in practice. A bimodal distribution of solutions similar to the RS thermodynamic prediction is observable more generally at high load, close to the parameterisations predicted by theory.

\subsection{Modulation schemes}

Simulations were also undertaken with respect to unmodulated codes for comparably sized systems to those presented. Results were comparable in the median, but had much more variability between samples, indicating that some finite size effects may be more pronounced in these systems. Unmodulated codes might be particularly sensitive to short loops, and variations in the mean bit transmitted $\<b_k\>$.

No firm evidence is uncovered for a preferential consideration of symmetric or asymmetric modulation patterns in this thesis, for sparse codes in the large system limit. The analysis of the linear stability of the RS bit-symmetric description indicates the same thermodynamic behaviour for both cases, and the noiseless analysis of Appendix~\ref{app:CDMAto1ink} indicates no distinction in decoding dynamics by unit clause propagation during the initial inference stage.

It seems that where RS is applicable, temperature is high, or MAI small, typical case equivalence of asymptotic properties may be correct. However, where correlations are strong, or with minor modifications to the ensemble, the asymptotic results may differ. In practice although the unmodulated code represents a simpler choice, it is noteworthy that in the limit of intensive connectivity $C=O(M)$ the unmodulated code fails dramatically. A further anomaly of the unmodulated codes is that with biased transmission of bits $\<b_k\>^2>0$, a typical case of the unmodulated code has an improved performance even when not including this prior in the model, a feature that may be exploited in practice.

The Gaussian modulation breaks some of the degeneracy apparent in the sparse finite connectivity ensemble. In terms of capacity this degeneracy plays an important role at low noise as shown in section~\ref{CDMA.marginals} and Appendix~\ref{app:CDMAto1ink}. However, performance is relatively poor with modest increases in noise or load, and one advantage of sparse codes, their concise specification, should be balanced against the gains achievable through amplitude modulation.

%% file: COMPOSITEMODEL.tex
\newcommand\newfigureyes[1]{#1}
\newcommand\newfigureno[1]{}
\newcommand\CUTFROMFINAL[1]{}
\newcommand\INCLUDEINFINAL[1]{#1}
\section{Introduction}
\label{composite.Introduction}
\begin{figure}[htb]
\includegraphics[width=1.0\linewidth]{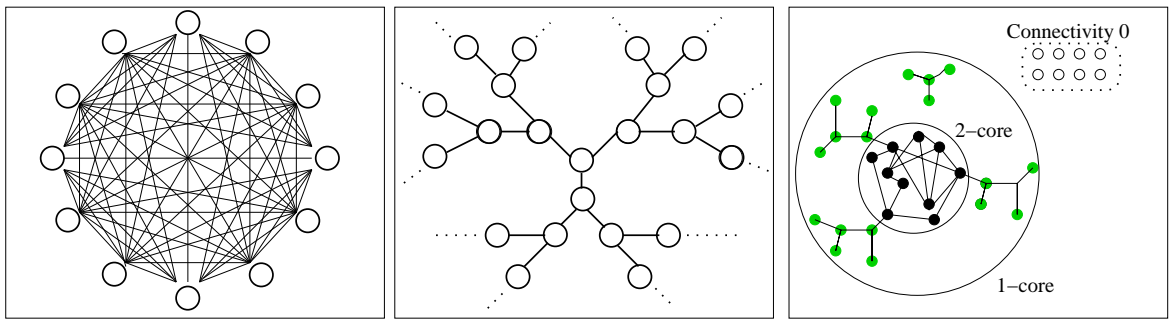}
\caption[Sparse and dense spin models.] {\label{fig:composite.2cores} Shown are the couplings (links) amongst a set of spin variables (circles), which describes graphically a particular quadratic Hamiltonian. Left figure: The fully connected graph is a special case of the dense graph describing the SK model, with $O(N)$ non-zero couplings per variable in the large system limit. Centre/Right figure: The VB model is defined with $O(1)$ non-zero couplings per variable in the large system limit. Centre figure: In the case of a regular connectivity random graph above the percolation threshold, there is an inhomogeneous structure on a global scale, but locally the structure is a Bethe lattice (regular tree). Right figure: In the case of a random graph with Poissonian connectivity the local structure is again tree like. Above the percolation threshold many trees of finite size, and unconnected variables exist, as well as a giant component containing $O(N)$ variables, and many loops~\cite{Bollobas:RG}. The 1-core contains all variables with at least one link, including the giant component above the percolation threshold. Addition structure within the giant component may be identified, including a 2-core, obtained by recursively removing leaves (singly connected variables) from the giant component.}
\end{figure}

Statistical physics methods for studying disordered spin systems have become well developed. Much of the development can be traced back to early work on mean-field models for disordered magnetic systems and the theory was strongly developed in spin-glass models~\cite{Fischer:SG,Mezard:SGT}. One problem in studying spin glasses and disordered media has been in appropriately modeling the inhomogeneity within tractable frameworks. Statistical descriptions of inhomogeneity are often realised by random coupling ensembles. Small systems described in this way may have strongly varying properties, but the ensemble may be chosen so that the macroscopic description is asymptotically well defined.

Both dense and sparse graphical models are useful in understanding a range of phenomena, such as neural networks~\cite{Hertz:ITNC}, information theory~\cite{Richardson:MCT} and other information processing~\cite{Nishimori:SP}, where spatial and dimensional constraints are often less rigid. Many complex systems have an inhomogeneous interaction structure that can be approached, if not exactly represented, by consideration of simple random graph ensembles. In this chapter spin glass models with couplings conforming to infinite dimensional Erd\"{o}s-R\'{e}nyi random graphs are considered~\cite{Bollobas:RG}. In the large system limit many equilibrium properties depend on the connectivity distribution, and how the number of couplings per variable scales with $N$, the system size. Dense graphs have a number of links per variable that in the large system limit is $O(N)$, whereas sparse ensembles have finite mean connectivity in this limit. Many topological features become well defined in these limits. Two standard sparse coupling distributions are considered, a description with regular user connectivity, and one with Poissonian user connectivity. The distinctions between these two sparse models and the limiting case of full connectivity are illustrated in figure~\ref{fig:composite.2cores}.

Some densely connected models may be analysed exactly for ensembles of uniform binary interactions, and certain random coupling models, most famously the Sherrington-Kirkpatrick (SK) model of spin glasses~\cite{Sherrington:SMSG}. Simplification of the analysis in the disordered case is often possible through noting the ability to describe large sets of interactions by central limit theorems~\cite{Ellis:ELD}. For sparse graphical models a locally tree like approximation (Bethe approximation) is often essential in simplifying analysis, central limit theorems again apply to certain objects, but not directly to the set of local interactions for any variable. Models which do not allow use of central limit theorems or locally tree-like approximations are normally significantly more difficult to analyse.

Frameworks in which an interplay between strong sparse and weak pervasive couplings might be proposed in a variety of areas. In nanotechnology for example, miniaturisation of classical components will preserve engineered short range interactions, but other accidental correlations may emerge not limited by the designed connectivity structure, and these may well be modeled by a mean-field (infinite connectivity) like interaction. A mixed connectivity may also be a designed feature. Neuronal activity is known to involve a combination short and long range information processing structures, this motivated a $1+\infty$ dimensional model of neuronal activity~\cite{Skantzos:1ID} discovering many novel properties. Another example of such an engineering application is CDMA, for which results are demonstrated in chapter~\ref{chapter:compositeCDMA}.

\begin{figure}[htb]
\begin{center}
\includegraphics[width=0.8\linewidth]{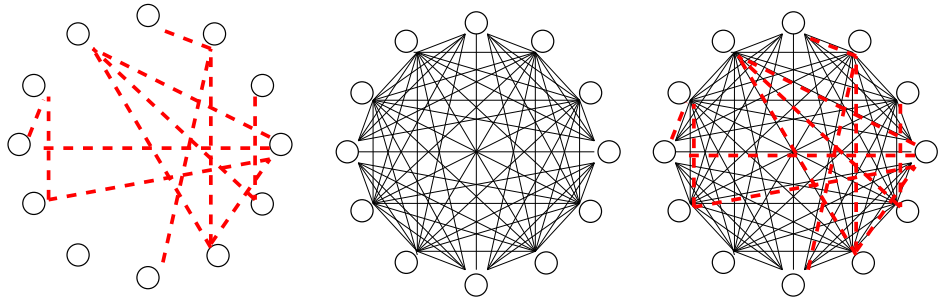}
\caption[Composite models.]{\label{fig:composite.COMPSYS} Left figure: A sparse model defined by some mean connectivity, describes couplings in the VB model. Centre figure: A fully connected model, describes couplings in the SK model. Right figure: A fully connected graph with a subset of strong sparse links, this is the composite model. The sparse subset of couplings are an order of magnitude stronger than the couplings on the other edges.}
\end{center}
\end{figure}

The work presented in this chapter considers the analysis of a composite model of $N$ densely connected Ising spins in which there are two scales of interaction, but no dimensionality constraints. A small subset $O(N)$ of the couplings are strong with the remainder of the couplings non-zero, but an order of magnitude weaker. The composite model~\cite{Raymond:OC,Raymond:CB,Hase:DA} is a new type of exactly solvable mean-field model. An illustration is given in figure~\ref{fig:composite.COMPSYS}.

To motivate a closely related study Hase and Mendes noted a possible application for theories of these structures~\cite{Hase:DA}. Consider the model with sparse anti-ferromagnetic couplings on a structure otherwise fully connected through ferromagnetic couplings. This composite model can be considered as one in which a ferromagnetic phase is maintained by a densely connected network, but with a small proportion of links attacked. Often only a small portion of a link structure is accessible to an attacker, so it is interesting to consider how the system response differs from weak attacks on all (or most) links.

The effect of an attack on a sparse subset may cause a transition away from the ordered phase, when sufficiently strong. It is possible that the nature of transitions away from the ordered state may differ from those with only a single interaction scale. The effect of disruption of networks by random attack, or frustrating interactions, is of importance in many practical network models~\cite{Albert:EA,Hase:DA}, the restriction to random topologies allows a focus on generic properties, in this case restricted to the issue of sparse and dense induced effects.

More generally a range of mean field behaviour, including spin glass like, may be supported by the dense sub-structure, combined with an arbitrary set supported by the sparse sub-structure. In so doing a wider variety of competitive phase behaviour are explored.

It may be expected that many of the results for composite systems will be similar to those for the limiting sparse and dense models. Four thermodynamic phases describe equilibrium properties of spin models with independent and identically distributed (i.i.d) couplings. A pure state with no macroscopic order, the paramagnetic phase; a pure state with macroscopic order aligned with some mean bias in the couplings, the replica symmetric (RS) ferromagnetic phase (F); a macroscopically aligned phase, but with some complicated phase space fragmentation, the mixed phase (M); and a phase with no macroscopic alignment and a complicated fragmentation of the phase space, the spin glass phase (SG). Within both the sparse and dense Ising spin models these phases are exhibited and many features are shared by the two models.

The main question investigated in this chapter is how phase behaviour and transitions differ in the composite model from the sparse and dense frameworks, and whether a simple interpolation is produced by the composite models. Attention is restricted to cases in which the sparse sub-structure is percolating, since in any other regime the long range coupling will be due solely to the dense links. A non-percolating sparse sub-structure would not test the effects of competing long range induced order, although some of the methods and results are inclusive of this scenario and appear to vary continuously (at finite temperature) across the percolation threshold corresponding to the sparse sub-structure.

\subsection{Summary of related results}

At the time my PhD began no substantial work existed on the
thermodynamic properties for a combination of sparse and dense random graphs. However, a recent paper by Hase and Mendes considered the stability of a mean field ferromagnetic model subject to a random attack by sparse anti-ferromagnetic couplings, acting between variables according to a sparse annealed structure~\cite{Hase:DA}. By contrast the author has studied a variety of models with a quenched interaction structure~\cite{Raymond:OC}~\cite{Raymond:CB}, many results being summarised in this chapter. Both these treatments involved a replica based analysis of the thermodynamic properties, which was solved under the RS assumption. In terms of iterative methods for constructing marginals a Belief Propagation (BP) method was constructed relating to the problem of composite CDMA~\cite{Mallard:BPDG}, this is examined in chapter~\ref{chapter:compositeCDMA}.

Results for both sparse and dense systems are very relevant. The SK model~\cite{Mezard:SGT,Sherrington:SMSG} is the appropriate benchmark as a densely connected model, whilst the sparse model is comparable to the Viana-Bray (VB) model~\cite{Viana:PD,Mottishaw:RSB}. These models exhibit continuous phase transitions between ferromagnetic, paramagnetic, mixed and spin-glass phases, with variation of temperature, external fields or disorder in the set of couplings. A triple point exists in both models where the phase boundaries, P-F-SG coincide. As temperature is lowered pure states are often susceptible to phase space fragmentation, which can be tested by local stability analyses~\cite{Rivoire:GM,Mottishaw:SRF,Almeida:SSK}. Some very general rigorous results are attainable in the high temperature paramagnetic phase~\cite{Ostilli:ISG}.

Properties of the spin glass phase may be exactly calculated in the dense model through the replica method, but many of the central limit theorems necessary for this analysis do not extend to sparse models, so that it is necessary to consider variational methods~\cite{Monasson:OP,Mezard:BLSG}. Much work has also been conducted studying exclusively ground state properties, the limit of zero temperature~\cite{Wong:GB,Mezard:MFT,Kanter:MFT}, though this limit is not considered in this chapter. The percolation threshold produces a novel transition in the sparse model absent from the dense model~\cite{DeDominicis:RSB}.

The effect of random external fields on densely connected models can be understood in the SK model through the AT line~\cite{Almeida:SSK}. A pure state phase to fragmented phase may occur with application of a strong random field, or vice-versa for an aligning field. In the sparse model trends are similar in response to uniform fields, the problems in understand SG and M phases are not solved except very near some high temperature transition points for uniform fields. Variational approximations must be considered away from these points~\cite{Hase:RSS}.

Generalisations of coupled Ising spin models include to systems with Potts spins, or continuous phase states, and also to systems with more than two point couplings, or without i.i.d couplings. A composite model with competing alignments in the sparse and dense parts may be created by introducing a random alignment for couplings in either the sparse or dense part~\cite{Mattis:SSS}. The properties for two misaligned sets of dense couplings can be understood through the Hopfield model~\cite{Amit:SG}, where a simple form of metastability arises in the case of two embedded alignments.

\subsection{Chapter outline and results summary}
\label{composite.CORS}

Section~\ref{composite.ensemble} outlines the ensemble of models studied in this chapter, with four special cases outlined, which form the basis for much of the specific analysis. Section~\ref{composite.RM} presents a replica analysis of the composite model.

Section~\ref{composite.RS} develops the RS solution to the replica method. A set of BP equations are developed and analysed, and a longitudinal stability analysis presented in the context of BP.

Section~\ref{composite.hightemp} presents a leading order solution to the composite system in terms of a simplified ansatz on the order parameter. It is shown that for some composite systems a local stability analysis is sufficient to determine properties of the paramagnetic phase and the leading order behaviour in the spin glass and ferromagnetic phases. The case of Poissonian connectivity in the sparse part is shown to lead to a high temperature thermodynamic solution identical to that of the SK model. A regular connectivity ensemble, by contrast, may undergo discontinuous ferromagnetic transitions, not characteristic of either the sparse or dense models.

Section~\ref{composite.leadingorderauxiliarysystems} demonstrates applications of some of these methods to some simple representative systems. It is shown that for many models near the triple point there is a transition from a spin glass phase to an RS ferromagnetic phase as temperature is decreased.

Section~\ref{composite.lowtemp} demonstrates the RS solutions for several composite models in the interesting range of parameters about the triple point in the phase diagram. This demonstrates departures from the leading order analysis, and some unanticipated low temperature transitions. A composite model is demonstrated that exhibits a low temperature transition to a mixed phase in spite of weak pervasive anti-ferromagnetic effects, which prevent ferromagnetic transitions at high temperature.

Section~\ref{composite.finite} presents hypotheses on the structure of the low temperature phases, which cause results to differ from comparable sparse and dense systems. Simulation results of BP and Metropolis-Hastings Monte-Carlo are presented for a model instance of a model with ferromagnetic dense couplings and anti-ferromagnetic sparse couplings.

\section{Composite ensembles}
\label{composite.ensemble}

The Composite model can be described by a Hamiltonian with coupling of $N$ spins
\begin{equation}
\Ham(\vS)= -\sum_{\ij} \left[J^D_{\ij} + J^S_{\ij}\right] S_i S_j - \sum_i \randomfield_i S_i\label{eq:composite.Ham}\;,
\end{equation}
where $\ij$ are an ordered set of variables. The couplings are labeled as dense ($D$) or sparse ($S$) and are sampled independently for each link according independent ensembles described shortly. The quenched variable abbreviation $\quenched$ indicates a sample of the couplings, and $\vS$ are the dynamic variables. The field vector $\vrandomfield$ is used only as a conjugate parameter to explore symmetries, the limit $\vrandomfield\rightarrow \vzeros$ (vector of zero fields) is always assumed throughout this chapter, although some physical quantities and insight are demonstrated using conjugate fields in Appendix~\ref{app:ConjugateFields}.

The equilibrium properties of the model are studied. The Hamiltonian implies a static probability distribution on the state space given by
\begin{equation}
P(\vS) = \frac{1}{Z(\beta,\quenched)} \exp\left\lbrace -\beta
\Ham(\vS)\right\rbrace\label{eq:PS}\;,
\end{equation}
where $\beta$ is the inverse temperature and $Z$ is the partition function.

The spin states of interest are the typical case equilibrium distribution, in the large system limit. Properties of these states are established through the mean free energy
\begin{equation}
\beta \safed(\beta) = -\lim_{N \rightarrow \infty} \frac{1}{N}\< \log \partitionfunction \>_{\quenched} \label{eq:composite.safed}\;,
\end{equation}
where $\ensemble$ is the ensemble parameterisation.

The model is fundamentally a fully connected one, the sparse component is realised as a subset of couplings that are an order of magnitude stronger. Due to this order of magnitude many results for standard densely connected spin models do not apply.

\subsubsection{Dense (SK) sub-structure}
\label{composite.SKSS}

The dense sub-structure fully connects the set of $N$ spin variables $\vS \in \left\lbrace \pm 1\right\rbrace^N$, with couplings sampled independently at random according to the Gaussian distribution parameterised by $J_0$ and $J$
\begin{equation}
P(\mJ^D)= \prod_\ij P(J^D_\ij) \;; \qquad P(J^D_\ij) = \frac{1}{\sqrt{2\pi/N}} \exp \left\lbrace -\frac{N}{2 J^2}\left(J^D_\ij - \frac{J_0}{N} \right)\right\rbrace
\label{eq:composite.SK}\;,
\end{equation}
with a necessary scaling of components included. This set of couplings has a statistical description corresponding to the SK model.

\subsubsection{Sparse (VB) sub-structure}
\label{composite.VBSS}

It is convenient to factorise the sparse couplings as
\begin{equation}
 J^S_\ij= A_\ij \modulationsymbol_\ij\;.
\end{equation}
The ensemble is described by a connectivity matrix, $\mA$, which is zero for all but a fraction $C/N$ of components, and a dense coupling matrix $\mxi$, with no zero elements. In the irregular ensemble each directed edge is present (non-zero) independently with probability $C/N$, $C$ is the mean variable connectivity, so that a prior for inclusion of an edge is
\begin{equation}
P(\mA)= \prod_\ij \left[\left(1-\frac{C}{N}\right)\delta(A_\ij) + \frac{C}{N}\delta(A_\ij -1)\right] \label{eq:composite.A}\;,
\end{equation}
this being the connectivity in a standard Erd\"{o}s-R\'{e}nyi random graph. The couplings in the non-zero cases are described by a distribution with finite moments, and are sampled independently according to
\begin{equation}
 P(\mxi)= \prod_\ij P(\modulationsymbol_\ij) \;; \qquad P(\modulationsymbol_\ij = x) = \phi(x) \label{eq:composite.phix}\;,
\end{equation}
in the general case. A practical distribution for analysis is the $\pm J$ distribution defined
\begin{equation}
\phi(x)=(1-p)\delta(x-J^S) + p \delta(x+J^S) \label{eq:composite.pmJ}\;,
\end{equation}
with two parameters, $p$ the probability that the link is anti-ferromagnetic, and $J^S$ the strength of coupling. Regular connectivity ensembles have each variable constrained to interact with exactly $C$ neighbours,
\begin{equation}
 P(\mA) \propto \prod_{i=1}^N \delta\left(\sum_j A_{i j} - C\right) \label{eq:composite.HamAux4}\;.
\end{equation}

\subsubsection{Representative parameterisations}
\label{composite.specialcases}

Four models are considered in greater detail owing to their simplicity and ability to make transparent a range of observed phenomena. The F-AF model includes ferromagnetic dense couplings ($J=0$, $J_0>0$ (\ref{eq:composite.pmJ})) and anti-ferromagnetic sparse couplings ($p=1$ (\ref{eq:composite.pmJ})), with connectivity $C=2$, and is described by
\begin{equation}
\Ham(\vS)= - \frac{\mB(\gamma,J^S)}{N} \sum_\ij S_i S_j + J^S \sum_{\ij} A_\ij S_i S_j \label{eq:composite.HamAux1}\;.
\end{equation}
The function $\mB(\gamma,J^S)/N$ is introduced to balance the ferromagnetic and anti-ferromagnetic tendency. Choosing $\mB(\gamma,J^S)$ as a positive, monotonically increasing function of the scalar parameter $\gamma$ the relative strength of the anti-ferromagnetic and ferromagnetic parts are kept in some intuitive balance. As $\gamma$ increases there is an increased tendency towards aligning spins within the Hamiltonian -- the ferromagnetic (ordered) state is promoted.

It is also interesting to consider the converse case, the AF-F model with a ferromagnetic sparse part ($p=0$) and anti-ferromagnetic dense model ($J=0$,$J_0<0$), with connectivity $C=2$,
\begin{equation}
\Ham(\vS)= - J_S\sum_{\ij} A_\ij S_i S_j + \frac{\mB(\gamma,J^S)}{N} \sum_\ij S_i S_j \;, \label{eq:composite.HamAux2}
\end{equation}
with $\mB$ being again some suitably re-scaled function, $J^S$ must also be defined.

These models can also be considered for the case of regular connectivity. The {\em regular} F-AF model (\ref{eq:composite.HamAux1}) and {\em regular} AF-F model (\ref{eq:composite.HamAux2}) are considered, but in each case with connectivity chosen to be $C=3$ (a minimal choice above the percolation threshold).

In the definitions of the sparse and dense sub-structures the
alignment of the order is equal in both parts, with $\vtau$ biased towards either $\vones$ or $-\vones$. It is interesting to consider the case that the alignment in the dense part is orthogonal to the alignment in the sparse part. This is achieved by sampling dense ensemble links $J^D_\ij$, according to (\ref{eq:composite.SK}), but with an additional modulation by $b_i b_j$. The quenched vector $\vb$ is sampled uniformly from $\left\lbrace -1,1\right\rbrace^N$. This embeds an alignment in a similar way to the Mattis model~\cite{Mattis:SSS}, which changes thermodynamic properties in the composite model since it applies only to one set of couplings, and not the other. Taking otherwise ferromagnetic couplings in the two parts, the F-F model is
\begin{equation}
\Ham(\vS) = - \gamma \sum_{\ij} A_\ij S_i S_j - (1-\gamma) \frac{1}{N} \sum_\ij b_i b_j S_i S_j \label{eq:composite.HamAux3}\;,
\end{equation}
where $p=0, J=1$ in the sparse part, and $J_0=1$ and $J \rightarrow 0$ in the dense part. The scalar parameter $\gamma$ controls the relative importance of the two parts.

\section{Replica method}
\label{composite.RM}

The replica method is used in both~\cite{Hase:DA,Raymond:OC} to study the composite system free energy in the limit of large $N$. The replica method is the most concise analytical method available, although many results presented herein can be developed through the cavity method with suitable assumptions. For convenience the fields $\vrandomfield\rightarrow 0$ (\ref{eq:composite.Ham}) in the calculation steps. Variations on this which are useful in establishing a number of system properties are explored in Appendix~\ref{app:ConjugateFields}.

In the replica approach the typical case behaviour is examined through the free energy density (\ref{eq:composite.safed}) averaged over the quenched disorder. That is to say typical samples from the ensembles are not expected to differ in the value of the order parameters and other extensive properties. The replica identity
\begin{equation}
\<\log \partitionfunction\>_{\quenched}= \left.\frac{\partial}{\partial n}\right|_{n=0} \repZ\;,
\end{equation}
allows the average over the logarithm to be replace by the partition sum of a replicated set of variables. This is by an analytic continuation of $n$ to the set of integers, giving a form for which the quenched averages may be taken. The properties of the free energy are constructed through the replicated partition function
\begin{equation}
 \repZ = \prod_{\alpha=1}^n \left[\sum_{\vS^\alpha}\right]
\< \prod_{\ij} \exp \left\lbrace \beta(J^D_\ij +
 J^S_\ij) \sum_\alpha S_i^{\alpha} S_j^{\alpha}\right\rbrace\>_\quenched\;,
\end{equation}
where the quenched averages and dynamic averages may be taken equivalently.

The exponent is factorised with respect to the quenched variables in the sparse and dense parts. The method in the sparse part is a simplification of that appropriate in chapter~\ref{chapter:sparseCDMA}. The average in the dense part involves an expansion to second order in $N$ of $J^D_\ij$. The leading order terms are described by $J_0$ and $J^2$ (\ref{eq:composite.SK}), and higher order terms are taken to be negligible in the large $N$ limit. The details of the averaging in the sparse and dense parts are left to Appendix~\ref{app:CompositeSystem_Replica}, including a derivation inclusive of the F-F model and non-Poissonian connectivity. The brief outline of the method in the remainder of this section applies for Poissonian connectivity only. The site dependence in the energetic part is factorised in general by introducing three classes of order parameters
\begin{equation}
\qal=\frac{1}{N}\sum_i S^\alpha_i \;; \qquad
\qalal=\frac{1}{N}\sum_i S^{\alpha_1}_i S^{\alpha_2}_i \;; \qquad \GENOP(\rvS) = \frac{1}{N}\sum_i \delta_{\rvS,\rvS_i} \label{eq:composite.GENOP}\;;
\end{equation}
where $\qal$ describes the homogeneous magnetisation, $\qalal$ describes the 2-replica correlations, and the generalised order parameter $\GENOP(\rvS)$ describes correlations of all orders, where the bold font vector notation is used to represent a vector labeled by replica indices rather than site indices.

The definitions of $\qal$ and $\qalal$ can be defined from the generalised order parameter in the Poissonian case
\begin{equation}
\qal = \sum_{\rvsigma} \GENOP(\vsigma) \sigma^\alpha \;; \qquad \qalal =\sum_{\rvsigma} \GENOP(\vsigma) \sigma^{\alpha_1} \sigma^{\alpha_2} \label{eq:composite.equi}\;.
\end{equation}
However, solving the saddle-point equations, by population dynamics in the RS description, is complicated without the redundant description (\ref{eq:composite.GENOP}), and the redundant description is necessary in the regular and F-F models. Furthermore having order parameters describing both dense and sparse parts is useful in discriminating effects due to sparse and dense interactions and the connection with the standard sparse and dense descriptions is also made transparent in the limiting cases: taking $\qal=\qalal=0$ to recover the thermodynamics of a sparse system; and $\GENOP(\rvsigma)=1$ to recover a purely dense thermodynamic description.

The original mixed topology problem is replaced by a site factorised (mean field) model - the complexity being encoded in a set of replica correlations encoded in the order parameters. The definitions of the order parameters may be transformed to an exponential form by introducing a weighted integral over conjugate parameters (denoted by hat). The exponential form allows a saddle-point method to be applied, an extremisation of the exponent allows the free energy to be identified as
\begin{equation}
\beta \safed = \lim_{n\rightarrow 0}\frac{\partial}{\partial n}\Extr_{\{\GENOP,\GENOPconj,\qal,\qhal,\qalal,\qhalal\}} \left\lbrace \Gone(\beta,\ensemble,\GENOP) + \Gtwo(\beta,\ensemble,\GENOPconj) + \Gthree(\GENOPconj,\GENOP) \right\rbrace \label{eq:composite.replicafreeenergy}\;,
\end{equation}
up to constant (ensemble parameter dependent) terms. The term $\Gone$ encodes an energetic term describing interactions, which in the absence of an external field is given by
\begin{equation}
\begin{array}{lcl}
 \Gone &=& - \frac{1}{2}\beta J_0 \sum_\alpha (\qal)^2 - \frac{1}{2}\beta^2 \! J^2 \sum_\alal (\qalal)^2 \\
&-& \frac{C}{2} \log \sum_{\rvS,\rvS'} \GENOP(\rvS)\GENOP(\rvS') \int \rmd x \phi(x) \exp\left\lbrace \beta x \sum_\alpha S^\alpha S'^\alpha \right\rbrace \label{eq:composite.G1}\;,
\end{array}
\end{equation}
where $\phi(x)$ is the coupling distribution in the sparse part (\ref{eq:composite.pmJ}). The term $\Gtwo$ is an entropic term coupling the sparse and dense order parameters
\begin{equation}
\Gtwo \!=\! - \log \sum_{\rvS}\exp\left\lbrace \!\sum_\alpha \qhal S_\alpha \!+\! \sum_{\alal} \qhalal S^{\alpha_1} S^{\alpha_2} \!+ C \GENOPconj (\rvS) \right\rbrace\;. \label{eq:composite.G2}
\end{equation}
The coupling between the order parameters and their conjugate forms is present in the term
\begin{equation}
\Gthree \!= \! C \sum_{\rvS} \GENOP(\rvS) \GENOPconj(\rvS) + \sum_\alpha\qal \qhal + \sum_{\alal} \qalal\qhalal \label{eq:composite.G3}\;.
\end{equation}

The free energy is used to calculate various self averaging properties of the system by taking derivatives with respect to conjugate parameter, as outlined in Appendix \ref{app:ConjugateFields}. The inverse temperature is conjugate to the energy, from which the entropy is calculated. Derivatives with respect to uniform fields conjugate to $\vones$ (and $\vb$ in the case (\ref{eq:composite.HamAux3})) can be used to test emergent ferromagnetic order. By inclusion of a random field of mean zero, the variance can be used to calculate correlation functions and susceptibility, by a comparable method to that of chapter~\ref{chapter:sparseCDMA} section \ref{CDMA.stability} section.

The order parameters, defined at the extrema of the saddle-point (denoted $*$), obey coupled saddle-point equations
\begin{equation}
\GENOP^*(\rvS)\! = \! \localreplicaprobability(\rvS); \qquad \qal^* = \sum_{\rvS} S^\alpha \localreplicaprobability(\rvS) ; \qquad \qalal^* =\sum_{\rvS} S^{\alpha_1} S^{\alpha_2} \localreplicaprobability(\rvS) \label{eq:composite.saddle1}\;,
\end{equation}
where
\begin{equation}
\localreplicaprobability(\rvsigma) \propto \exp\left\lbrace C \GENOPconj^*(\rvsigma)\! + \!\sum_\alpha \qhal^* \sigma^\alpha \!+\! \sum_\alal \qhalal^* \sigma^{\alpha_1} \sigma^{\alpha_2} \right\rbrace \label{eq:composite.saddlePrvsigma}\;,
\end{equation}
is a normalised probability distribution on the replicated state space.

The conjugate parameters are determined by equations without coupling between the sparse and dense parts
\begin{equation}
\GENOPconj^*(\vsigma) \propto \sum_{\rvtau} \GENOP^*(\rvtau) \< \exp \left\lbrace \beta x \sum_\alpha \tau^\alpha \sigma^\alpha \right\rbrace\>_x \;; \qquad \qhal^* = \beta J_0 \qal^* \;; \qquad \qhalal^* = \beta^2 J^2 \qalal^* \;; \label{eq:composite.saddle2}
\end{equation}
with $x$ distributed according to $\phi(x)$ (\ref{eq:composite.phix}). From these six equations it is possible to eliminate the conjugate parameters (\ref{eq:composite.saddle2}) to leave a fixed point defined without the conjugate parameters.

\section{Replica symmetric formulation and message passing}
\label{composite.RS}

\subsection{The RS saddle-point equations}
\label{composite.RSsaddle-point}

The order parameters are defined by the standard sparse and dense RS forms
\begin{equation}
\GENOP^*(\vsigma) = \int \rmd h \RSOP(h) \prod_{\alpha=1}^n
\frac{\exp\left\lbrace h \sigma^\alpha \right\rbrace}{2 \cosh
 h}\;;\qquad \qal^* = m \;; \qquad \qalal^* = q \;;
\end{equation}
with the variational aspects captured by the normalised distribution on the real line ($\RSOP$) and two scalar parameters ($m,q$).

The saddle-point equations can then be written for the general case, inclusive of F-F and regular connectivity models, as
\begin{equation}
\RSOP(h) \propto \int \<\prod_{c=1}^{c_e} \left[\rmd h_c \rmd x_c \RSOP(h_c) \phi(x_c) \right] \delta\left(h - h^{RS} \right)\>_{b,c_e,\lambda}\label{eq:composite.RSsaddle1}\;,
\end{equation}
where
\begin{equation}
 h^{RS} = b m + \lambda \sqrt{q} + \sum_{c=1}^{c_e} \atanh\left(\tanh(\beta x_c)\tanh(h_c) \right) \label{eq:composite.hRS}\;,
\end{equation}
and $c_e$ is distributed according to the excess connectivity distribution, $b=1$ except in the case (\ref{eq:composite.HamAux3}) where $b= \pm 1$ with equal probability. The integration variable $\lambda$ is normally distributed. The dense parts are defined similarly
\begin{equation}
m = \int \<\prod_{c=1}^{c_f} \left[\rmd h_c \rmd x_c \RSOP(h_c) \phi(x_c) \right] \delta\left(h - h^{RS}\right) b\tanh(h)\>_{b,{c_f},\lambda}\label{eq:composite.RSsaddle2}\;,
\end{equation}
and
\begin{equation}
q = \int \<\prod_{c=1}^{c_f} \left[\rmd h_c \rmd x_c \RSOP(h_c) \phi(x_c) \right] \delta\left(h - h^{RS}\right)\tanh^2(h) \>_{b,{c_f},\lambda} \label{eq:composite.RSsaddle3}\;,
\end{equation}
but with the averages in ${c_f}$ being with respect to the
full connectivity distribution.

These equations can be solved by a method of population dynamics~\cite{Mezard:BLSG} as used in the previous chapter subject to two additional recursions on scalar quantities (\ref{eq:composite.RSsaddle2})-(\ref{eq:composite.RSsaddle3}).
\subsection{Composite belief propagation equations}
\label{composite.BP}

Composite BP can be interpreted in the context of the composite system as a heuristic method of determining marginals of the static probability distribution (\ref{eq:composite.P}), given a quenched sample. Whereas an exhaustive calculation requires $O(2^N)$ operations to construct a marginal, BP is guaranteed to produce an estimate in a number of operations that scales only linearly with the number of edges.

The equations from factors to nodes are trivial in the case of binary factors, so iterations on variable messages alone can be composed. Defining two directed messages for every link $\ij$, which are log-posterior ratios for spins on some cavity graphs
\begin{equation}
h^{(t+1)}_{i \rightarrow j} =\frac{1}{2 \beta} \sum_{b_i} b_i \log {\hat P}^{(t+1)}(S_i=b_i | G_{i \rightarrow j}) = \frac{1}{\beta}\sum_{k \setminus \{i,j\}} \atanh\left(\tanh(\beta h^{(t)}_{k \rightarrow i})\tanh(\beta J_{\< i k \>})\right) \label{eq:composite.BP}\;,
\end{equation}
where ${\hat P}$ is used to denote an approximated probability distribution. The cavity graph is a factor graph rooted in variable $i$ with the coupling $J_\ij$ set to zero. The assumption underlying the probabilistic recursion is the independence of log-posterior ratios, which allows them to be used accurately as priors in each step, so that the recursion is equivalent to that on a tree.

BP can be iterated from some initial condition. If correlations between messages are sufficiently weak then the messages will converge to correctly describe the probabilities. From these marginal properties, such as the magnetisation at equilibrium, can be constructed. A log-marginal may be estimated by
\begin{equation}
H^{(t+1)}_i = \frac{1}{2\beta} \sum_\tau \tau \log {\hat P}^{(t+1)}(S_j=\tau | G) = \frac{1}{\beta}\sum_{j \setminus i} \atanh\left(\tanh(\beta h^{(t)}_{j \rightarrow i})\tanh(\beta J_\ij)\right) \label{eq:composite.H}\;.
\end{equation}

The condition of sufficiently weak correlations is closely related to the notion of a pure state is statistical mechanics~\cite{Mezard:SGT}. The assumption of independent messages applies only when the log-posteriors (\ref{eq:composite.BP}) reflect the distribution in a pure state, the similarity with (\ref{eq:composite.hRS}) is not coincidental. Pure states act as local attractors of the BP dynamics, and it is only when there is a competition between these attractors that dynamics is expected to fail. With BP initialised sufficiently close (globally) to a pure state, or in the case of a unique attractor, convergence to the pure state can be anticipated.

\subsubsection{Simplification of dense messages}
\label{composite.SDM}
\begin{figure*}
\begin{center}
\includegraphics[width=\linewidth]{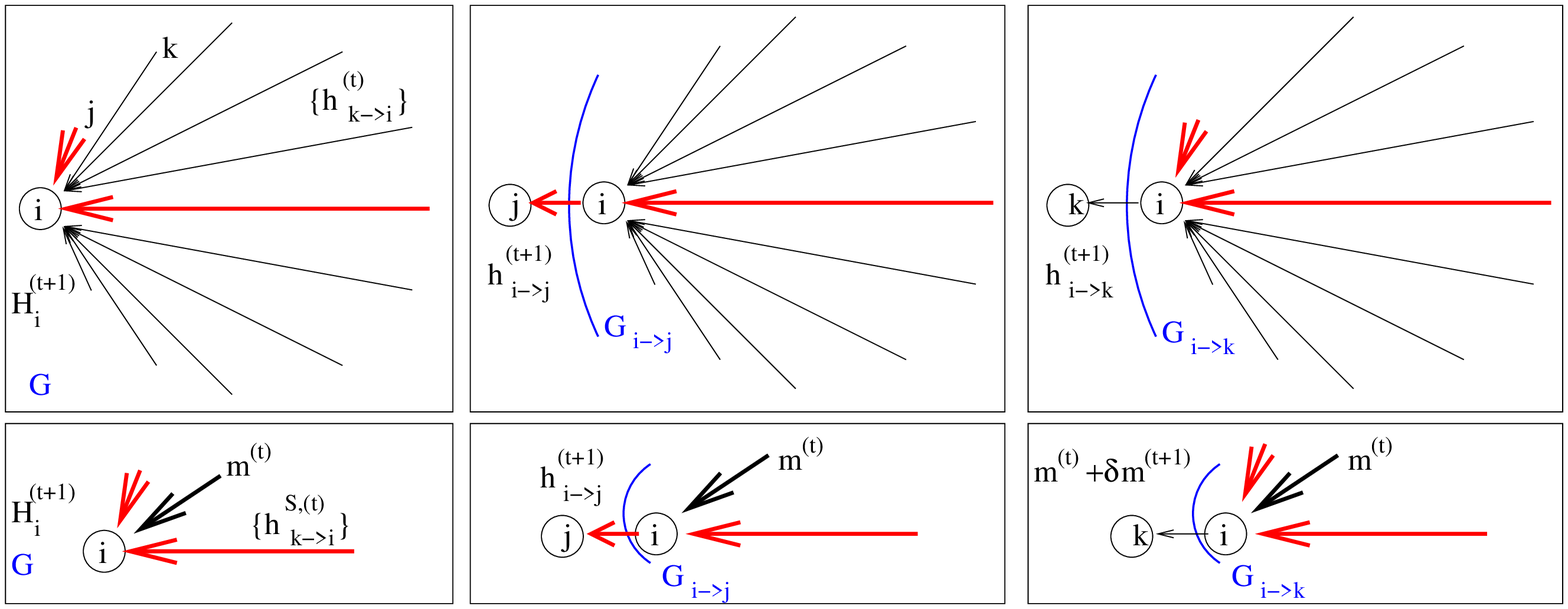}
\caption[Composite BP]{\label{fig:composite.BP} BP constructs an estimate of the posteriors by message passing, each message is a log-posterior estimate for some variable, as in the top sub-figures. In the lower two sub-figures the central limit is applied to the messages on dense links and in some cases only a single parameter is then required to represent the $O(N)$ dense messages. A related approximation is implicit in the derivation of the RS free energy.}
\end{center}
\end{figure*}

Assuming the messages to be independent, then each message can be considered as a random object determined by the couplings in the cavity graph. The messages are therefore i.i.d. and the sum over many messages will converge to a Gaussian random variable. To leading order the messages may be rewritten incorporating this insight
\begin{equation}
h^{(t+1)}_{i \rightarrow j} = m^{(t)} + \sqrt{q^{(t)}}\lambda^{(t)}_{i \rightarrow j} + \frac{1}{\beta}\sum_{k \in \{ \partial_i \setminus j\}} \atanh\left( \tanh(\beta h^{(t)}_{k \rightarrow i}) \tanh(\beta J_\ij)\right) \label{eq:composite.gaussian}\;,
\end{equation}
where $m^{(t)}$ is the mean and $q^{(t)}$ the variance, and term $\partial_i$ is used to denote variables connected to $i$ through strong couplings. The distribution over reweighed messages $\lambda_{*\rightarrow*}$ will be asymptotically Gaussian if the approximation is correct. The value of the message for a particular instance of the quenched disorder is given by
\begin{equation}
m^{(t)} + q^{(t)}\lambda^{(t)}_{i \rightarrow j} = \frac{1}{\beta}\sum_{k \setminus \{ \partial_i \cup j\}} \atanh\left( \tanh(\beta h^{(t)}_{k \rightarrow i}) \tanh(\beta J_\ij)\right)\label{eq:composite.lambda_ij}\;.
\end{equation}
The Gaussian statistics are defined by analogy with the RS thermodynamic quantities, to leading order in $N$
\begin{equation}
m^{(t)} = \beta J_0 \frac{1}{N}\sum_{i=1}^N \tanh(\beta
 H_i^{(t)}) \;; \qquad q^{(t)} = \beta^2 J^2
\frac{1}{N}\sum_{i=1}^N \tanh^2(\beta H_{i}^{(t)})\;,
\end{equation}
for any dense set of couplings~\cite{Kabashima:PB}. The log-posterior ratios for the spin states on the full graph are approximated as
\begin{equation}
\beta H^{(t+1)}_{j} = m^{(t)} + \sqrt{q^{(t)}}\lambda^{(t)}_{j} + \sum_{k \in \partial_i} \atanh\left( \tanh(\beta h^{(t)}_{k \rightarrow i})\tanh(\beta J_\ij) \right) \label{eq:composite.Hgaussian}\;.
\end{equation}
The term $\lambda_i$ is closely related to $\lambda_{i \rightarrow j}$, up to a correction of order $1/N$, by removing the restriction on the sum in $j$ from (\ref{eq:composite.lambda_ij}).

In the case that $J\neq 0$ it is necessary to evaluate $\lambda_{i}$ for each link, still requiring $O(N^2)$ evaluations as in the original algorithm. To reduce computational complexity it may be valuable to marginalise over this if $J \ll J_0$ or if the sparse couplings dominate dynamics, but if $J=0$ it is sufficient to take $\lambda_{i}^{(t)}=0$ and algorithm complexity is reduced to $O(N)$, as illustrated in figure~\ref{fig:composite.BP}.

\subsection{Stability analysis}
\label{composite.Stability}

If the replica description correctly describes a single pure state, then this implies the spin glass susceptibility is not divergent in the thermodynamic limit. The average connected correlation function can be calculated in the thermodynamic analysis by applying an infinitesimal field to each variable in the Hamiltonian, determining the derivative with respect to this field, and taking the limit of small field at the end of the calculation. It was shown in section~\ref{CDMA.stability} that the stability of the RS description is equivalent to a test of the local stability of the order parameter, and such methods may also apply to this model.

The local stability of the saddle-point equations is in fact an equivalent condition to stability of the BP equations on a typical graph in the limit $N\rightarrow\infty$~\cite{Kabashima:PB}. Stability of the BP equations is therefore explored for a typical sample. Assuming a linear perturbation $\{\delta h^{(t)}_{i\rightarrow j}\}$ about some fixed point $\{ h^{(t)}_{i\rightarrow j}\}$ of the BP equations (\ref{eq:composite.BP}), implies an independent recursion on the perturbations that may be written at leading order
\begin{equation}
 \delta h^{(t+1)}_{j\rightarrow k} = \sum_{i \setminus \{j,k\}} \delta h^{(t)}_{i\rightarrow j} \frac{(1-\tanh^{2}(\beta h^{(t)}_{i \rightarrow j}))\tanh(\beta J_\ij)}{1-\tanh^2(\beta h^{(t)}_{i\rightarrow j}) \tanh^2(\beta J_\ij)}\label{eq:composite.BPstab}\;.
\end{equation}
In the dense part the fluctuations may again be represented by a Gaussian random variable of mean and variance
\begin{equation}
J_0 \<\delta h_{i\rightarrow j}^{(t)} (1-\tanh^2(\beta
h_{i\rightarrow j}^{(t)}))\>\;; \qquad \<(\delta h_{i\rightarrow j}^{(t)})^2 (1-\tanh^2(\beta h_{i\rightarrow j}^{(t)}))^2\>\;;
\end{equation}
since the couplings are assumed to be uncorrelated with the perturbations in BP, the average is with respect to all perturbations and fields incident on $j$. An expansion of $h_{i \rightarrow j}$ in terms of $H_i$ is possible so that the statistics can be shown to be identical at leading order for all $j$~\cite{Kabashima:PB}, therefore the perturbations evolve according to quantities which are time but not site dependent
\begin{equation}
\delta m^{(t)} = J_0 \<\delta H_i^{(t)}\left( 1 - \tanh^2(\beta H_i^{(t)})\right) \>\;; \qquad \delta q^{(t)} = J^2 \<(\delta H_i^{(t)})^2 \left(1-\tanh^2(\beta H_i^{(t)})\right)^2 \> \;;\label{eq:composite.perturbations}
\end{equation}
where $\delta H_i^{(t)}$ are the perturbations in the log-posteriors, which are equal to $\delta h^{(t)}_{i \rightarrow j}$ at leading order whenever $J_\ij$ is not a strong coupling term.

A final approximation is to assume $H_i$ is uncorrelated with $\delta H_i$. In this case the statistics can be written only in terms of $q^{(t)}$, $\<\delta H_i\>$ and $\<(\delta H_i)^2\>$. However, this is not true at leading order when a sparse component is present. Variables with larger connectivity in the sparse part, are described by a field distribution of greater variance, and the perturbations scale similarly. Instead the pair of correlation functions (\ref{eq:composite.perturbations}) determines the evolution of perturbations.

Evolution of the perturbations can be undertaken in parallel with BP; to each message is attached a representative statistic for, or a distribution over, perturbations. It is sufficient to consider a distribution of perturbations characterised by a mean $\bar{\delta h}_{i \rightarrow j}^{(t)}$, and variance $\bar{\delta h^2}_{i \rightarrow j}^{(t)}$, attached to each macroscopic field. If these parameters decay exponentially, in expectation, then this is an indication of fixed point stability.

Assuming that there is no linear instability, the equation determining $\bar{\delta h^2}_{i \rightarrow j}^{(t)}$ is
\begin{equation}
\bar{\delta h^2}_{i \rightarrow j}^{(t+1)} = \delta q^{(t)} + \sum_{i \in \partial_j \setminus k} \bar{\delta h^2}_{i \rightarrow j}^{(t)} \left(\frac{(1-\tanh^{2}(\beta h^{(t)}_{i \rightarrow j})) \tanh(\beta J_\ij)}{1-\tanh^2(\beta h^{(t)}_{i\rightarrow j}) \tanh^2(\beta J_\ij)} \right)^2 \label{eq:composite.VarianceProp}\;,
\end{equation}
with a similar equation applicable to the case of a linear perturbation.

The BP equations can be interpreted as a recursive instantiation of the RS saddle-point equations (\ref{eq:composite.RSsaddle1})-(\ref{eq:composite.RSsaddle3}) except in the explicit site dependence, so that quenched disorder specific correlations may accumulate over several updates. Assuming a negligible feedback process in BP, or a modified problem without loops or with annealed disorder, the macroscopic properties established by BP will depend only on the steady state distribution of messages on sparse links and the mean and variance of dense messages. Objects analogous to a histogram estimate to $\RSOP$ (\ref{eq:composite.RSsaddle1}), and scalar parameters $m^{(t)}$ and $q^{(t)}$ in the saddle-point method. However, at the level of the mapping of individual points in the RS description (\ref{eq:composite.hRS}) it is possible that local fluctuations of the messages on fields are unstable, despite stability in the distribution. Whereas divergence in $\<\bar{\delta h}\>$ might be observed in a macroscopic instability in the first moment of $\RSOP$, an instability of the mapping in $\<\bar{\delta h^2}\>$ will not be realised in any macroscopic moment of the distribution. It is this instability in the mapping which is probed by the BP stability analysis. In the absence of a linear instability it is assumed divergence in $\<\bar{\delta h^2}\>$ is a necessary condition for any local instability.

The fluctuations on sparse messages are represented fully in this framework, whereas dense messages are summarised under approximation. The stability is a self-consistent (longitudinal) test of stability, but is known not to probe all possible instabilities and so provides only a sufficient criteria for instability. The SK model is an example where the longitudinal stability of the ferromagnetic phase, as derived through a BP framework~\cite{Kabashima:PB}, does not capture correctly the spin glass transition at low temperature, as shown in figure~\ref{fig:composite.StandardDiag}. Since the models investigated in detail later have inhomogeneity in the sparse sub-structure only ($J^2=0$), it is felt the test of stability as applied in this paper may be a more accurate reflection of true local stability towards replica symmetry breaking.


An analytic framework entirely within the replica method might also be constructed to test spin-glass susceptibility. As in chapter \ref{chapter:sparseCDMA} section \ref{CDMA.stability} a connection can be made between the particular instability in the order parameter and the divergence of the physical quantity, spin glass stability, within the RS framework. This identity is not pursued within this chapter.

\section{Exact high temperature formulation}
\label{composite.hightemp}

In the limit $\beta\rightarrow 0$ the paramagnetic solution $\GENOP=1,\qal=0,\qalal=0$ is the only stable solution, but becomes unstable as temperature is decreased. This process can be investigated by considering the moments of $\GENOP$ through a moment expansion representation
\begin{equation}
\GENOP(\rvsigma) = 1 + \sum_\alpha \qbal \sigma^\alpha + \sum_\alal \qbalal \sigma^{\alpha_1} \sigma^{\alpha_2} + \sum_{L=3} \sum_{\orderedL{\alpha}} {\bar q}_\orderedL{\alpha} \sigma^{\alpha_1}\ldots \sigma^{\alpha_L} \label{eq:composite.GENOPexpanded}\;.
\end{equation}
The saddle-point equations can be solved in each moment $\left\lbrace {\bar q}\right\rbrace$, and stability tested in some subset of the moments.

In the sparse sub-structure both the excess and full connectivity distributions are Poissonian, the saddle-point equation (\ref{eq:composite.saddle1}) can be expanded, using the identity (\ref{eq:composite.equi}), as
\begin{equation}
P(\rvsigma) = \prod_{L=1}^\infty \left[\prod_{\orderedL{\alpha}} \left[\cosh( X_L {\bar q}_{\orderedL{\alpha}}) (1 + \sigma^{\alpha_1} \cdots \sigma^{\alpha_L} \tanh(X_L {\bar q}_{\orderedL{\alpha}}) \right]\right] \label{eq:composite.P}\;,
\end{equation}
eliminating the conjugate parameters (\ref{eq:composite.saddle2}). The terms
\begin{equation}
X_1 = \beta J_0 + T_1\;; \qquad X_2 = \beta^2 J^2 + T_2\;; \qquad X_i = T_i\; \hbox{ if } \; i>2 \label{eq:composite.X}\;,
\end{equation}
determine transition properties where
\begin{equation}
T_i = C \int \rmd x \phi(x) \tanh^i(\beta x)\;.
\end{equation}
The saddle-point equations can be written in terms of an equation for each moment
\begin{equation}
\bq_\orderedL{\alpha} \! = \! \tanh(X_L \bq_{\orderedL{\alpha}}) \!+ \! \frac{(1-\tanh^2(X_L \bq_{\orderedL{\alpha}}))\< S^{\alpha_1} \ldots S^{\alpha_L}\>_{\sim \bq_{\orderedL{\alpha}}} X_L}{1 \! + \< S^{\alpha_1} \ldots S^{\alpha_L}\>_{\sim \bq_{\orderedL{\alpha}}} \tanh(X_L \bq_{\orderedL{\alpha}})}\;,\label{eq:composite.saddleexpanded}
\end{equation}
where the notation $\<\>_{\sim x}$ indicates an average with respect to (\ref{eq:composite.P}), but with $x=0$. A solution is apparent which is the paramagnetic solution with $z=\<\sigma^{\alpha_1} \ldots \sigma^{\alpha_L}\>$ and $q_*$ equal to zero. This is the only solution when $X_L\rightarrow 0$, the high temperature limit.

At lower temperature a solution may emerge in one of the moments. It is only necessary to show that some component $\bq$ allows a non-zero solution. The second term in (\ref{eq:composite.saddleexpanded}) is zero at leading order in $\bq$ in the moments of the distribution, there is no coupling of the moments at leading order. Hence any solution which emerges continuously from the solution must do so with equality at leading order between the first term of the right hand side and the left hand side. This leads to a criteria $X_L=1$ for the existence of a continuous transition.

For a discontinuous transition to occur in some component, without $X_i > 1$, requires the derivative of the second part with respect to $\bq$ to be a convex function of $\bq$ in some range of the parameter (\ref{eq:composite.saddleexpanded}). However, the derivative is a concave function of $\bq$, so that unless $X_i>1$ for some component there can be no solution other than the paramagnetic one.

\subsection{High temperature phase transitions}
\label{composite.HTPT}

The existence of non-paramagnetic order is determined from (\ref{eq:composite.saddleexpanded}) as:
\begin{equation}
\begin{array}{lcl}
 X_1 &>& 1 \qquad\hbox{1-spin / Ferromagnetic (F) order}\;;\\
 X_2 &>& 1 \qquad\hbox{2-spin / Spin Glass (SG) order}\;;\\
 X_L &>& 1 \qquad\hbox{L-spin order}\;.
\end{array}\label{eq:composite.HTT}
\end{equation}
In each case the solution which emerges may be estimated by an expansion in the right hand side of (\ref{eq:composite.saddleexpanded}) up to some order. Cubic order can be considered as a minimum to obtain the continuously emerging solution. To allow the limit $n \rightarrow 0$ an assumption on the correlations is required, RS being the simplest, the order parameters may then be determined. Depending on the order of solution required coupling between moments is relevant, and it is necessary to solve a set of coupled equations.

The emergence of a ferromagnetic phase is realised in a continuous transition towards non-zero values of $\qbal$. Through coupling of the moments $\bq_\orderedL{\alpha}$ become non-zero at order $O(X_L(\qbal)^L)$.

The emergence of a spin glass phase is realised in a continuous transition towards non-zero values of $\qbalal$, while $\qbal=0$. Other even moments become non-zero through coupling.

The transition towards an L-spin order is not relevant to the high temperature analysis, by consideration of (\ref{eq:composite.X}) it is clear that $X_L \leq X_2$ for all $L>2$, with equality only in pathological cases, therefore the transition can only be towards a ferromagnetic or spin glass phase.

In the case that $X_1\!=\!X_2$ at the high temperature transition both orders may emerge simultaneously and in competition. This case can be understood at leading order through an SK auxiliary model.

\subsection{SK auxiliary system}
\label{composite.auxiliary}
\begin{figure*}
\centering
 \includegraphics[width=\linewidth]{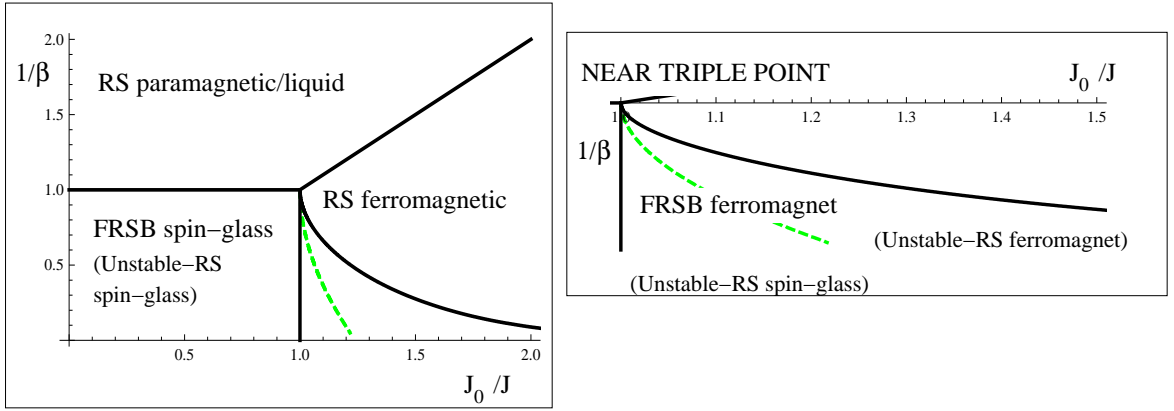}
\caption[SK phase diagram.]{\label{fig:composite.StandardDiag} The phase diagrams for disordered spin glass systems often exhibit a phase behaviour similar to the SK model. Left figure: The phase transitions are indicated by solid dark lines. As temperature is lowered there is a transition from an RS paramagnetic phase ($m=q=0$) to either an RS ferromagnetic ($m>0$) or spin glass ($q>0$,$m=0$) phase. As temperature is lowered in the ferromagnetic phase there is also an RS to Full-Replica Symmetry Breaking (FRSB) transition. Under the RS assumption the longitudinal instability measures calculated in the context of BP coincides with the F-SG transition in the RS description (dashed line). The instability of the ferromagnetic phase is not correctly predicted, the result is a lower bound in temperature for the replica instability in the ferromagnetic phase (towards a mixed phase).}
\end{figure*}

In either the case of a ferromagnetic order, or spin glass order, the behaviour is described at leading order about the paramagnetic phase by the terms $\left\lbrace \qal\right\rbrace$ and $\left\lbrace \qalal\right\rbrace$. The free energy can be written in these cases as a function of only these two types of order parameter. After elimination of conjugate parameters the free energy can be written up to constant terms as
\begin{equation}
\beta \safed \!=\! \lim_{n \rightarrow 0} \frac{\partial}{\partial n} \!\left(\!-\! \log \!\!\sum_{\rvS} \! \exp \!\left\lbrace \! X_1 \!\sum_\alpha q_\alpha S^\alpha \! + \! X_2\!\!\! \sum_\alal \qalal S^{\alpha_1} S^{\alpha_2}\!\!\right\rbrace \!+\! \frac{X_1}{2}\!\! \sum_\alpha q_\alpha^2 \!+\! \frac{X_2}{2}\!\! \sum_\alal \!\! \qalal^2 \!\right) \label{eq:composite.SKauxiliary}\;.
\end{equation}
This is the replica formulation of the SK model free energy~\cite{Sherrington:SMSG}. Therefore at leading order the high temperature phases are equivalent to the SK model, up to the $\beta$ dependence of the energetic coupling terms. Instead of the standard term $\beta J_0$ there is $X_1$, and instead of $\beta^2 J^2$ there is $X_2$.

For every composite system of Poissonian connectivity there exists an auxiliary SK model with an equivalent leading order behaviour at high temperature. By mapping the composite parameterisations to the SK model all the leading order high temperature transition properties must carry over, including the nature of Replica Symmetry Breaking (RSB) and the stability of the RS description.

Let $A$ denote the parameterisation $(J^A_0,J^A,\beta^A)$ of an SK model with an equivalent high temperature behaviour to some composite system at the high temperature transition. This parameterisation is redundant, there are only two independent parameters and so $J^A=1$ is chosen. The standard phase diagram for an SK model under these parameterisations is demonstrated in figure~\ref{fig:composite.StandardDiag}.

The auxiliary parameterisation is determined by the mapping equilibrating the coefficients in the free energy (\ref{eq:composite.X})
\begin{equation}
\beta^A J^A_0 = X_1 \;; \qquad (\beta^A)^2 = X_2
\label{eq:composite.AuxiliaryMapping}\;.
\end{equation}
Where this mapping is continuous it is possible to consider how the auxiliary system parameterisation responds to variation of temperature (or some other parameter) in the composite system.
Variation of $\beta$ in the composite model is realised as a trajectory in the auxiliary model parameter space given by
\begin{equation}
\frac{\partial J_0^A}{\partial \beta^A} = 2\frac{J_0 - J_S C
(1-\tanh^2(\beta J^S)}{J^S C \tanh(\beta J^S)(1-\tanh^2(\beta
J^S))} - \frac{1}{\beta^A} \label{eq:composite.partialJ0ApartialbetaA}\;.
\end{equation}

In the case that the couplings to higher order moments are small ($X_L \ll 1$ for $L>2$), then the mapping may be applied with some confidence to lower temperature. Such a scenario will occur when the $X_1$ and $X_2$ are dominated by the dense sub-structure terms, or when $C$ is large in the sparse sub-structure.

\subsection{Beyond leading order}
\label{composite.BLO}

The leading order approximation to the composite system differs from the SK model in the anomalous dependence of energetic components on $\beta$. This observation alone is sufficient to account for many of the novel features of composite models reported at high temperature.

About the ferromagnetic transition the term $\qal$ appears at leading order to provide a thermodynamic description. The magnitude of $(\qal)^2$ is proportional to $\Delta_1=X_1-1$ at leading order and at $L^{th}$ order the value is dependent on moments of the distribution up to $\bq_{\orderedL{\alpha}}$. The set of non-linear coupled equations can be solved in parallel at each order. The ferromagnetic phase is at leading order an RS phase so an expansion with simple RS components will be stable at leading order. The full description of the ferromagnetic phase differs from the auxiliary system description at third or fourth order.

The spin glass phase does not include any non-zero odd moments, and is described at leading order by $\Delta_2=X_2-1$, and at second order includes the term $\qbalalalal$. This term arises from the sparse sub-structure and so behaviour deviates from the auxiliary model at second order. However, since even moments have positive coefficients, all with a monotonic dependence on $\beta$, phenomenological properties may not differ significantly from the VB model which has been frequently studied (e.g.~\cite{Mottishaw:RSB}).

In the vicinity of the triple point, where both $\Delta_1$ and $\Delta_2$ are positive the terms $\qbalalal$ and $\qbalalalal$ are relevant at second order. The literature developed in studying the VB model is sufficient to describe RS properties, and stability about the triple point~\cite{Viana:PD,Mottishaw:SRF}. The leading order behaviour gives a transition from an RS ferromagnet to a spin glass according to a balance in the components $\Delta_1= \Delta_2/2$. The second order term in the sparse model indicates the existence of a mixed phase, with a refinement of the transition line.

The AT line is sufficient to describe stability of an RS solution in the dense model at all temperatures~\cite{Almeida:SSK}. In order to correctly describe transitions in the sparse or composite models it is necessary to consider a wider range of eigenvalues~\cite{Mottishaw:SRF}, which cannot be evaluated other than numerically, except at the percolation threshold (absent in the composite model) or as a polynomial expansion truncated at some order.

In~\cite{Raymond:OC} a stability analysis considering moments up to fourth order was presented. The stability analysis considers an RS description with inclusion of second order effects $\{\qbalalal,\qbalalalal\}\neq 0\}$, but with an analysis of instabilities restricted to variation in $\{\qbal,\qbalal\}$. This predicts a comparable splitting of the line $\Delta_1=\Delta_2/2$ to those found for the VB model, but for some ranges of parameters a stable spin glass phase is incorrectly identified. Since only a restricted set of eigenvalues is considered this is not unreasonable, but demonstrates a weakness in the method.

\subsection{Transitions in non-Poissonian composite systems}
\label{composite.reg}

The derivations of this section so far beginning from (\ref{eq:composite.saddleexpanded}) onwards have been specific to the case of Poissonian connectivity (\ref{eq:composite.G2}), and do not necessarily extend to composite systems with non-Poissonian connectivity or to non-coherent embedded alignments (the F-F model). The equivalent theories, and some contrasting results are demonstrated.

\subsubsection{The F-F model at high temperature}

In this case there are two distinct one spin orders described by $\qal$ and $\qbal$, describing macroscopic ferromagnetic ordered aligned with $\vb$ or $\vones$ respectively, but still the single spin glass order parameter $\qbalal=\qalal$. The derivations of previous sections remains very similar~\cite{Raymond:OC}.

At leading order all three terms are uncoupled, so that the
emergence of a ferromagnetic order, or a spin glass order remains valid. The first transition of (\ref{eq:composite.HTT}) must be modified, there are two possible one spin orders, one of which dominates so that the criteria
\begin{equation}
 X_1 \rightarrow \max\left(T_1, \beta J_0\right) > 1 \qquad \hbox{(Ferromagnetic order)} \label{eq:composite.HTT2}\;.
 \end{equation}
In the case that the maximum is in the first term the emergent phase is characterised by ($\qbal^2>0$,$\qal=0$); the spins have a macroscopic alignment along $\vones$. In the opposite case of a large second term the phase has ($\qbal^2>0$,$\qal=0$), with a macroscopic alignment of spins along $\vb$.

Each phase is a simple RS spin ferromagnet at high temperature. The case in which the critical behaviour emerges simultaneously along both alignments as temperature is lowered, the degenerate solution to (\ref{eq:composite.HTT2}), may be understood by contrast with a comparable fully connected model, the Hopfield model~\cite{Amit:SG}. The prediction is that at leading order the high temperature behaviour should be symmetric, but as temperature is lowered about the triple point P-F-F the thermodynamically favoured phase corresponds to a dense alignment, by contrast with the exact symmetry in the Hopfield model.

\subsubsection{Regular connectivity}

The replica theory is developed along similar lines to previous sections in Appendix~\ref{app:CompositeSystem_Replica} to be inclusive of the regular connectivity ensemble. The 1-spin and 2-spin dense sub-structure order parameters are determined by (\ref{eq:composite.GENOP}) and take zero values in the paramagnetic phase. The sparse sub-structure order parameter is different from (\ref{eq:composite.GENOP}) to be inclusive of non-Poissonian connectivity, but in general takes a value $1$ in the paramagnetic solution, and may be expanded as a set of moments (\ref{eq:composite.GENOPexpanded}). However, with the new definition $\qal\neq\qbal$ and $\qalal\neq\qbalal$ in general. Each of these order parameters corresponds to distinct physical quantities: $\qal (\qalal)$ are related to the mean magnetisation (2-spin correlation), whereas $\qbal,\qbalal$ correspond to these quantities weighted by connectivity in the sparse sub-structure, as indicated in Appendix~\ref{app:ConjugateFields}.

Along similar lines to the previous analysis it is possible to consider the emergence of order by treatment only of the leading order behaviour about the paramagnetic solution.
The 1-spin order terms are coupled at leading order in the saddle-point equations, thus there is no decoupled description describing emergence of spin glass and ferromagnetic order in general. The criteria for a ferromagnetic solution to emerge continuously from the paramagnetic solution as temperature is lowered is determined by the point at which
\begin{equation}
\left(\begin{array}{c} \qal \\ \qbal \end{array}\right) = \left(\begin{array}{cc} \beta J_0 & T_1 \tanh(\beta x)\\
\beta J_0 & \frac{(C-1)}{C} T_1 \end{array} \right) \left(\begin{array}{c} \qal \\ \qbal \end{array}\right) \label{eq:composite.stability}\;;
\end{equation}
if such a point exists. Existence requires the principal eigenvector of the matrix to be one. However, the existence of a solution point in the coupled equations is not guaranteed, and there exist a range of parameters in which decreasing temperature results in a pair of complex conjugate eigenvalues which exceed one in modulus.

The right hand side of (\ref{eq:composite.stability}) represents the leading order 1-spin terms in the saddle-point equations (\ref{eq:composite.saddle1}), after elimination of the conjugate parameters. In the case of Poissonian connectivity the existence of a continuous transition is necessary for the existence of a ferromagnetic or spin glass phase (\ref{eq:composite.saddleexpanded}). This is due to the concavity of the saddle-point equation, which is assumed to hold also for the regular connectivity composite system.

However, in the composite system it is necessary only for the modulus of (\ref{eq:composite.stability}) to be positive for a non-zero solution to exist. Parameterisations leading first to complex modulus one eigenvalues as temperature is lowered do not characterise a local instability in the paramagnetic solution, but the modulus $1$ criteria is sufficient for the existence of a solution distinct from the paramagnetic solution.

When the modulus of the principal eigenvalue exceeds one the assumption of weak coupling with other order parameters ceases to be valid. The criteria that the modulus in the leading order expansion is greater than one corresponds to a set of criteria
\begin{equation}
\begin{array}{lcllr}
\frac{1}{2}\left| \left(\beta J_0 + \frac{C-1}{C}T_1\right) \pm \sqrt{\left(\beta J_0 + \frac{C-1}{C}T_1\right)^2 + 4 \frac{\beta J_0 T_1}{C}} \right| &>& 1 &&\hbox{1-spin order}\;; \\
\frac{1}{2}\left| \left(\beta^2 J^2 + \frac{C-1}{C}T_2\right) \pm \sqrt{\left(\beta^2 J^2 + \frac{C-1}{C}T_2\right)^2 + 4 \frac{\beta^2 J^2 T_2}{C}} \right| &>& 1 &&\hbox{2-spin order}\;; \\
\frac{C-1}{C} T_L &>& 1 &&\hbox{L-spin order}\;.
\end{array} \label{eq:composite.HTT3}
\end{equation}
The potential exists for the modulus to exceed one whilst the discriminant is less than zero, when either $T_1$ or $\beta J_0$ are negative. This phenomena absent in the VB and SK model is contingent on one set of couplings being anti-ferromagnetic on average. In spite of a similar coupling in the spin glass term, the transition from a paramagnet to a spin glass is always a continuous one, since the discriminant is always non-negative.

\INCLUDEINFINAL{

The complex eigenvalues imply complex conjugate eigenvectors. Where the eigenvalues are real it is possible to test the stability of the equilibrium solution by inclusion of a conjugate field in proportion to the eigenvector components (see Appendix \ref{app:ConjugateFields}).
However, where the eigenvalue is complex such a field is not physical and is not consistent with assumptions made in the development of the equilibrium solution.

Attention is restricted to real valued perturbations, of the order parameters, which can be associated to real valued conjugate fields. A local instability in the paramagnetic solution is only anticipated towards a ferromagnetic phase when the real part of the principal eigenvalue is larger than one, or towards a spin-glass solution when criteria (\ref{eq:composite.HTT3}) is met. If the paramagnetic solution is stable with respect to an infinitesimal term conjugate to the magnetisation in the Hamiltonian then the paramagnetic solution will be recovered continuously as the conjugate field approaches zero. This is equivalent to the criteria that the linearised saddle-point equations are convergent to the zero solution.  Linear instability is apparent when the real-part of the eigenvalues exceed one. However, since the perturbation is not coincident with an eigenvector there is no leading order solution to the linearised equations when the external field is added. The instability in the paramagnetic solution is towards a discontinuously emerging solution.

The discontinuously emerging solution from the paramagnetic instability might be a locally stable (thermodynamic or metastable) solution across a wider range of temperatures than that indicated by the local stability analysis of the paramagnetic solution. In limited simulations, comparable in size to those described in section~\ref{composite.finite}, the behavior observed at temperatures close to (but below) the modulus one criteria (\ref{eq:composite.HTT3}) is consistent with the hypothesis of two locally stable solutions.
One solution describes the thermodynamic phase, and the other a metastable solution, with decreasing temperature a discontinuous thermodynamic transition is anticipated.

%

The case of large $\gamma$ allows only for a transition from a paramagnetic to ferromagnetic phase, and this may be discontinuous. As well as a thermodynamic solution, several dynamical transitions may describe changes in local stability criteria of the solutions; these local instabilities may dominant aspects of dynamics, and in general will not be coincident with thermodynamic transitions.

At intermediate $\gamma$ values the paramagnetic solution may be locally unstable first towards a spin glass solution as temperature is lowered. The presence of another metastable or thermodynamic ferromagnetic phase may change the properties of this transition by comparison with the standard continuous case.
}

\CUTFROMFINAL{
The discontinuously emerging solution may be only metastable, and it is difficult to establish whether it is physically reasonable, but in limited simulations, comparable in size to section~\ref{composite.finite}, behaviour observed is consistent with the hypothesis of two locally stable solutions similar to chapter~\ref{chapter:sparseCDMA}.

The presence of a pair of ferromagnetic and paramagnetic locally stable solutions can be anticipated across a range of parameterisations. One solution describing the thermodynamic phase, and the other a metastable solution, with variation of temperature a discontinuous thermodynamic transitions between the two is possible. With variation of temperature or another parameter the real-part of the conjugate eigenvalues may come to exceed one, but because the eigenvectors are not orthogonal the paramagnetic solution must remain locally stable. A local instability is only anticipated when the imaginary solutions to the eigenvalue problem disappear, or when the spin-glass solution criteria is met (\ref{eq:composite.HTT3}).

The case of large $\gamma$ allows only a transition from a paramagnetic to ferromagnetic state, and this may be discontinuous. As well as a thermodynamic solution, several dynamical transitions may describe changes in local stability criteria of the solutions, the metastable to thermodynamic transitions may dominant aspects of dynamics on these systems. These transitions may be expected not to be coincident in the regular connectivity case.

At small $\gamma$ the paramagnetic solution may be local instability first towards a spin glass solution as temperature is lowered. The presence of another metastable or thermodynamic ferromagnetic phase may change the properties of this transition by comparison with the standard continuous case.
}

In the limit of large $C$ a simplified description is possible in the transition criteria in the regular connectivity case. With a sensible scaling of the moments of $\phi(x)$ so that $T_1$ and $T_2$ remain finite as $C$ becomes large, the final term in the discriminant becomes negligible and a simple transition criteria is recovered, consistent with the Poissonian system
\begin{equation}
 \beta J_0 + T_1 > 1\;; \qquad
 \beta^2 J^2 + T_2 > 1 \;. \label{eq:composite.reglimHTT}
\end{equation}
This is also the result that would be obtained in naively applying the dense system method, using only a mean and variance of link strengths, to the two scale system. Examples of discontinuous high temperature transitions are examined in section~\ref{composite.DHTT}, with a clear departure from (\ref{eq:composite.reglimHTT}).

\section{Leading order predictions for phase behaviour}
\label{composite.leadingorderauxiliarysystems}

\subsection{The F-AF model}
\label{composite.FAFauxiliarysystem}
\begin{figure*}[!htbp]
\centering{
 \includegraphics[width=0.8\linewidth]{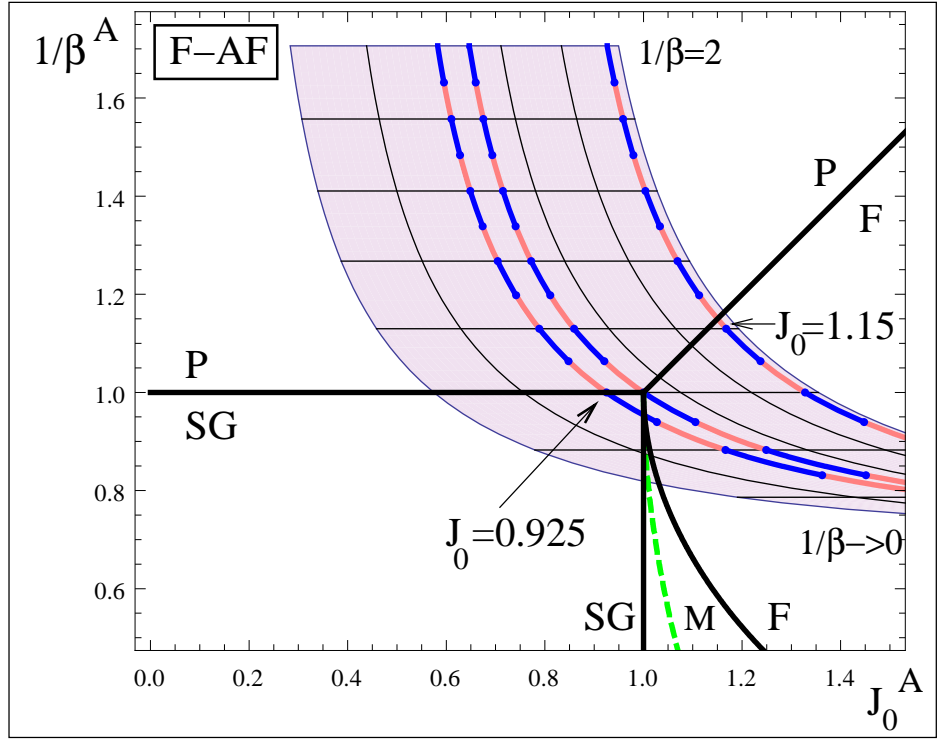}
\caption[Composite F-AF auxiliary system phase diagram.]{\label{fig:F-AF} The F-AF models (\ref{eq:composite.HamAux1}) in a parameter range ($\gamma=[0.75,1.25],1/\beta=(0,2]$) are mapped through (\ref{eq:composite.AuxiliaryMapping}) to auxiliary SK models parameterised by ($J^A_0/J^A$,$1/\beta^A$). These models are equivalent about the high temperature transition lines, and elsewhere equivalent when constraining higher than second order moments to zero (\ref{eq:composite.GENOPexpanded}). Horizontal isobars indicate constant $\beta$, and the near vertical isobars indicate constant $\gamma$, in the composite model parameter space. The set of transition lines for the SK model are shown, the upper most solid lines describing the high temperature phase transition. The SK auxiliary model predicts that as temperature is lowered in the composite models behaviour converges towards a mean field ferromagnetic behaviour. For small $\gamma$ the prediction is that a spin glass phase transforms through a mixed phase to an RS ferromagnet behaviour as temperature is lowered. Decreasing temperature about the triple point ($\gamma=1$) there is only an RS ferromagnetic behaviour. The three highlighted isobars correspond to composite systems from left to right parameterised by $\gamma=0.952$ ($J_0^{A}=0.925$ at $\beta_C$), $\gamma=1$ ($J_0^{A}=1$ at $\beta_C$) and $\gamma=1.23$ ($J_0^{A}=1.15$ at $\beta_C$), across a range of temperatures. }
}
\end{figure*}

The SK auxiliary model can be used to predict trends as temperature or some other parameter is varied in the F-AF model about the high temperature transition points. Using the mapping (\ref{eq:composite.AuxiliaryMapping}) combined with an exact (FRSB) description of the transitions and phases of the SK model at high and low temperature, the trajectories implied by the mapping can be used as a leading order predictor of phase behaviour.

Choosing the F-AF models (\ref{eq:composite.HamAux1}) such that
\begin{equation}
\mB = \gamma \;; \qquad J^S = \atanh(1/\sqrt{C})\;; \label{eq:composite.mB1}
\end{equation}
a class of models parameterised by $\gamma \in [0,\infty)$ is created. The disorder in couplings decreases with $\gamma$ from a typical spin glass set to an ordered ferromagnetic set. These models are characterised by a high temperature spin glass transition at $\beta_C=1$ when $\gamma<1$, and a high temperature ferromagnetic transition at a temperature $\beta_C^{-1}=\gamma$ when $\gamma>1$. There is a triple point in the parameter space at $\gamma=1,\beta=1$. Phase transitions between ferromagnetic and spin glass phases are possible where $\beta\gtrsim 1$ and $\gamma\sim 1$.

Near the triple point model parameterisation ($\gamma=1$) a decrease in temperature results in a competition between ferromagnetic and spin glass solutions. A graphical answer to which solution dominates is provided by figure~\ref{fig:F-AF}, for a range of high temperature transition properties. If only leading order moments are considered in the free energy then all composite systems evolve towards an RS ferromagnetic behaviour with decreasing temperature. Thus unusual transitions away from FRSB spin-glass phases towards RS ferromagnetic phases is predicted as temperature is lowered.

In the vicinity of the triple point the prediction is an accurate one at leading order about the high temperature transition. The prediction at leading order is that spin-glass to ferromagnetic solutions are possible. The derivative describing the line of RS instability in the SK model is strictly vertical at the triple point, whereas the trajectory of the composite model in the auxiliary model space (\ref{eq:composite.partialJ0ApartialbetaA}) is positive as temperature is lowered. Therefore some models exhibit a transition towards first an RSB spin-glass phase with decreasing temperature, then towards an RS ferromagnetic phase; this does not preclude transitions back to RSB at lower temperature.

In the F-AF model a spin glass phase with zero magnetisation can not be a sufficient description at low temperature. This is because the spins disconnected from the sparse sub-structure can evolve independently and undergo an independent phase transition induced by the dense sub-structure. The results at leading order are in agreement with this observation.

\subsection{The AF-F model}
\label{composite.AFFauxiliarysystem}
\begin{figure*}[!htbp]
\centering{
 \includegraphics[width=0.8\linewidth]{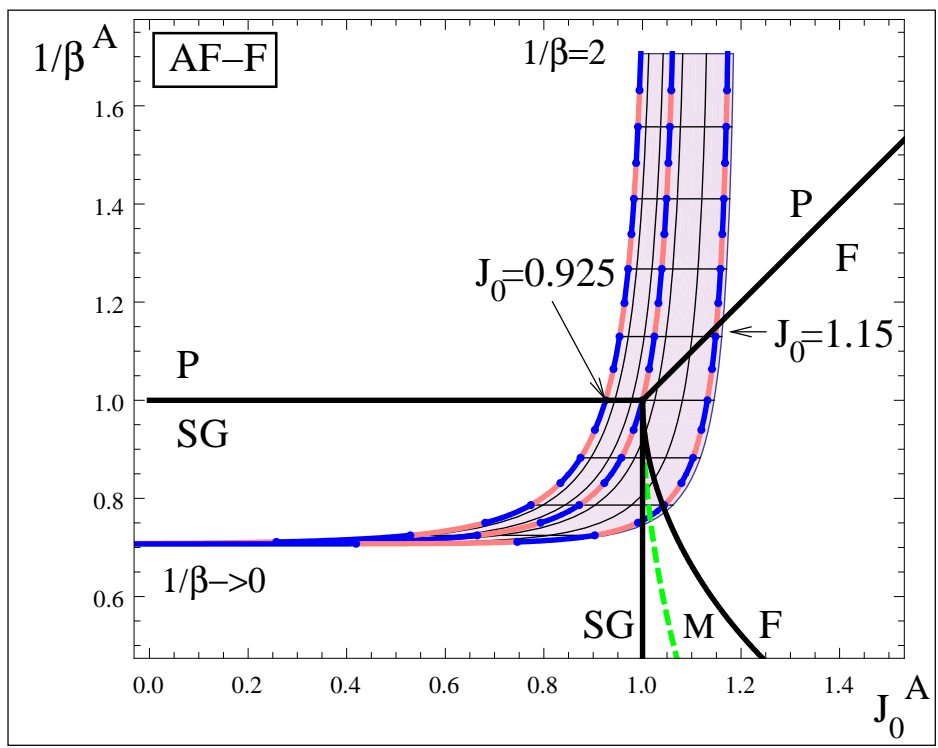}
\caption[Composite AF-F auxiliary model phase diagram.]
{\label{fig:AF-F}. The AF-F model (\ref{eq:composite.HamAux2}) as parameterised in $\gamma-\beta$ space ($\gamma=[0.75,1.25],1/\beta=(0,2]$) is mapped (\ref{eq:composite.AuxiliaryMapping}) to an auxiliary dense model parameter space. The auxiliary model prediction is that the magnetic order parameter ($m^2$) goes to zero in all composite models as temperature is lowered, a FRSB spin glass phase describes the zero temperature limit. The three highlighted systems correspond to systems with $\gamma=0.746$ ($J_0^{A}=0.925$ at $\beta_C$), $\gamma=1$ ($J_0^{A}=1$ at $\beta_C$) and $\gamma=1.23$ ($J_0^{A}=1.15$ at $\beta_C$).}
}
\end{figure*}

Consider the choice
\begin{equation}
\mB = \gamma (1 - C \tanh(J_S/\gamma))\;, \qquad J^S = \atanh(1/\sqrt{C}) \label{eq:composite.mB2}\;,
\end{equation}
as applied to the AF-F model (\ref{eq:composite.HamAux2}), with $\gamma \in [0,J_S/\atanh\left(1/C\right)]$. Again $\gamma$ describes the amount of order in couplings. Larger $\gamma$ can be considered, but these correspond to systems with small ferromagnetic couplings in the dense part rather than anti-ferromagnetic ones.

The predictions based on a leading order representation of the order parameters are shown in figure~\ref{fig:AF-F}. Composite systems are predicted to evolve towards spin glass phases as temperature is lowered, lowering temperature at large $\gamma$ results first in transitions to a stable RS ferromagnetic phases then towards a mixed phase before finally a spin glass phase. The auxiliary model predicts that at lower temperature the magnetic moment is suppressed, for all $\gamma$ up to the maximum value $J_S/\atanh\left(1/C\right)$, so that in the low temperature limit all systems are in a phase equivalent to a "finite temperature" spin-glass phase in the SK model. As temperature is lowered RS states are unstable towards RSB, which is the scenario normally observed in dense or sparse spin glass models.

The prediction that all systems converge towards a finite temperature spin glass is a consequence of the limited moment description. The spin glass behaviour is a residual effect of the sparse couplings, and at low temperature depends strongly on higher order moments which are absent in the auxiliary model. The spin glass phase is not induced by the dense anti-ferromagnetic couplings.

\subsection{Regular connectivity models}
\label{composite.DHTT}
\newfigureno{
\begin{figure}[!htbp]
\begin{center}
\includegraphics[width=\linewidth]{FIGURES/regularcomptransitions.eps}
\caption[Regular F-AF and AF-F composite model phase diagrams.] {\label{fig:composite.Regularcomptransitions} The assumption that real eigenvectors describe the dominant perturbations towards 1-spin order about the paramagnetic solution breaks down for some regular connectivity composite models (shaded region). The horizontal thick line indicate the P-SG instability, below which a spin glass solution emerges continuously. The dashed diagonal line indicates the P-F instability under an assumption the mean and variance of couplings strengths describe the transition (\ref{eq:composite.reglimHTT}). The solid line indicates the points in parameter space where the modulus of the eigenvectors (\ref{eq:composite.HTT3}) is one, below which a ferromagnetic solution may exist. Left figure: In the AF-F model continuous high temperature transitions are expected everywhere. (a) A spin glass transition is found. (b) A triple point is observed, with a single dominated ferromagnetic orientation described by a real principal eigenvector $(\qbal,\qal)$. (c) Continuous ferromagnetic transitions are observed at large $\gamma$. Under the mapping (\ref{eq:composite.mBreg}), only systems up to $\gamma\sim 2.5$, just beyond the triple point are valid AF-F models. Larger $\gamma$ describe ferromagnetic rather than anti-ferromagnetic dense couplings. Right figure: In the F-AF model a discontinuous transition to the ferromagnetic phase may be anticipated for intermediate $\gamma$. (a) The high temperature transition is to a spin glass phase. (b) The modulus (\ref{eq:composite.HTT2}) exceeds one, and so a solution in competition with the paramagnetic phase is expected, but a ferromagnetic solution does not appear continuously. The paramagnetic solution is unstable first towards a spin-glass phase as temperature is lowered. (c) A discontinuous transition from a paramagnetic phase to ferromagnetic phase is anticipated, with metastability in a range of parameters. (d) With strong dense ferromagnetic couplings a continuous high temperature transition is observed towards an RS ferromagnetic state.}
\end{center}
\end{figure}
}\newfigureyes{
\begin{figure}[!htbp]
\begin{center}
\includegraphics[width=\linewidth]{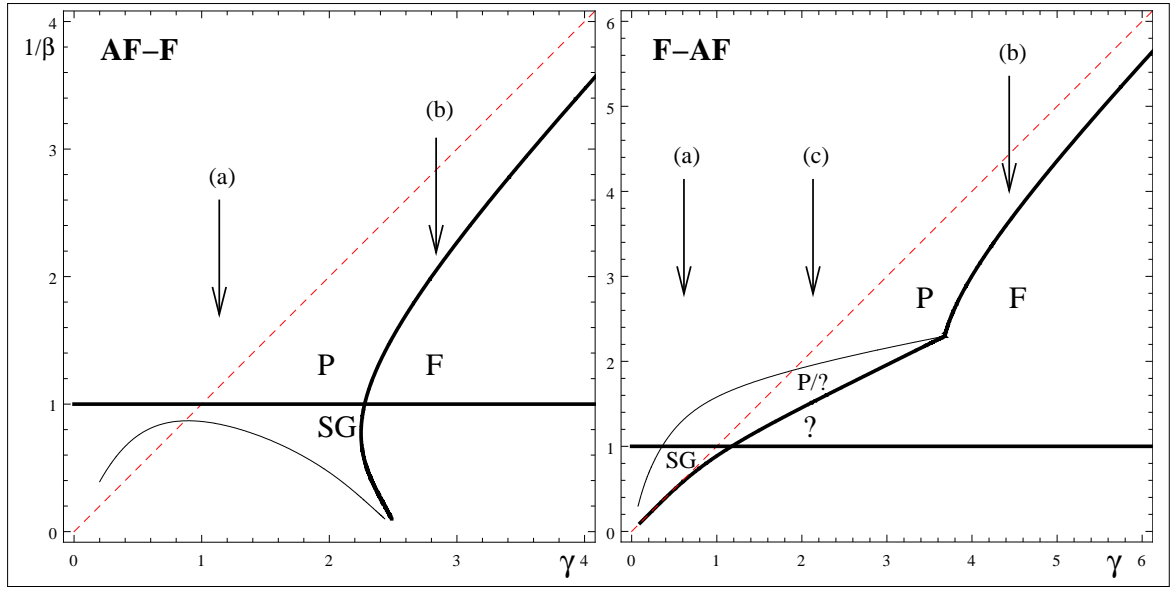}
\caption[Regular F-AF and AF-F composite model phase diagrams.]
{\label{fig:composite.Regularcomptransitions}
The phase diagrams based on high temperature perturbative analysis for regular connectivity models with connectivity $C=3$. In this figure the horizontal line indicates a high temperature instability in the paramagnetic solution towards two spin order. Other lines indicate instabilities towards 1-spin order: the straight diagonal line is assuming (\ref{eq:composite.reglimHTT}), the thick and thin lines are the points where the real part or modulus respectively of the principal eigenvector(s) equal one. Left figure: In the AF-F model decreasing temperature results in either a continuous  spin glass or ferromagnetic transition. (a) At small $\gamma$ a spin glass phase emerges continuously with decreasing temperature. (b) At large $\gamma$ eigenvectors describing 1-spin order are real, a continuous ferromagnetic transition is found.  Right figure: In the F-AF  model continuous and discontinuous transitions occur, no continuous transition triple-point exists. (a) At small $\gamma$ eigenvectors describing the 1-spin order are complex, but a spin glass high temperature transition is dominant. (b) At large $\gamma$ a continuous transition occurs described by a real eigenvector. (c) An instability in the paramagnetic solution in the first moment is anticipated at the lower (thick) line for intermediate $\gamma$, the properties of the discontinuously  emerging solution (labeled ?) cannot be established by a linearised approach. The thin line indicates instability in the modulus for the linearised system, which is speculated to relate to the existence of the non-paramagnetic solution.
}
\end{center}
\end{figure}}

Figure~\ref{fig:composite.Regularcomptransitions} demonstrates the limitations on the parameter range consistent with unique locally stable RS solutions, in the case of regular connectivity systems. The two figures correspond to the systems (\ref{eq:composite.HamAux1}) and (\ref{eq:composite.HamAux2}), but with regular couplings (\ref{eq:composite.HamAux4}). The coupling scaling is
\begin{equation}
\mB = \gamma (1 - C \tanh(J_S/\gamma)) \;; \qquad J^S = \atanh(1/\sqrt{C-1}) \label{eq:composite.mBreg}\;.
\end{equation}
The choice of $J^S$ ensures that everywhere temperature $\beta=1$ corresponds to a spin glass instability in the paramagnetic solution. The choice of scaling means that under the approximated ferromagnetic transition scheme (\ref{eq:composite.reglimHTT}), the critical temperature implying local instability in the paramagnetic solution towards ferromagnetism increases linearly with $\gamma$,, denoted by the dashed line in figure~\ref{fig:composite.Regularcomptransitions}. If the transition were predicted by (\ref{eq:composite.reglimHTT}) then a triple point would occur at $1$: for $\gamma<1$ all high temperature transitions would be of a spin glass type; and for $\gamma>1$ transitions would be of a ferromagnetic type.

With (\ref{eq:composite.mBreg}) a range of $\gamma$ allow complex eigenvalues describing the ferromagnetic instability. In the AF-F regular model this is relevant at small $\gamma$, as shown in figure ~\ref{fig:composite.Regularcomptransitions}. Although the principal eigenvalue\newfigureno{ is complex in this region the relevant high temperature transition is towards a spin glass phase, since the modulus one result is everywhere below the high temperature spin glass transition.}\newfigureyes{(s) may exceed one in modulus in some range of temperature at small $\gamma$, it is the spin glass instability that controls transition behaviour.} At larger $\gamma$ (equivalently $J_0$) a triple point is reached but here the eigenvalues are real, and a continuous ferromagnetic transition may be expected, with a leading order behaviour comparable to the SK model.

In the F-AF regular model complex eigenvalues occur in a parameter range relevant to the high temperature transition. When $J_0$ is sufficiently large a continuous high temperature ferromagnetic transition is observed, and at small $\gamma$ there is a continuous spin-glass transition. There exists a broad range of $\gamma$ between these regimes where the ferromagnetic solution can not emerge continuously from the paramagnetic solution and two locally stable solutions are anticipated. There is no triple-point in this model suitable for a perturbative analysis.

In a small number of Metropolis-Hasting Monte-Carlo simulations~\cite{Landau:GMC} two attractors corresponding to paramagnetic and ferromagnetic type configurations were found in these parameter ranges, though no systematic analysis was undertaken.

\section{Replica symmetric solution of low temperature behaviour}
\label{composite.lowtemp}

In figures~\ref{fig:J085},\ref{fig:J1} and~\ref{fig:J115} stability measures and magnetisations for the composite models, equivalent at $\beta_C$ to SK models with $J_0^{A}\!=\!1$, $J_0^{A}\!=\!1.15$ and $J_0^{A}=0.925$, are presented at various temperatures below the $1/\beta_C$. The trends found are compared to those predicted by the auxiliary model in the vicinity of the transitions, as shown in figures~\ref{fig:F-AF} and~\ref{fig:AF-F}, and also RS solutions to dense (SK) and sparse (VB) models with equivalent high temperature properties.

\subsection{Numerical evaluation of the saddle-point equations}
To work beyond a perturbative approach the RS saddle-point equations are solved by population dynamics~\cite{Mezard:BLSG}. The results are presented based on samples from a single run of a population dynamics algorithm. In population dynamics machine numbers are used for $m$ and $q$ and the distribution $\RSOP$ is represented by an order-parameter histogram ($\Histogram$) of $N$ components
\begin{equation}
\RSOP \rightarrow \Histogram = \left\lbrace h_1, \ldots, h_N \right\rbrace \label{eq:composite.Histogram}\;.
\end{equation}
The saddle-point equations (\ref{eq:composite.RSsaddle1})-(\ref{eq:composite.RSsaddle3}) are treated as a mapping with integrals and summations replaced by random samples. This implies a random map from the histogram to itself. Updating Histograms recursively by a large number of random maps, from a random initial condition, leads to an accurate description of the fixed point $\RSOP$.

The random sampling is done in such a way as to reduce fluctuations in the variance of the Gaussian distributed samples, and mean of the Poissonian distributed samples, to $O(1/N)$. A single iteration includes an update of every field in the histogram $\Histogram$ with either parallel or random sequential order. Given that anti-ferromagnetic couplings play a role in the dynamics, there is a risk that an invalid macroscopic anti-ferromagnetic state could be amplified by parallel updates. This scenario does not form a problematic point in the analyis undertaken, but was relevant to work undertaken in~\cite{Raymond:OC}, and carefully avoided. In order to control finite size effects a scheme of microcanonical sampling was employed with respect to $\Histogram$, so that each field in generation $(t)$ is involved in forming exactly $C$ fields in generation $(t+1)$.

A histogram of $65556$ floating point fields run for $1024$ iterations appears to resolve all statistical quantities of interest down to a temperature of $\sim 1/(10 \beta_C)$, with great precision, even in the vicinity of phase transitions. At lower temperature there is a rapid decrease in the resolution of statistical quantities, which is uniform across tested systems and probably related to numerical precision limitations in the representation for hyperbolic functions. Based on the converged set of order parameters samples are taken in the following $256$ iterations to determine robust system statistics.

The initial condition for the order parameters $m^2$, $q$ and $\Histogram$ are chosen as paramagnetic, combined with a small systematic bias towards spin-glass and ferromagnetic configurations with small, but non-zero values to the dense sub-structure moments ($m^2=q$), elements of $\Histogram$ are sampled according to a Gaussian ${\cal N}(m,q)$ such that the mean and variance of the histogram values are $m+O(1/N)$ and $q=O(1/N)$. Other initial conditions were also tested to ensure that dynamical bias was not implied by initial conditions, the suggested scheme converged effectively and systematically.

\subsubsection{Numerical evaluation of the stability equations}

The longitudinal stability is tested by initialising a fluctuation histogram $\delta \Histogram$
\begin{equation}
\delta \Histogram = \{ (\chi^2)^{(t)}_1, (\chi^2)^{(t)}_2,
\ldots,(\chi^2)^{(t)}_N \}\;,
\end{equation}
where each component corresponds to a distinct field in the histogram $\Histogram$ (\ref{eq:composite.Histogram}). Each component represents a topology free measure of $\bar{\delta h^2}_{i \rightarrow j}^{(t)}$, each of which is evolved according to (\ref{eq:composite.VarianceProp}), with the site dependent fields and parameters replaced by a sample of fields from $\Histogram$ and other quenched disorder determined as in the field update. Cases in which $J^2=0$ ($q^(t)=0$), without linear perturbations are considered. The stability exponent is
\begin{equation}
\lambda^{(t)} = \log \frac{ \sum_l (\chi^2)_l^{(t)}}{ \sum_l (\chi^2)_l^{(t-1)}} \label{eq:composite.lambda}\;,
\end{equation}
and is negative if BP is convergent in expectation. This is averaged over many generations, alongside renormalisation of $\delta \Histogram$ to prevent numerical precision problems.

\subsection{The F-AF and AF-F models}
\begin{figure*}
\centering{
 \includegraphics[width=0.8\linewidth]{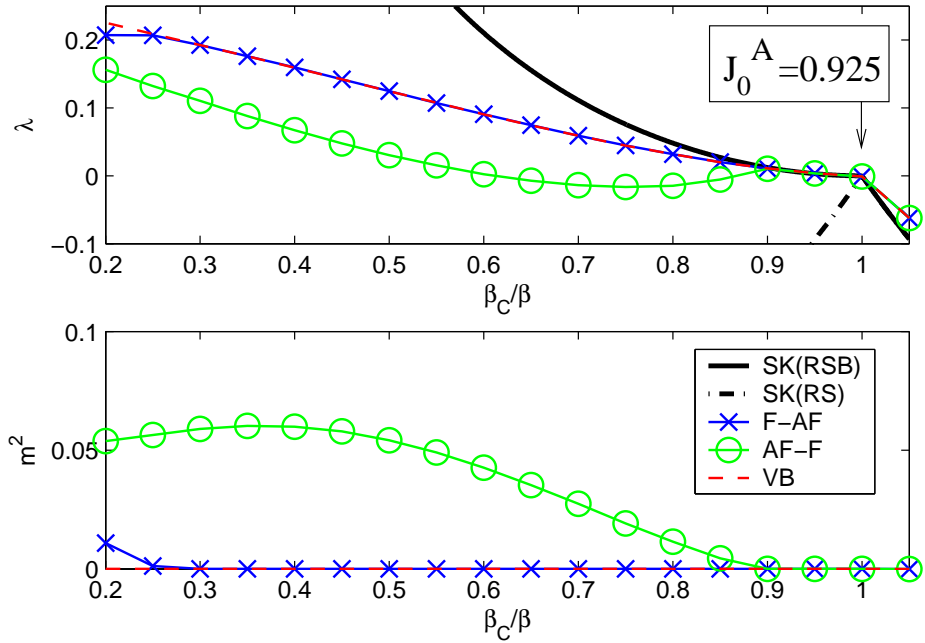}
\caption[Composite system behaviour lowering temperature from a SG-P phase transition.]
{\label{fig:J085} A comparison of the stability exponent and magnetisation for the F-AF (circles),AF-F (crosses), VB (dashed line) and SK (solid line) models under the RS assumption. Every model is equivalent at the high temperature spin glass transition point to an SK model parameterised by $J^{A}_0=0.925$, and temperature variation is considered on the rescaled interval $\beta_C/\beta=[0.2,1.05]$. In the top figure two stability exponents are given for the SK model, a longitudinal measure SK(RS) and a latitudinal measure SK(RSB). In the lower figure the sparse and dense models show similar trends with $\lambda>0$ and $m^2=0$. Composite models behave as sparse spin-glass models whenever $m^2=0$, but there is a departure in both models at low temperature. In all models as temperature decreases $\lambda>0$, except the F-AF model which is negative over an intermediate temperature range. Both the composite models attain a non-zero magnetic moment at low temperature, which is not seen in the VB or SK models. The F-AF model is in approximate agreement with figure~\ref{fig:F-AF} at high temperature. However, the behaviours observed in the composite models at low temperature are not anticipated by the auxiliary model.}
}
\end{figure*}

Results for VB, SK, F-AF and AF-F models are shown. The VB model presented for comparison is of connectivity $2$, the same as the sparse sub-structures for F-AF and AF-F models, and has a balance of anti-ferromagnetic and ferromagnetic interactions described by a PMJ model (\ref{eq:composite.pmJ}). Figure~\ref{fig:J085} demonstrates the results for the set of systems equivalent at the high temperature transition point to a dense model with $J_0^{A}=0.925$.  In all systems there is a high temperature transition that is $P-SG$ at $\beta_C=1$, behaviour is examined for relative temperature $\beta_C/\beta$ in the interval $(0.1,1.05)$.

The stability exponent ($\lambda$) and magnetisation ($m^2$) are identical in all the models very close to the transition, the phase is a spin glass ($m=0$,$q>0$) and the RS description is unstable ($\lambda>0$). The F-AF model becomes unstable towards a mixed (unstable RS ferromagnetic) phase at relatively high temperature. This is qualitatively similar to the prediction based on the auxiliary model of the composite system (see figure (\ref{fig:F-AF}), and the transition temperature is comparable to what would be predicted by the auxiliary model.

When the magnetisation is zero (the spin glass solution) only the even moments of the distribution in the composite models contribute to the composite system behaviour. These include only sparse model dependent parts for F-AF, AF-F so that these models are described by a saddle-point solution identical to the sparse model.

In the AF-F model the ferromagnetic order parameter is suppressed down to a temperature $\beta_C/\beta \approx 0.25$ where it acquires a small value. This is close to the point where $q$ reaches a maximum value, saturation is reached before $q=1$ due to the disconnected component in the sparse sub-structure. This low temperature transition must have a strong dependence on higher order moments since it is in strong contrast with the auxiliary model prediction (figure~\ref{fig:AF-F}).

\begin{figure*}
\centering
 \includegraphics[width=0.8\linewidth]{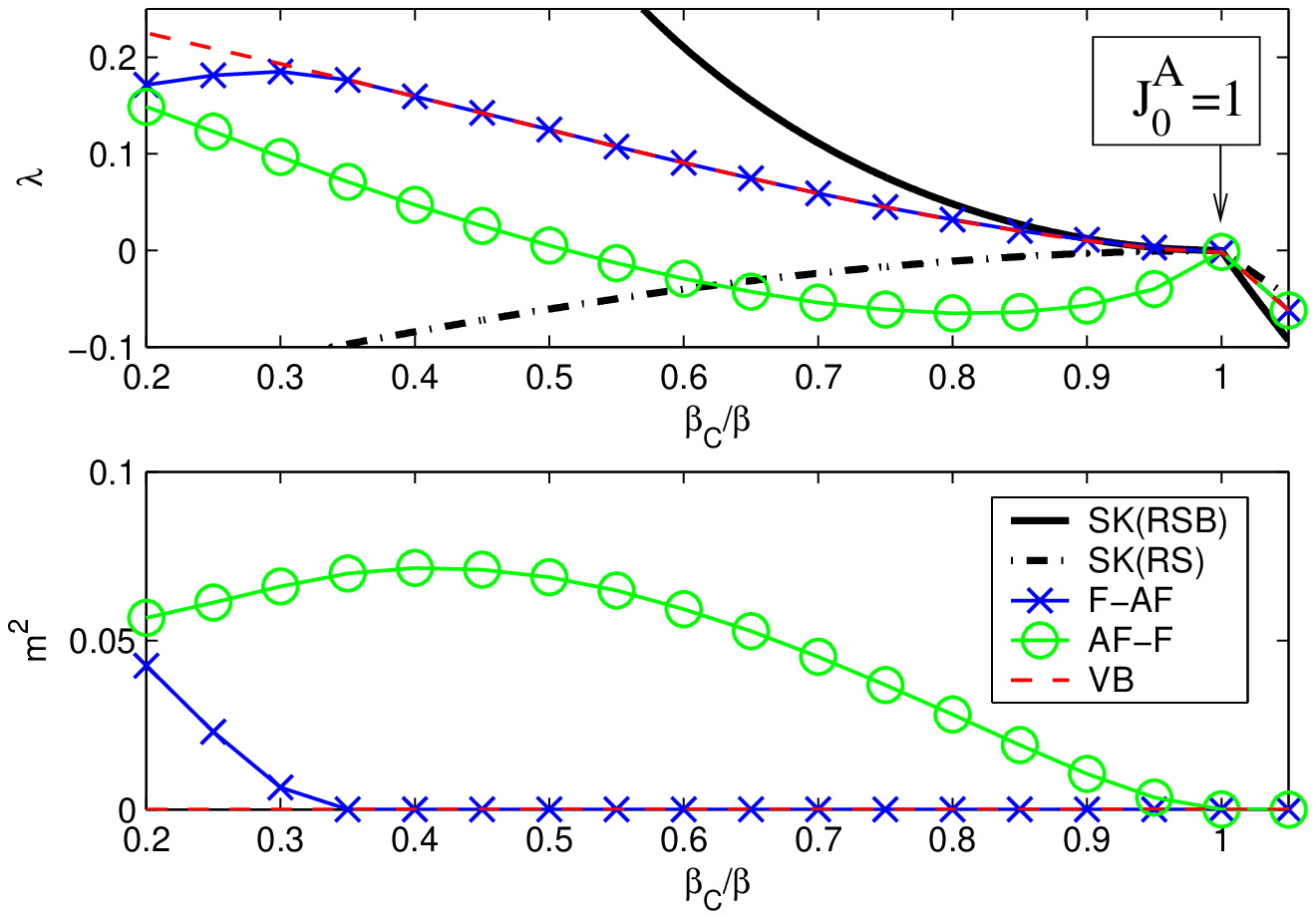}
\caption[Composite system behaviour lowering temperature from a triple-point.]
{\label{fig:J1} A comparison of the longitudinal stability and magnetisation for the F-AF (circles),AF-F (crosses), VB (dashed line) and SK (solid line) models under the RS assumption. Every model is equivalent at the high temperature transition to an SK model with $J^{A}_0=1$, coincident with the triple point in the phase diagram. Temperature variation is considered on the rescaled interval $\beta_C/\beta=[0.2,1.05]$. Trends differ in F-AF from figure~\ref{fig:F-AF} in that the magnetisation acquires a maximum value, and the stability exponent tends towards a positive value at sufficiently low temperatures. Trends differ in AF-F from figure~\ref{fig:AF-F} in the appearance of a magnetic moment at low temperatures.}
\end{figure*}

Figure~\ref{fig:J1} demonstrates results for the same models and temperature range, but for cases in which the models have a high temperature triple-point transition. In this figure the F-AF model has a behaviour clearly distinct from the other three models. As temperature is lowered a ferromagnet phase is found rather than a spin glass phase in the other cases, in agreement with figure~\ref{fig:F-AF}. At lower temperatures a maximum magnetisation is reached and a small decrease in magnetisation is discernable at the lowest values in the temperature range. With $\beta_C/\beta<0.5$ the RS ferromagnetic phase becomes unstable to a mixed phase.

Initially, at high temperatures, the AF-F model is described by a spin glass phase. With the continuous emergence of a ferromagnetic moment at low temperature there is a decrease in the stability exponent.

\begin{figure*}
\centering
 \includegraphics[width=0.8\linewidth]{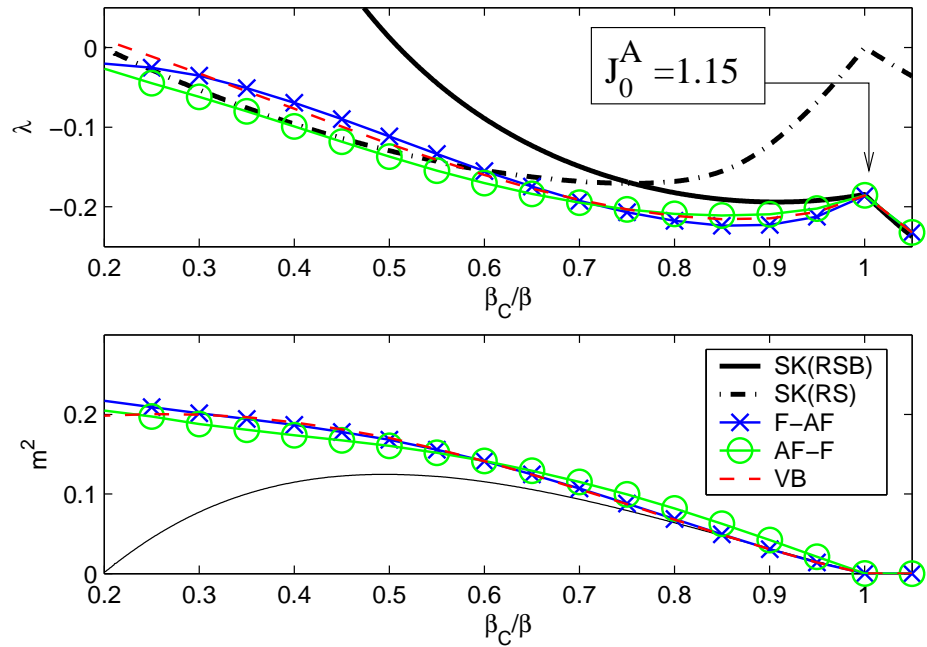}
\caption[Composite system behaviour lowering temperature from a F-P phase transition.]{\label{fig:J115} A comparison of the longitudinal stability and magnetisation for the F-AF (circles),AF-F (crosses), VB (dashed line) and SK (solid line) models under the RS assumption. Every model is equivalent at the high temperature ferromagnetic transition point to an SK model with $J^{A}_0=1.15$, and temperature variation is considered on the rescaled interval $\beta_C/\beta=[0.2,1.05]$. Two stability exponents are given for SK. The marginal stability at the Paramagnetic-Ferromagnetic transition point ($\beta_C/\beta=1$) is with respect to a linear instability, which is captured by the longitudinal instability exponent [SK(RS)], but not by the other non-linear stability exponents. F-AF properties display features of the VB model rather than the auxiliary model predictions (figure~\ref{fig:F-AF}). Trends also differ in AF-F from figure~\ref{fig:AF-F}, instability is not realised until much lower than the predicted temperature, properties are again closer to the VB model.}
\end{figure*}

In figure~\ref{fig:J115} the behaviour of systems exhibiting a high temperature ferromagnetic transition are shown, systems with auxiliary models defined by $J_0^{A}=1.15$ at the high temperature transition. In this regime reentrant behaviour is seen in the SK model, but not in the VB or composite models. The two composite models follow very closely the behaviour of the VB model, although at $\beta_C/\beta\sim 0.3$ there appears to be a modification of the trend in the stability exponent for the AF-F model absent in the F-AF and VB models.

The ferromagnetic moment is largest in the AF-F model at high temperature, and the F-AF model at low temperatures. There are also several such cross overs in the stability exponent. The RS solutions are stable for the composite systems and VB over the full temperature range presented.

\section{Reentrant behaviour and structure in finite systems}
\label{composite.finite}
\subsection{BP and Monte-Carlo simulation}
\begin{figure}[htb]
\begin{center}
\includegraphics[width=\linewidth]{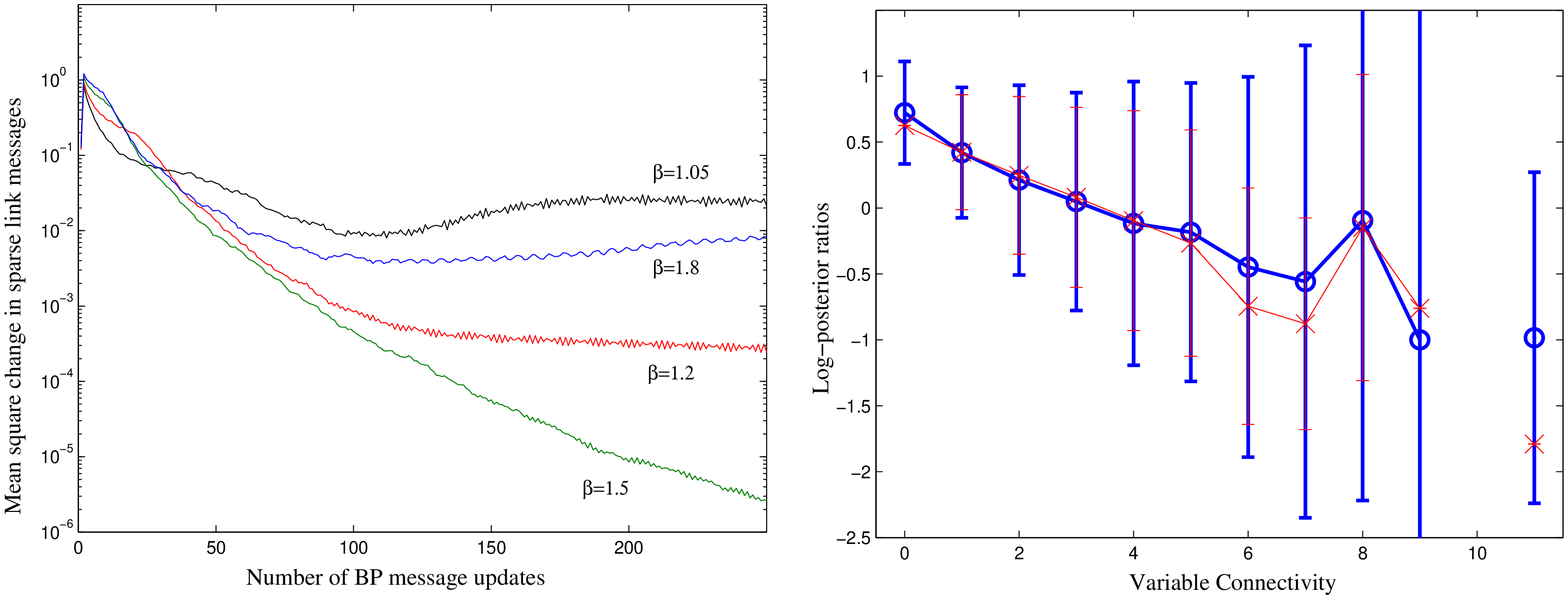}
\caption[Reentrant stability behaviour in BP and monte-carlo simulations.]
{\label{fig:composite.BPresult} Results in applying BP and Monte-Carlo simulation to an F-AF model of size $5000$ spins, and $\gamma=1$ for various temperatures. Left figure: Iteration of BP on a sample graph from various initial conditions is convergent for this sample of quenched disorder at intermediate temperature only, as indicated by the exponential decay in the stability measure. Right: For the case that BP converges $\beta=1.5$ the mean and variance in the field distribution are demonstrated as a function of variable connectivity in the sparse sub-structure. Thick lines (circles) demonstrate the results of Metropolis-Hastings Monte-Carlo simulation. Thin lines (crosses) demonstrate the estimates of BP. These are in agreement except at high variable connectivity. The magnetism of the system is supported by the alignment of low connectivity variables, with variables of high connectivity in the sparse sub-structure being magnetised in an opposite sense.}
\end{center}
\end{figure}

Some testing of thermodynamic results was undertaken in samples of $N \!= \!O(100)-\!\!O(8000)$ spins by sampling through a Metropolis-Hastings algorithm~\cite{Landau:GMC}, and estimating log-posterior ratios by Belief Propagation. These studies verified qualitatively the outcomes of the thermodynamic analysis at high temperature. The paramagnetic phase was observed to transform continuously into either a ferromagnetic, spin glass, or mixed (unstable ferromagnetic) phase as temperature was decreased. The ferromagnetic state is assumed to be described by a connected phase space up to finite size effects. Stability of the BP algorithm was measured through the mean square change in BP log-posterior estimates (\ref{eq:composite.H})
\begin{equation}
\lambda^{(t)} = \frac{1}{N} \sum_{i=1}^N \left(H_i^{(t)}-H_i^{(t+1)}\right)^2 \;,
\end{equation}
this being a new definition of $\lambda$ related to (\ref{eq:composite.lambda}), but distinguished by the algorithmic context.

Figure~\ref{fig:composite.BPresult} demonstrates a simulation of an F-AF model with $5000$ spins. This demonstrates that the non-monotonic behaviour seen in the RS solution of the F-AF model, and predicted by the leading order expansion, can be realised in finite systems also. The second part of the figure demonstrates the structure of the magnetic phase in the F-AF model. The macroscopic magnetisation is supported primarily by spins coincident with the disconnected component in the sparse sub-structure.

\subsection{Structure of phases and transitions}

In the F-AF model the inhomogeneity in magnetisations, with the disconnected component being the most strongly aligned set of variables, seems an intuitive and necessary feature in a model with such a stark contrast in coupling types.

The disconnected component appears to play an even more vital role in the AF-F model. In the magnetic phase of this model all the disconnected components are observed in Monte-Carlo and BP experiments to be anti-correlated with the macroscopic magnetisation, which is an intuitive result. Whereas almost all other variables, connected through the sparse sub-structure take values aligned with the macroscopic order. In the large system limit there should be some discrimination in the topology within the sparse-substructure. Some important topological features of sparse Poissonian graphs are outlined in figure~\ref{fig:composite.2cores}.  In general the highly connected spins may take one alignment, the disconnected component an opposite alignment, with other variables intermediate.

The inhomogeneity in the structure must also be vital in allowing continuous transitions between various phases, and in the dynamics of models. The continuous emergence of a magnetic phase as temperature is lowered in the AF-F model is presumably by a nucleation process, whereas in the F-AF model the ferromagnetic part can emerge first in the disconnected component and percolate inwards to the core of the sparse sub-structure. The absence of sufficient inhomogeneity in the regular connectivity models is responsible for the metastability found in some parameter ranges.

%% file: COMPOSITECDMA.tex
\KEEPNOTE{CHECK HORIZONTAL SPARSEDENSECOMPOSITEFIGURE FOR USE OF N}
\section{Introduction}
\label{compositeCDMA.introduction}
Code Division Multiple Access (CDMA) is an efficient method of bandwidth allocation, employed in many to one wireless communication channels~\cite{Verdu:MD}. Schematically, each source (user) is allocated a code by which to modulate some source bit across the bandwidth. The signal arriving at a sink (base station) is a superposition of the user signals and channel noise; with carefully chosen codes, the source bits may be robustly inferred. The problem addressed in this chapter is one of multiuser detection, in which the bandwidth access patterns for different users are random and not correlated in such a way as to prevent, or reduce optimally, Multi-Access Interference (MAI).

The base station must extract information from the relevant parts of the bandwidth in order to decode for a particular user. It is convenient to consider two spreading paradigms. In the first, each user transmits on the full bandwidth allocating a small amount of power to each section. Alternatively, the user may have power concentrated on one or several small sections of the bandwidth. In the former case the code is said to be dense, and in the latter, sparse. The case in which bandwidth access patterns are random and uncoordinated between users~\cite{Tanaka:SMA,Yoshida:ASS,Montanari:BPB,Guo:MDSS,Raymond:SS} is considered alongside a simple case of coordination between users on the (microscopic) level of bandwidth access. Coordination between the users allows opportunities to reduce MAI, thus producing an improved performance.

The process of wireless multi-user detection is idealised as a linear vector channel subject to Additive White Gaussian Noise (AWGN), with the transmission between each user and the base station being subject to perfect synchronisation and power control. In other words there is no unknown fading or scattering of the transmitted signals, and user power and transmission timings may be synchronised under the coordination of the base-station. A bit interval is considered, which is a bandwidth interval on which each user transmits exactly one bit.

The bandwidth is discretised as $M$ Time-Frequency blocks (chips), so that a vector describes the spreading pattern across the bandwidth. In the detection problem the set of chips are synonymous with the set of factor nodes in a factor graph, whereas the users are synonymous with variable nodes. Each user (labeled by $k=1\ldots K$) is assigned a modulating code ($\vs_k$) for transmission/detection of a random bit, $b_k=\pm 1$ sent to/from a base station. The channel load is $\load=K/M$, which is finite. Consider the transmission case where the base station has knowledge of all codes in use. A superposition of the user transmissions, along with noise ($\vomega$) arrives at the base station
\begin{equation}
 \vy = \vomega + \sum_k b_k \vs_{k} \label{eq:compositeCDMA.channel}\;.
\end{equation}
In order to allow good decoding the base station may coordinate the amplitude of codes so that in expectation the received signal to noise ratio is uniform for all users, which is a special case of power control. For example, users at greater distances (suffering greater fading) in a practical wireless phone network will be instructed to use a higher transmission power to mitigate this effect. We assume such a determination of transmission power levels and timing has been achieved, so that codes may be taken as normalised $\vs_k \cdot \vs_k = 1$. A suitable power scale is determined by the ratio of user transmission power to the noise variance, so the choice of $1$ in the model system is without loss of generality.

In the dense case, bits of information may be transmitted at a near optimal rate using pseudo-random dense spreading codes~\cite{Verdu:MD}, which are amongst the best understood CDMA systems. These codes may be generated randomly on a user by user basis, and may be quickly decoded by a matched filter or modified message passing methods under standard operating conditions. A more recent interest has been in the sparse analogue of these codes, in which performance is comparable, but decoding is based on sparse iterative methods such as Belief Propagation (BP)~\cite{Montanari:BPB}. There exists enough latitude in parameters and channel properties encountered in real systems to anticipate that each method may be optimal in different applications and operating conditions.

The composite code, like sparse and dense codes from which it is composed, has a structure that is suitable to detailed mean-field type analysis in the spread-spectrum limit, and as will be shown, can outperform sparse and dense analogues in some reasonable parameterisations of the linear vector channel. This system represents an extension of the binary coupling, zero field, composite model considered in chapter~\ref{chapter:composite}.

\subsection{Summary of related results}
The majority of results presented in this chapter form part of the paper~\cite{Raymond:CC}. Most other related literature exists in research focusing on dense or sparse coding methods, for which many results were outlined in chapter~\ref{chapter:sparseCDMA} section~\ref{CDMA.summaryrelatedresults}. Of additional relevance to the algorithmic approaches of this chapter is the work by Kabashima~\cite{Kabashima:SMA}, which formulated the dense BP algorithm in a manner suitable for low (algorithmic) complexity detection.

One model that considers a combination of sparse and dense inference structures in the linear vector channel was studied by Mallard and Saad~\cite{Mallard:BPDG}. In these works there is a consideration of a BP method for a composite CDMA detection problem, including both sparse and dense access patterns.

The work of chapters~\ref{chapter:composite} and~\ref{chapter:sparseCDMA} are relevant to this chapter, although more so the later. The case of zero field was considered in chapter~\ref{chapter:composite}, and the sparse substructure does not have a comparable local interaction structure, so no non-trivial phenomena appear to be directly transferrable.

\subsection{Chapter outline and results summary}
Section~\ref{compositeCDMA.ensemble} outlines the detection model used and describes the ensembles of composite codes, which will be studied in this chapter.

Section~\ref{compositeCDMA.RM} solves the general case of composite CDMA by the replica method. The final results are reformulated for the case of replica symmetry. Necessary modifications to the standard population dynamics and survey dynamics algorithms are proposed. An efficient composite algorithm based on BP is presented~\cite{Mallard:BPDG}, alongside other iterative detection methods. Combining a transformations of the dense messages~\cite{Kabashima:SMA} with the BP algorithm of Mallard and Saad~\cite{Mallard:BPDG} an algorithm for composite codes of complexity comparable to the quickest dense graph detectors is produced.

Section~\ref{compositeCDMA.thermodynamics} solves the saddle-point equations by population dynamics to determine optimal performance at the Nishimori temperature for the various composite systems, as well as establish the properties of the meta-stable states, which are expected to dominate detector dynamics. The meta-stable behaviour is shown to be less prevalent in composite coding methods.

Section~\ref{compositeCDMA.finitesystems} considers the algorithmic performance of the composite algorithm to standard methods in a variety of finite composite system samples. Where the power ratio is balanced between the sparse and dense part performance is relatively poor, and in some cases unstable at high MAI. Codes with a regular sparse chip access pattern are introduced and a regime in which composite codes outperform either the sparse or dense codes at equivalent power is identified in the equilibrium analysis. Composite BP is found to achieve the predicted bit error rate in moderately sized samples.

\section{Composite ensembles}
\label{compositeCDMA.ensemble}

\begin{figure}[!htbp]
\begin{center}
\includegraphics[width=\linewidth]{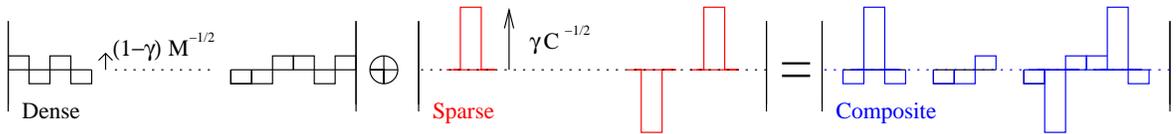}
\caption[Sparse and dense code
composition.]{\label{fig:composite.sd} The upper figure shows a standard random BPSK code for a dense system. The middle figure shows the sparse ensemble where all power is concentrated on a few ($C=3$) chips at higher power on each chip. The composite system is a superposition of these systems, the power in the sparse system is normalised to $\gamma$ and in the dense code to $1-\gamma$. The codes intersect on a small number of points, which has a negligible effect as $M \rightarrow \infty$.}
\end{center}
\end{figure}

The codes used for transmission are generated according to the sum of random sparse and dense codes (sub-codes) drawn from
independent ensembles
\begin{equation}
\vs_k = \sqrt{\gamma} \vs^S_k + \sqrt{(1-\gamma)} \vs^D_k \label{eq:compositeCDMA.s}\;.
\end{equation}
where superscripts indicate sparse and dense respectively in the right part. A schematic is shown in figure~\ref{fig:composite.sd}. If the sparse and dense sub-codes are normalised the new code will be normalised, up to a small ($O(1/M)$) factor, which is not important in the large $M$ limit considered in the equilibrium treatment. In the algorithmic analyses this is corrected for to reduce finite size effects.

The difference between composite and dense codes is in the hierarchical nature of the modulation sequences, all chips are transmitted on, but with two power scales of transmission (provided $\gamma \gg 1/M$). In terms of detector performance the subset of chips transmitted on in the absence of the dense sub-code ($\partial_k$), representing only a fraction $O(1/M)$ of the total, but remains thermodynamically relevant even as $M$ becomes large. This is not the case for standard, finite variance modulation patterns (dense codes).

\subsubsection{The detection model and statistical analysis}

A probabilistic detection model is appropriate to quantify the uncertainties in channel noise and source bits. This might be presented for a general CDMA ensemble as in section~\ref{CDMA.probability}. The principles of the statistical treatment for composite CDMA remain the same, to determine optimal detector performance the spectral efficiency is determined given the set of codes $\SEmath = I(\vb,\vy; \ms)/M$, a measure of mutual information between the source bits and signal given the code, also called the capacity or spectral efficiency. This quantity may be concisely determined through a statistical mechanics methodology. For source (bit sequence) detection purposes a noise model, and prior knowledge on the source bit sequences, can be introduced. Marginalising over the assumed noise distribution gives an estimate of the joint signal and bit probability distribution given the model (\ref{eq:compositeCDMA.channel})
\begin{equation}
P(\vb,\vy) = P(\vb)\prod_\mu \left[\int \rmd\nu_\mu
\delta\left(y_\mu -\sum_{k=1}^K s_{\mu k} b_k - \nu_\mu\right)
P(\omega_\mu=\nu_\mu)\right]\label{eq:compositeCDMA.Pvbvy}\;,
\end{equation}
where $P$ are determined by model parameters, and are assumed to match the true generative probability distributions in this section (the Nishimori temperature is used). The more general case of incorrect estimation under the AWGN model was examined in chapter~\ref{chapter:sparseCDMA}, and results of this section generalise in a comparable way.

A good model for a noisy channel of some assumed power spectral density, would be an AWGN model. The distribution may be parameterised by a variance, which in the thermodynamic formulation is equivalent to a temperature, $\beta^{-1}$. The prior estimate of bits is taken to be uniform. The detection properties are then determined from an Ising spin model with Hamiltonian
\begin{equation}
\Ham(\vtau) = \sum_{\mu=1}^M \left(y_\mu -\sum_{k=1}^K s_{\mu k} \tau_k\right)^2\label{eq:compositeCDMA.Ham}\;,
\end{equation}
at inverse temperature $\beta$. The low energy configurations of the dynamical variables $\left\lbrace \tau_k\right\rbrace$ approximate the encoded bit sequence $\vb$, according to the quenched variables: evidence $\vy$ and code $\ms$. The spectral efficiency is affine to the free energy density for this model, and takes an upper bound of $\load$ bits.

The typical case of the free energy is assumed to be representative of the set of ensembles under consideration
\begin{equation}
\beta \safed\! = \!\lim_{K \rightarrow \infty} \<-\frac{1}{K}\log \partitionfunction\>_\quenched \;; \qquad Z = \sum_{\vtau}\exp \left\lbrace -\beta \Ham(\vtau)\right\rbrace\;;
\end{equation}
where $Z$ is the partition function, $\epsilon$ the ensemble parameterisation, and $\quenched$ the weighted set of samples (codes and noise) drawn from the ensemble. From a functional form for this quantity the information theoretic properties of the channel can be extracted.

\subsubsection{The dense sub-code ensemble}

In standard dense CDMA a code is assigned to each user so that on any chip the signal transmitted is modulated according to $s^D_{\mu k}$, which is non-zero for all, or some large fraction of, chips. The standard Binary Phase Shift Keying (BPSK) random ensemble takes for each user a normalised code sampled uniformly at randomly from $\left\lbrace 1/\sqrt{M}, -1/\sqrt{M}\right\rbrace^M$, each chip is transmitted on by a user with identical power. It is convenient for analytical purposes to separate the scaling in $M$ from the modulation pattern, defining $s^D_{\mu k}\rightarrow \frac{1}{\sqrt{M}} \modulationsymbol^D_{\mu k}$, so that each modulation pattern $\modulationsymbol^D_{\mu k}$ is sampled uniformly and independently from $\{-1,1\}$.

In terms of the equilibrium analysis for large system size, all uncorrelated shift keying pattern distributions are equivalent provided the mean is $0$, variance scales as $1/M$, and some reasonable criteria are met in higher order moments, as a consequence of a central limit theorem. In finite size samples (of the size presented in this thesis) the differences amongst keying patterns is found to be quite modest and attention is restricted to the BPSK case.

\subsubsection{The sparse sub-code ensemble}
Several code ensembles were presented in chapter~\ref{chapter:sparseCDMA} and the same set are of interest in this chapter.

In a standard definition of sparse CDMA there is no transmission by user $k$ except on some finite subset ($\partial_k$) of $C_k$ chips ($\mu_1\ldots\mu_{C_k}$). Let $C$ be the mean connectivity of users in the ensemble, which is small and finite for the ensembles we study. Then the user connectivity distribution is parameterised by a distribution $P_C$. Similarly it is possible to consider the chip connectivity distribution, defined $P_L$, where $L$ is the mean chip connectivity in the sparse sub-code. In the case of no constraints on these distributions, a sparse connectivity matrix $\mA$, where $A_{\mu k}=1$ if user $k$ transmits on chip $\mu$ and zero otherwise, encodes the sparseness through a prior of the form
\begin{equation}
P(A_{\mu k})= (1-L/K)\delta_{A_{\mu k},0}+ L/K \delta_{A_{\mu
k},1} \label{eq:compositeCDMA.prior}\;.
\end{equation}
In the absence of further constraints this implies a Poissonian distribution for chip and user connectivity.

Amongst the simplest ensembles is the user regular ensemble, in which the number of accesses for all users is identical $C_k=C$, with the set of chips accessed by each user sampled independently and uniformly from the set of ${M \choose C}$ possible chip combinations. In the large $M$ and $K$ limits users become homogeneous in terms of the local connectivity profile. This homogeneous profile represents an extreme scenario amongst choices $P_C$, minimising the excess degree distribution for example. The excess degree distribution is expected to play a (non-trivial) role in information recovery, so that the homogeneous case might be optimal in the restricted set of codes parameterised only by marginal distributions.

The ensemble description in terms of $P_C$ and $P_L$ implies a distribution on the sparse connectivity matrix given by
\begin{equation}
P(\mA|P_C,P_L) \propto \prod_k \<\frac{{c_f}!}{C^{{c_f}}}\delta\left({c_f} - \sum_{\mu=1}^M A_{\mu k} \right) \>_{c_f} \prod_\mu \<\frac{{l_e}!}{L^{l_e}} \delta\left({l_e} - \sum_{k=1}^K A_{\mu k} \right)\>_{{l_e}} \prod_{\mu,k} P(A_{\mu k})\;,
\end{equation}
where ${c_f}$ and ${l_e}$ are distributed according to $P_C$ and $P_L$. The form of the pre-factors are motivated in Appendix~\ref{app.Sparsematrix}, but can be interpreted as reweighing, according to the multiplicity of the $\delta$ function and the sparse prior distribution (\ref{eq:compositeCDMA.prior}). The factors can be derived by Bayes' law.

\subsubsection{The modulation pattern for sparse codes}

BPSK is assumed to be the modulation method applied, so that $s^S_{\mu k}$ is sampled uniformly from $\left\lbrace \sqrt{1/C},-\sqrt{1/C}\right\rbrace^{C}$ for $\mu \in \partial_k$, and is otherwise $0$. Defining a quenched matrix $\mxi^S$ of modulation patterns on $\left\lbrace -1,1\right\rbrace^{M\times K}$, the sparse sub-code may be decomposed as $s^S_{\mu k}\rightarrow \frac{1}{\sqrt{C}} A_{\mu k}\modulationsymbol_{\mu k}$ to allow a convenient separation of the power, connectivity and modulation effects. The dense ensemble is recovered when $C\rightarrow M$.

For the sparse ensemble there is the possibility of strongly varying performance depending on the details of the modulation sequence, even in the absence of correlated modulation patterns. However, the room for optimisation with respect to the marginal sequence seems small and, on a practical note, the overhead in storing and processing complicated amplitude patterns in detection algorithms is an undesirable feature of any non-uniform modulation method. As shown in chapter~\ref{chapter:sparseCDMA} section~\ref{CDMA.marginals} BPSK in any case outperforms, across a range of noise levels, a Gaussian modulation pattern under independent chip analysis.

Since the sparse ensemble has many more parameters than the dense case, results are not so general, but most phenomena highlighted are expected to generalise, except possibility in the relative intensity and importance of finite size effects.

\subsubsection{Bit sequence ensemble}
\label{compositeCDMA.bits}

The bit sequence is sampled uniformly at random from the set of all possible bit sequences. This corresponds to maximum rate transmission. Since BPSK is used in this chapter, for purposes of analysis it is possible to gauge the bit codes from the Hamiltonian (\ref{eq:compositeCDMA.Ham}), $\vb=\vones$, without loss of generality.

\subsubsection{Channel noise ensemble}
An AWGN source is assumed on each chip. Using normalised codes for each user the Signal to Noise Ratio (SNR) per bit is identical for all chips and defined as
\begin{equation}
 \SNRmath = \beta_0/2 \;,
\end{equation}
where $\beta_0^{-1}$ is the variance of the noise per chip, $\<\omega_\mu^2\>$. The overall channel signal to noise ratio, power spectral density, increases linearly with $\load$.

\section{Replica method}
\label{compositeCDMA.RM}
The replica method evaluates the free energy by averaging over all samples of quenched variables, subject to the ensemble description. The replica method gives a site factorised analytical description of the free energy in the limit $K\rightarrow\infty$. In this way the random free energy determined from a sample of quenched variables
\begin{equation}
\quenched = \left\lbrace \vb, \mxi, \mxi^D, \mA, \vomega \right\rbrace \label{eq:compositeCDMA.quenched}\;,
\end{equation}
according to the ensemble ($\ensemble$) is replaced by a non-random free energy dependent on the ensemble parameterisation
\begin{equation}
\ensemble = \left\lbrace \left\lbrace \gamma, P(b_k), P_L(L_\mu), P(\omega_\mu), P(\modulationsymbol_{\mu k})\right\rbrace, P_C(C_k) \right\rbrace \label{eq:compositeCDMA.ensemble}\;,
\end{equation}
with the parameterisation broken into those parts with, and without, a chip dependence.

Whereas the model takes a prescribed form, determined by the AWGN assumption, parameterised by variance $\beta^{-1}$, the ensemble details at the level of (\ref{eq:compositeCDMA.ensemble}) are to a large extent flexible. For brevity and generality it is easiest to write expressions with some marginalisations unevaluated $\<\cdots\>$, so that the broadest class of cases is represented. This may include averages that can be computed only numerically.

The self averaged free energy is analysed by the replica trick
\begin{equation}
\beta \safed = \lim_{K \rightarrow \infty} - \frac{1}{K} \lim_{n \rightarrow 0} \frac{\partial}{\partial n} \repZ \;.
\end{equation}
The power over partition sums $\partitionfunction^n$ can be analysed for $n$ integer. The problem is then described by $n$ replicas of $K$ dynamical variables, all replicas subject to the same set of quenched variables. For each of the replicated partition functions a Gaussian integral identity may be applied to reduce the square in the exponent to a linear form
\begin{equation}
\exp \left\lbrace\! -\beta/2\left(y_\mu - \sum_k s_{\mu k} \tau^\alpha_k\right)^2\right\rbrace = \int \rmd \lambda^\alpha \frac{1}{\sqrt{2\pi}} \exp\left\lbrace -(\lambda^\alpha)^2/2 \right\rbrace \exp \left\lbrace \sqrt{-\beta}\lambda^\alpha \left(y_\mu-\sum_k s_{\mu k}\tau^\alpha_k\right) \!\right\rbrace \label{composite.HS}\;.
\end{equation}
By representing $y_\mu$ as a function of the quenched variables (\ref{eq:compositeCDMA.quenched}) and then separating those parts of order $1/\sqrt{M}$ in the exponent (due to the dense sub-code), the decomposition
\begin{equation}
y_\mu-\sum_k s_{\mu k}\tau^\alpha_k = \omega_\mu +
\sqrt{\gamma/C}\sum_{k} A_{\mu k} \modulationsymbol_{\mu
k}(1-\tau^\alpha_k) + \sqrt{(1-\gamma)/M} \sum_{k}
\modulationsymbol^D_{\mu k}(1-\tau^\alpha_k)
\label{eq:compositeCDMA.ybreakdown}\;,
\end{equation}
is possible. The sparse and dense code parts are now factorised in the exponent and the quenched averages may be made independently according to standard sparse and dense methodologies~\cite{Monasson:OP,Tanaka:SMA}. Separate order parameters are defined to describe statistical properties due to the sparse and dense factor nodes, as viewed from a particular user, and these encode a rich set of possible replica symmetries in the general case, which is undertaken in Appendix~\ref{app:compositeCDMAequi}.

Under the assumption of RS it is found that the replica variables in the site factorised form evolve independently conditioned on a set of correlations, which are a function of $\sum_\alpha \sigma^\alpha$ only. The dependence, as relevant to the dense sub-code interaction, is characterised by a Gaussian distribution, ${\cal N}(m,q)$, with mean $m$ and variance $q$. The sparse order parameter may be written in a general form
\begin{equation}
\GENOP(\rvsigma) \rightarrow \GENOP\left(\sum_\alpha \sigma^\alpha\right) = \int \rmd h \RSOP(h) \prod_{\alpha=1}^n \frac{\exp \left\lbrace h \sigma^\alpha \right\rbrace }{2\cosh(h)} \label{eq:compositeCDMA.RSOP}\;.
\end{equation}
The distribution over real valued fields $\RSOP$ encodes the set of correlations.

The free energy has only been numerically evaluated for the case of replica symmetry, the variational form for the free energy within this approximation is determined by an extremisation problem
\begin{equation}
 \beta \safed \!\propto\! \Extr_{\{\RSOP,\RSOPconj,q,{\hat q},m,{\hat m}\}} \left[\gone + \gtwo + \gthree\right]\label{eq:compositeCDMA.safed}\;.
\end{equation}
The conjugate order parameters are denoted by hat. The first term $\gone$ in the maximisation problem includes those parts that are dependent on {\em chip} ensemble parameters (\ref{eq:compositeCDMA.ensemble})
\begin{equation}
 \load \gone \!= \! - \!\log\left(I \beta_0\right)-\!\< \int \prod_{l=1}^{l_e} \left[\rmd h_l \RSOP(h_l) \sum_{\tau_l} \frac{\exp\left\lbrace h_l \tau_l \right\rbrace}{2 \cosh(h_l)}\right] \exp{\frac{1}{2 I}\left(\sum_{l=1}^{l_e} \modulationsymbol_l (1-\tau_l) + \sqrt{I'}\lambda\right)^2} \>_{\left\lbrace \modulationsymbol_l \right\rbrace,\lambda,{l_e}} \;.
 \end{equation}
The averages are with respect to a Gaussian distributed variable $\lambda$, marginal coupling distribution $\modulationsymbol_l=\pm 1$ and ${l_e}$ distributed according to $P_L$. The quantity $I$ combines the uncertainty due to the channel noise with an additional uncertainty from incomplete determination of the bit sequences, it is a signal to noise plus interference ratio (SINR)
\begin{equation}
\begin{array}{lcll}
 I \!&=&\! 1/\beta + \load (1-\gamma) (1 - q)\qquad & \hbox{Assumed SINR}\;; \\
 I'\!&=&\! 1/\beta_0 + \load (1-\gamma) (1 - 2 m +q) \qquad & \hbox{True SINR}\label{eq:compositeCDMA.SINR}\;.
 \end{array}
\end{equation}
In the free energy the true uncertainty $I$ differs from model uncertainty $I'$, except at the Nishimori temperature $\beta=\beta_0$, when $m=q$ as shown in Appendix~\ref{app:Nishimori}. At the Nishimori temperature the free energy is correctly described by the RS assumption~\cite{Nishimori:CO}.

The free energy also contains a part dependent on site ensemble parameters
\begin{equation}
\gtwo \!=\! \frac{{\hat q}}{2} - \int \< \prod_{c=1}^{c_f} \rmd u_c \RSOPconj(u_c) \log\frac{2 \cosh(\sum u_c + {\hat m} + \sqrt{{\hat q}}\lambda)}{\prod 2\cosh(u_c)}\>_{\lambda,{c_f}} \;, 
\end{equation}
where ${c_f}$ is sampled from $P_C$, and $\lambda$ is marginalised with respect to a Normal distribution. The final part of the free energy couples the two classes of order parameter
\begin{equation}
\gthree \!=\! - \frac{q{\hat q}}{2} + m{\hat m} + C \int \rmd h \RSOP(h)\rmd \RSOPconj(u) \log \left( \frac{1 + \tanh(u)\tanh(h)}{2} \right)\;.
\end{equation}
In the large or small $\gamma$ limit either the sparse or dense order parameters become negligible in determining thermodynamic properties and the usual expressions are recovered for sparse and dense ensembles~\cite{Raymond:SS,Tanaka:SMA}.

At the Nishimori temperature an equivalence of several parameters is apparent ($q=m$, so that ${\hat q} = {\hat m}$ and $I=I'$). The order parameters satisfying the extremisation condition of (\ref{eq:compositeCDMA.safed}) must give partial derivatives of the free energy evaluating to zero. This leads to the set of saddle-point equations in the sparse order parameters. For the RS case the variables $\left\lbrace m, {\hat m} \right\rbrace$ can be eliminated so that the derivatives with respect to ${\hat q}$ and $q$ imply the constraints:
\begin{equation}
q \!=\! \int \rmd \vu \< \prod_{c=1}^{{c_f}} \RSOPconj(u_c) \tanh^2\left(\sum_{k=1}^{c_f} u_k + {\hat q}+\sqrt{\hat q}\lambda\right)\>_{\lambda}\;; \qquad {\hat q} \!=\! \load (1-\gamma)(1-q) \label{eq:compositeCDMA.RSsaddle1}\;;
\end{equation}
and the derivatives with respect to the sparse order parameters imply:
\begin{equation}
\begin{array}{lcl}
\RSOP(x) \!&=&\! \int \rmd\vu \<\prod_{c=1}^{c_e} \RSOPconj(u_c)\delta\left(x -\sum_{c=1}^{{c_f}} u_c + {\hat q} + \sqrt{{\hat q}}\>_{c_e}\lambda\right)\;;\\
\RSOPconj(u) &=& \int \rmd\vh \< \prod_{l=1}^{{l_e}} \RSOP(h_l)\delta\left(u - \frac{1}{2}\sum_{\tau_0} \tau_0 \log \localpartitionfunction_{{l_e}}(\tau_0) \right) \>_{{l_e},\vmodulationsymbol,\lambda} \label{eq:compositeCDMA.RSsaddle2}\;; \end{array}
\end{equation}
where the subscript $e$ in the connectivity averages ($c_e$,${l_e}$) implies an average with respect to the marginal excess connectivity distributions for variables and chips. The quantity $\localpartitionfunction_{l_e}$ is a type of mean-field partition function
\begin{equation}
\localpartitionfunction_{{l_e}}(\tau_{0})\! \!=\! \sum_{\vtau\! \setminus\! \left\lbrace \tau_{0}\right\rbrace} \!\exp\! \left\lbrace\!\sum_{l=1}^{{l_e}} \!x_l \!\tau_l \!-\! \frac{1}{2}\!\left(\!\lambda \!+\! \sum_{l=0}^{L}\! \frac{\modulationsymbol_l}{\sqrt{I}}\! (1\!-\!\tau_l) \!\right)^2\!\right\rbrace \;.
\end{equation}

From the free energy at the saddle-point, by application of small conjugate field against the terms $\sum_\orderedL{i} \tau_{i_1} \ldots \tau_{i_L}$, it is possible to identify $P(H)$ with the distribution of log-posterior ratios on source bits in typical instances of the quenched model. Let
\begin{equation}
H_k = \frac{1}{2}\sum_{b=\pm 1} b \log \left(P(b_k=b)\right) \;,
\end{equation}
then the quantity
\begin{equation}
P(H) = \lim_{K\rightarrow\infty} \frac{1}{K}\sum_{k=1}^K \delta(H - H_k) = \int \rmd u \rmd h \RSOPconj(u) \RSOP(h) \delta(H - (u+h)) \label{eq:compositeCDMA.PH}\;.
\end{equation}
once an analytic continuation is taken in the sum. From this observation the bit error rate is defined by the integral
\begin{equation}
\BERmath = \int_{-\infty}^0 \rmd H P(H)  \label{eq:compositeCDMA.BER} \;,
\end{equation}
the spectral efficiency, with regards typical case of (\ref{eq:compositeCDMA.Pvbvy}), is attained by an affine transformation of the free energy. The entropy is also a simple function of the free energy, since at the Nishimori temperature the energy is $\frac{1}{2\load}$.

\subsection{Decoding: multistage detection and belief propagation}
\label{compositeCDMA.algorithms} The idealised achievable performance is calculated in the limit of large $M$ under the RS assumption. In practice one must deal with finite systems, and the finite size effects tend to degrade performance relative to the ideal. However, for reasonable size systems ($M \gtrsim 100$) and $\gamma\gg 1/M$ the properties of composite codes in decoding, based on suitably constructed heuristics, become distinguishable from the performance through sparse or dense decoding methods, and approach in many cases the solutions predicted by the equilibrium analysis.

Two algorithms are analysed: BP and Multistage detection (MSD). The MSD algorithm~\cite{Verdu:MD} involves iteration of a vector approximation to the source bits
\begin{equation}
H_k^{(t+1)} = \sign\left[\vs_k \cdot \vy - \sum_{k'\setminus k} Y_{k k'} H_{k'}^{(t)}\right] \label{MSD}\;,
\end{equation}
using a matrix of interference factors
\begin{equation}
Y_{k k'} = \vs_k \cdot \vs_{k'}\;,
\end{equation}
to adjust an initial matched filter estimate. MSD is a heuristic method~\cite{Verdu:MD}, which works well in dense codes and simple noise models, provided MAI is not too large. BP is based on passing of conditional probabilities (real valued messages) between nodes in a graphical representation of the problem~\cite{Kschischang:FG}.

BP involves passing conditional probabilities and marginalising of probabilistic dependencies. The most time consuming step in BP is marginalisation, a naive approach in the dense case requires $O(2^M)$ floating point operations for every interaction (chip). However, due to the central limit theorem the dependence on the weakly interacting bits, not connected strongly through the sparse code, is equivalent to a Gaussian random variable and the marginalisation is replaced by an exact Gaussian integral. This reduces algorithm complexity asymptotically to $O(M^2)$, as shown in Appendix~\ref{app:compalg.marginalisation}.

The approximation leads to a more concise form for the {\em evidential} messages (passed from factor nodes to variable nodes):
\begin{eqnarray}
u^{(t)}_{\mu \rightarrow k} &=& \sum \tau_k \frac{1}{2 \beta}\log\left( \partitionfunction_{\mu \rightarrow k} (\tau_k) \right) \label{cavbiasmessS}\;;\\
\partitionfunction_{\mu \rightarrow k}(\tau_k) \!&\doteq&\! \prod_{l \in \partial_\mu \setminus k} \left[\sum_{\tau_l}\exp \left\lbrace \beta h^{(t)}_{l\rightarrow \mu} \tau_l \right\rbrace\right] \nonumber\\
&\times& \exp\left\lbrace - \frac{1}{2 I^{(t)}_{\mu k}}\left(y_\mu - \sum_{l \in \partial_\mu} s_{\mu l}\tau_l - \sum_{l\setminus \partial_\mu} s_{\mu l} \tanh(\beta H^{(t)}_l)\right)^2\right\rbrace\label{eq:compositeCDMA.Zmuk}\;; \\
I_{\mu k}^{(t)} &=& \frac{1}{\beta} + \sum_{l\setminus \{k,\partial_\mu\}} s_{\mu l}^2 \tanh^2 (\beta h^{(t)}_{l \rightarrow \mu}) \!\doteq \! \frac{1}{\beta} + \load (1-\gamma)\left(1 - \frac{1}{K}\sum_{l=1}^K \tanh^2 (\beta H^{(t)}_{l})\right) \label{eq:compositeCDMA.Amuk}\;;
\end{eqnarray}
where a further simplification is possible for messages passed along dense links, using an expansion to leading order in $s_{\mu k}= O(1/\sqrt{M})$,
\begin{equation}
u^{(t)}_{\mu \rightarrow k} \!\doteq\! \frac{1}{\beta I_{\mu k}^{(t)}} s_{\mu k}\left(y_\mu - \sum_{i\setminus \{k,\partial_\mu\}} s_{\mu i} \tanh (\beta H^{(t)}_i) - \sum_{l \in \partial_\mu} s_{\mu k} \tanh (\beta h^{(t)}_{l \rightarrow \mu})\right)\label{cavbiasmessD}\;,
\end{equation}
as constructed in~\cite{Mallard:BPDG}. In these expressions the notation $\doteq$ indicates those equations where some $O(1/M)$ corrections have been eliminated, the most critical being the replacement of the full marginalisation over densely connected variables in (\ref{eq:compositeCDMA.Zmuk}) by a Gaussian integral that is taken analytically. At termination time a bit estimate is determined by decimating all fields to their nearest bit value, $\vtau^{BP} = \sign(\vH^{(T)})$. Evidential messages may be combined in a standard way to give marginal log-posterior estimates for the source bits
\begin{equation}
H^{(t+1)}_k = \frac{1}{2 \beta} \sum_b b \log P(b_k=b| \vy) = \sum_{\mu=1}^M u^{(t)}_{\mu \rightarrow k} \;,
\end{equation}
and {\em variable} messages (passed from variable nodes to factor nodes)
\begin{equation}
 h^{(t+1)}_{k \rightarrow \mu} = H^{(t+1)}_k - u^{(t)}_{\mu \rightarrow k} \label{cavfieldmess}\;.
\end{equation}
The algorithm remains $O(M^2)$ comparable to matched filter or MSD (\ref{MSD}) but with a large multiplicative factor; however, the expression may be manipulated without introducing any additional errors at leading order in $M$ to an algorithm with dense messages eliminated, as outlined in Appendix~\ref{app:compalg.elimination}. The manipulation is an application of methods proposed in~\cite{Kabashima:SMA} for a dense inference problem. The removal of the $O(K\times M)$ dense messages is a substantial improvement on the algorithm, reducing memory requirements as well as improving the speed by a large factor.

Composite BP is applied as a heuristic algorithm based on an unbiased initialisation of the messages, in the hope that the various simplifications on the algorithm do not produce strong finite size effects. BP exactly describes the marginal probability distributions only if the messages (\ref{cavbiasmessS}), (\ref{cavbiasmessD}), (\ref{cavfieldmess}) converge to a unique fixed point, since in this case $\vH$ describes the log-posterior ratios. There are two scenarios to be concerned about, either the BP messages fail to converge, or they converge to an incorrect fixed point - both scenarios occur in different decoding regimes for CDMA. The requirements for standard BP to successful decode are closely related to the assumption of RS, hence the similarity of the minimisation process for the functions $\{\RSOP,\RSOPconj\}$ (\ref{eq:compositeCDMA.RSsaddle2}) and the BP equations.

In decoding samples of finite size two message update schemes are considered for MSD and BP. The first is a parallel update scheme where all variables are updated such that the values of the current generation of messages ($t+1$) are conditionally independent given the previous generation of messages $t$. The second schemes is a random stochastic update method, the updates are applied to all messages in the population, but in a random order. As soon as a variable is updated it is made available to subsequent updates, the messages in a single generation ($t+1$) are then not conditionally independent given the previous generation ($t$). The sequential update method is slower to implement, but helps to suppress oscillations observed in some parallel update schemes, that can lead to oscillating dynamical attractors. This was not found to be a significant problem as load ($\load$) increased.

A measure of convergence for BP and MSD is the mean square change in variable estimates as determined by a log-posterior in the case of BP, and bit estimates in the case of MSD,
\begin{equation}
 \lambda^{(t)}=\frac{1}{K}\sum_{i=1}^K \left(H^{(t)}_i-H^{(t-1)}_i\right)^2 \label{eq:lambda}\;.
\end{equation}
An exponential decay in this quantity, or an evaluation to zero, would be characteristic of a converging, or converged, iterative method.

\subsection{Properties of decoders in finite systems}

MSD is an iterative method which works very well in systems with small load $\load$ and mixing parameter $\gamma$. In the first iteration the achieved result is equivalent to a matched filter. In subsequent iterations the estimates are updated, but because the information is rather crudely used the consequence can be instability of the iterative procedure when MAI is large. Since MSD is based on filtering it is not so successful for composite ensembles as for dense ones, and its reliability in dense codes improves as system size increases.

The critical scenario in which BP is guaranteed to produce the correct marginal posteriors is that the graphical model is tree like. However, BP often produces a reasonable performance in loopy models including sparse~\cite{Montanari:BPB} and dense~\cite{Kabashima:SMA} CDMA. A failing regime in BP often corresponds to large $\load$ for sparse and dense codes, but in the composite code a strong dependence on $\gamma$ is also apparent. The composite algorithm proposed works less effectively with intermediate $\gamma$.

In many of the cases studied it was found that BP converged in the marginal log-likelihood ratios, this was the case for systems at small $\load$, and/or high SNR. In other cases the fields did not converge, and instead a steady state was reached -- remembering the BP equations describe a dynamic algorithm, which does not obeying detailed balance, this might be expected. The steady state is one in which the distribution of messages converges up to finite size effects, but the individual messages do not converge. Steady states were characteristic of systems initiated with messages unbiased towards the source bits at high $\load$. The estimates determined from the distribution of messages in the steady state typically correspond to high BER estimates.

In regimes where message passing is unstable the detectors may still be used to provide an estimate, subject to some termination criteria. Variations on MSD and BP involving heuristic tricks may avoid some of these effects, but some of the standard methods may be unsuitable to the composite model. Experimentation with the update scheme demonstrated improved results in MSD for example.

The dynamics of numerically solving the saddle-point equations (\ref{eq:compositeCDMA.RSsaddle1})-(\ref{eq:compositeCDMA.RSsaddle2}) are very closely related to the dynamics employed in BP. The saddle-point dynamics (figure~\ref{metatimeseries}) appear smooth and systematic even at large $\load$ based on a numerical solution involving $10000$ points. However, in addition to the use of a large system size, control was exercised over finite size effects through selective sampling from the integration variables so as to reduce finite size effects in the mapping (\ref{eq:compositeCDMA.RSsaddle2}), which is not possible for the analogous quenched variables in BP (\ref{cavbiasmessS}). Therefore any realisation of the problem in BP, even at an equivalent system size, is not expected to produce such smooth effects. However, many qualitative features such as the speed of convergence in the vicinity of dynamical transition points appear to be reproduced in some finite size realisations.

Finite size effects are significant for the size of model investigated, but the trends presented appeared consistent across a range of system size from $O(100)$ to $O(1000)$ chips. No structured attempt is made to calculate these effects, or to distinguish the contributions due to the different $O(1/M)$
approximations in the algorithm, and other instabilities implicit to BP. Working with a sufficiently large graph to study these effects is restricted by the storage and manipulation of a $K \times M$ dense sub-code, the algorithm complexity is asymptotically $O(M^2)$ rather than linear as in a sparse system.

\section{Statistical physics results}
\label{compositeCDMA.thermodynamics}
\subsection{Parameters considered}

The model constructed is already quite simple, avoiding many idiosyncracies of real channels and making no attempt to optimise composite ensembles to account for finite size effects. However, even with these simplifications the channel produces interesting behavior. In order to demonstrate the equilibrium properties of composite codes in such a way as to produce strong contrast between the composite, dense and sparse ensembles samples parameterised by $C=3$ and $\load$ between $3/5$ and $2$ are used. Except in section (\ref{compositeCDMA.regular}) all results correspond to the user regular sparse sub-code ensemble -- the ensemble in which codes are independently sampled for every user.

Analysis of the sparse code ($\gamma=1$) is for this range of parameters a loopy inference problem, but is sufficiently far from the percolation transition for a giant graph component to exist in the sparse part in every sample. At the same time $C=3$ is sufficiently small to allowing quick decoding, and produces a contrast with the dense code. It has been noticed since the first studies on sparse codes that the mean connectivity of the sparse code ensemble $C$ need not be very large for results to become indistinguishable from the dense code~\cite{Yoshida:ASS,Montanari:ABP}.

A lower bound to the achievable bit error rate in all ensembles is given by the single user Gaussian channel (SUG) result over a bit interval
\begin{equation}
\hbox{SUG}= \int_{-\infty}^{0} \rmd \nu \frac{\sqrt{\beta_0}}{\sqrt{2\pi}} \exp\left\lbrace -\beta_0 (\nu-1)^2/2 \right\rbrace\;,
\end{equation}
which is the complementary error function of SNR. In the absence of MAI this lower bound can be achieved if spreading patterns are coordinated so as to be orthogonal. On the vector channel this orthogonality is possible only if $K \leq M$, unavoidable MAI at higher loads strictly degrades performance.

The saddle-point equations (\ref{eq:compositeCDMA.RSsaddle1}) - (\ref{eq:compositeCDMA.RSsaddle2}) are solved by population dynamics~\cite{Mezard:BLSG}, an iterative method using a histogram approximation to the distribution $\RSOP$ ($10000$ points are sufficient to attain our results). Evolving the order parameters from initial conditions that correspond to low and high BER finds either the unique solution or a pair of locally stable solutions.

A distinction is made between a good solution and a bad solution. A good solution has low BER, less than $10^{-2}$, which is a strongly aligned state. The bad solution has higher BER, so that good and bad are qualitative statements of detector performance. Many features in detection undergo changes in behaviour at about $\SNRmath=6-10 $dB, which, amongst other effects, may be detected as a cusp in the strengths of correlations. In the case that solutions are unique then this cusp in some sense discriminates the good and bad solutions, although there is not a technical transition (discontinuity in any moment) as the transition occurs. In the case that metastable solutions exist, they occur as locally stable complementary (bad/good) solutions to the stable thermodynamic good/bad solution. In regimes without unique solutions there are both dynamical and thermodynamic transitions between the solutions.

\subsection{Equilibrium behavior of unique saddle-point solutions}
\begin{figure}[!htbp]
\centering{
\includegraphics[width=0.7\linewidth]{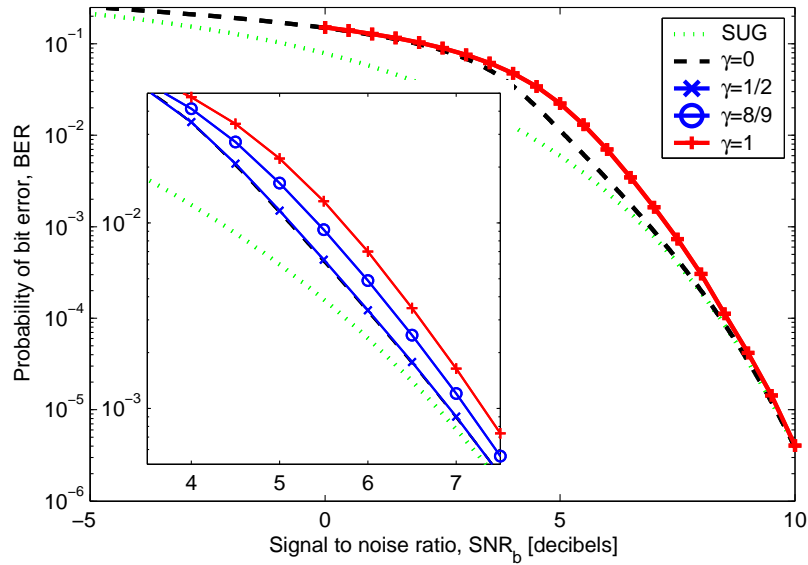}
\caption[Composite CDMA RS results, single solution
regime.]
{\label{fig:composite.normalregime} The figure demonstrates the BER determined from the order parameters at the equilibrium solution of the free energy for various SNR and $\load=1$. The curves represent different ensembles ($\gamma$), with the single user Gaussian (SUG) channel lower bound also displayed (dotted line) for comparison. Error bars are significantly smaller than symbol size for BER above $10^{-4}$, and are excluded for clarity. The lower bound is approached for the CDMA codes at large and small SNR, the dense code is best amongst the random codes. The code with an even power distribution between the sparse and dense parts ($\gamma=1/2$) is not easily distinguishable in thermodynamic performance from the dense code, even where the spread of codes is greatest (inset).}}
\end{figure}

Generally with $\load \lesssim 1.5$ there is a unique solution of the saddle-point equations with a smooth transition between bad and good solutions as SNR is increased. The population dynamics equations require few iterations to converge and results can be achieved with relatively fewer points in the histogram. The normal working range of CDMA is often by design one with a relatively small load ($\load\!<\!1$) and so falls into this class of behaviour.

The equilibrium values for BER with $\load\!=\!1$ are demonstrated in figure~\ref{fig:composite.normalregime}. The dense code ensemble achieves a smaller bit error rate than the sparse code ensemble, and the composite code ensembles interpolate between these. With $\gamma\!=\!0.5$ the curve is indistinguishable at this magnification from the dense curve, performance resembles the dense code with evenly distributed power in the two codes. At intermediate SNR there is a large gap in BER between the composite codes and the single user channel performance, which narrows in the limits of high and small SNR. Trends in the free energy follow a similar monotonic pattern -- the dense code has the highest spectral efficiency everywhere.

\subsection{Metastable solutions of the saddle-point equations}
\begin{figure}[!htbp]
\includegraphics[width=1\linewidth]{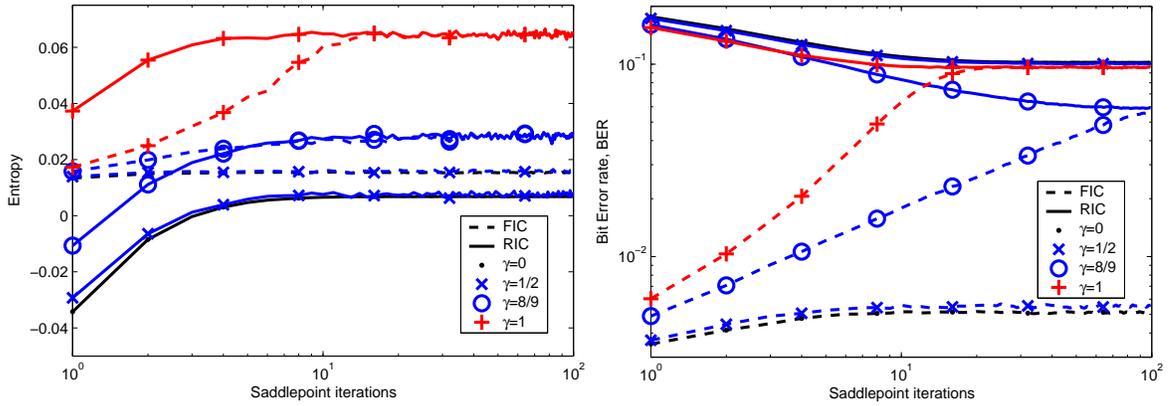}
\caption[Saddle-point equation dynamics for composite CDMA in the vicinity of a dynamical phase transition.] {\label{metatimeseries} The dynamics of the order parameters determined by iteration of the saddle-point equations is shown for $\load=5/3$ and $\SNRmath=6$dB, with a large histograms of $10^6$ points to represent the distribution $\RSOP$ (\ref{eq:compositeCDMA.RSsaddle1}). Evolving the saddle-points from either Ferromagnetic or Random Initial Conditions (FIC/RIC) discovers either the unique solution ($\gamma\!=\!1$ or $8/9$), or two locally stable solutions ($\gamma\!=\!0$ or $1/2$). Left figure: The maximum free energy is determined by the system of maximum entropy at the Nishimori temperature. In cases of small $\gamma$ there are two candidate solutions. The fluctuations are visible in some curves and are due to the sampling method, these fluctuations are not sufficient to escape the local solutions in the cases of metastability. Right figure: BER demonstrates a clear performance contrast between solutions, for small $\gamma$ the thermodynamic solution is the good solution in this example. At larger $\gamma$ there is a unique solution of BER between the metastable and thermodynamic results at small $\gamma$.}
\end{figure}

\begin{figure}[!htbp]
\includegraphics[width=1\linewidth]{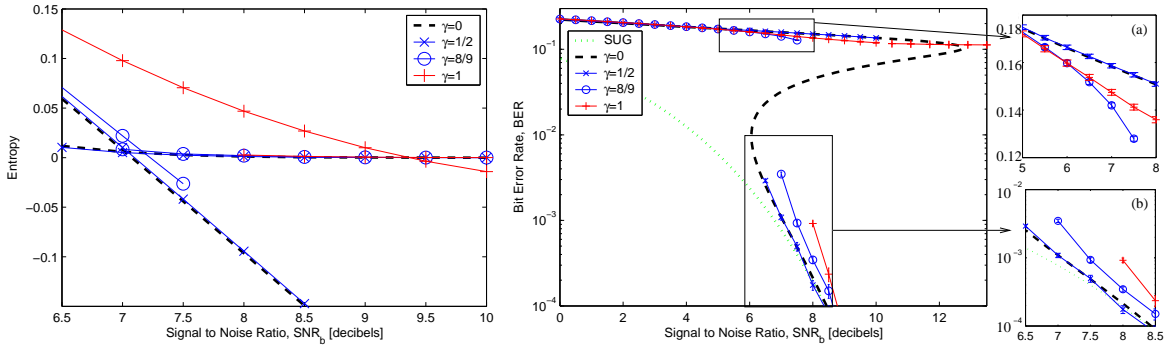}
\caption[Metastable sparse CDMA regimes.]
{\label{metaregime} The figure covers the same range of parameters as figure~\ref{fig:composite.normalregime}, but with a load $\load=2$. Two locally stable solutions are found by minimisation of the RS saddle-point equations in a range of SNR for all $\gamma$. Left figure: The entropy indicates a second order transition between the good and bad solutions for each ensemble. At SNR greater than the thermodynamic transition point metastable solutions evolve towards a freezing point ($\sasd=0$) and a regime of negative entropy. The thermodynamic transition point is at significantly greater SNR in the sparse ensemble than the composite ensembles. The range of SNR for which metastability exists is minimised in composite systems with $\gamma\!\approx\! 8/9$. Error bars are everywhere much smaller than symbol size. Right figure: The two saddle-point solutions are distinguishable in BER everywhere, a discontinuous transition occurs in BER at the thermodynamic transition. The properties of the good and bad solutions change smoothly about the thermodynamic transition and freezing (negative entropy) point. Right figure inset (a): The bad solution has high BER even at large SNR and becomes locally unstable at lower SNR for ensembles at intermediate $\gamma$, as shown for $\gamma\!\approx\!8/9$. Right figure inset (b) Good solutions with smaller $\gamma$ have lower BER and exist at smaller SNR.}
\end{figure}
The regime of high $\load$ is of greater theoretical interest in multi-user detection since this is where MAI causes results to differ substantially from single user models. As $\load$ is increased beyond $1.5$ a spinodal point may be reached beyond which there are multiple locally stable solutions to the saddle-point equation.

In regimes with a competition between locally stable attractors, or with one marginally stable attractor convergence of the saddle-point equations is slower; one such scenario is shown in figure~\ref{metatimeseries}. At $\load\!=\!5/3$ there is a unique solution for some of the sparse and composite ensembles, but not for the dense ensemble. In this example the composite code solution is superior to the sparse solution, and the dense metastable (bad) solution. The best solution is the dense thermodynamic (good) solution. As shown in figures~\ref{metaregime} and~\ref{metatimeseries}, the entropy is positive for all the thermodynamic solutions. However, at larger $\load$ and higher SNR the metastable solutions can have negative entropy, indicating an inadequacy in the RS description.

The saddle-point solutions for our ensembles with load $\load\!=\!2$ at a range of SNR is shown in figure~\ref{metaregime}. For this load metastability is present at all $\gamma$ values. Where the solution is not unique the correct and metastable solutions can be distinguished from the free energy (equivalently entropy at the Nishimori temperature). At the Nishimori temperature there is a second order transition, the energy is equal to $1/(2\load)$ in both solutions, which is realised as a discontinuous transition in the BER. In the metastable regimes the entropy evolves towards a negative value as SNR increases, the correct metastable state in the negative entropy regime is described by the phase space at the freezing point, where entropy first becomes negative.

Up to 7dB the bad solution is the thermodynamic solution in all ensembles. Close to the transition the best performing codes are composite ones with $\gamma\!\sim\! 8/9$, but at lower SNR the regular code ensemble appears best. The composite systems displayed all have thermodynamic transitions near 7dB, the entropy and free energy of the sparse bad solution is much larger, so that thermodynamic transition does not occur until about 9.5dB. This entropy gap might be a manifestation of the local configurational freedom available in some neighbourhoods in the sparse inference problem, but absent in the composite and dense structures. In the case of a regular sparse part, with a more homogeneous interaction structure, the gap in entropy and thermodynamic transition point are significantly reduced~\cite{Raymond:SS}. Amongst the good solutions, in contrast to the bad solutions, both the ensemble entropy and BER appear to be ordered by $\gamma$ for all SNR.

The metastable solutions appear to be qualitatively similar in the composite ensemble to the sparse and dense ensembles~\cite{Kabashima:SMA,Raymond:SS}. What is interesting in the metastable regime is that the positioning of the composite ensemble performance is not a simple interpolation between the sparse and dense ensemble results. In the example shown the metastable solutions for composite codes are at lower BER than either the sparse or dense metastable solutions. Furthermore, for $\gamma\!=\!8/9$ there is a unique solution beyond 8dB in spite of the persistence of metastable solutions in the sparse and dense ensembles at significantly larger SNR.

The microscopic stability of the metastable solutions for the composite system were not tested, but this should be possible, in part, by a local stability analysis of the RS description. It is expected that at, and above, the Nishimori temperature ($\beta\!<\!1$) the RS description will be locally stable even for the metastable states, as was found for the dense~\cite{Kabashima:SMA} and sparse ensembles~\cite{Raymond:SS}.

The composite codes exhibit a thermodynamic behaviour most strongly contrasting with sparse and dense codes when $\gamma \lesssim 1$, and close to the thermodynamic transition of the dense code. The effect of distributing power mostly in the sparse code appears to destabilise the bad solution in some marginal cases. The instability of metastable solutions for the sparse code to the inclusion of a small, but $O(1)$, dense component, occurs across a wide range of SNR, including regimes far from dynamical transition points so that the phenomena can not be a numerical artefact.

To understand the origins of this instability requires a more detailed investigation of the stability of the RS metastable solutions, and possibly an RSB type treatment. The combination of an external field with a sparse code might be expected to produce a comparable behaviour to the composite system, and this might be one way in which to understand the origins of reduced metastability in this system.

\section{Algorithm results}
\label{compositeCDMA.finitesystems}

Algorithm have been tested on representative sample sizes for systems of between $O(100)$ and $O(1000)$ users, and a variety of ensembles. In all the figures presented each sample involves an independent generation of Gaussian noise, a dense matrix and a sparse matrix, with the different sub-structures being rescaled appropriately by $\gamma$ and SNR. In order to fairly sample the sparse sub-codes a method has been developed and is outlined in Appendix \ref{app:RandomGraphSamples}.

\subsection{User-regular code ensembles}
\label{compositeCDMA.poissonian}

\begin{figure}[!htbp]
\centering{
 \includegraphics[width=0.9\linewidth]{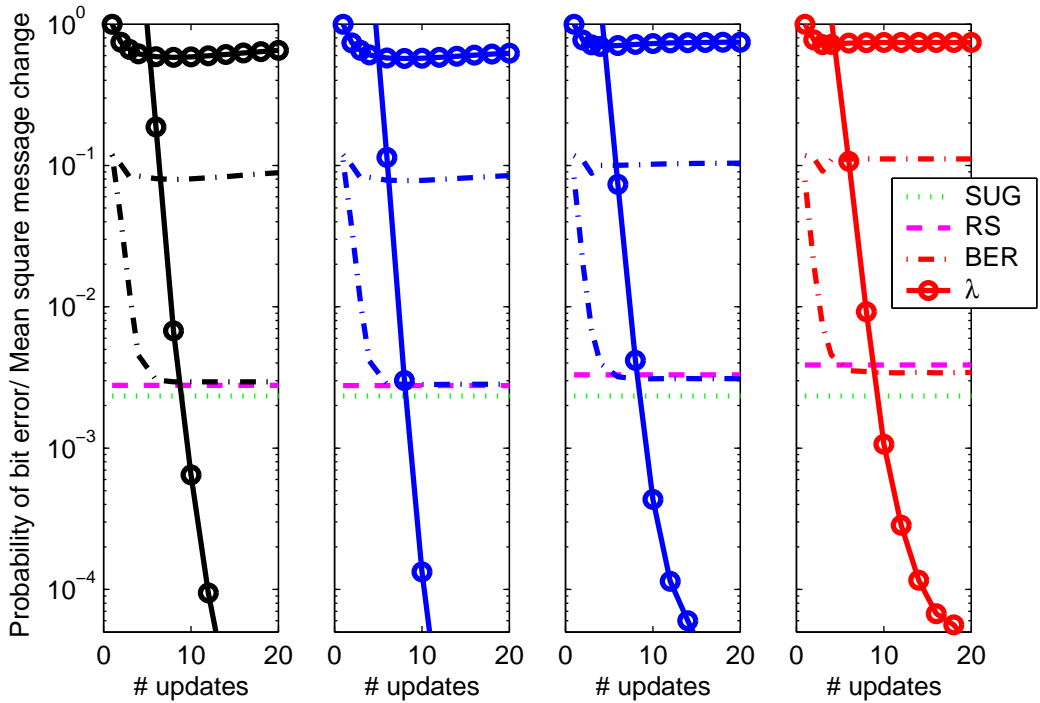}
\caption[BP and MSD detection of sparse CDMA in a regime with one solution.]
{\label{normal_decoding} Mean BER (dashed line) and $\lambda$ (solid line) are shown for different ensembles, $\gamma=\{0,1/2,8/9,1\}$ from left to right, as a function of the number of variable estimate updates for BP and MSD implemented with parallel updates. $\SNRmath=6$dB and $\load=3/5$ ($M=1000$, $K=600$): for each point $300$ independent sparse and dense connectivity profiles were sampled and combined in proportion to $\gamma$, with channel noise randomly sampled from a Gaussian distribution. The convergence measure $\lambda$ (\ref{eq:lambda}) indicates exponential convergence in BP and non-convergence of MSD for all ensembles. The RS result is approached after $10$ updates by the simulation average, but with some systematic error due to finite size effects. The MSD result does not improve beyond about five updates.}
}
\end{figure}
\begin{figure}[!htbp]
\centering{
 \includegraphics[width=0.6\linewidth]{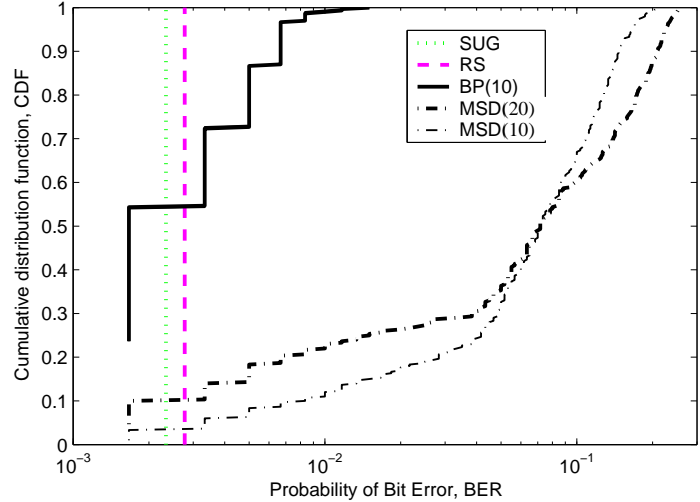}
\caption[BP and MSD detection of sparse CDMA in regime with one solution.]{\label{normal_decoding2} The cumulative distribution function for the decoding at $\gamma=0$ of the $300$ samples taken, as in figure~\ref{normal_decoding}, is typical in structure of all composite systems. The BER found by BP has converged for all samples taken within 10 updates. The BER found by MSD continues to evolve between 10 and 20 updates, with increasing BER for some subset of the samples. The median of the samples decoded by BP is close to the RS thermodynamic prediction of BER (vertical line), but the cumulative distribution function is not yet approaching a tight Gaussian and finite size effects are thus important. Some percentage of samples obtain a zero bit error rate which accounts for a small density range absent on the logarithmic scale.}}
\end{figure}

\begin{figure}[!htbp]
\centering{
 \includegraphics[width=0.8\linewidth]{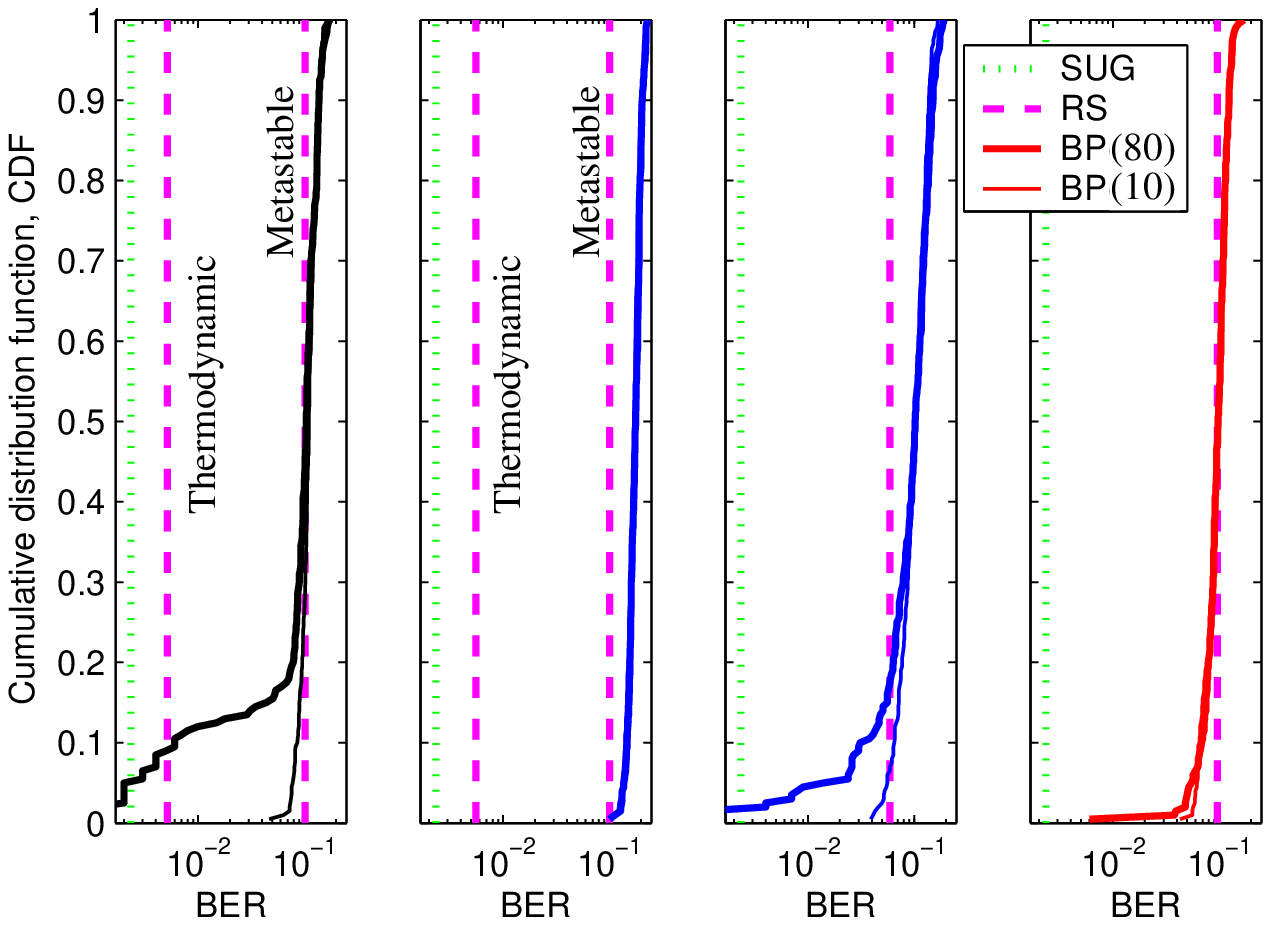}
\caption[BP and MSD detection of sparse CDMA in regime with high load.]
{\label{fig:decodingbadly} For $\SNRmath=6$dB and $\load=5/3$ ($M=600$, $K=1000$) the decoding performance of algorithms is presented as cumulative distribution functions in BER based on $200$ runs. The histograms from left to right represent mixing parameter values $\gamma=\{0,1/2,8/9,1\}$. The sparse code samples (far right) converge in most cases after $10$ iterations, and the median performance is close to the unique RS solution. The dense ensemble (far left) is after $10$ iterations close to the median performance for the metastable RS solution (right RS solution). A subset of samples evolve further, towards or beyond the thermodynamic RS solution (left RS solution), as can be seen in a discrepancy in the distributions correspond to posteriors at ($10$) and ($80$) updates. For $\gamma=1/2$, and $\gamma=8/9$, BER is larger than the asymptotic RS predictions in most samples.}}
\end{figure}

In assemblies with $\load \lesssim 1$ the equilibrium results are achievable by iteration of BP equations, this was established previously for the dense case in~\cite{Kabashima:SMA}. Such an example is shown in Figure~\ref{normal_decoding} with $\load=3/5$. The performance of MSD is poor, although initially the achieved bit error rate is improving with each iteration, over many iterations a destructive oscillation emerges. For systems of higher SNR and/or decreased $\load$ the MSD result is found to be very close to BP and the theoretical result. The BP algorithms reproduce the equilibrium result to within a small error after only a few iterations, even in systems with only $600$ users and $1000$ chips ($\load=3/5$), across a range of $\gamma$. Where unique saddle-point solutions were predicted by the equilibrium analysis decoding by BP normally produced a stable fixed point. The MSD results are not shown in subsequent figures, but are suboptimal with respect to BP in all cases.

A histogram of BERs for the BP decodings is demonstrated in figure~\ref{normal_decoding2}. In the large system limit the cumulative distribution functions is expected to converge towards a step function, which is the self-averaging assumption, in the metastable regime there may initial be convergence on two values (two steps), but with one solution dominating asymptotically. It is clear that for the sample sizes considered the distributions are far from a step function. BP converges quickly towards results of very low or zero BER. The MSD algorithm works very well, but more slowly than BP, for a subset of examples. In many other samples the performance deteriorates as MSD is iterated, the initial approximation (matched filter) is not significantly improved upon.

If a similarly sized system of $600$ chips and $1000$ users is considered with $\load=5/3$, the corresponding asymptotic result predicts metastability in the dense code, but not in the sparse code. The final BER achieved in $300$ samples for various systems is shown as a cumulative probability distribution in figure~\ref{fig:decodingbadly} after 10 iterations and after 80 iterations. The sparse system is uni-modal, with fast convergence in most systems. The dense ensemble is multi-modal as expected, the convergence time towards the low BER solutions are very slow, and the majority of achieved solutions are close to the high BER metastable solution. Random initial conditions tend to produce steady states characterised by the bad solution, even if this corresponds to the metastable (probabilistically suboptimal), rather than equilibrium, solution. The composite system equilibrium solution is unique for $\gamma=8/9$. For $\gamma=8/9$ some $40\%$ of samples improve between iteration $10$ and iteration $80$, but $40\%$ also worsen, the median performance is quite far from the equilibrium prediction. The equilibrium results for $\gamma=0.5$ are not closely approximated in the decoding experiments, the performance in BER is worse everywhere than the equilibrium prediction, and also significantly worse than if power were distributed on only the sparse ($\gamma=1$) or dense sub-codes ($\gamma=0$). For large $\load$ it appears the finite size effects are more limiting in the case of the composite codes, particularly at intermediate values of $\gamma$. However, it is noteworthy that even without elimination of dense messages, performance with the proposed update schemes are poor at intermediate $\gamma$. The ordering of updates may also be important, and other sensible schemes might be consider. For example to iterate only the sparse messages until convergence (a fast process) between updates of dense message dependent quantities.

The composite systems shown in figure~\ref{fig:decodingbadly} does not come close to the performance of even the bad solution in either the median or mean for this system size except for large or small $\gamma$. The ability of the composite BP algorithm is more limited in achieving the equilibrium result for intermediate $\gamma$ than for comparable methods applied to sparse and dense code ensembles for systems of this size. A quantitative comparison of the equilibrium and finite size systems in the metastable regime with bulk statistics such as the mean is difficult due to the multi-modal nature of the distributions.

The decoder performance for systems of size $O(1000)$ seem to provide mean values for the BER, which are quite far from the theoretical values and unable to realise the asymptotic advantages of some composite codes predicted by the equilibrium analysis. There are many approximations made at $O(1/M)$ in construction of the BP algorithms, some specific to the composite codes. It is likely these systematic and random fluctuations are at the root of the BP instability for intermediate values of $\gamma$. In BP without the various leading order approximations the algorithm relies on the assumption of negligible correlations between messages, although this assumption breaks down for the loopy graphs considered, it is not clear that the assumption is weaker for intermediate $\gamma$.

\subsection{Regular code ensembles}
\label{compositeCDMA.regular}
\begin{figure}[!htbp]
\centering{
\includegraphics[width=0.6\linewidth]{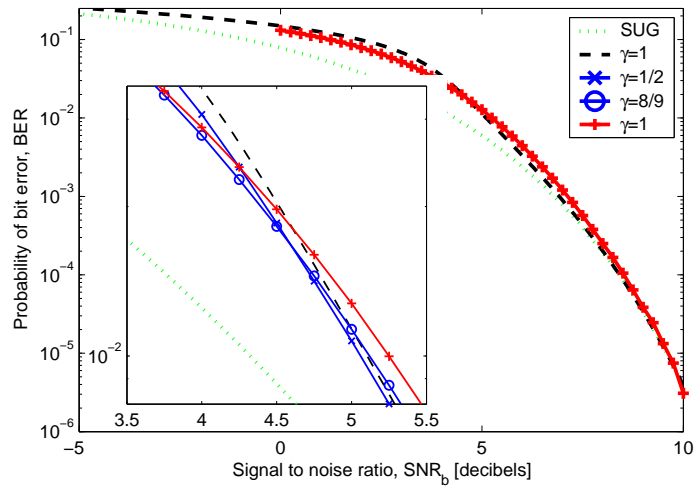}
\caption[RS equilibrium results for regular composite CDMA.]
{\label{fig:composite.EnsembleComparison} Shown is the optimal performance for $\gamma=\{0,1/2,8/9,1\}$, with a regular sparse ensemble as a component in the composite system. At high SNR the performance decreases with $\gamma$, at low SNR the performance increases with $\gamma$. For a small range of SNR, inclusive of the inset range, the composite codes outperform both the sparse and dense codes.}}
\end{figure}

Alternative composite system involving correlated sampling of user codes, so as to reduce MAI or inhomogeneity, represent interesting cases for study. A scenario in which the user-regular sparse sub-codes are sampled so that the number of accesses per chip ($L_\mu$) is uniform for all chips is one example, the ensemble of codes may be described as regular. This requires global coordination of user codes, without the restriction that user codes should be sampled independently there are significantly more options available allowing code optimisation. However, the regular code is interesting because it shares many of the topological features of the sparse user-regular ensemble, but has very slightly lower MAI, the mean square code overlap is reduced by a factor $(L-1)/L$ in the sparse sub-code (as shown in section~\ref{CDMA.marginals}). There are also some finite size effects removed from the composite BP algorithm with this choice. The reduced MAI has the effect that at low values for SNR the unique stable solutions for the regular ensemble is superior in BER to the dense ensemble. With this ensemble it is possible to demonstrate a statistically significant result, in decoding by BP, for which the composite code ensemble outperforms the corresponding sparse and dense sub-codes in BER.

The equilibrium behaviour of the regular sparse ensemble was analysed in~\cite{Raymond:SS}, and in chapter~\ref{chapter:sparseCDMA}. In the analysis it is found that the composite code BER interpolates the sparse and dense performance in low and high SNR regimes. However, in an intermediate range of SNR, where the BER is approximately equal in the sparse and dense models, the unique solution of the composite ensembles has an improved BER over both the dense and sparse solutions. The performance of several ensembles is shown in figure~\ref{fig:composite.EnsembleComparison}.

\begin{figure}[!htbp]
\centering{
\includegraphics[width=0.8\linewidth]{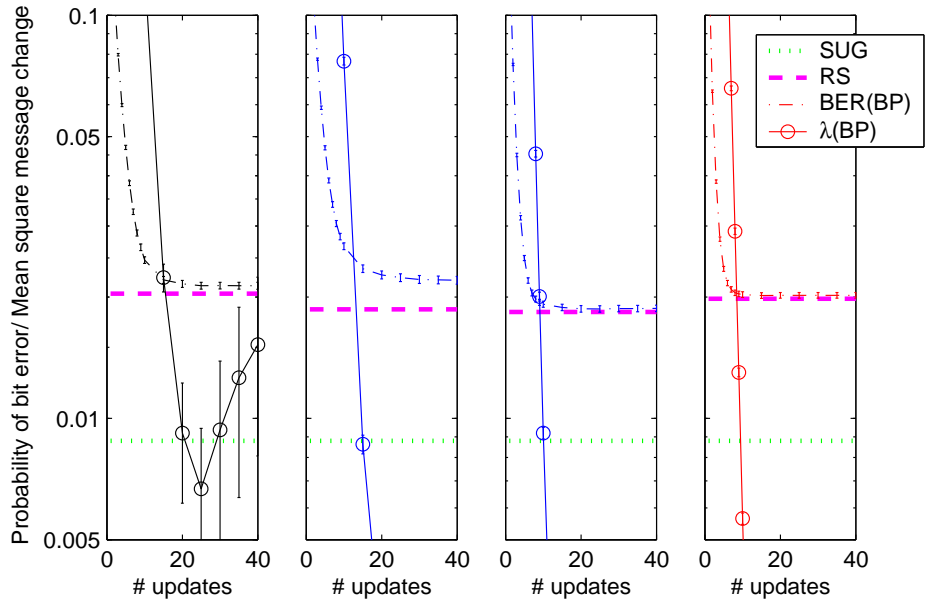}
\caption[BP detection results for regular composite CDMA.]
{\label{fig:composite.finallysuccess} At $\SNRmath=4.5$ the mean results of $500$ decoding experiments using $\gamma=\{0,1/2,8/9,1\}$ (from left to right) with a chip regular sparse component in each composite system. The BP equations converge except for $\gamma=0$, where some samples were unstable. A similar effect is manifested in the $\gamma=\frac{1}{2}$ ensemble after about 45 iterations, but not within the scale of the figure. Each set of samples produced a BER in decoding close to the RS prediction except for $\gamma=1/2$, where decoding performance was substantially poorer than the prediction. The BER by the RS result and decoding experiment is best amongst ensembles with $\gamma\approx 8/9$.}}
\end{figure}

Working with a simulation of $1000$ users and $1000$ chips it is possible to demonstrate that the mean performance of several composite codes exceed the performance of the sub-codes re-scaled to an equivalent SNR, as shown in figure~\ref{fig:composite.finallysuccess}. The results for $\gamma=\{0,1/2,8/9,1\}$ ensembles are close to the large system limit prediction, to within the error bars. The composite code ($8/9$) achieves the lowest bit error rate in expectation amongst the codes, and has convergence properties interpolating between the sparse and dense ensembles. However, as in previous experiments on the user regular code, the performance for $\gamma=1/2$ is much poorer than the large system limit prediction.

As can be seen the dense code fields are initially converging in a similar way to figure~\ref{normal_decoding}. However, at later time the estimates begin to diverge slightly, at least within a significant fraction of simulations. This instability is most apparent in the dense ensemble and absent in the sparse ensemble, and might be an indication of the inaccuracy of the Gaussian BP approximation (\ref{eq:compositeCDMA.Zmuk}) when BER in decoding becomes very small. Similar trends are seen in some of the composite codes, often the messages do not converge exactly, but only to within some fixed variability.

\section{Discussion}

The equilibrium analysis demonstrates that in regions of metastability the composite coding structure, comprised of a sparse and densely connected component, might have some interesting and valuable properties. When power is approximately equal in the two parts performance is very close to the dense ensemble, but with only a small amount of power in the dense code properties are strongly distinguishable. At the same time it has been shown that in reasonably sized samples the BP approaches, based on $O(1/M)$ approximations in the dense part, work relatively poorly when MAI is large. This instability in some composite codes can persist even in scenarios where the equilibrium analysis predicts a unique RS solution.

The failure of the composite BP algorithm is likely to be in part due to the Gaussian approximation in marginalisation over states (\ref{eq:compositeCDMA.Zmuk}), which may be a poor approximation when messages become strongly biased. If this is the case then the problem may be avoided or mitigated by standard algorithmic tricks such as annealing or damping. When the messages become very biased, replacing the full marginalisation by one considering only a truncated set of states might be a viable polynomial time alternative to using the analytical Gaussian approach. In small realisations of composite systems many heuristics might be employed.

In the final section results are presented with a chip-regular sparse sub-codes, which improves performance, but requires coordinated sampling of user codes. Where coordination is possible there would be some value in considering either an ordered (optimised) sparse code combined with a random dense code or vice-versa. The ordered sparse code might provide a method for detection under ideal channel conditions, whereas the dense code provides a contingency and some of the advantages of the spread spectrum approach, such as multi-path resolution. This might be a practical application of composite codes.

The composite code presents an interesting dichotomy in its suppression of metastable behaviour, but greater apparent instability in simulation. Aside from standard convergence measures in simulations a concrete way to probe the origins of this instability, specific to the mixed topology, has not been envisaged. Some further insight on the stability issues might be found by probing more thoroughly the properties of the metastable solution in the equilibrium analysis.

A finite size scaling of the algorithm results would be valuable, unfortunately the need to manipulate an $M$ by $K$ dense spreading matrix prevents moving to larger scales. The scales we have presented, and error bounds, are chosen subject to this restriction in such a way as to demonstrate the breadth of behavior. Many results in the cited papers go much further in dealing with the question of finite size effects in cases of sparse and dense random codes.

The method developed is applicable where the sparse sub-code defines a connected graph, above the percolation threshold. If a composite code is used with a sparse sub-code below the percolation threshold a more fruitful analysis may be possible working with the sparse trees as the microscopic states, connected through a homogeneous (dense code) interaction. A similar decomposition may allow some algorithm simplifications.

%% file: Conclusion.tex
\section{Summary}
\label{conclusion.summary}

This thesis has addressed theoretical problems relating to satisfiability in the random one in k satisfiability model, novel phase behaviour and transitions in composite systems, and the problem of source detection in a linear vector channel using sparse and composite random codes. Each of these problems may be constructed as an inference problem on a large random graph.

Random graphical models have played important roles in the development of many fields. In the case of disordered systems random graphs form the natural basis for encoding unstructured correlations amongst interacting variables, and so are essential to capture uncertainty. In other applications, such as channel coding or neural network, random graph structures may be deliberately engineered features, capable of achieving some robust performance in typical case. Finally random graph structures can be used as a simplification of an intractable model, allowing certain features to be probed through exact or variational methods.

In the problem of one in k satisfiability the random graph ensembles studied are minimal descriptions of an interaction structure. In this way an inference problem including complicated correlations may be studied with minimal assumptions on the structure, rather than examining worst case structures typical inference properties can be established. These results benchmarks by which to test the scalability of algorithms and heuristic methods, and owing to the simple ensemble description it is often possible to identify generic features of graphs and constraints that lead to algorithmic hardness.

The work outlined in chapter~\ref{chapter:1inkSAT} sought an understanding of the dichotomy between the algorithmically easy symmetric one in k satisfiability ensemble, and the algorithmically more challenging Exact Cover ensemble. An algorithmic method based on branch and bound could be formulated analytically to study this problem, and so demonstrate a range of transitions in algorithmic hardness in the large system limit. One result indicated that the unit clause algorithm, a local search method, could work exactly in a part of the phase diagram for which a fragmented solution space applied. Results of this kind are essential in developing theories on the nature of typical case algorithmic hardness.

In the study of multi-access channels, methods based on random code division multiple access are of increasing interest in theory and application. Work on dense random codes has become developed to the stage that it is used in many wireless networks. Utilising the bandwidth through a randomised structure offers a numbers of practical advantages. The sparse code ensemble has slightly different properties that have also come to the attention of researchers in the field, these include the potential for optimal and fast detection based on message passing methods.

Chapter~\ref{chapter:sparseCDMA} investigates the typical case properties of some ensembles of sparse random codes. These are found to produce trends in performance comparable to the dense codes, with detection by Belief Propagation a viable method. The phase space is shown to be a simple one. A variety of different code ensembles were proposed and each was shown to have relative strengths and weaknesses in analysis and optimal and practical detection performance.

In developing models of physical or information systems it is often possible to make an approximation to the interaction structure by a fully connected graph. However, the lack of locality for variables in this model means that some features are not captured, so that sometimes a sparse random graphical model is more appropriate. Both of these models have a form suitable for exact analytical methods. The stability of these models is often studied with respect to self consistent perturbations, and it is often impossible to generalise to tractable models involving several scales of interactions, or different kinds of topology.

In chapter~\ref{chapter:composite} a new form of exactly solvable graphical model was proposed, the composite model. In spin systems with simple couplings it was found that the competition between sparse and dense effects produced unusual behaviour both in the vicinity of the paramagnetic phase, where the model is exactly solvable, and at lower temperature. The structure of ferromagnetic phases were shown to have an interesting structure which causes a significant difference in results between models with regular connectivity and models with inhomogeneities.

Composite models may be useful in application. Many information structures allow a choice between dense and sparse graphical frameworks, and the possibility exists to use both in combination through a composite structure. Chapter~\ref{chapter:compositeCDMA} proposes such a code division multiple access method involving both the sparse and dense spreading paradigms. The key finding was that, at least in the large system limit, there exist regimes where the achievable bit error rate in detection from independently sampled coding is substantially improved by spreading power between a sparse and dense transmission protocol, rather than relying on a single type to convey information. It is also shown that efficient detection algorithms can be formulated for the composite inference framework.

\section{Some future directions}
\label{conclusion.future}

There remains much work to do with respect to each of the topics studied, but several are of particular interest to the author.
\begin{itemize}
\item[1] An analysis of the properties of the noiseless sparse codes was undertaken using dynamical features of unit clause propagation, and this analysis may be extended to be inclusive of the binary erasure channel and more practical code ensembles. In cases where unit clause propagation terminates without reducing the inference problem to a tree like structure, as occurs in some overloaded regimes, the properties of the residual inference problem require a complementary analysis and understanding. Just as the success of unit clause propagation in some low load or low connectivity regimes indicates easy detection regimes that may generalise to problems with noise, understanding the high load residual inference problem, once the unit clause propagation from initial conditions has terminated, might provide insight on algorithmic hardness at large load. Understanding in greater detail how the embedding of a bit sequence solution effects the detection solution space would be an outcome of this analysis, clearly the solution space differs substantially from the superficially related one in k satisfiability ensemble.
\item[2] The composite model with a non-Poissonian sub-structure has not yet been extensively analysed by methods other than perturbation about the high temperature transition, and it would be interesting to consider an equilibrium analysis of this model in the regime of metastability. Topological features differ between the sparse regular and Poissonian graphs, and the solution spaces supported were demonstrated in this thesis to be very different when these sub-structures underpin a composite system. It would be interesting to identify which features are most important in determining low temperature equilibrium and dynamical properties for composite models. An extension of the equilibrium methods, or development of new methods, to understand the zero temperature limit would also be interesting.
\item[3] Exploring the dynamics of composite systems represents an interesting research direction. Testing whether there exist significant differences between the local stability of thermodynamic and metastable solutions, and the nature of their dynamical attractors as a function of the coupling type through which they are sustained might highlight new difference in the robustness sparse and dense induced order. The F-F model with unaligned ferromagnetic orders represents a particularly simple model by which to begin such a consideration.
\item[4] The composite code has been shown to demonstrate reduced metastability in the equilibrium analysis, but in simulation for moderate system sizes the modified BP algorithms perform very poorly. Identifying the finite size effects or dynamical features of the algorithm implementation responsible for this breakdown is essential if composite inference algorithms are to be practical. Understanding the physical origin of reduced metastability in the composite code ensemble is also an unfulfilled ambition.
\end{itemize}

%% file: APPENDICES.tex
\cut{
\chapter{Notation and Mathematical Idenities}
\section{Notation}
\label{app:Notation}
\input{APPENDICES/Notation.tex}
}
\chapter{Mathematical identities}
\label{app:Identities}
\input{APPENDICES/Identities.tex}

\chapter{Analytic bounds for UCP}
\label{app:AnalyticBounds}
\input{APPENDICES/Analytic_bounds.tex}

\chapter{Exact results for sparse CDMA}

\section{Nishimori temperature for multi-user detection}
\label{app:Nishimori}
\input{APPENDICES/Nishimori.tex}

\section{Noiseless CDMA}
\label{app:CDMAto1ink}
\input{APPENDICES/Ucp_for_cdma.tex}

\chapter{Replica calculations}

\section{Replica method for sparse connectivity matrices}
\label{app:SparseMatrixAnalysis}
\input{APPENDICES/Sparse_matrix_analysis.tex}

\section{Conjugate field methods}
\label{app:ConjugateFields}
\input{APPENDICES/Conjugate_fields.tex}

\section{Composite model replica method}
\label{app:CompositeSystem_Replica}
\input{APPENDICES/Composite_system_replica.tex}

\section{Composite CDMA replica method}
\label{app:compositeCDMAequi}
\input{APPENDICES/Composite_cdma_replica.tex}

\chapter{Sampling of sparse random graphs}
\label{app:RandomGraphSamples}
\input{APPENDICES/Random_graph_samples.tex}

\chapter{Composite belief propagation for CDMA}
\label{app:compositeCDMAalg}
\input{APPENDICES/Composite_cdma_algorithm.tex}

%% file: APPENDICES/Notation.tex
\cut{
In the section a brief explanation of tensor and set notation is given followed by a list of some the symbols used to represent various quantities. Finally a list of abbreviations is provided.

\subsection{Tensor notation}

Column vectors are denoted by a vector symbol $\vx$, except in the case of components in replica space where the notation $\rvx$ applies, matrices have a notation $\mx$, always capitalised. Tensors are everywhere written with at most $2$ implicit indices, so that a three index tensor may be written as for example $x_{\mu k}^{t}$, or $\vx_{\mu k}$ or $\mx^{t}$. Normal multiplication and summation applies to vector and matrix quantities but the Einstein summation convention is not assumed anywhere, so that for example $x_i y_i$ does not imply an inner product. The inner/dot product $\sum_i x_i y_i$ may be abbreviated $\vx \cdot\vy$ and the determinant of a matrix is denoted $\determinant{\mx}$.

\subsection{Set notation}

A set of labeled or unlabeled elements is denoted by $\Set$. The notation $\orderedL{x}$ indicates an ordered set $i_1<i_2<\ldots<i_L$. By contrast $(i_,\ldots,i_L)$ indicates the set of unordered, but distinct, pairs. Where x is a node in a factor graph $\partial_x$ indicates the set of nodes attached to $x$. Since factor graphs are bi-partite factors have variable neighbours and vice-versa. The exception for $\partial_x$ notation is chapter \ref{chapter:composite} where $\partial_x$ is the set of variables coupled to variable $x$. The size, or modulus, of a set $x$ is denoted $|x|$.

The notation
\begin{equation}
{N \choose C} = \frac{N\factorial}{(N-C)\factorial C \factorial} \label{appeq:NchooseC}\;,
\end{equation}
is the standard $N$ choose $C$ combinatorial factor, defined as $1$ when $C=0$ or $C=N$, and $0$ if $C<0$ or $C>N$.
}

\subsection{Acronyms}
A list of acronyms in alphabetical order are presented. These acronyms are included as standard abbreviations to well known concepts, to abbreviate statements in a short section of text, or because the phrases are standard within the wider relevant literature.
\begin{itemize}
\item[\oRSB] One step of Replica Symmetry Breaking.
\item[\ASK] Amplitude Shift Keying.
\item[\BER] Bit Error Rate.
\item[\BP] Belief Propagation.
\item[\CDMA] Code Division Multiple Access.
  divergence between functions.
\item[DP,DPLL] Davis-Putnam,Davis-Putnam-Logemann-Loveland, types of algorithm.
\item[\UCP] Unit Clause Propagation.
\item[\EC] Exact Cover.
\item[\FDMA] Frequency Division Multiple Access.
\item[\FRSB] Full Replica Symmetry Breaking.
\item[\HUCP] Heuristic driven Unit Clause Propagation.
\item[\KL] Kullback-Leibler, specifies a measure of divergence between functions.
\item[\kSAT] k-SATisfiability.
\item[\MAI] Multi-Access Interference.
\item[\MAP] Maximum A Posteriori.
\item[\MPM] Marginal Posterior Mode, or Maximiser of Posterior Marginals.
\item[\PSD] Power Spectral Density (equal to load times \SNR).
\item[\RS,\RSB] Replica Symmetric,Replica Symmetry Breaking.
\item[\SAT] SATisfiability.
\item[SNR,SINR] Signal to Noise Ratio, Signal to (Interference plus) Noise Ratio.
\item[\SK] Sherrington-Kirkpatrick, specifies a dense spin model.
\item[\TDMA] Time Division Multiple Access.
\item[\VB] Viana-Bray, specifies a sparse spin model.
\item[\UCP] Unit Clause Propagation.
\item[\UNSAT] UNSATisfiability

\end{itemize}

\subsection{Common symbols}
A list of Greek, Latin, calligraphic and composite symbols representing standard quantities are presented in approximately chronological order within the thesis. Many definitions are common across all chapters, and where possible are consistent with wider literature,
\begin{itemize}
\item[$K$] Number of users in a multi-access channel, number of variable nodes in a factor graph.
\item[$M$] Number of discrete elements in the bandwidth, number of factor nodes in a factor graph.
\item[$N$] Number of spins/System size.
\item[$n$] Number of replica.
\item[$\load$] Channel Load, ratio of users to bandwidth.
\item[$\Reals$] The real line.
\item[$\Complexs$] The complex plane.
\item[$\ensemble$] The set of ensemble parameters.
\item[$\quenched$] The set of quenched variables, a sample from the ensemble.
\item[$\freeenergy,\freeenergydensity,\safed$] The free energy, free energy density, and free energy density averaged over quenched disorder.
\item[$\SpinGlassSusceptibility$] The spin-glass/non-linear susceptibility.
\item[$\SEmath,\SEmath_{U},\SEmath_{U}$] The spectral efficiency, and (U)pper and (L)ower bounds.
\item[$P_C$] The marginal variable/user connectivity distribution.
\item[$P_L$] The marginal factor/chip connectivity distribution.
\end{itemize}

%% file: APPENDICES/Identities.tex
A number of transformations are required in calculating quantities through the replica method. Many of these transformations allow analytic continuations of discrete quantities essential to the method, or factorisation of dependencies. A brief overview is provided.

\subsection{The Fourier transform}
\label{app.fourier}
The Fourier transform of a function on the real numbers is a representation of a function ($\arbitraryfunction$) in reciprocal space, the representation is formed through the transformation with an integral in the complex plane
\begin{equation}
 \arbitraryfunction(s) \propto \int_{-\rmi\infty}^{\rmi\infty} \rmd \lambda \exp \left\lbrace \lambda s\right\rbrace
\arbitraryfunction(s)\label{appeq:Fourier}\;.
\end{equation}
The constant of proportionality is not significant in establishing properties of interest.

In this thesis the Fourier transform is frequently applied with a scaling ($\lambda \rightarrow N \lambda$), where $N$ is the system size. The scaling with $N$ reflects the physical intuition of an extensive entropy, and is also necessary in scalable solutions of the saddle-point equations.

\subsection{Cauchy's integral formula}
\label{app.cauchy}
Cauchy's integral formula is useful for representation of identity functions on a discrete space by an analytic form. Constraints on the sums of discrete quantities form an important part of the analysis in the replica method. A convenient way to represent these constraints is through the Cauchy Integral theorem, transformations of the form
\begin{equation}
 \delta\left(\sum_i x_i - L\right) = \frac{1}{2 \pi i}\oint \rmd Z \frac{\prod_{i} Z^{x_i}}{Z^{L+1}} \label{appeq:CauchyIntegral}\;,
\end{equation}
are used in the thesis. The integral is along a closed curve in the complex plane about the origin, which may be taken as the unit circle. Equivalent representations of the delta function, such as Fourier series, also allow the factorisation critical to the calculations.

\subsection{The Hubbard-Stratonovich transform}
\label{app.HS}
The Hubbard-Stratonovich transform~\cite{Ellis:ELD} can be applied to factorisable quadratic forms in an exponent, the quadratic form may be encoded in a matrix ($\QuadraticForm=\RotationMatrix^T\RotationMatrix$). The dimensional dependence can be factorised within a Gaussian weighted integral
\begin{equation}
\exp\left\lbrace-\vS^T \QuadraticForm \vS/2\right\rbrace = \sqrt{\frac{\determinant{\QuadraticForm}}{2 \pi }} \int \rmd\vZ \exp\left\lbrace - \vZ\cdot\vZ/2\right\rbrace \exp \left\lbrace \rmi \vZ^T \RotationMatrix \vS \right\rbrace\label{appeq:Hubbard}\;.
\end{equation}
In most of the calculations undertaken $\QuadraticForm$ is not a matrix but a scalar, in which case the expression is substantially simplified to
\begin{equation}
\exp\left\lbrace - \ScalarForm \sum_i (S_i)^2/2 \right\rbrace = \int \left[\rmd Z \exp\left\lbrace - Z^2/2 \right\rbrace\right] \exp\left\lbrace\sqrt{-\ScalarForm} Z \sum_i S_i\right\rbrace \;. \label{ScalarHubbardStratonovich}
\end{equation}
The Gaussian weighted integral $[\cdots]$ is in many sections abbreviated to $\rmD Z$.

\subsection{Laplace's method}

Many integrations in this thesis involve an exponential form with an $N$ dependent exponent, where $N$ is system size. The integral over such a form in the case of a real valued exponent may be accurately approximated in the limit of large $N$. Assuming the exponent $N \arbitraryfunction(\lambda)$, $\arbitraryfunction$ being an arbitrary real smooth function, and $\lambda$ a scalar or vector argument, has some unique maxima which is not on the boundary of the integration range then
\begin{equation}
\lim_{N \rightarrow \infty} \frac{1}{N}\log \int \rmd \lambda \exp\{N \arbitraryfunction(\lambda)\} \doteq \arbitraryfunction(\lambda^*)\;,
\end{equation}
where $\lambda^*$ are the integration parameters maximising the exponent, the result applies to the leading order in $N$. The maxima may be determined by assuming the first derivatives with respect to $\lambda$ of the exponent are zero. The set of equations defining the first derivatives as zero form a closed set of equations called in this thesis the saddle-point equations. The second derivatives may be checked to test whether the fixed point is a local maxima. Different local maxima may be compared to determine the global maxima, and hence the correct solution. In the case of degenerate maxima, or maxima differing only at $O(N)$ a sum of maxima must be considered.

\section{The saddle-point method and physical interpretation}
\label{app:saddle-pointmethod}

The saddle-point method is an extension of Laplace's method of integral approximation to integrals on the complex plane. In this thesis the integrals found in the replica method involve an application of the Fourier transform (\ref{appeq:Fourier}), and definitions of order parameters defined on the complex plane, so that although the exponent scales as $N$ the integrals are over a complex domain. In the saddle-point method the integral is dominated not by a maxima on the real line/volume, but by a saddle-point in the complex plane/volume. However, it is assumed that any physical solution must be dominated by real valued arguments of the integration parameters, and hence a real valued saddle-point. Laplace's method for determining extrema (rather than maxima) is used assuming a real valued space of integration variables.

The rigorous justification of this assumption may be approached in at least two directions. Since the free energy, and its derivatives with respect to physical perturbations, must be real valued, an application of conjugate fields in the Hamiltonian may be used to demonstrate that the integration parameters maximising the free energy must lie on the real access, as demonstrated for replica symmetric solutions in Appendix~\ref{app:ConjugateFields}. Alternatively, a transformation of the integrals on the complex line by a rotation of the complex line into the real line, combined with an application of Cauchy's residue formula, may also be used in some situations.

The saddle-point method is necessary in determining a tractable functional form for the free energy within the replica method~\cite{Mezard:SGT,Talagrand:SG}, but as applied in this thesis involves many implicit assumptions that are not rigorously justified. The principle justification for use of the method as presented is in the self-consistency of results obtained, the success of the method in the wider literature, and consistency of results with both experiments on finite size samples and known rigorous results.

Physical insight plays a role in the application and evaluation of the saddle-point method for several reasons. Firstly, the method of population dynamics used in determining extrema provides no guarantee that all, or even a unique, extrema may be determined in general. However, population dynamics used to search for solutions are often analogous to some physical dynamics, providing insight into the properties of optimisation methods for example. Secondly, degeneracy in the extrema is often related to a discontinuous phase transition, or else to some exact symmetry, a symmetry that may be broken in physically meaningful systems. Thirdly, sub-optimal extrema may be worth evaluating by comparable methods to the global optima, since these often provide valuable information on metastable solutions, as opposed to the thermodynamic solution. Finally, it is in many cases necessary to search for an extrema only in a subspace of possible replica correlations, such as the replica symmetric space, and without testing local and/or global perturbations physical insight is essential. The symmetry assumptions are often analogous to some intuitive physical structure to the phase space.

%% file: APPENDICES/Analytic_bounds.tex

\begin{figure}[htb]
\centering{
\includegraphics[width=0.8\linewidth]{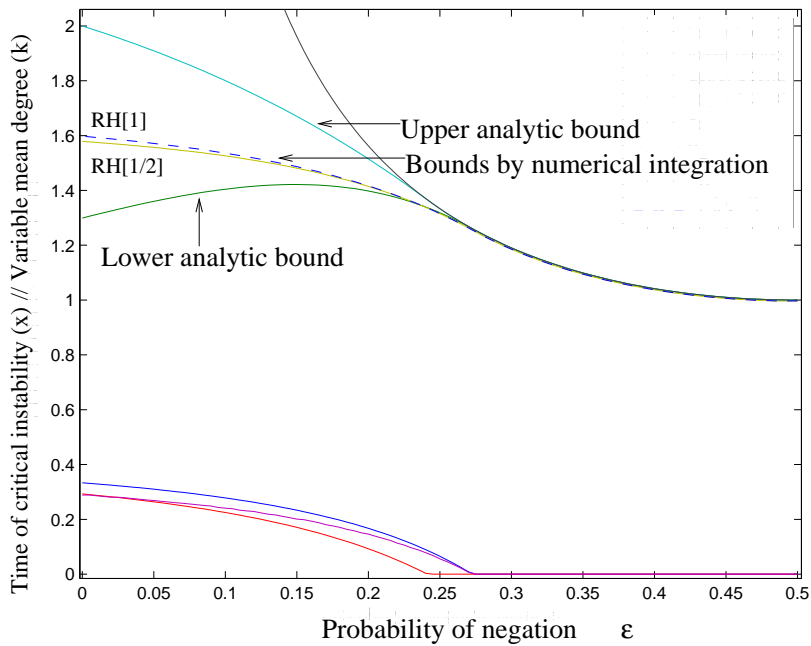}
\caption[Algorithmic bounds on the SAT/UNSAT transition for
$\eotSAT$.]{\label{fig:app.lowerupperbounds} The critical values in variable connectivity below which the samples are Easy-SAT are attained by solving a set of non-linear coupled partial differential equations RH[1] (central dashed) and RH[1/2] (central solid) are presented with their bounds. The bounds take $\UCPtransitioner(x,\epsilon)$ to be constant (labeled curves). The accuracy of the bounds depend on the critical algorithm time ($x^*$) at which branching process is strongest (lower curves). At large $\epsilon$ the bounds are tight to the integration result and identify correctly the maxima $x^*=0$. At small $\epsilon$ where the position of the criticality in algorithm time $x^*$ is non-zero, the bounds worsen.}}
\end{figure}

The functions $C_k(x)$ can be determined exactly (\ref{eq:1inkSAT.c_k}) given an initial condition, provided the variables are selected for decimation independently of their multiplicity. The number of clauses of size $2$ to $k-1$ can only be determined by numerical integration, with the extent of the non-linearity encoded through $\UCPtransitioner$ (\ref{UCPcalF}). $\UCPtransitioner$ determines the rate at which $(i\!-\!1)$-clauses are generated from $i$-clauses independently of the creation rate for unit clauses (subject to conserved mass in expectation).

Although $\UCPtransitioner$ is a complicated non-linear function that must be calculated recursively in general, some monotonic properties may be used. The value of $\gamma$ for which criticality occurs is a non-decreasing function of $\UCPtransitioner$ -- since increasing $\UCPtransitioner$ increases the number of clauses at every stage in the algorithm. Furthermore, $\UCPtransitioner$ is bounded in the interval $[\frac{1}{2},1]$. As such it should be possible to set $\UCPtransitioner$ as a function of $x$ within these bounds and attain a variational approximation, which can be used to bypass the numerical integration.

It is in fact true that $\UCPtransitioner$ is not only bounded, but is a monotonically decreasing function of $x$ if the heuristic rules RH$[\frac{1}{2}]$ or SCH are used. Reduction of the largest clauses, where mass is concentrated at $x=0$, produces $\boldsymbol{m}$ biased towards $\boldsymbol{\epsilon}$. The $k-1$ unit clauses generated by reducing a $k$-clause directly is biased towards production of negative unit clauses for any $\epsilon<\frac{1}{2}$. Conversely reduction of smaller clauses produces unit clauses with less bias in the literals, and exactly balanced in the case of reduced 2-clauses.

Given this knowledge two values produce intuitive upper and lower bounds on $\UCPtransitioner$ that can be used to bound the results obtained by numerical integration. The first value is $\UCPtransitioner(x,\epsilon)=\UCPtransitioner(0,\epsilon)$, which maximises the rate of $i$-clause production and so indicates an upper bound to the principal eigenvalue, and a lower bound to the Hard-SAT phase. An opposite bound is attained by setting $\UCPtransitioner(x,\epsilon)=\frac{1}{2}$, which underestimates clause production even at $x=0$. The important bound is obtained in the first case; this gives an analytically determined lower bound in $\gamma$ to the Hard-SAT regime.

In this framework the equations describing clause dynamics (\ref{eq:1inkSAT.ciroundSC}) becomes solvable in a closed form. Figure~\ref{fig:app.lowerupperbounds}
 demonstrates the result for $k=3$ and RH[p] with a comparison to the quantity determined by numerical integration. The upper and lower bounds coincide with the integration result for large $\epsilon$. At smaller $\epsilon$ bounds diverge from the integration result, since $x^*$, the critical algorithm time, is greater than $0$ and hence there is some departure during the heuristic stage of the algorithm in $\UCPtransitioner$ that causes a significant fluctuation $c_2(x)$. The upper curve overestimates the instability, the lower curve underestimates the instability. The difference between the bounds becomes worse with increasing $x$. The full numerical integration improves upon the estimates obtained using the bounding methods, but the bounds are useful verification of these results removing some uncertainty from the numerical methods. 

%% file: APPENDICES/Nishimori.tex
At the Nishimori temperature many properties of sparse and composite CDMA may be determined by exact methods~\cite{Nishimori:CO}. Derivations are demonstrated for the zero field case $z\rightarrow 0$ and unbiased sources for brevity.

\subsection{Energy}
The internal energy density for the arbitrary spreading sequences is one such quantity
\begin{equation}
\saed = \frac{\partial}{\partial \beta} \lim_{K \rightarrow \infty} - \frac{1}{K}\<\log \partitionfunction(\beta,z,\quenched)\> \label{appeq:e1}\;,
\end{equation}
using the definition of the free energy this gives
\begin{equation}
\saed = \lim_{K \rightarrow \infty}\<\frac{1}{Z(\beta,\vy,\ms) K} \sum_{\vtau} {\cal
H}(\vtau,\vy,\ms) \exp\left\lbrace -\beta {\cal H}(\vtau)
\right\rbrace \>_{\vy,\ms} \label{appeq:e2}\;.
\end{equation}
The average with respect to $\vy$, for a generic function $\arbitraryfunction(\vy)$, can be decomposed
\begin{equation}
\<\arbitraryfunction(\vy)\>_\vy = \int \rmd\vy \rmd\vomega \left[\sum_{\vb} P(\vy| \vomega,\vb) P(\vomega) P(\vb)\right] \arbitraryfunction(\vy)\;.
\end{equation}
Substituting the exact expression for the likelihood term, and marginalising with respect to the Gaussian noise gives
\begin{equation}
\int \rmd\vomega P(\vy | \vomega,\vb) P(\vomega) = \left(\frac{\beta_0}{2\pi}\right)^{\frac{M}{2}} \exp \left\lbrace -\frac{\beta_0}{2}\sum_\mu\left(y_\mu - \sum_{k} s_{\mu k} b_k\right)^2\right\rbrace \;.
\end{equation}
This is equal to the partition sum when $\beta=\beta_0$ and without an external field. In the case of an external field that matches $P(\vb)$ a similar cancellation occurs. This parameterisation is the Nishimori temperature. Taking the partial derivative leaves, in the case of a uniform prior on bits,
\begin{equation}
\saed \propto \int \frac{\rmd\vy}{\sqrt{2\pi}^M} \sum_{\vtau} \frac{1}{2 K}\sum_\mu \left(y_\mu -\sum_k s_{\mu k} \tau_k \right)^2 \exp \left\lbrace - \sum_\mu \frac{\beta}{2}\left(y_\mu -\sum_k s_{\mu k} \tau_k \right)^2\right\rbrace\;,
\end{equation}
and the energy finally evaluates to $1/(2\load)$. The constant of proportionality exactly cancels the partition function denominator (\ref{appeq:e2}), once the final averages are taken.

\subsection{The sufficiency of replica symmetry}
A more interesting result involves the correlations between two estimates/states sampled according to the Hamiltonian - these are real replicas of the system. The average overlap ($q$) between reconstructed bits sequences is described by
\begin{equation}
{\hat P}(q| \ms,\vy) = \sum_{\vtau,\vsigma} \delta\left(q - \frac{1}{K}\sum_k b_k b'_k \right){\hat P}(\vb'| \ms, \vy) {\hat P}(\vb| \ms, \vy) \label{Nishimori:q}\;.
\end{equation}
This can be compared to the magnetisation, where $\vb$ is the quenched random variable encoding the bit sequence
\begin{equation}
{\hat P}(m | \ms,\vy) = \sum_{\vb'} \delta \left(m - \frac{1}{2}\sum_k b_k b'_k \right) {\hat P}(\vb'|\vy,\ms)\;.
\end{equation}
This is a random quantity with respect to the signal, but self-averaging can be assumed to apply with respect to the bit sequence. Averaging over realisations of the bit sequence distributed with a uniform prior and conditioned on $\vy$ and $\ms$, gives
\begin{equation}
\<{\hat P}(m | \ms,\vy)\>_{\vb | \vy, \ms} = \sum_{\vb} {\hat P}(m | \ms,\vy) P(\vb | \vy,\ms) \label{Nishimori:m}\;.
\end{equation}
Therefore the random variable, which describes the overlap of two replicas of the system (\ref{Nishimori:q}) is identical to the magnetisation in the large system limit, provided the estimate and generative probability distributions are identical.

The consequence of this latter result is that the many replica correlation function is a simple one, allowing a connected description of the phase space. The RS assumption will give a correct description of behaviour provided the self-averaging assumption applies. The result is exact in the case $\beta=\beta_0$, but since the system properties might be expected to change smoothly with respect to small changes in the estimation probability model, so that the RS assumption may apply in a range of $\beta$. 

%% file: APPENDICES/Ucp_for_cdma.tex
In this Appendix it is shown that the inference problem for a variety of loopy ensembles can be reduced to tree like inference problems, so that an optimal decoding can be efficiently determined. The proof that a probabilistically optimal decoding configuration can be determined easily has consequences for algorithm development in noisy regimes; the spectral efficiency and attainable bit error rate at zero noise is of course a limitation to the spectral efficiency in noisy systems. To demonstrate the inference problem is computationally easy an equivalent Constraint Satisfaction Problem (CSP) is reduced to a solvable CSP on a tree by Unit Clause Propagation (UCP).

In the case of noiseless CDMA with uniform amplitude BPSK a solution is sought to the set of chip constraints
\begin{equation}
\vy = \ms \vb \label{appeq:vymsvb}\;.
\end{equation}
The transmission amplitude of users is zero or of fixed amplitude ($C^{-\frac{1}{2}}$) when BPSK or unmodulated codes are used (\ref{eq:CDMA.BPSK}), and it is convenient to consider the code and signal rescaled to integer values in this Appendix so that every term in (\ref{appeq:vymsvb}) is integer valued. For typical bit sequences, and well chosen codes, the solution is unique, but the inference structure is a loopy graph in the case of ensembles with reasonable load, and an efficient method is not known for worst case transmission scenarios.. However, in worst case the problem, even in the case of a sparse matrix, is NP-complete, so that there may be no practical way to determine the optima.

By contrast the case of random Gaussian amplitude shift keying does not correspond to an integer valued inference problem (\ref{appeq:vymsvb}) and is easily solved on a chip by chip basis. However, BPSK is believed to be comparable to Gaussian modulation in noisy systems, or provably better in some marginal properties (section~\ref{CDMA.ScalarChannel}), so the insight into the noiseless limit of this system is also valuable. The ambiguity introduced by using uniform amplitude modulation is intuitively a better reflection of the ambiguity relevant, for all modulation patterns, in noisy systems. Since the leading order properties of the unmodulated and BPSK codes (\ref{eq:CDMA.BPSK}) are found to be described by the same UCP dynamics, therefore little further reference is made to the choice of modulation.

\subsection{Sparse noiseless CDMA as a constraint satisfaction problem}

The rescaled signal takes integer values. If the number of variables attached to a chip is $L_\mu$ then the set of values for $y_\mu$ is $\{-L_\mu + 2 i\}$, where $i=0,\ldots L_\mu$. Only the chip-regular ensemble (\ref{eq:CDMA.chipregular}) is examined in detail, with $L_\mu=L=3$ and Poissonian variable connectivity (\ref{eq:CDMA.chipregular}). Comparable methods can be developed for other ensembles~\cite{Raymond:OD}, but in the context of this thesis it is informative to consider just the single ensemble at various $\load$, which is comparable to the $\otSAT$ ensemble of chapter~\ref{chapter:1inkSAT}.

With $K$ source bits and $K/\load$ clauses, the duality $\{-1,1\}\rightarrow\{\True,\False\}$ may be assumed in the source bits. A variable ($k$) is included in a clause ($\mu$) if the corresponding component in the connectivity matrix $A_{\mu k}=1$, otherwise it is absent from the clause. The variable appears as a positive literal if the corresponding modulation pattern ($\modulationsymbol_{\mu k}$) is positive, and as a negative literal otherwise. Each chip implies a logical constraint
\begin{equation}\begin{array}{lclcl}
 y_\mu&=&-3 (3) &\qquad&\hbox{All three literals are
$\True$ ($\False$)}\;; \\
y_\mu&=& -1 (1)&\qquad&\hbox{Two in three literals are
$\True$ ($\False$)}\;.
\end{array}\end{equation}
Therefore clauses include the all in 3 type clause, which is equivalent to 3 unit clauses (trivial logical statements), and the 2 in 3 clause. Finally all 2 $\True$ in 3 clauses may be transformed to 1 in 3 clause by negating all the literals in a clause.

The problem is formulated as a $\otSAT$-type ensemble, combined with a set of unit clauses. Within this structure there is no correlation between the distribution of $\otSAT$ and 3 in 3 SAT clauses and the marginal probability for a literal to be positive is $1/2$. Therefore the solution space might be expected to be related to the ($\epsilon=\frac{1}{2}$) $\eotSAT$ ensembles studied in chapter~\ref{chapter:1inkSAT}, but squeezed randomly by the unit clauses.

Although the analogy with $\eotSAT$ is apparent, the distribution of literals is not independent of the embedded sequence. In both the case of symmetric and unmodulated BPSK modulation, true ($-1$) variables are twice as likely to appear as negative literals than as positive literals. This causes some modifications to the branching processes demonstrated in chapter~\ref{chapter:1inkSAT}. In combination with the extensive number of unit clauses in the initial condition, a substantially different analysis is required.

\subsection{UCP applied to sparse chip-regular CDMA}

In the large system limit a sparse code always provides some free information (unit clauses). Incorporating this deterministically ensures that any bit implied only by logical deduction must coincide exactly with the source bits, and form part of an optimal detection. By iteratively decimating (assigning) variables, and modifying the inference problem structure, the original inference problem can be substantially reduced. Each decimation modifies only the clauses in the graph in which it is a literal, and the mean dynamics of these processes can be studied. A literal decimation, consistent with the embedded sequence, is twice as likely to be false than true in a 3-clause (one in 3 clause). When this literal is decimated, with probability $1/3$ the clause is covered implying the other two literals to be false (unit clauses), otherwise the 3-clause is reduced to a 2-clause containing one positive and one negative literal. Let $C_3(X)$ be the population of clauses of length $3$. After $X$ variable decimations the expected change in the population after one further assignment is in expectation
\begin{equation}
 \Delta C_3(X) = C_3(X+1) - C_3(X) = - \frac{3}{K-X}C_3(X) \;,
\end{equation}
where the coefficient is the probability that a variable selected at random is in the clause.

The population of 2-clauses, clauses of the type "1 in 2 literals are true", are absent from the initial condition, but if variables are decimated at random a population
 of 2-clauses is created from reduced 3-clauses. If the distribution of variables within the 3-clauses is independent and uncorrelated then the same will hold true for the 2-clauses generated by decimation of 3-clauses. At the same time the population is reduced when a decimated literal is coincident with a 2-clause. The dynamics of two clauses population $C_2$ evolves, in expectation, according to
\begin{equation}
\Delta C_2(X) = C_2(X+1) - C_2(X) = - \frac{2}{K-X}C_2(X) + \Delta C_3(X)(1-\UCPtransitioner) \;,
\end{equation}
where $\UCPtransitioner=1/3$ is the probability (2) unit clauses are created given a 3-clause decimation. Finally the creation of unit clauses from either 3 or 2 clauses will be an i.i.d process given the number of 2 and 3-clauses reduced, and the number of variables left in the problem ($K-X$). If at each time step a large set of uncorrelated unit clauses exist, and one variable (one variable, rather than one unit clause) in the set is selected at random, the population decreases by at least one. Variables, rather than unit clauses (which may be degenerate in a variable), are selected at random from the set. Let $U_1(X)$ be the number of variables that are unknown, either because they are not in the decimated set, or not represented in the set of unit clauses at decimation time $X$. When new unit clauses are created, some of these are coincident with this ambiguous set, and so the quantity is always reduced
\begin{equation}
U_1(X+1) - U_1(X) = - \frac{U_1(X)}{K-X} \left( \frac{6\UCPtransitioner}{K-X} C_3(X) + \frac{2}{K-X} C_2(X) \right) \label{appeq:U_1}\;,
\end{equation}
while $U_1(X)>K-X$, where the quantity \[\cdots\] is the number of new unit clauses created in expectation. Unit clauses created from $2$ and $3$ clauses are equally likely to describe new variables.

The set of equations describe the mean so long as the number of unit clauses is non-zero at all decimation times, this implies the condition $K-X < U_1(X)$. If populations are large and quantities concentrate on their mean, these equations should also describe typical dynamics. This is observed in experiments~\cite{Raymond:OD}.

The clause populations are extensive at $X=0$, and provided this situation is maintained an analytic continuation to rescaled parameters is reasonable $x=X/K$, $C_i(X)=K c_i(x)$ $U_1(X)=K U_1(x)$, which allows a differential description of dynamics~\cite{Wormald:DE}. The clause dynamics are
\begin{equation}
\frac{\rmd}{\rmd x} c_3(x) = \frac{-3 c_3(x)}{1-x} \;;\qquad c_2(x) = \frac{3(1-\UCPtransitioner) c_3(x)}{1-x} - \frac{2 c_2(x)}{1-x}\;,
\end{equation}
and are exactly solvable in the both cases
\begin{equation}
c_3(x)= c_3(0)(1-x)^3 \;; \qquad c_2(x) = \left[c_2(0) + 3 (1-\UCPtransitioner) c_3(0)\right] x (1 - x)^2 + c_2(0)(1-x)^3\;,
\end{equation}
according to the initial condition on the clause populations are $c_3(0)=3/(4\load)$ and $c_2(0)=0$. The expression in the number of unit clauses
\begin{equation}
\frac{\rmd u_1(x)}{\rmd x} = - \frac{u_1(X)}{1-x} \left[ \frac{6\UCPtransitioner}{1-x}c_3(X) + \frac{2}{1-x}c_2(x) \right] \label{appeq:u_1}\;,
\end{equation}
may also be solved exactly using the initial conditions $u_1(0)=1-\exp(-3/(4\load))$, in a form with a quadratic exponent dependence on $x$.

By determining numerically the critical decimation time where the algorithm ceases to operate $x^*=\argmin(u_1(x)=1-x | x>0)$, it is possible to establish how many variables are set deterministically from the initial condition. If the remaining problem is tree-like it is efficiently solvable even in worst case. If it remains loopy then higher level inference than UCP may be required to obtain an optimal solution, although a heuristic driven UCP approach may be successful in typical case by analogy with chapter \ref{chapter:1inkSAT}.

\subsubsection{Results}

\begin{figure}
\centering{
\includegraphics[width=0.8\linewidth]{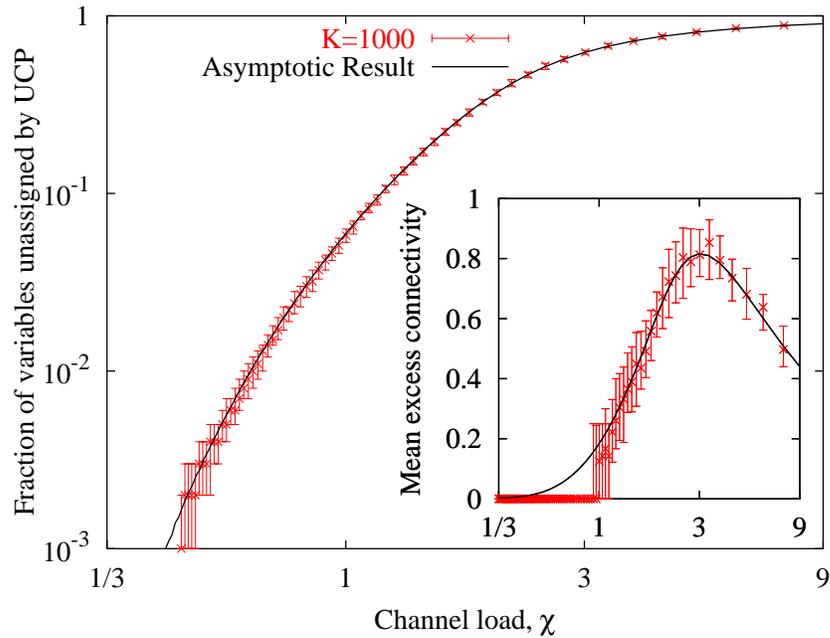}
\caption[UCP in noiseless CDMA.]{\label{fig:UCPvariablessetCDMA} The figure demonstrates the inference problem properties at $x^*$ (when UCP runs out of deterministically implied unit clauses) for typical sparse chip regular ($L=3$) BPSK ensembles at various load. The main figure is the fraction of initial variables remaining in the inference problem. The inset shows the mean excess connectivity for the residual inference problem. The median, lower and upper quartiles for the two quantities are represented as bars, $100$ samples were taken, each sample applying to $1000$ user. The continuous lines demonstrate the mean quantities calculated through $u_1$,$c_2$ and $c_3$. At low load almost all variables are set deterministically, at large load few are determined. However, in both cases the residual problem has an excess connectivity less than 1, indicating that few loops remain (no loops asymptotically). Thus the residual inference problem is computationally simple.}}
\end{figure}

Results of detection by UCP are demonstrated in figure
\ref{fig:UCPvariablessetCDMA}. Experimental results are in excellent agreement except where finite size effects are clear at small load, and concentration on the analytic result increases quickly with $K$. As load increases towards the percolation threshold of the inference problem ($\log_3 \load = 2$) fewer variables are implied by UCP. In all problems, even those where few variables are set, the excess connectivity of the residual inference problem is less than one, indicating that asymptotically loops are absent from the residual inference problem when UCP halts. Therefore, even at high loads the problem is computationally easy, this is in contrast to dynamical load limitations well understood in the dense random CDMA ensemble~\cite{Tanaka:SMA}, and approached by the sparse case as $L$ increases.

A change in the performance characteristics might be anticipated as the load $\load\sim 1.6$, corresponding to an upper bound in spectral efficiency (\ref{eq:CDMA.noiselessseU}), before the percolation threshold. A point of inflexion is observed on a linear scale near this point, but no sharp features in the number of variables inferred by UCP is observed. Finally it is noted that in experiments with BPSK and unmodulated codes, with $K=1000$ and $K=10000$, no qualitative differences are visible in the parameter range of figure~\ref{fig:UCPvariablessetCDMA}, and the deviation in the quartiles from the analytic result decreases substantially at $K=1000$.

\subsection{UCP applied to other ensembles}

The case of chip regular ensemble is generalised in a straightforward manner to incorporate the irregular ensemble. Additional correlations must be considered in handling the user-regular user connectivity ensembles~\cite{Raymond:OD}, and in principle a calculation for ensembles constrained in both chip and user connectivity may be possible.

When noise is present some of the information from chips becomes unreliable. One way in which to consider such noisy effects, but allow an exact UCP analysis, is to consider the binary erasure channel~\cite{Richardson:MCT,Luby:EE}. This is straightforwardly incorporated as an initial condition change for each of the ensembles. In the case of the irregular ensemble variation of the fraction of chips erased is analytically equivalent to a decrease in $C$. UCP might be used as a heuristic component within algorithms for more complicated noise models, especially in regimes where the SNR is very large and the number of users finite.

In the case of chip-regular $L=3$ ensembles, the algorithm terminates and all residual problems are tree-like and hence algorithmically easy, although the solution space is degenerate it is easy to determine as many solutions as are required, for example by heuristically driven UCP. The more interesting case occurs at larger $L$, where UCP may halt with the residual problem being loopy~\cite{Raymond:OD}. The inference problem remaining takes a form described by a distribution on clause types, functionally determined by residual occupancy and signal $(L_\mu,y_\mu)$, alongside variations in variable connectivity. For different ensembles and terminating stages a wide variety of ground (solutions) and excited (possibly dynamically dominant) state distributions may be relevant. The complexity equivalence of unmodulated and BPSK ensembles does not extend to these structures necessarily.

In spite of the wide of clauses and variable connectivity distributions possible, some commonalities are apparent, such as the {\em locked} nature of all clause types~\cite{Zdeborova:LC}. These clause types are known to lead to a fragmented distribution of low energy solutions in many cases, leading to algorithmic complications. One feature common to the $\eotSAT$ problem (chapter~\ref{chapter:1inkSAT}) and the high load CDMA noisy inference problem (chapter~\ref{chapter:sparseCDMA}) is the appearance of a dynamical transition, with variation in variable connectivity. A parameter range exists for which determining the optimal solution is difficult, at least by UCP type algorithms. It seems likely that the origins of the noisy metastability when using optimal detection of CDMA is rooted in some equilibrium or dynamical properties also relevant and more easily analysed by UCP in noiseless model.

%% file: APPENDICES/Sparse_matrix_analysis.tex
\subsection{Sparse connectivity matrix}
\label{app.Sparsematrix}

This subsection gives an analytic representation for a marginal connectivity distribution allowing the quenched average, with respect to the sparse part of a disordered ensemble, to be carried out in all models.

The factor graphs considered are described by a fixed ratio of the number of variable nodes to number of factor nodes, or equivalently the ratio of the mean factor node connectivity to mean variable node connectivity
\begin{equation}
 \load =\frac{K}{M} = \frac{L}{C} \label{appeq:chi}\;.
\end{equation}
For analysis purposes all probability distributions, including priors, are taken to be conditioned on these global parameters, these are written explicitly and selectively only for priors.

Let $P(\vL,\vC)$ describe the joint probability distribution for the connectivity of factors (chips) and variables (users) in a graphical model. Cases in which the coupling distributions are conditionally independent given a fixed number of edges in the graphical model can be described by
\begin{equation}
P(\vL,\vC|P_L,P_C) = \frac{1}{\cal N}\prod_{\mu=1}^M P_L(L_\mu) \prod_{k=1}^K P_C(C_k) \delta\left(\sum_{\mu=1}^M L_\mu - \sum_{k=1}^K C_k\right) \label{appeq:PLCgivenPlPc}\;,
\end{equation}
where $P_C$ is the variable node connectivity distribution conditioned on a mean connectivity of $C$, and $P_L$ is the factor node connectivity distribution conditioned on a mean value of $L$. Both $L$ and $C$ are finite in this analysis, much less than $K$ and $M$ respectively, the limit of $\infty$ is taken in these latter quantities to determine asymptotic results. The final constraint balances the number of edges leaving factor nodes, in expectation $M L$, and leaving variable nodes, in expectation $K C$. In the large system limit the final constraint (\ref{appeq:PLCgivenPlPc}) is assumed to be negligible at leading order in calculation of typical case properties.

This form constrained by both $P_L$ and $P_C$ can be relaxed in many derivations. For example $P_{L=2}=\delta_{L_\mu,2}$, which describes a graph with factor nodes of connectivity two, might be combined with the weak global constraint (\ref{appeq:chi}). However, within the analysis presented it is equivalent to consider a Poissonian distributions ($P_C$ in this case) to describe weakly constrained scenarios.

An understanding of the probability distribution over sample graphs is quantified in a distribution over the connectivity/adjacency matrix $\mA$. Each matrix describes a distinct labeled instance of a sparse factor graph
\begin{equation}
A_{\mu k} =\left\lbrace \begin{array}{cc}
 1 &\hbox{If a link exists between factor $\mu$ and variable $k$}\;;\\
 0 &\hbox{otherwise}\;.
\end{array}\right.
\end{equation}
A prior distribution on the edges for sparse models is
\begin{equation}
P(A_{\mu k}| L,K)= \left(1 -\frac{L}{K}\right)\delta_{A_{\mu k},0} + \frac{L}{K}\delta_{A_{\mu k},1}\;.
\end{equation}
However, the quantity of interest is the distribution on the matrix $\mA$, given the probability distributions
\begin{equation}
P(\mA|P_C,P_L) = \sum_{\vL,\vC} P(\mA|\vC,\vL)P(\vC,\vL|P_C,P_L)\;.
\end{equation}
The first term may be rewritten by Bayes' Theorem
\begin{equation}
P(\mA|\vC,\vL) = \frac{P(\vC|\mA)P(\vL|\mA) }{P(\vC,\vL|L,K,\load)}  \prod_{\mu,k} P(A_{\mu k}|L,K) \label{appeq:APLPC}\;.
\end{equation}
The posterior distributions over factor and variable connectivity may be factorised. Similarly the denominator takes a form (\ref{appeq:PLCgivenPlPc}), and is assumed to be factorised by approximation as two Poissonian distributions. Therefore the $\vC$ dependent part of (\ref{appeq:APLPC}) is
\begin{equation}
\sum_{\vC} P(\vC|\mA)P(\vC|P_C) = \prod_{k} \<\frac{1}{P(C_k|L,K)} \delta\left(\sum_\mu A_{\mu k} - C _k \right)\>_{C_k}\;,
\end{equation}
and the $\vL$ dependent part is
\begin{equation}
\sum_{\vL} P(\vL|\mA)P(\vL|P_L) = \prod_{\mu} \< \frac{1}{P(L_\mu|L,K)} \delta\left(\sum_k A_{\mu k} - L_\mu \right) \>_{L_\mu}\;.
\end{equation}

In the limit of large $K$, the conditional probabilities normalising the distributions are simple Poissonian factors
\begin{equation}
\left({\cal N}_\mu\right)^{-1} = \lim_{K\rightarrow \infty} P(L_\mu | L,K) = \frac{L^{L_\mu}\exp\{-L\}}{L_\mu!} \label{appeq:distL}\;,
\end{equation}
and similarly
\begin{equation}
\left({\cal N}_k\right)^{-1} = \lim_{M\rightarrow\infty} P(C_k | C,M) = \frac{C^{C_k}\exp\{-C\}}{C_k!}
\label{appeq:distC}\;.
\end{equation}
The most convenient form for the posterior (\ref{appeq:APLPC}) as used in the main test is
\begin{equation}
P(\mA|P_L,P_C) \propto \<\prod_{k} \left[\frac{C_k!}{C^{C_k}}
\delta\left(\sum_\mu x_{\mu k} - C_k \right)\right] \prod_{\mu} \left[\frac{L_\mu!}{L^{L_\mu}}\delta\left(\sum_k A_{\mu k} - L_\mu \right)\right] \>_{\vL,\vC} \prod_{\mu,k} P(A_{\mu k}) \label{GENFORM}\;,
\end{equation}
absorbing some constant terms in a global normalisation constant. The results in this Appendix are developed using the full forms ${\cal N}_k$ and ${\cal N}_\mu$, without a global normalisation.

In terms of calculating ensemble averages an edge factorised form is required, the $\delta$ functions must be replaced by analytic forms. This is achieved with the Cauchy integral formula (\ref{appeq:CauchyIntegral}), for the factor constraint
\begin{equation}
\delta\left(\sum_k A_{\mu k} - L_\mu \right) =\frac{1}{2\pi \rmi} \oint_\Complexs \rmd Y_\mu\frac{1}{Y_\mu^{L_\mu+1}} \prod_k Y_\mu^{A_{\mu k}}\label{eq:CDMA.Ymuintegral}\;,
\end{equation}
and variable constraint
\begin{equation}
\delta\left(\sum_\mu A_{\mu k} - C_k \right) =\frac{1}{2\pi \rmi} \oint_\Complexs \rmd Z_k \frac{1}{Z_k^{C_k+1}} \prod_\mu Z_k^{A_{\mu k}}\;,
\end{equation}
where the integrals are around unit circles in the complex plane ($\Complexs$). All non-site dependent normalisations are easy to establish retrospectively and will be dropped until the final expression.

This allows the posterior to be written
\begin{equation}
P(\mA|P_L,P_C) \propto \prod_k\left[\frac{C_k!}{C^{C_k}}\exp(C) \oint\frac{\rmd Z_k}{Z^{C_k+1}} \right]\prod_\mu \left[ \exp(L) \frac{L_\mu!}{L^{L_\mu}} \oint \frac{\rmd Y_\mu}{Y_\mu^{L_\mu+1}} \prod_k P(A_{\mu k}|L)\left[ Y_\mu Z_k \right]^{A_{\mu k}}\right] \label{appeq:AnPALC}\;.
\end{equation}
Thus a factorised form with respect to $\mA$ is obtained subject to two sets of complex integrals.

\subsection{General average with respect to factor connectivity}
\label{app.GAmu}

A sufficient general case of averaging with respect to
(\ref{appeq:AnPALC}) is considered, of the function $\arbitraryfunction$ with some site dependence in $k$ and also a factorised dependence on $\mu$, but dependence on connectivity only through an exponent
\begin{equation}
\arbitraryfunction = \prod_{k,\mu}\left[\arbitraryfunctiontwo_k \arbitraryfunctiontwo_\mu \right]^{A_{\mu k}},
\end{equation}
Aside from a more complicated form for $\arbitraryfunction$, involving additional marginalisations over quenched parameters or auxiliary variables, the process of averaging in factor connectivity in all chapters follows closely this Appendix. The quenched marginalisations can be introduced after first taking the connectivity average.

The average over $\arbitraryfunction$ is
\begin{equation}
\begin{array}{lcl}
\< \arbitraryfunction \>_\mA &=& {\cal N}\prod_k \left[\<\frac{C_k!\exp\{C\}}{C^{C_k}} \oint \frac{\rmd Z_k}{Z^{C_{k}+1}}\>_{C_k} \right] \\
&\times&\prod_\mu \<\left[ \frac{L_\mu!\exp\{L\}}{L^{L_\mu}} \oint \frac{\rmd Y_\mu}{Y_\mu^{L_\mu+1}} \prod_k \left[ \sum_{A_{\mu k}} P(A_{\mu k})\left(Y_\mu \arbitraryfunctiontwo_\mu Z_k \arbitraryfunctiontwo_k\right)^{A_{\mu k}} \right] \right]\>_{L_\mu} \label{appeq:Exi}\;.\end{array}
\end{equation}

Considering the inner most set of brackets, which is factorised with respect to $\mu$, evaluation of the sum gives a product of $k$ in terms of the type $[(1-L/K) + L/K Z_k Y_\mu \arbitraryfunctiontwo_k]$.
Inverting the Cauchy integral (\ref{eq:CDMA.Ymuintegral}) selects only the $L_\mu^{th}$ term in the expansion of $k$
\begin{equation}
\left[\cdots\right] = \left\lbrace\begin{array}{ccc}
L_\mu \factorial \sum_{\<k_1 \ldots k_{L_\mu}\>} \prod_{l=1}^{L_\mu} \left[\frac{1}{K} \arbitraryfunctiontwo_\mu Z_{k_l} \arbitraryfunctiontwo_{k_l}\right] &\hbox{if} &{L_\mu>1}\;; \\
\left(\arbitraryfunctiontwo_\mu \frac{1}{K} \sum_{k=1}^K Z_k \arbitraryfunctiontwo_{k}\right)^1 &\hbox{if} &{L_\mu=1} \;;\\
\left(\arbitraryfunctiontwo_\mu \frac{1}{K}\sum_{k=1}^K Z_{k} \arbitraryfunctiontwo_{k}\right)^0 &\hbox{if}&{L_\mu=0}\label{appeq:orderparamexp}\;;
\end{array}\right.
\end{equation}
where $\<k_1 \ldots k_{L_\mu}\>$ is the ordered set of indices, $k_1 <\ldots < k_{L_\mu}$. A definition is convenient to extract the $k$ dependence, in the simplest case
\begin{equation}
 1 = \int \rmd\GENOP \delta\left(\GENOP - \frac{1}{K}\sum_{k=1}^K \arbitraryfunctiontwo_k Z_k\right) \label{appeq:GENOP}\;.
\end{equation}
Since the identity is complex, the integral is over the complex plane. The general order parameter used in chapter~\ref{chapter:sparseCDMA} is
\begin{equation}
1 = \int \rmd\GENOP \delta\left(\GENOP(b,\rvsigma) - \frac{1}{K}\sum_{k=1}^K \arbitraryfunctiontwo_k \delta_{b,b_k} Z_k \prod_\alpha \delta_{\sigma^\alpha,\tau^\alpha_i}\right) \label{appeq:GENOP2}\;.
\end{equation}
introduced for every $(b,\rvsigma)$ where $\arbitraryfunctiontwo_k$ is one.

The part of (\ref{appeq:Exi}) factorised in $\mu$ may be expanded,
\begin{equation}
\prod_\mu \<\left[ \cdots\right]\>_{L_\mu} = \prod_\mu \<(\GENOP \arbitraryfunctiontwo_\mu)^{L_\mu} + O\left(\frac{1}{K}\right)\>_{L_\mu}\;,
\end{equation}
when working with (\ref{appeq:GENOP}), but easily extended to (\ref{appeq:GENOP2}). The $O(1/K)$ terms are taken to be negligible. Assuming $\arbitraryfunctiontwo_k$ is dependent on some quenched parameters these can be averaged over, and the sum replaced by $M$. This final form can be written
\begin{equation}
\<\prod_\mu \left[\cdots\right] \>_{\{\arbitraryfunctiontwo_\mu\}} = \exp \left\lbrace \sum_\mu \log\left( \<(\GENOP \<\arbitraryfunctiontwo_\mu\>)^{L_\mu} \>_{L_\mu}\right) \right\rbrace = \exp \left\lbrace - K \Gone(\GENOP) \right\rbrace \;.
\end{equation}

\subsection{General average with respect to user connectivity}
\label{app.GAk}

The average over the $k$ dependent terms in (\ref{appeq:Exi}), can be taken once a form factorised with respect to $k$ is derived. The simpler order parameter (\ref{appeq:GENOP}) is considered but a product over order parameters (\ref{appeq:GENOP2}), suitable for all chapters, may also be processed through the method of this Appendix. The expression of interest given the average over the factor connectivity of section \ref{app.GAmu} is
\begin{equation}
\< \arbitraryfunction \>_\mA = {\cal N} \int \rmd\GENOP \< \prod_k \left[\frac{C_k!}{C^{C_k}} \exp\{C\}\oint \frac{\rmd Z_k}{Z^{C_{k}+1}} \right] \delta\left(\GENOP - \frac{1}{K}\sum_{k=1}^K \arbitraryfunctiontwo_k(k) Z_k\right) \>_{\vC} \exp \left\lbrace - K \Gone(\GENOP) \right\rbrace
\label{appeq:Exi2}\;.
\end{equation}
Each of the delta functions may be represented by a Fourier
transform (\ref{appeq:Fourier}), the resulting integral is again in the complex plane
\begin{equation}
\delta\left(\GENOP - \frac{1}{K}\sum_{k=1}^K \arbitraryfunctiontwo_k(k) Z_k\right) \propto \int \rmd\GENOPconj \exp \left\lbrace - (C K) \GENOP \GENOPconj \right\rbrace \exp \left\lbrace (C K)\frac{1}{K} \sum_{k=1}^K \arbitraryfunctiontwo_k Z_k \GENOPconj \right\rbrace \label{appeq:Ktransformpart}\;.
\end{equation}
It is convenient to include the constant factor $C$, by contrast with (\ref{appeq:Fourier}). In the thermodynamic limit it can be assumed that $\GENOPconj$ is proportional to $K$, this is necessary for scalable solutions of the saddle-point equations. The $\vZ$ dependence is factorised so it is finally possible to take the integral with respect to $\vZ$
\begin{equation}
 \prod_k \<\frac{C_k!}{C^{C_k}} \exp{C} \intZ{{C_k}}\exp\left\lbrace C_k \arbitraryfunctiontwo_k Z_k \GENOPconj \right\rbrace\>_{C_k} = \prod_k\< \left[\arbitraryfunctiontwo_k \GENOPconj\right]^{C_k}\>_{C_k} \;,
\end{equation}
which can be calculated on a marginal basis with respect to $k$, each integral picks out the $C_k^{th}$ component. Averaging over the quenched disorder associated to $\arbitraryfunctiontwo_k$ all topology is removed and an exponential form is apparent
\begin{equation}
\<\arbitraryfunction\>_{\{\arbitraryfunctiontwo_k\},\{\arbitraryfunctiontwo_\mu\},\mA} \!=\! {\cal N}\! \int \!\rmd \GENOP \!\rmd \GENOPconj\!\exp\left\lbrace \!K C \!+\! K \log\<\!\left[\GENOPconj \<\arbitraryfunctiontwo_k\>\! \right]^{C_k} \!\>_{C_k} \!-\! K C \GENOP \GENOPconj \!+\! K/\load \log \< (\<\arbitraryfunctiontwo_\mu\> \GENOP)^{L_\mu}\>_{L_\mu} \!\right\rbrace\;.
\end{equation}

\subsection{Ensemble and order parameter normalisation}
\label{app.normalisation}

The case of $\arbitraryfunction=\arbitraryfunctiontwo_k \arbitraryfunctiontwo_\mu=1$ can be considered to establish the global normalisation constant. The remaining problem is a simple saddle-point problem,
\begin{equation}
1 =  {\cal N} \int \rmd \GENOP \rmd \GENOPconj \exp \left\lbrace K C + K \log\<\left[\GENOPconj \right]^{{c_f}} \>_{c_f} - K C \GENOP \GENOPconj + K/\load \log \< \GENOP^{{l_e}}\>_{l_e} \right\rbrace\;.
\end{equation}
The integral is dominated at a saddle-point where the first derivatives with respect to $(\GENOP,\GENOPconj)$ are zero. The derivative with respect to $\GENOPconj$ is
\begin{equation}
\GENOP= \frac{1}{C}\frac{ \<{c_f}\left[\GENOPconj\right]^{{c_f}-1} \>_{{c_f}}}{\< \left[\GENOPconj\right]^{{c_f}} \>_{c_f}}\;,
\end{equation}
and with resect to $\GENOP$
\begin{equation}
\GENOPconj = \frac{1}{C\load} \frac{\<{l_e} \GENOP^{{l_e}-1} \>_{{l_e}}}{\< \GENOP^{{l_e}}\>_{l_e}}\;.
\end{equation}
In general a consist solution is $\GENOPconj=1$ and $\GENOP=1$, due to the choice of scaling for $\GENOPconj$ (\ref{appeq:Ktransformpart}) the form is parameter independent. The same normalisations apply to the constant part when more complicate forms of $\arbitraryfunction$ are considered. The global normalisation constant is ${\cal N} = 1$ as expected.

%% file: APPENDICES/Conjugate_fields.tex
An interpretation for some parameters can be gained by consideration of derivatives of the free energy with respect to $\beta$, and simple random external fields $\vrandomfield$. This may also be used to prove the consistency of some method assumptions in the case of replica symmetry. The choice of a random field is primarily to allow a concise inclusion within the variational free energy description. It is equivalent to work directly with fields conjugate to quantities such as $\sum_\ij \tau_i \tau_j$, or with annealed random fields in some cases.

\subsection{Energy and entropy in sparse CDMA}
A derivative of the free energy density $\beta \safed$, with respect to $\beta$, gives the average of the Hamiltonian, the energy density. In chapter~\ref{chapter:sparseCDMA} the free energy is (\ref{eq:CDMA.freeenegyreplica}) and a partial derivative with respect to $\beta$ is
\begin{equation}
\begin{array}{lcl}
\saed &=& \frac{\partial}{\partial \beta} (\beta f) \propto \left. \frac{\partial}{\partial n} \right|_{n=0} \Bigg\langle\prod_{l=1}^{l_e} \left[\sum_{b_l,\rvsigma_l} \GENOP^*_{b_l}(\rvsigma_l)\right]\\
&\times&\left<\frac{1}{2}\left( \omega +\sum_l \frac{\modulationsymbol_l(b_l - \sigma^\alpha_l)}{\sqrt{C}}\right)^2 \exp \left\lbrace \frac{-\beta}{2}\left(\omega + \sum_l \frac{\modulationsymbol_l (b_l - \sigma^\alpha_l)}{\sqrt{C}} \right)^2\right\rbrace \>_{\left\lbrace \modulationsymbol_l\right\rbrace,\omega} \Bigg\rangle_{l_e}\;.
\end{array}
\end{equation}
For complicated noise distributions ($P(\vomega)$) energy may be determined given the saddle-point solution for the order parameters. Where $P(\vomega)$ is Gaussian the expression can be evaluated explicitly, in agreement with the known exact result at the Nishimori temperature (Appendix~\ref{app:Nishimori}). The entropy of the model, which measures the size of the phase space, the number of states determining equilibrium properties, is determined straightforwardly from the Helmholtz relation
\begin{equation}
\sasd = \beta(\saed-\safed)\;.
\end{equation}
A negative self-averaged entropy is often an indication of failure in the saddle-point approximation method (replica symmetric in most of this thesis), or a more fundamental failure of the self-averaging assumption.

\subsection{Random external field analysis in sparse CDMA}
\label{app:physicsfromconjfields}
The random external field ($\vrandomfield$) is introduced in all chapters as a means to break symmetries or to evaluate properties of interest. In chapter~\ref{chapter:sparseCDMA} various external fields are useful. The external field is introduced in the Hamiltonian (\ref{eq:CDMA.Hamiltonian}), but taken to be zero in the free-energy derivation. Assume instead that the field is non-negligible and the $k$ dependence in $\randomfield_k$ takes one of two forms, either uniform or random
\begin{equation}
\randomfield_k = z^D b_k\;\qquad\hbox{or}\;\qquad\randomfield_k = \sqrt{\frac{z^D}{\beta}} \zeta_k \;. \label{appeq:randomfield}
\end{equation}
$\zeta_k$ is a quenched variable sampled uniformly from $\{-b_k,b_k\}.$

The derivative in the first case gives the bit error rate (magnetisation), when applied to the un-replicated expression for the self-averaged free-energy density
\begin{equation}
\BERmath = m =\<\! \left(\frac{1}{K}\sum_i \! b_i \!\tau_i
\!\right)\>_{\quenched,\vtau} = \frac{\partial
}{\partial z^D} \safed \label{appeq:CDMA.mag}\;,
\end{equation}
whereas in the latter case the standard definition of the linear susceptibility~\cite{Fischer:SG} is found
\begin{equation}
\Susceptibility + \frac{1}{2}(1-m^2) = \frac{1}{2 K} \<\left(\sum_{i} b_i \tau_i\right)^2 - K^2 m^2\> = \frac{\partial}{\partial z^D} \safed \label{appeq:CDMA.susceptibility}\;.
\end{equation}

The modifications required to the replica method to incorporate these external fields are realised in a change of the variable centric term $\Gtwo$ (\ref{eq:CDMA.G_2})
\begin{equation}
\Gtwo = - \log \sum_{\rvsigma}\< \<\left[\GENOPspconj\right]^{c_f}\>_{{c_f}} \exp \left\lbrace \beta \randomfield_k \sum_\alpha \sigma_\alpha \right\rbrace \>_{\randomfield_k,b}\label{appeq:CDMA.G_2}\;.
\end{equation}
With this the derivative (\ref{appeq:CDMA.mag}) may be evaluated given the saddle-point solution ($*$) as
\begin{equation}
m \propto \lim_{n\rightarrow 0} \frac{\partial}{\partial n} \sum_{\rvtau} \< b \sum_\alpha \tau^\alpha \< \left[\GENOPconj^*_{b}(\rvtau) \right]^{c_f} \>_{{c_f}} \>_b \label{appeq:CDMA.BER}\;.,
\end{equation}
and the linear susceptibility (\ref{appeq:CDMA.susceptibility}) is given by
\begin{equation}
\Susceptibility + \frac{1}{2}(1-m^2) = \frac{\partial}{\partial n} \sum_{\rvtau} \left(\sum_\alpha \tau^\alpha\right)^2 \< \left[ \GENOPconj^*_{b}(\rvtau) \right]^{c_f} \>_{{c_f,b}} \label{appeq:CDMA.replicaSGS}\;,
\end{equation}
keeping only those terms relevant to the limit in $n$.

A failure of the RS description is often found through the spin glass (or non-linear) susceptibility
\begin{equation}
\SpinGlassSusceptibility=\frac{1}{N}\sum_\ij \left(\<b_i b_j \tau_i \tau_j \> - \<b_i \tau_i \>\<b_j \tau_j \>\right)^2,\label{appeq:spinglasssusceptibility}
\end{equation}
a measure of correlation strength, and divergence in this quantity is an indication of method pathology, either due to a phase transition or an incorrect symmetry assumption. In the case of zero magnetisation, not relevant specifically to the CDMA problem, an important spherical symmetry is broken by this term, which is not broken by the susceptibility. The spin glass susceptibility can be probed by considering two sets of spin variables evolving independently given the same instance of quenched disorder (real replica), but with non-identical weak random external fields. The joint Hamiltonian for the CDMA system may be
\begin{equation}
\Ham(\rvsigma,\rvtau)= \Ham(\rvsigma) + \Ham(\rvtau) + z \sum_{k} \zeta_k \left(z_1\sigma_k + z_2 \tau_k \right)\;,
\end{equation}
where $\zeta_k,z_1,z_2$ are quenched random variables sampled independently from $\{+1,-1\}^K$, $z$ is an infinitesimal non-negative field. When $z$ is zero the free energy is twice that of each uncoupled model. An expansion of the free energy in $z$ gives aside from dependence on constants at $O(z^2)$ and $O(z^4)$, a term dependent on (\ref{appeq:spinglasssusceptibility}) at order $O(z^4)$. This term must be non-divergent in order for the RS description to be locally stable.

As shown in section (\ref{CDMA.stabilityb}) a field dependent on the quenched interaction structure, probing the linear stability in the example explored, might be transformed into a test of stability on the order parameter. Testing divergence of spin glass stability in the limit of small $z$ can then be formulated as a test of stability in the order parameters at the saddle-point for $z=0$. Analogous stability tests on the order parameters can be motivated through either a consideration of the cavity method, utilising the sparse graph structure~\cite{Rivoire:GM}, or a consideration of the stability of BP equations~\cite{Kabashima:PB}.

\subsection{Physical constraints on order parameters in composite models}
An assumption of the saddle-point method used to evaluate the exponential term describing the free energy is that only real valued integration parameters (order-parameters) need be considered. The arguments of \ref{app:saddle-pointmethod} are developed here for the composite model of chapter~\ref{chapter:composite} following in the replica calculations the scheme of Appendix~\ref{app:CompositeSystem_Replica}, to demonstrate that any physical solution must have order parameters real valued in some moments.

Consider the composite model with a quenched parameter dependence in the field
\begin{equation}
\Delta\Ham(\vtau) = \sum_i z_i \tau_i \;;\qquad z_i = \eta_{(i,j)} \sum_{(i,j)} \left( z^S J^S_{i,j} + z^D J^D_{i,j}\right) \label{appeq:randomfield2}\;.
\end{equation}
Unordered matrices are used in (\ref{appeq:randomfield2}) to describe their ordered counterparts, so that $(i,j)$ is $\ij$ or $\<ji\>$ as ordering dictates, for each ordered pair only one quenched parameter exists. Each of $\eta_\ij$  are assumed to be exactly zero (a default), uniform ($1$) or quenched variables independently samples from $\{-1,1\}$, with $\{z^S,z^D\}$ being infinitesimal positive real fields.

As in Appendix~\ref{app:physicsfromconjfields} a more standard choice for the fields involve a variable node dependence $\zeta_k=1$ with derivatives corresponding to magnetisations. Similarly the derivatives with respect to $z^S$ or $z^D$ when $\eta_\ij=1$ probes alignments of variables with couplings, again giving a physical measure that can distinguish a ferromagnetic phase from a paramagnetic one. The more complicated quantities probed involve $\eta_\ij=\{-1,1\}$, for example
\begin{equation}
\left.\frac{\partial}{\partial \left[\beta(z^D)^2\right]}\right|_{z^D=0} \safed = \frac{1}{2 N} \<\left(\sum_{(i,j)} J^D_{(i,j)} S_i \right)^2\> - \frac{1}{2 N} \<\left(\sum_{(i,j)} J^D_{(i,j)} S_i \right)\>^2\label{appeq:thisistheend}\;,
\end{equation}
which determines a type of linear susceptibility. These quantities are necessarily real-valued at a saddle-point.

The free energy in the replica formulation, with inclusion of these possible fields involves a modification of the factor-centric ($\Gone$) term (\ref{eq:composite.G1}) in the free energy. Following Appendix~\ref{app:CompositeSystem_Replica}, and relabeling non-zero sparse couplings $\ij$ by $\mu$
\begin{equation}
\begin{array}{lcl}
\Gone &=& - \sum_\alpha \frac{1}{2}\beta J_0 (\qal + 2 z^D\<\eta_{(i,j)}\>)\qal - \frac{1}{2} \! \sum_\alal \beta^2 J^2 \left(\qalal + 2(z^D)^2\<(\eta_{(i,j)})^2\>\right)\qalal \\
&-& \frac{C}{2} \log \sum_{\rvS,\rvS'} \GENOP(\rvS)\GENOP(\rvS') \int \rmd x \phi(x) \< \exp\left\lbrace \sum_\alpha \beta x \left(S^\alpha + z^S \eta_\mu \right) S'^\alpha \right\rbrace \>_{\eta_\mu} \label{appeq:composite.G1}\;.
\end{array}
\end{equation}

Now consider the analogous quantity to (\ref{appeq:thisistheend}), the term found is up to ensemble dependent constants
\begin{equation}
\left.\frac{\partial}{\partial \left[\beta (z^D)^2\right]}\right|_{z^D=0} \safed = \left.\frac{\partial}{\partial n}\right|_{n=0} J^2 \sum_\alal q^*_\alal \label{appeq:keepgoingalmostthere}\;.
\end{equation}
The derivative with respect to $z^S$ gives by contrast
\begin{equation}
\left.\frac{\partial}{\partial \left[\beta (z^D)^2\right]}\right|_{z^D=0} \safed = \left.\frac{\partial}{\partial n}\right|_{n=0}\sum_{\rvS} \GENOP^*(\rvS) \left(\sum_\alpha S^\alpha\right)^2 \int \rmd x \phi(x) \frac{C}{2\beta^2}\tanh^2(\beta x)
 \label{appeq:itwillbeworthit}\;,
\end{equation}
keeping only the relevant terms in $n$. These two expressions constrain the sum of dense and sparse order parameters, in replica space, to real values in the second moments. Similar real-valued constraints apply to the first moments. In the case of an RS assumption the saddle-point solution is therefore constrained to real valued moments, and the search for a saddle-point may justifiably be restricted to this space.

Some analysis is added to the arguments of Appendix~\ref{app:saddle-pointmethod} for the restriction of saddle-point analysis to the real axis. In the dense case two moments describe the RS solution and these are shown to be real, in the sparse case there are many higher order moments for which complex solutions are difficult to rule out analytically. The question of $L^{th}$ order moments and higher in the RS sparse order parameter ($\GENOP$) seems possible to address analytically if the Hamiltonian is analytically continued to the complex plane, or some other field description. Using for example $\zeta_k = \{\exp\{2\pi \rmi/l\}|l=1\ldots L\}$ with $L=4$ describes the fourth moment of the generalised order parameter and can be associated to a physical quantity involving 4-spin correlations through a derivative. In order to establish properties for inter-replica correlations, as are relevant in RSB formulations, it is likely that real-replica must be considered as in Appendix \ref{app:Nishimori}, where the Nishimori temperature result is derived for sparse CDMA ensembles.

%% file: APPENDICES/Composite_system_replica.tex
\subsection{Modifications to the saddle-point equations}
The saddle-point equations can be written down for the general case (\ref{appeq:composite.G2gen}), the generalisation of (\ref{eq:composite.saddle1}) in the sparse order parameter is
\begin{equation}
\GENOP(\rvsigma) \propto \<\left[\GENOPconj(\rvsigma)\right]^{c_e}\exp \left\lbrace \sum_\alpha b \qhal \sigma^\alpha + \sum_\alal \qhalal \sigma^{\alpha_1}\sigma^{\alpha_2}\right\rbrace\>_{b,c_e} \label{appeq:composite.GENOP_saddle_gen}\;,
\end{equation}
where the average of $c_e$ is with respect to the excess variable connectivity distribution. The dense order parameters are determined through the recursions
\begin{equation}
\qal = \sum_\rvsigma \sigma_\alpha \localreplicaprobability(\rvsigma) \;; \qquad \qalal = \sum_\rvsigma \sigma^{\alpha_1} \sigma^{\alpha_2} \localreplicaprobability(\rvsigma) \label{appeq:composite.q_saddle_gen} \;;
\end{equation}
involving a normalised distribution
\begin{equation}
 \localreplicaprobability(\rvS) = \<\left[\GENOPconj(\rvS)\right]^{c_f} \exp\left\lbrace \!\sum_\alpha b \qhal S_\alpha \!+\!
\sum_{\alal} \qhalal S^{\alpha_1} S^{\alpha_2}
\right\rbrace \>_{{c_f}}\;,
\end{equation}
with an averages according to the full variable connectivity distribution. The conjugate saddle-point equations are unchanged in form (\ref{eq:composite.saddle2}).

The replica method involves both a sparse and dense average. In order to connect the sparse description to those of previous chapters it is useful to redefine the sparse matrix in terms of a factor graph representation. Labeling each edge by $\mu$, an adjacency matrix with factor (edge) and variable labeling, $A_{\mu k}=\{0,1\}$, may be defined. With uniform connectivity $C$ the number of edges is $C N/2$ therefore in the absence of other constraints, the probability distribution is defined
\begin{equation}
P(\mA) = \prod_{\mu=1}^{C N/2} \left[{N \choose 2}^{-1} \delta\left(\sum_k A_{\mu k} - 2 \right) \right]\;.
\end{equation}
This is a micro-canonical description of interactions, but formulations with the number of edges not strictly fixed (to $C N/2$) are possible. This describes a Poissonian connectivity distribution in the variable connectivity. Both Poissonian and regular connectivity are given as special cases of
\begin{equation}
P(\mA) \propto \prod_{\mu=1}^{C N/2} \left[\frac{1}{2} \delta\left(\sum_k A_{\mu k} - 2 \right)\right] \prod_{i=1}^{N} \< \frac{c_f\factorial}{C^{c_f}} \delta\left(\sum_\mu A_{\mu k} - {c_f}
\right) \>_{c_f}\prod_{\mu,k} P(A_{\mu k}) \;,
\end{equation}
the average is with respect to the marginal variable connectivity distribution of mean variable connectivity $C$, and $P(A_{\mu k})$ is a sparse prior
\begin{equation}
P(A_{\mu k}) = \left(1- \frac{2}{N}\right)\delta_{A_{\mu k}} + \frac{2}{N}\delta_{A_{\mu k},1}\;.
\end{equation}

The Hamiltonian may be written in a form
\begin{equation}
\Ham = \frac{1}{2} \sum_\mu J^S_\mu \left[ \left( \sum_k A_{\mu k} \tau_k \right)^2 - 2\right] + \sum_{\ij} J^D_\ij b_i b_j S_i S_j \;,\label{appeq:composite.Ham}
\end{equation}
where the representation of the dense part is unmodified from (\ref{eq:composite.Ham}), $J^S_\mu$ is the sparse random coupling sampled from to $\phi$ (\ref{eq:composite.phix}), in the self-averaged free energy an average over instances ($J^S_\mu=x$) applies. The replicated partition function is
\begin{equation}
\begin{array}{lcl}
\repZ &=& \prod_\alpha\left[\sum_{\vS^\alpha}\right] \<\prod_\mu \<\exp\left\lbrace \frac{\beta}{2} x \sum_\alpha \left[\left(\sum_k A_{\mu k} S^\alpha_k \right)^2-2\right] \right\rbrace \>_{x} \>_{\mA} \\
&\times& \<\prod_\ij \<\exp \left\lbrace \beta J^D_\ij \sum_\alpha S_i^\alpha S_j^\alpha \right\rbrace\>_{J^D_\ij}\>_{\vb,\mJ^D}\;.
\end{array}
\end{equation}
Since the Hamiltonian is factorised with respect to the sparse and dense quenched variables, these averages may be taken independently.

In the sparse part it is useful to linearise the squared components with a Hubbard-Stratonovich transform (\ref{appeq:Hubbard}) for each factor node and replica index pair
\begin{equation}
\< \cdots \>_{\mA} = \int \prod_{\mu,\alpha} \left[ \rmD_{1}\lambda_\mu^\alpha\right]  \prod_\mu \< \exp \left\lbrace-\beta J_\mu n \right\rbrace \prod_k \left[\exp \left\lbrace \sqrt{\beta x} \sum_\alpha \lambda_\mu^\alpha S^\alpha_k \right\rbrace \right]^{A_{\mu k}}\>_{x}\;,
\end{equation}
with $\rmD_1 x$ is a Gaussian weighted integral of variance $1$. The form is now factorised with respect to connectivity so that the average with respect to $\mA$ can be taken according to Appendix~\ref{app:SparseMatrixAnalysis}, with minor modifications. Having taken the average in $\mA$ the Hubbard-Stratonovich transform may be inverted to give
\begin{equation}
\begin{array}{lcl}
\< \cdots \>_{\mA} &=& \int \prod_{\rvsigma}\rmd \GENOP(\rvsigma) \prod_{k}\< \frac{C_k!}{C^{C_k}} \intZ{{C_k}} \prod_{\rvsigma} \delta\left(\GENOP(\rvsigma) - \frac{1}{K} \sum_k Z_k \delta_{\rvS_k,\rvsigma} \right)\>_{C_k} \\
&\times& \prod_\mu \left[ \sum_{\rvtau,\rvsigma} \GENOP(\rvtau)\GENOP(\rvsigma)\<\exp \left\lbrace \beta x \sum_\alpha \tau^\alpha \sigma^\alpha \right\rbrace \>_{x} \right] \end{array}\;,
\end{equation}
where the average is with respect to the coupling distribution $\phi(x)$ (\ref{eq:composite.phix}).

The dense part of the Hamiltonian can be expanded to second order, including the possibility of a Mattis type disorder $\vb \neq \vones$,
\begin{equation}
\<\cdots\>_{\vb,\mJ^D} = \prod_\ij \left[ 1 + \beta \frac{J_0}{N} \sum_\alpha b_i S_i^{\alpha} b_j S_j^{\alpha} + \frac{\beta^2 J^2}{2 N} \sum_\alal S_i^{\alpha_1} S_i^{\alpha_2} S_j^{\alpha_1} S_j^{\alpha_2} \right]\;.
\end{equation}
Defining the dense order parameters
\begin{equation}
\qal = \frac{1}{N} \sum_i b_i S^\alpha_i \;;\qquad \qalal = \frac{1}{N} \sum_i S^{\alpha_1}_i S^{\alpha_2}_i \;;
\end{equation}
where $b_i$ is a quenched variable ultimately marginalised over. Introducing these definitions
\begin{equation}
\begin{array}{lcl}
\<\cdots\>_{\vb,\mJ^D} &=& \<\prod_\alpha \delta\left(q_\alpha - \frac{1}{N}\sum_k b_k S_k^\alpha\right)\>_\vb \prod_\alal \delta\left(\qalal - \frac{1}{N}\sum_k S_k^{\alpha_1} S_k^{\alpha_2}\right) \\
&\times& \exp \left\lbrace \frac{N \beta J_0}{2} \sum_\alpha q_\alpha^2\right\rbrace \exp \left\lbrace \frac{N\beta^2 J^2}{2} \sum_\alal q_{\alal}^2\right\rbrace\;.
\end{array}
 \end{equation}
The definitions of $\qal$, $\qalal$ and $\GENOP$, introduced as $\delta$-functions may be Fourier transformed introducing conjugate parameters, and the integral with respect to $Z$ taken as in Appendix~\ref{app.GAk}, the scaling of the fourier transform exponent by $C N$ applies in the sparse part, the scaling in the dense part is assumed to be $N$. A factorisation of the dependence in $\vb$ allows the final quenched dependence to be removed. The trace over replicated spins is finally taken to give an expression for free energy (\ref{eq:composite.replicafreeenergy}), composed of terms (\ref{eq:composite.G1}),(\ref{eq:composite.G3}) and ($\Gtwo$). The final term is (\ref{eq:composite.G2}) in the Poissonian variable connectivity and (\ref{appeq:composite.G2gen}) for general connectivity and Mattis disorder.

\subsection{Modifications to the variational free energy}

Following the previous appendix modifications are required in the factor term of the free energy in the case of constrained variable connectivity or embedded disorder $\vb\neq 0$, the adjustment effects (\ref{eq:composite.G2}) in the main text. A form sufficient for the general case is given up to constant terms by
\begin{equation}
\Gtwo \!=\! - \log \<\sum_{\rvS} \left(\GENOPconj(\rvS)\right)^{c_f} \exp\left\lbrace \!\sum_\alpha b \qhal S_\alpha \!+\! \sum_{\alal} \qhalal S^{\alpha_1} S^{\alpha_2} \right\rbrace \>_{{c_f},b} \label{appeq:composite.G2gen}\;,
\end{equation}
where the average over ${c_f}$ is with respect to the marginal variable connectivity distribution, and $\vb$ is averaged according to the alignment. The special case (\ref{eq:composite.G2}) is recovered when $b=1$ and ${c_f}$ is Poisson distributed.

%% file: APPENDICES/Composite_cdma_replica.tex
The Appendix determines the average replicated partition function for the composite system Hamiltonian (\ref{eq:composite.Ham}). Following a Hubbard-Stratonovich (H-S) transform (\ref{appeq:Hubbard}) and replacement of the
signal $\vy$ by $\vb$ and $\vomega$ (\ref{eq:compositeCDMA.ybreakdown}) the replicated partition function is
\begin{equation}
\begin{array}{lcl}
\repZ &=& \prod_\alpha \left[\sum_{\vsigma^\alpha}\right] \prod_{\mu} \vintlambda{{\mathbb I}} \<\prod_{\mu} \prod_k\left[\exp \left\lbrace \sqrt{-\beta\gamma/C} \modulationsymbol_{\mu k} \sum_\alpha \lambda^\alpha (1- \tau^\alpha_k) \right\rbrace \right]^{A_{\mu k}} \right.\\
&\times& \left.\prod_\mu\left[\prod_k\left[\exp \left\lbrace \sqrt{-\beta(1-\gamma)/N} \modulationsymbol^D_{\mu k} \sum_\alpha \lambda^\alpha (1- \tau^\alpha_k) \right\rbrace \right] \exp \left\lbrace \sqrt{-\beta}\omega_\mu \sum_\alpha \lambda^\alpha \right\rbrace\right] \>_{\quenched} \label{appeq:compositeCDMA.repZ}\;,
\end{array}
\end{equation}
where $\rmD_{x}$ abbreviates a Gaussian weighted integral described by a covariance matrix (or scalar) $x$, and the bold-font notation implies a vector in replica space.

The form of the replicated partition function within the integral is factorised with respect to the sparse and dense sub-code quenched parameters, the averages in each part may be taken separately. The averages over the sparse sub-codes are the same as in chapter~\ref{chapter:sparseCDMA}. The dependence on the sparse connectivity matrix takes a factorised form ($\prod(\arbitraryfunction)^A_{\mu k}$) within each integral and so the analysis of section~\ref{app.GAmu} applies. The order parameter contains a $k$ dependence through the replicated variables, which is captured in the identity
\begin{equation}
1=\int \rmd\GENOP(\rvsigma) \delta\left(\GENOP(\rvsigma)-\frac{1}{K}\sum_k Z_k \prod_{\alpha=1}^n \delta_{\tau^\alpha_k,\sigma^\alpha} \right)\;.
\end{equation}
With this definition the term factorised in $\mu$, top line of (\ref{appeq:compositeCDMA.repZ}), becomes
\begin{equation}
 \< \prod_l^{l_e}\left[\sum_{\vsigma_l} \GENOP(\vsigma_l)\<\exp \left\lbrace \sqrt{-\beta\gamma/C} \sum_l \modulationsymbol_l \sum_\alpha \lambda^\alpha (1- \sigma_l^\alpha) \right\rbrace \>_{\modulationsymbol_l}\right] \>_{{l_e}} \;. \label{appeq:compositeCDMAenergeticSP}
\end{equation}
the exceptional case ${l_e}=0$ evaluates to one.
In the model studied $\modulationsymbol_l$ are distributed uniformly on ${-1,+1}$, and ${l_e}$ is distributed according to the chip/factor connectivity of the sparse sub-code, in the analysis the bit sequence can be gauged to the code modulation, so $\vb=1$ can be considered in general.

The quenched averages in the dense sub-code are simplified by an expansion in terms of $O(\sqrt{N}^{-1})$. Taking a second order expansion two more order parameters are identified
\begin{equation}
\qal = \frac{1}{K} \sum_k b_k \sigma_k^\alpha \;;\qquad
q_{\alpha_1\neq\alpha_2} = \frac{1}{K} \sum_k \sigma_k^{\alpha_1} \sigma_k^{\alpha_2} \;;\label{appeq:compositeCDMA.denseOP}
\end{equation}
where $q_{\alpha_1 \alpha_2}=q_{\alpha_2 \alpha_1}$ by definition. The form for the dense sub-code dependent part of the second line in (\ref{appeq:compositeCDMA.repZ}) becomes
\begin{equation}
 \exp \left\lbrace - \frac{\beta \load (1-\gamma)}{2} \left[\sum_\alpha (1-q) (\lambda^\alpha)^2 + \sum_{\alpha_1,\alpha_2} \lambda_{\alpha_1}\lambda_{\alpha_2} (1-q_{\alpha_1}-q_{\alpha_2}+ q_{\alpha_1,\alpha_2}) \right] \right\rbrace \label{appeq:composite.CDMAenergeticSP}\;,
\end{equation}
defining $q_{\alpha\alpha}=q$ as an auxiliary argument in the interval $[0,1)$. The exponent is a quadratic form in $\rvlambda$ and so, in combination with (\ref{appeq:compositeCDMAenergeticSP}) the exact integration of $\rvlambda$ is possible.

It is convenient to define the second term in the exponent as $-(\RotationMatrix\vlambda)^T \RotationMatrix \vlambda/2$, where $\RotationMatrix$ is a matrix combining the eigenvalues and eigenvectors of the quadratic form. This term can be transformed using the H-S transform, so that (\ref{appeq:composite.CDMAenergeticSP}) becomes
\begin{equation}
 \int \rmD_{{\mathbb I}} \xi \exp \left\lbrace - \frac{\beta \load (1-\gamma)(1-q)}{2} \sum_\alpha (\lambda^\alpha)^2 + \sqrt{-1}\sum_\alpha \left[\sum_{\alpha_1} \xi^{\alpha_1} \RotationMatrix_{\alpha_1 \alpha}\right] \lambda^{\alpha} \right\rbrace \label{appeq:composite.reform}\;,
\end{equation}
with ${\mathbb I}$ the identity matrix. The remaining quadratic form in $\vlambda$ is a simple diagonal one. Including the exponential terms from $D_1\lambda^\alpha$ the integral over $\lambda$ is with respect to a quadratic term $\beta I \sum_\alpha (\lambda^\alpha)^2$, where
\begin{equation}
I = 1/\beta + \load(1-\gamma)(1-q)\;,
\end{equation}
which quantifies a signal to interference plus noise ratio.

Integrals of $\lambda^\alpha$ can be taken, combining (\ref{appeq:composite.CDMAenergeticSP}) and (\ref{appeq:composite.reform})
\begin{equation}
\begin{array}{lcl}
\load \Gone &=& -\Big[ \left[\beta I \right]^{-1/2} \int \rmD\xi_{{\mathbb I}} \prod_{l=1}^{l_e} \left[\frac{1}{2}\sum_{\modulationsymbol,\rvsigma} \GENOP(\rvsigma_l)\right]\\
&\times&\<\exp\left\lbrace - \frac{1}{2 I} \left( \nu + \sqrt{\gamma/C}\sum_l \modulationsymbol_l (1-\sigma_l^\alpha) + \left[\sum_{\alpha_1} \xi^{\alpha_1} \RotationMatrix_{\alpha_1 \alpha}\right] \right)^2\right\rbrace \>_{{l_e},\nu}\Big]\;,
\end{array}
\end{equation}
where $\nu$ is distributed according the channel noise.

Introduction of the order parameter identities in (\ref{appeq:compositeCDMA.repZ}) allows a straightforward evaluation, with terms comparable to Appendix~\ref{app:CompositeSystem_Replica}. Introducing a Fourier transform for each of the identities the remaining two contributions to the free energy are
\begin{equation}
\Gtwo = - \log \sum_{\rvtau} \left[\<\left[\GENOPconj(\rvtau) \right]^{c_f}\>_{{c_f}} \exp \left\lbrace \sum_\alpha \qhal \tau^\alpha + \sum_{\alpha_1\neq\alpha_2}{\hat q}_{\alpha_1,\alpha_2} \tau^{\alpha_1}\tau^{\alpha_2} \right\rbrace\right]\;,
\end{equation}
and
\begin{equation}
\Gthree = C \sum_{\rvsigma} \GENOP(\rvsigma) \GENOPconj(\rvsigma)\;,
\end{equation}
so that the free energy is, as usual, determined by an
extremisation problem
\begin{equation}
 \beta \safed \!\propto\! \Extr_{\{\GENOP,\GENOPconj,q_\alal,{\hat q}_\alal,q_\alpha,{\hat q}_\alpha\}} \left.\frac{\partial}{\partial n}\right|_{n=0}\left[\Gone + \Gtwo + \Gthree\right]\label{appeq:compositeCDMA.safed}\;.
\end{equation}

\subsection{RS solution}

With a particular ansatz on the form of correlations the treatment is simplified. In the case of an RS ansatz, taking the standard set of definitions (\ref{eq:compositeCDMA.RSOP}), $\{q_\alpha=m,q_\alal=q_{\alpha\alpha}=q\}$, the matrix $\RotationMatrix$ is determined by only one non-zero eigenvalue (column), and the results of chapter~\ref{chapter:compositeCDMA} are obtained. The RS free energy (\ref{eq:compositeCDMA.safed}) is dependent on the linear terms in $n$ in the exponent, defined
\begin{equation}
\gi = \lim_{n\rightarrow 0} \frac{\partial}{\partial n} \Gi\;.
\end{equation}

%% file: APPENDICES/Random_graph_samples.tex
Algorithmic generation of random bi-partite (factor) graphs defined by the arbitrary marginal connectivity distributions ($P_L$,$P_C$) used in analysis is in general computationally difficult. Although generating some sample is simple for sparse ensembles, intuitive methods of sampling or rewiring graphs for experimental purposes may produce unintended bias in results, unless uniform sampling of the space is guaranteed at a statistically significant level. This Appendix explains how graph samples were generated to obtain experimental results presented or referred to in various chapters by a method sampling in an unbiased way asymptotically. The case with marginal Poissonian connectivity in either factor or variables is first addressed, followed by the more complicated case of regular connectivity in variables and factors. By introducing some additional processes the method may be generalised to some non-regular non-Poissonian connectivity distribution pairs.

\section{Poissonian connectivity in factors or variables}

To generate an unbiased sample of a labeled sparse matrix ($\mA$) constrained only in the mean factor/variable connectivity ($L$/$C$) and number of factor/variable nodes ($M$/$K$) it is possible to sample components independently, setting $A_{\mu k}$ to one with probability $C/M$, and to zero otherwise. This algorithm requires a reliable random number generator, which can be approximated by pseudo-random method in practical situations~\cite{Press:NRC}. The samples constructed have marginal factor and variable connectivity distributions converging towards Poissonian.

In this thesis the Poissonian connectivity distribution appears frequently. This distribution reflects an asymptotic outcome of unconstrained sparse connectivity. In a finite system the Poissonian is not realised, instead it is a Binomial distribution that is the appropriate analogue. When considering one row of the matrix, with unconstrained occupation subject to a marginal probability $C/M$, a variable connectivity distribution is given by
\begin{equation}
P_{M,C/M}(C_k) = \frac{M\factorial}{(M-C_k)\factorial C_k\factorial} \left(\frac{C}{M}\right)^{C_k} \left(1-\frac{C}{M}\right)^{M-C_k}\;\label{appeq:Binomial}
\end{equation}
which is asymptotically Poissonian. An asymptotically correct sampling for a Poissonian distribution in a finite system is therefore an unconstrained one. Suppose a constrained distribution exists for the factor nodes ($P_L$), but the desired variable connectivity is Poissonian (Binomial) of mean $C=LM/K$. For this ensemble it is possible to generate a random graph by sampling $L_\mu$ independently from $P_L$, for each row and then to choose these elements within a row uniformly at random. This will generate a Binomial distribution of variable connectivities as intended. Thus it is possible where one set of connectivity constraints is Poissonian to generate an unbiased sample efficiently.

\section{Regular connectivity in factors and variables}
A regular connectivity ensemble is defined by marginals $P_L({l_f})=\delta_{L,{l_f}}$ and $P_C({c_f})=\delta_{C,{c_f}}$, and providing $M$ and $K$ are much greater than $C$ and $L$ many graphs exist for this ensemble. For finite systems it is possible to developing an efficient iterative sampling method that is unbiased at leading order in the system size, and seems to produce reasonable finite graphs. The iterative method is applicable to an arbitrary variable connectivity distribution $P_C({c_f})$ assuming a maximum sample connectivity of $M$.

To generate a random sample of $\mA$, of dimension $M \times K$, first a vector of variable connectivities ($\vC$) is sampled according to $P_C$, but consistent with the fixed number of edges $L K$ implied by the regular distribution on the factors.  Elements of $\mA$ are assigned on a row by row basis, according to a decomposition
\begin{equation}
P(\mA|\vC,L,M) = P(\mA\setminus\vA_\mu|\vA_\mu,\vC,L,M) P(\vA_\mu|\vC,L,M)\label{appeq:PmAvC}\;.
\end{equation}
The sample $\vA_\mu$ will be constructed accurately according to the latter probability as explained latter. Having achieved this sampling the remaining sampling problem is equivalent to one of size $M-1 \times K$, with a modified set of variable connectivities.. Introducing the notation $\mA^M$ to describe the matrix with rows labeled $1$ to $M$
\begin{equation}
P(\mA^M|\vC,\vA_{M},L,M) \equiv P(\mA^{M-1}|\vC-\vA_{M},L,M-1) \;.
\end{equation}
This probability may again be decomposed as (\ref{appeq:PmAvC}) so that by iteration up to $M=1$ a matrix is generated.

Sampling a vector $\vA_\mu$ is more problematic. By Bayes' rule the second expression in (\ref{appeq:PmAvC}) is
\begin{equation}
P(\vA_\mu|\vC,L,M) \propto P(\vC|\vA_\mu,L,M) P(\vA_\mu|L,M)\;,
\end{equation}
where
\begin{equation}
P(\vA_\mu|L,M) = {K\choose L}^{-1}\delta\left(\sum_k A_{\mu k} - L\right)\label{appeq:PvAmuL}\;,
\end{equation}
and the likelihood term can be constructed by a marginalisation over the residual matrix
\begin{equation}
P(\vC|\vA_M,L,M) = \sum_{\mA^{M-1}} P(\vC|\mA^M,L,M) P(\mA^{M-1}|L,M-1) \;.
\end{equation}
An approximation to the prior is described by a factorised form, using the marginal for a single row to approximated the set of coupled rows
\begin{equation}
P(\mA^{M-1}|L,M-1)=\prod_k \sum_{C_k} \left[P_{M-1,L/K}(C_k) \delta\left(\sum_{\mu} A_{\mu k} - C_k \right) \right] \label{appeq:PmAL}\;,
\end{equation}
where $P_{M,L/K}$ is the Binomial distribution (\ref{appeq:Binomial}), which converges to a correct description of the joint probability when $M,K$ become large. The approximation (\ref{appeq:PmAL}) does not seem to produce obvious pathological features for graphs of the size experimented with in this thesis.

Finally this allows a factorised form for the row sample likelihood, carrying out the marginalisation
\begin{equation}
P(\vC|\vA_M,L,M) \propto \prod_k \left[\left( P_{M,L/K}(C_k) \right)^{-1}\left\lbrace \! P_{M-1,L/K}(C_k-1) \delta\left(A_{M k} - 1\right) \!+\! P_{M-1,L/K}(C_k)\delta\left(A_{M k}\right) \right\rbrace\! \right]\label{appeq:PvCvAmuL2}\;,
\end{equation}
where $P_{M,L/K}(x)$ is defined as zero for $x$ outside the interval $[0,M]$.

According to (\ref{appeq:PvAmuL}) exactly $L$ non-zero elements must be sampled, and according to (\ref{appeq:PvCvAmuL2}) this set ($\Set$) must include all variables $k$ for which $M=C_k$, and no values for which $C_k=0$. The remaining elements of $\Set$ are selected by a rejection sampling method, which is possible due to the factorisation  and efficient because the matrix is sparse. While the set size $|\Set|<L$, select some $k$ not in $\Set$ uniformly at random from all elements with $C_k>0$. Let the maximum variable connectivity not equal to $M$ be $C_{max}$. Sample uniformly a random number $\randomnumber \in [0,1]$, and evaluate the expression
\begin{equation}
P_{M-1,L/K}(C_{max}-1)\left[P_{M,L/K}(C_{max})\right]^{-1} \randomnumber< P_{M-1,L/K}(C_k-1)\left[P_{M,L/K}(C_k)\right]^{-1} \label{appeq:rejectionmethod}\;.
\end{equation}
If the constraint is met set $A_{\mu k}=1$, add $k$ to $\Set$, and repeat the process. Otherwise repeat the process without adding to the set. In this way all elements in the set are determined. The acceptance rate for a column is given by the ratio of the marginal probability of acceptance to the highest marginal probability of acceptance, $C_k/C_{max}$. In other words variables are sampled in proportion to their connectivity.

Every sample used in the thesis is generated by this method, no local rewiring procedure is applied, each sample is generated independently. Coincident (hyper-)edges (matrices with two identical rows or columns) are always removed to prevent pathological effects. This was achieved in a dynamical manner in comparing rows, and at a matrix level comparing columns. Partially overlapping hyper-edges (irrelevant for binary couplings) were not excluded. 

%% file: APPENDICES/Composite_cdma_algorithm.tex
\section{BP equations}
\label{app:compalg.BP}
Estimation of marginal probability distributions can be achieved by BP for a fixed spreading code $\ms$ consisting of a dense and sparse parts. A scalable algorithm is developed for the cases where $K=\chi M$, with $\chi\sim 1$ and $M$ large. A prior (external field) may be included in the equations, but is left absent for brevity.

A set of perfectly normalised codes is considered, defined by
\begin{equation}
 \ms = \sqrt{{1-\gamma}} \ms^D + \sqrt{\gamma} \ms^S \;,
\end{equation}
with the dense and sparse codes given by matrices $\ms^D$ and $\ms^S$ with components
\begin{equation}
s^D_{\mu k} = \sqrt{\frac{1-\gamma}{M-C}} \modulationsymbol^D_{\mu k}(1 - A_{\mu k}) \; ; \qquad s^S_{\mu k} = \sqrt{\frac{\gamma}{C}} A_{\mu k} \modulationsymbol^S_{\mu k}\;.
\end{equation}
In this form, a small variation on (\ref{eq:compositeCDMA.s}), links transmitted on with power $O(1)$, the strongly connected component, is separated from that part with weak transmission power. However, the algorithm is identical at leading order in the large system limit for the two cases. The strongly connected component is determined by the sparse connectivity matrix $\mA$, which contains a fraction $C/M$, of non-zero components. The dense code is defined as zero on all components that include a sparse transmission. The matrices $\mxi^*$ are random dense modulation matrices with components $\pm
1$ in the case of BPSK.

A self consistent marginal probability distribution can be
constructed based on the probabilistic relations amongst
log-likelihood ratios. These define the $2 M \times K$ BP
equations (two for each link in the factor graph) based on
variable (log-posterior) messages
\begin{equation}
h^{(t+1)}_{k \rightarrow \mu} \!=\! \frac{1}{2 \beta} \sum_b b
\log P^{(t+1)}(b_k = b | \vy \setminus y_\mu) =
\sum_{\nu\setminus\mu} u^{(t)}_{\nu \rightarrow k}
\label{appeq:composite.hkmu}\;,
\end{equation}
and evidential (log-likelihood) messages
\begin{equation}
u^{(t)}_{\mu \rightarrow k} \!=\! \frac{1}{2 \beta} \sum_b b \log P^{(t)}(y_\mu | b_k\!=\!b, \vy \setminus y_\mu) \!=\! \frac{1}{2 \beta}\sum_{\tau_k} \tau_k \!\log(\localpartitionfunction_{\mu k}^{(t)}(\tau_k))
\label{appeq:composite.umuk}\;.
\end{equation}
Defining
\begin{equation}
\localpartitionfunction_{\mu k}^{(t)}(\tau_k) \!=\! \sum_{\vtau\setminus \tau_k} \!\exp\left\lbrace - \frac{\beta}{2} \left(y_\mu - \sum_l s_{\mu l} \tau_l\right)^2 + \!\sum_{l\setminus k} \beta h^{(t)}_{l \rightarrow \mu} \tau_l\right\rbrace \label{appeq:composite.Z}\;,
\end{equation}
as the partition function for a single bit variable in the cavity graph with all factors (including prior factors) removed, except $\mu$. An estimate to the log-posterior ratio for bits is given by
\begin{equation}
H^{(t+1)}_{k} \!=\! \frac{1}{2 \beta} \log \frac{P^{(t)}(b_k = 1 | \vy)}{P^{(t)}(b_k = -1 | \vy)} = \sum_{\nu} u^{(t)}_{\nu \rightarrow k} \label{appeq:composite.Hk}\;.
\end{equation}

The messages form a self-consistent set of probabilistic relations, if the messages incident on a site are independent. In the fully connected case considered, the messages must be weakly correlated in order for BP to apply. In the large system limit it may be that correlations perturb estimates only at $O(1/K)$, so that the BP equations are exact at leading order. This might be expected to occur at parameterisations accurately described by connected pure states, as exist at the Nishimori temperature for example.

\section{Marginalisation over states}
\label{app:compalg.marginalisation}
In deriving the following assumptions the superscripts are attached to edge ($\mu k$) dependent quantities to distinguish strong and weak types: the Dense (D) edges, $s^D_{\mu k}$ and evidential messages $u^D_{\mu \rightarrow k}$ are $O(1/\sqrt{M})$, by contrast with Sparse (S) edges and evidential messages, and all variable messages. This is used to motivate some simplifications.

The algorithmic complexity for the complete BP equations is dominated by marginalisation in (\ref{appeq:composite.Z}), an evaluation of the evidential messages is not feasible with such term. For the composite system complexity is reduced by assuming independence of messages and making a Gaussian approximation~\cite{Kabashima:SMA}, the identity
\begin{equation}
\prod_{k \setminus \partial_\mu} \left[\sum_{\tau_k} \exp\left\lbrace \beta h^{(t)}_{k \rightarrow \mu} \tau_k\right\rbrace\right] = \int \rmd X \prod_{k\setminus \partial_\mu} \left[\sum_{\tau_k} \exp\left\lbrace \beta h^{(t)}_{k \rightarrow \mu} \tau_k\right\rbrace\right] \delta\left(X - \sum s^D_{\mu k} \tau_k\right)\;.
\end{equation}
can be introduced into (\ref{appeq:composite.Z}), and is simplified to a Gaussian integral in auxiliary mean ($m^D$) and variance ($v^D$) parameters
\begin{equation}
\int \rmd X \int \rmd\lambda \exp\left\lbrace - \lambda^2/2\right\rbrace \delta\left(X-(m^D+\sqrt{v^D}\lambda) \right)\;,
\end{equation}
in the large system limit. Taking the Gaussian integral explicitly, (\ref{appeq:composite.Z}) becomes
\begin{equation}
\localpartitionfunction_{\mu k}^{(t)}(\tau_k) \propto \prod_{i \in \partial_\mu \setminus k} \left[\sum_{\tau_i}\right] \!\exp\left\lbrace - \frac{1}{2 I^{(t)}_{\mu k}} \left({\hat y}_{\mu k} - \sum_l s^S_{\mu l} \tau_l\right)^2 + \!\sum_{l\setminus k} \beta h^{S,(t)}_{l \rightarrow \mu} \tau_l\right\rbrace \label{appeq:composite.Z2}\;,
\end{equation}
with an effective signal ${\hat y}_{\mu k}$ and noise variance $I^{(t)}_{\mu k}$. The estimated signal to noise ratio ($\beta$) is modified to a kind of signal to interference ratio with inclusion of ($v^D$)
\begin{equation}
I^{(t)}_{\mu k} = \beta^{-1} + \sum_{l \setminus \left\lbrace \partial_\mu\cap k\right\rbrace} (s^D_{\mu l})^2 (1 - \tanh^2 (\beta h^{(t)}_{l\rightarrow \mu})) \label{appeq:Imuk}\;,
\end{equation}
and the edge dependent signal is modified subject to the dominant (mean) estimate to the dense bit sequence ($m^D$)
\begin{equation}
{\hat y}^{(t)}_{\mu k} = y_\mu - \sum_{l \setminus \left\lbrace \partial_\mu \cap k \right\rbrace} s^D_{\mu k} \tanh( \beta h^{(t)}_{l \rightarrow \mu}) \label{appeq:yhat}\;.
\end{equation}

Calculation of each of these components requires only $O(K)$ operations per factor node per time step. Updating of all variable messages and evidential messages in a time step requires $O(K^2)$ operations. Explicit marginalisation is still required with respect to variables connected through the sparse sub-structure.

\section{Further leading order approximations}

The evidential messages can be simplified based on an expansion of the exponent (\ref{appeq:composite.Z}) in the small $s^D_{\mu k}$ terms
\begin{equation}
u^{(t)}_{\mu \rightarrow k} = ({I_{\mu k}^{(t)}})^{-1} s^D_{\mu k} \left({\hat y}_{\mu k} - \sum_{l \in \partial_\mu} s^S_{\mu k} \tanh(\beta h^{S,(t)}_{l \rightarrow \mu})\right)\label{appeq:composite.usimp}\;.
\end{equation}
Furthermore, an expansion of $I_{\mu k}$ in terms of the marginal magnetisations is possible using
\begin{equation}
\tanh(\beta h^{(t)}_{k\rightarrow \mu}) = m^{(t)}_k + \left(1-(m^{(t)}_k)^2\right) u^{(t-1)}_{\mu \rightarrow k}\;; \qquad m^{(t)}_k = \tanh(\beta H^{(t)}_k)) \label{appeq:composite.tanhh}\;;
\end{equation}
and keeping only leading order terms in $M$ gives
\begin{equation}
I^{(t)}_{\mu k} \rightarrow I^{(t)} = \beta^{-1} - \chi(1-\gamma) \left(1 - Q^{(t)} \right) \label{appeq:composite.Isimple}\;,
\end{equation}
assuming a mean square value for the dense modulation pattern of $(1-\gamma)/M$, with
\begin{equation}
Q^{(t)} = \frac{1} {K} \sum_{l=1}^K \tanh^2 (\beta H^{(t)}_{l})\;.
\end{equation}
So that no site dependence at leading order remains in (\ref{appeq:composite.Isimple}). These two observations allow a more concise algorithmic form~\cite{Mallard:BPDG}, although algorithm complexity remains $O(K^2)$. The corrections to $I_{\mu k}$ (\ref{appeq:composite.Isimple}) for all ensembles are $O(1/\sqrt(M))$, and assuming $\modulationsymbol^D_{\mu k}$ is uncorrelated with these corrections the variable messages will be unaffected at leading order. Similarly sized corrections, relative to $u$, apply to the expansion (\ref{appeq:composite.usimp}), and are assumed to be negligible.

\section{Elimination of dense BP messages}
\label{app:compalg.elimination}

The possibility to eliminate dense variable message dependence in the algorithm exists through use of the expansion (\ref{appeq:composite.tanhh}). When applied to (\ref{appeq:composite.umuk}) the sparse evidential messages become conditionally independent of the dense messages given $\vH^{(t)}$, the dependence is given through ${\hat y}_\mu^{(t)}$ which can be taken to be in the sparse part
\begin{equation}
{\hat y}^{(t)}_\mu = y_\mu - \sum_{l=1}^K s^D_{\mu l} \tanh(\beta H^{(t)}_l) \label{appeq:composite.haty}\;.
\end{equation}
The recursion on dense evidential messages can be written in terms of auxiliary quantities $R$ and $U$
\begin{equation}
u^{(t)}_{\mu \rightarrow k} = (\beta I^{(t)})^{-1} \left( R^{(t)} s^D_{\mu k} y_\mu - U^{\mu (t)}_{k} + (s^D_{\mu k})^2 m^{(t)}_k \right)\;,
\end{equation}
obeying recursive equations without site dependence in the case of $R$
\begin{equation}
R^{(t)} = (\beta I^{(t)})^{-1}\left(1 + \chi (1-\gamma)(1-Q^{(t)})R^{(t-1)} \right) \label{appeq:composite.R}\;,
\end{equation}
and with a dependency in $U$ given by
\begin{equation}
U_{k}^{\mu,(t+1)} = (\beta I^{(t)})^{-1}\left(\left(\sum_{l} s^D_{\mu k} s^D_{\mu l} (1 - \delta_{k,l}) m^{(t)}_{l}\right) + \chi (1-\gamma)(1-Q^{(t)}) U_{k}^{\mu,(t)} \right)\;.
\end{equation}
Determination of the magnetisations is possible with respect to $U^{(t)}_k = \sum_\mu U_{k}^{\mu,(t)}$. Therefore assuming an interest in only the magnetisation, which is sufficient to determine a bit approximation, the dense evidential messages can be removed and replaced by the recursions on $R$ and $\vU$. The total algorithm can be written down as a recursion in the dense part
\begin{equation}
\begin{array}{lcl}
\vU^{(t)} &=& \frac{1}{\beta I^{(t)}}\left( \mW \vm^{(t)} + \chi (1-\gamma)(1-Q^{(t)})\vU^{(t-1)}\right)\;; \\
\vH^{d,(t)} &=& \left( R^{(t)} \vy^{(t)} \ms^D - \vU^{(t)} + \frac{(1-\gamma)}{I^{(t)}} m^{(t)}_k \right) \;; \\
H_k^{(t+1)} &=& H_k^{d,(t)} + \sum_{\mu \in \partial_k} u_{\mu \rightarrow k} \;;\\
W_{k l} &=& \sum_\mu s^D_{\mu k} s^D_{\mu l} \left(1 - \delta_{k,l} \right)\;;
 \end{array}
\end{equation}
combined with (\ref{appeq:composite.R}), and standard BP equations on the strongly connected parts, subject to modified components (\ref{appeq:composite.Isimple}) and (\ref{appeq:composite.haty}). The final algorithm complexity is $O(K^2)$, but a large constant factor is removed as well as a large burden on memory, even in a distributed system, with the elimination of dense messages. The message passing on the sparse subsystem given ${\hat y}_\mu^{(t)}$ and $I^{(t)}_\mu$ remains of complexity $O(K)$, as in standard sparse BP.

Variable messages are assumed to be unbiased in the first step, therefore $h^{(0)}_{*\rightarrow*}=0$. This results in an initial condition for the new estimations of $R^{(0)}=1$ and $\vU^{(0)}=\vzeros$.